# Laboratory X-ray Microscopy Using Fresnel Zone Plate: Method and Practice

by

Yongshuai Ge, Yuhang Tan

This book is partially supported by
the NSFC funding

Funding Number

(12027812)

at the

SIAT, CAS

September 10, 2026

*To see or not to see.*

# Acknowledgments

First and foremost, we sincerely acknowledge the support from the National Natural Science Foundation of China (NSFC). Without this support (funding number: 12027812), majority of the content presented in this book would not have been possible.

Moreover, we would like to express our sincerest gratitude to Prof. Peiping Zhu, who initiated this research project and inspired many of the hereafter discussions.

In addition, we would like to thank Dr. Jiecheng Yang for his significant contributions at the early stage of this project, particularly in the development of key components of the numerical simulation platform, such as X-ray wave propagation and system geometry design.

Finally, we we would like to thank Prof. Tiqiao Xiao, Prof. Yangchao Tian (and his team members), Prof. Yifang Chen (and his team members), Prof. Biao Deng, Prof. Ke Liu, Prof. Qingxi Yuan, Dr. Qili He, Dr. Wenbing Yun (and his coworkers at Sigray Inc.) and Dr. Guibin Zan for their valuable supports and insightful discussions.

# Contents

# Tables

# Figures

# Abstract

This book was written to share our experience in developing a Fresnel zone plate (FZP) based X-ray microscope. The research began in September 2020, when we were going to build such a scientific imaging instrument from scratch in our laboratory. At that time, we could not find a textbook that systematically explained the entire development process step by step. We therefore had to make progress through trials and mistakes. After nearly five years of hard effort, we finally managed to build such an X-ray microscope in April 2025. Through this experience, we came to realize how challenging it can be to develop a new FZP-based X-ray microscope without systematic guidance. We therefore believe that it is worthwhile to share the learned lessons with readers who may be interested in developing their own FZP-based X-ray microscopes in the future.

We firmly believe that no technology can thrive if it remains inaccessible to the broader professional community. Without widespread discussion, understanding, and participation, a technology is unlikely to achieve broad adoption or benefit from continuous improvement. In turn, the development of better products will be constrained, and the technology itself may eventually lose its capacity to evolve. For any technology, isolation from its community is ultimately a form of self-destruction. Our goal, therefore, is not to prevent X-ray microscope companies from commercializing their products. On the contrary, we believe that broader access to the X-ray microscope knowledge can stimulate innovation and competition, enabling the development of more competitive and advanced products.

I would encourage readers to treat this book as a practical manual rather than a traditional textbook. Readers are not expected to be experts in X-ray imaging. Instead, a basic background in college-level physics should be sufficient to understand most of the material. Inevitably, the current edition has limitations and may contain omissions or inaccuracies due to the authors' limited experience and perspective. I sincerely welcome comments, suggestions and discussions from readers, please feel free to contact me at geyongshuai1987@qq.com. Your feedback will be greatly appreciated and will help make future editions of this book more complete.

Yongshuai Ge<br>
Shenzhen, 13 Jan 2026

# 1 Introduction

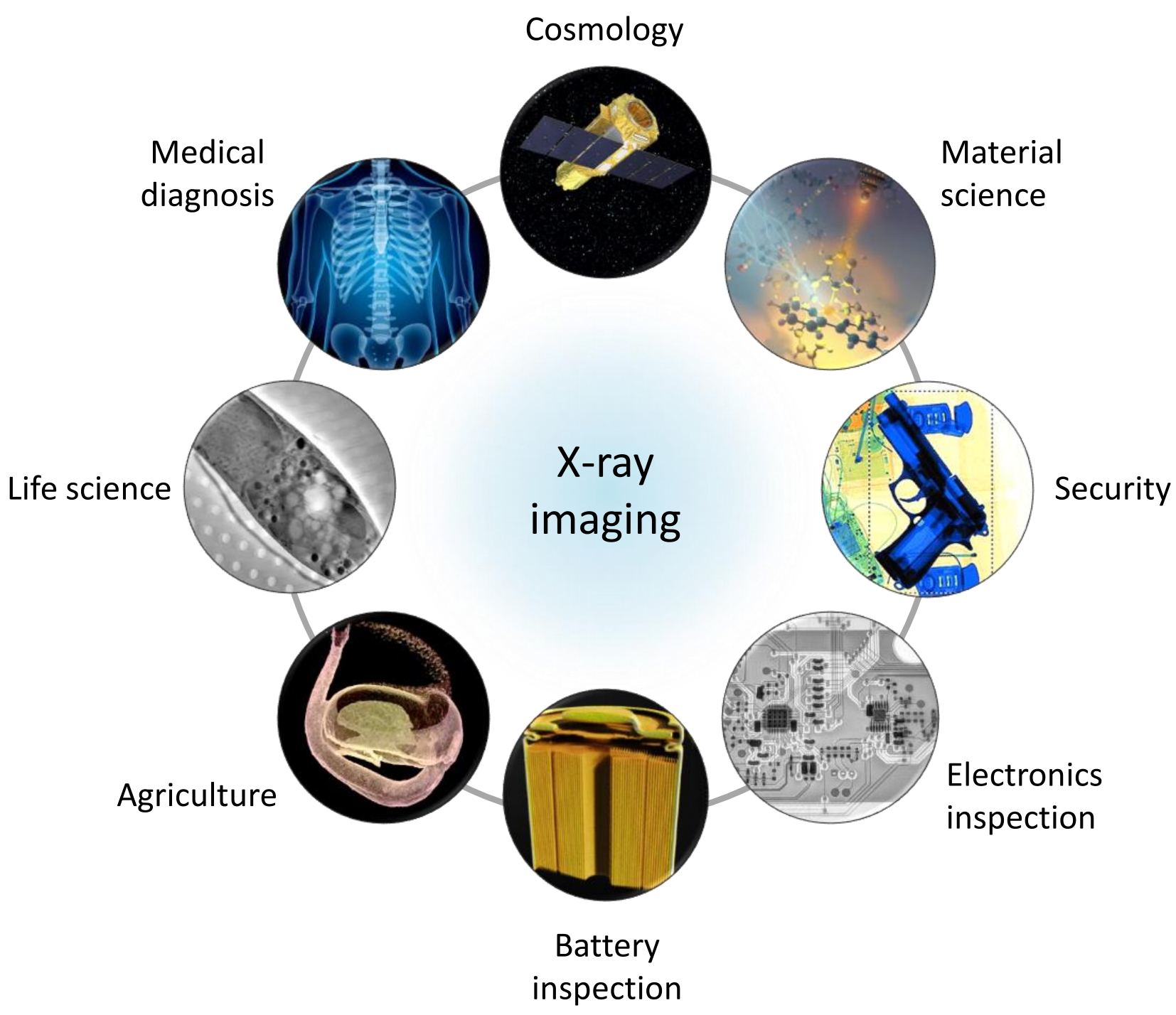

Figure 1.1: The application fields of X-ray imaging.

## 1.1 X-ray imaging

Very accidentally, the historical discovery of X-ray[1] in 8 November 1895 by Dr. Wilhelm Conrad Röntgen came along with the world's first X-ray image of human's hand. Since then, X-rays unveiled a whole new era for the entire twentieth century and after. Since X-rays are not directly visible to the human eyes, therefore, recording an image, corresponding to a certain type of mathematical representations, became the most convenient way to visualize it. However, no one realized that this indicated its future representation form in modern science such as physics, chemistry, biology, and medicine. Like the invention of photography, X-ray imaging since then began to play very important roles in unveiling numerous scientific findings in human history. For instance, the crystal lattice structure[2], the DNA double helix[3], the computed tomography (CT)[4–6], and the protein crystallography. Nowadays, X-ray imaging has been widely utilized in a number of areas, see Fig. 1.1.

## 1.2 X-ray microscopy

X-ray microscopes[7, 8] are specialized instruments developed to resolve fine internal structures of objects that cannot be seen by the human eye. Together with light microscopes[9, 10] and electron microscopes[11–13], they form the most dominant microscopy techniques available for scientific researches, see Table 1.1.

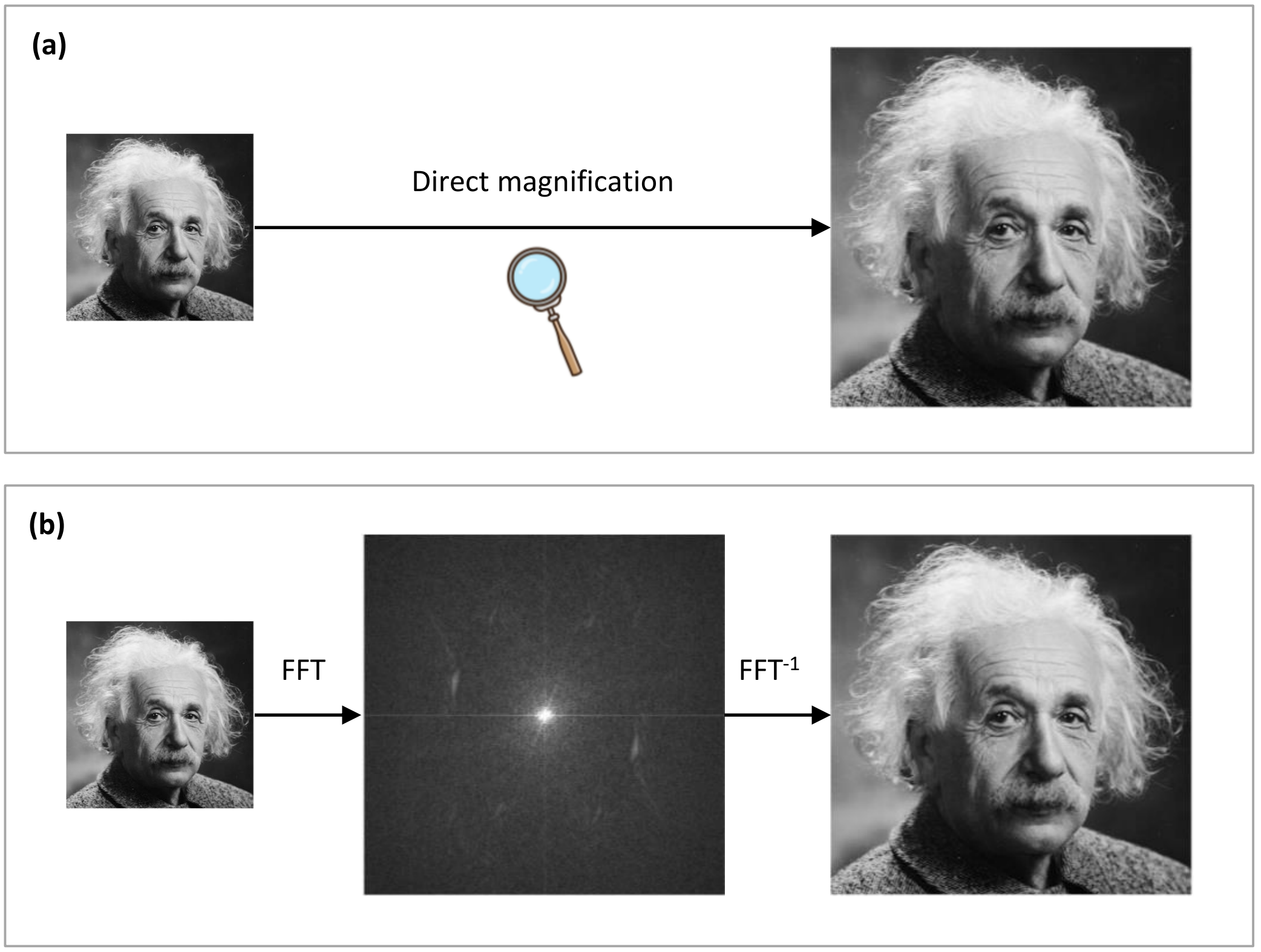

Figure 1.2: Two different typical approaches to obtain X-ray microscope images. (a) In the spatial domain, the nanometer scale structures are obtained via direct magnification. (b) In the transform domain, the transformed representations of nanometer scale structures are obtained, from which the X-ray microscope image can be retrieved. Herein, the most conventional Fast Fourier Transform (FFT) is assumed.

Table 1.1: Main properties of light microscope, X-ray microscope and electron microscope. Note that the provided values are only for orientation.

| **Tpye** | **Light microscope** | **X-ray microscope** | **Electron microscope** |
|---|---|---|---|
| Wavelength (nm) | 400-800 | 0.03-10 | 0.06-3 |
| Resolution (nm) | 200 | 20 | 2 |
| Imaging dimension | Inner/Surface | Inner | Surface |
| Sample types | All | All | Metallized |
| Imaging modes | Full field/Scanning | Full field/Scanning | Scanning |

In general, the X-ray microscope imaging information can be recorded and generated via two different domains: the spatial domain and the transform domain (usually the Fourier frequency domain[14, 15]). In particular, the spatial domain provides direct magnification or representation of the micro structures down to nanometer scale, see Fig. 1.2(a), and the transform domain reveals the transformed representations of nanometer scale structures, from which the X-ray microscope image can be retrieved, see Fig. 1.2(b).

## 1.3 A brief history review

### 1.3.1 In 1890s-1920s

The origins of X-ray microscopy can be traced all the way back to 1895, right after the discovery of X-rays. Röntgen himself ever attempted to focus X-rays with any sort of lens, but without success[16]. As a consequence, unfortunately, the possibility of inventing the first X-ray microscope was dismissed by Röntgen.

Regardless, X-rays were being utilized to inspect the fine internal structures within a few months of Röntgen's discovery via the contact microradiography, termed by Goby[17] in 1913. The resolution can be down to a few microns. Burch and Ranwea applied it to botanical specimens[18, 19] in 1896, and Heycock & Neville applied it to alloys[20] in 1898. Note that the contact microradiography approach does not require the focusing of X-rays. Its fundamental idea is reducing the large focal spot blurring effect by putting the object close to the photographic emulsion of very fine grain, while enlarging the region-of-interest (ROI) selected under a high-power optical microscope photographically[21], see the illustration in Fig. 1.3. The key relies on the ultra-fine-grain emulsions, which can be treated as flat-panel-detectors (FPDs) with ultra-small pixel dimensions. Afterwards, the development of contact microradiography was slow and mainly to biological applications. Until late 1930s, this technique started to be widely used in metallography due to the commercial availability of ultra-fine-grain emulsions.

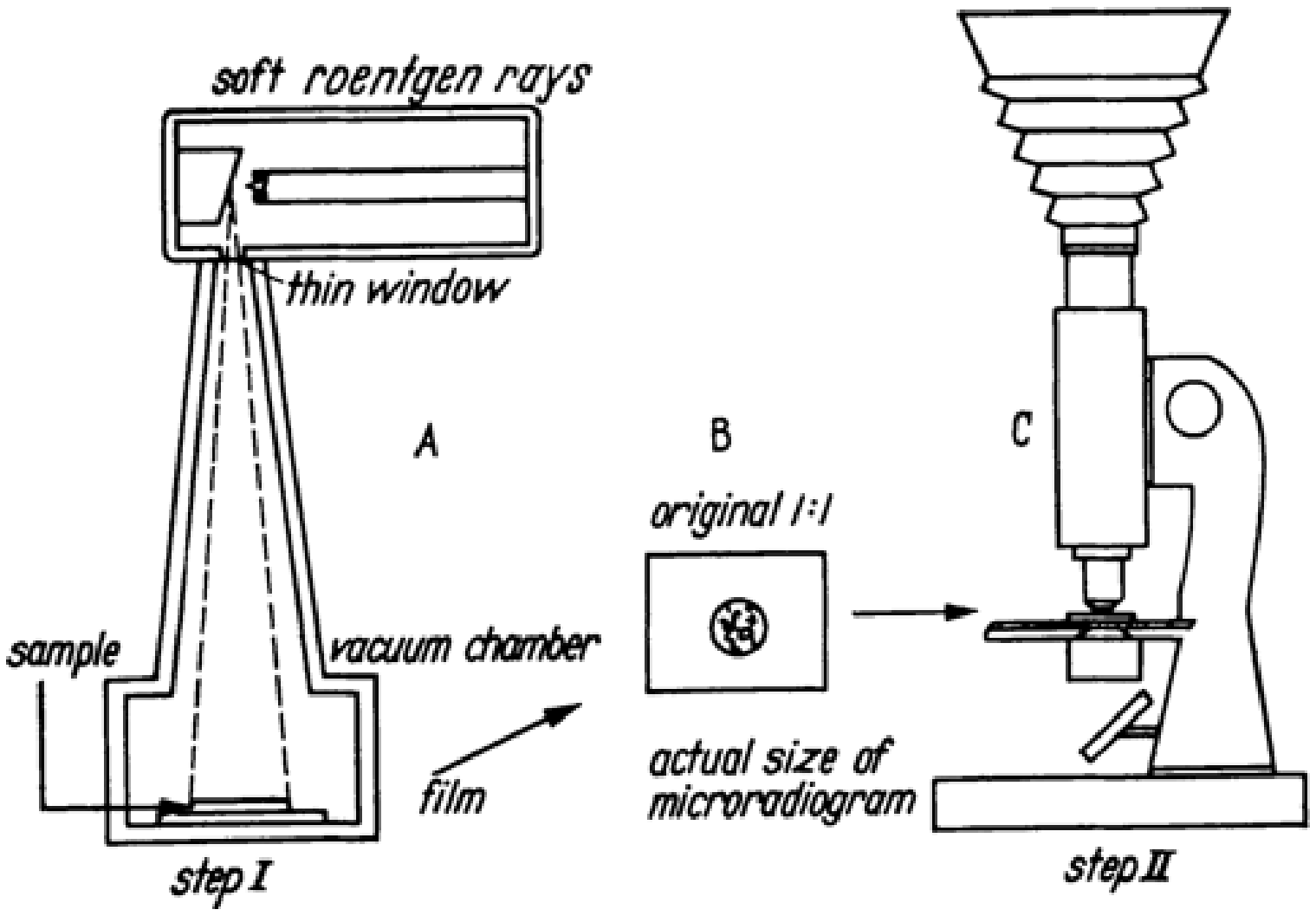

Figure 1.3: Illustration of the contact microradiography approach in generating high resolution X-ray images. In the first step, the sample is placed on the surface of film, which is made by ultra-fine-grain emulsions. In the second step, the recorded high resolution information is retrieved using a high-power optical microscope.

### 1.3.2 In 1930s

Instead of using ultra-fine-grain emulsions to record the high spatial resolution structures, the other simplest way of forming enlarged images with X-rays is to place the object close to a point source. Essentially, the resolution depends primarily on the diameter of the X-ray source. To obtain a point source, a limiting-pinhole is placed in front of a large focal spot in the early time, see Fig. 1.4. Early attempts were disappointing due to the low intensity of the X-ray tubes[22], but the method got improved with modern types of X-ray tubes[23].

In 1933, the use of a fine focal spot X-ray source was proposed by Malsch[24] to replace the above limiting-pinhole approach. In 1939, Ardenne[25] and Marton[26] independently suggested that concentrating an electron beam into a 1 $\mu$m or less size is possible with the newly developed electron optical techniques. Note that this approach does not require the focusing of X-rays either. In addition to the small focal spot size, a source of high intensity is also desirable to allow high resolution imaging.

In fact, the point projection method with an ultra-fine focused electron beam has become the standard selection in modern micro-CT ($\mu$CT) imaging nowadays. On the other hand, interestingly, the earlier strategy of placing a limiting-pinhole in front of the tube has been evolved as a standard operation in modern times to measure the shape and distribution of the focal spot[27].

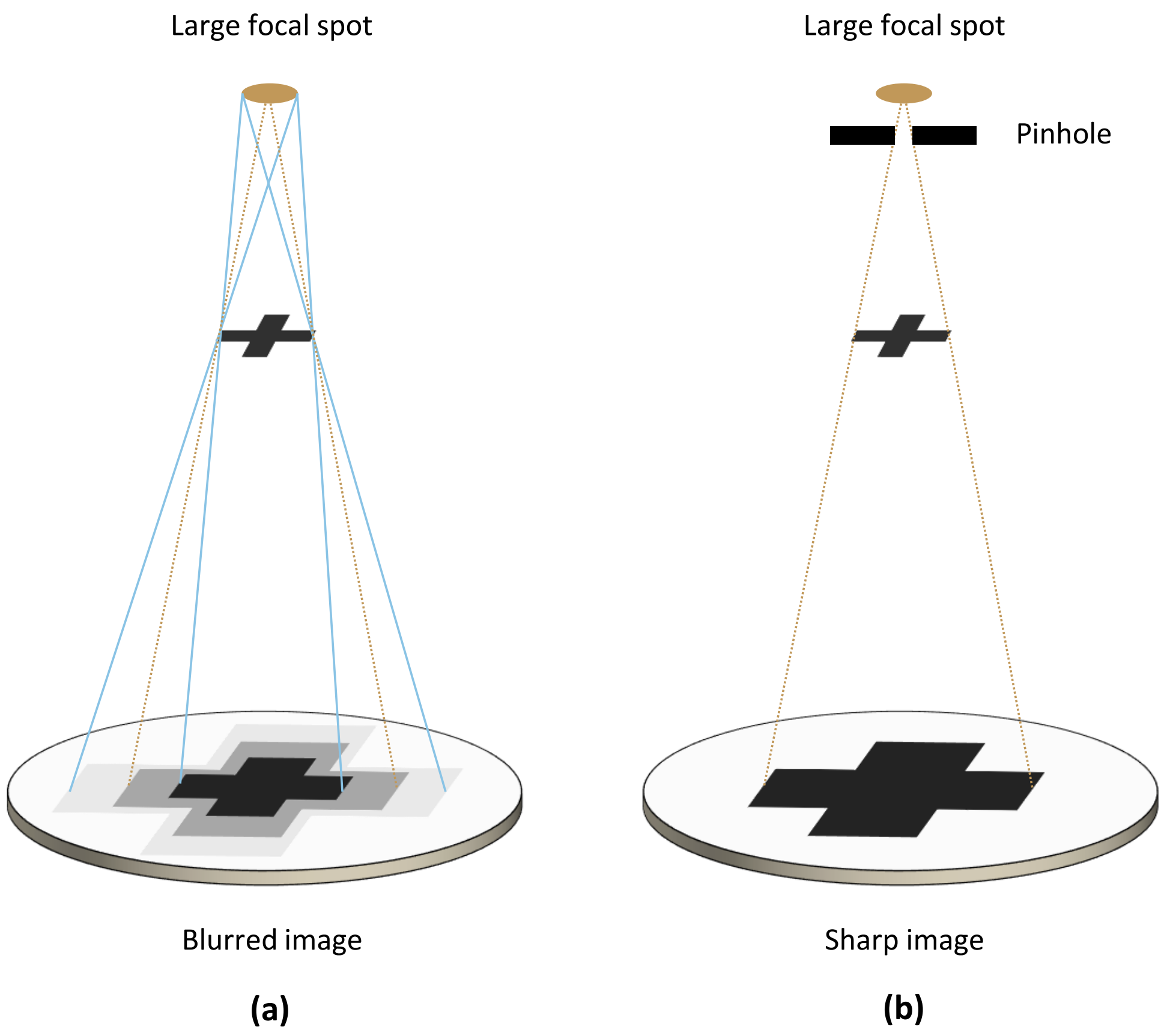

Figure 1.4: Illustration of the point projection method, in which a limiting-pinhole is positioned in front of a large focal spot, for high resolution X-ray imaging. With pinhole, the object can be positioned close to the focal spot to form a large geometric magnification ratio.

### 1.3.3 In 1940s-1950s

The X-ray microscope imaging was continuously evolved from the 1940s onwards at two distinct strategies: the projection method and the reflection method.

In 1940, Cosslett confirmed in experiments that an ultra-fine focused electron beam with severely limited tube power was viable to generate magnified X-ray images. The first successful X-ray microscope, a device capable of imaging small objects with resolution approaching optical microscope levels, was built in 1951 by Coslett & Nixon[28, 29]. That instrument is considered the first practical

X-ray microscope based on a point-focus (1-2 $\mu$m) X-ray tube, see Fig. 1.5(a). In 1954, Sterling Newberry and Selby Summers[30] at the General Electric Company designed the first commercial X-ray microscope using a micro focus X-ray tube, and achieved a spatial resolution comparable to a light microscope.

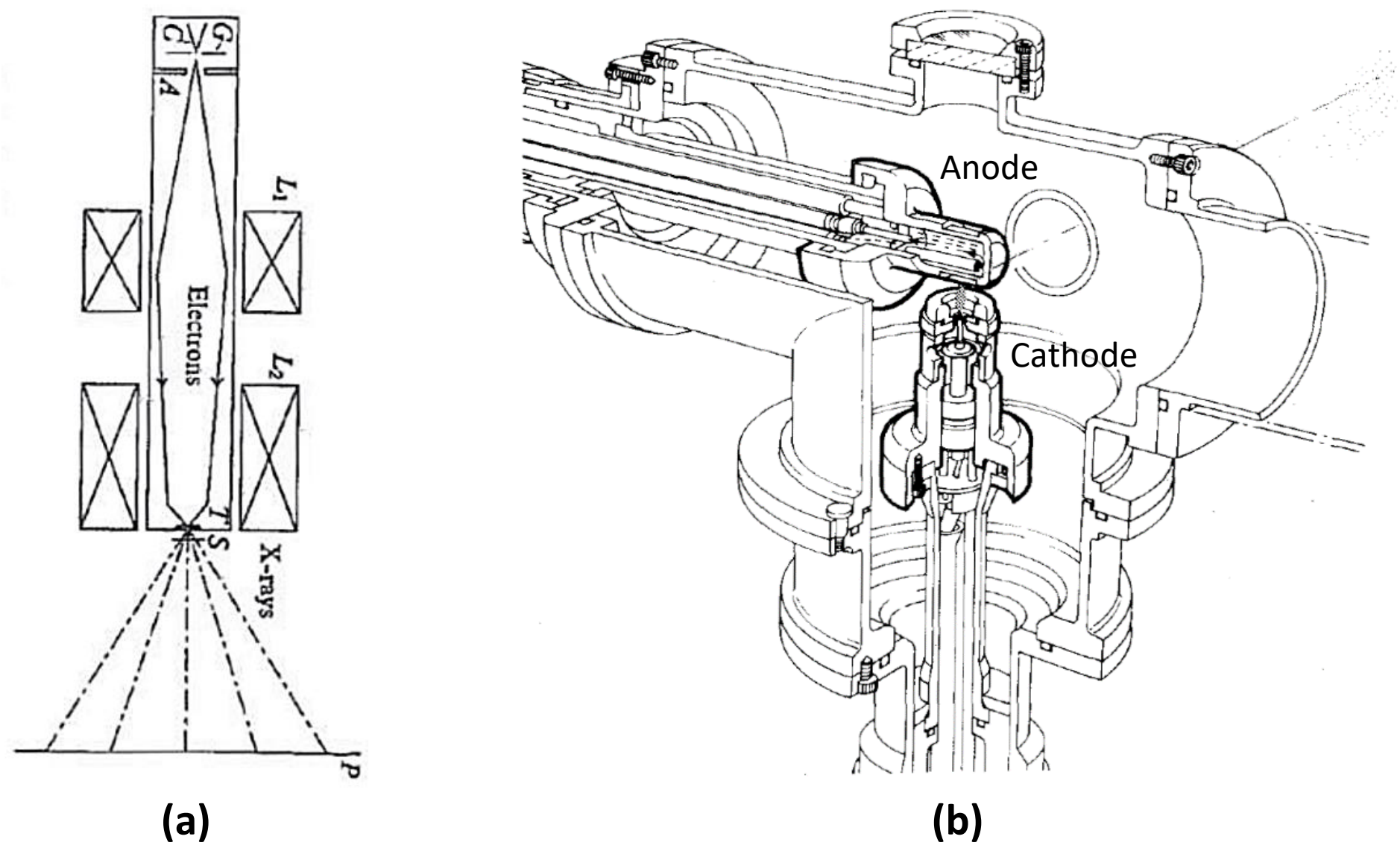

Figure 1.5: Inner structures of different micro-focus X-ray tubes. (a) The X-ray micro-focus tube with a transmissive anode. The electron lenses $L_1$ and $L_2$ form a reduced image at T of the cathode C, and the X-rays emitted from T project an image of a specimen S on to the screen (or plate) P. (b) The X-ray micro-focus tube with a reflective anode. The anode is in the horizontal direction, and the cathode is in the vertical direction.

The X-ray reflection techniques developed very slowly after Röntgen. For this approach, the fundamental issue is how to focus X-rays. In 1918, Albert Einstein[31] already pointed out that the refractive index for X-rays should be close to 1 but in general different from 1. The critical angle is between 89° and 90°, so that the glancing-angle for total reflection is of older 30'. This indicates that it would be very difficult to reflect X-rays in practice. However, it is obvious that for almost grazing incidence on a well polished surface there must be detectable total reflection for X-rays. This was investigated by Jentzsch[32] in 1929. The practical realization of a reflection X-ray microscope was delayed for another 20 years until Paul Kirkpatrick and Albert Baez[33] found an effective means to mount two parabolic curved mirrors at right-angle in overcoming the strong astigmatism effect, see Fig. 1.6. Different from Röntgen, Kirkpatrick and Baez changed the spherical focusing surfaces into cylindrical focusing surfaces. This is because X-ray focusing is very much stronger in the plane of incidence than in that of normal to it. In fact, until now, the development of the reflection method is hampered more by practical fabrications than by any theoretical limitation.

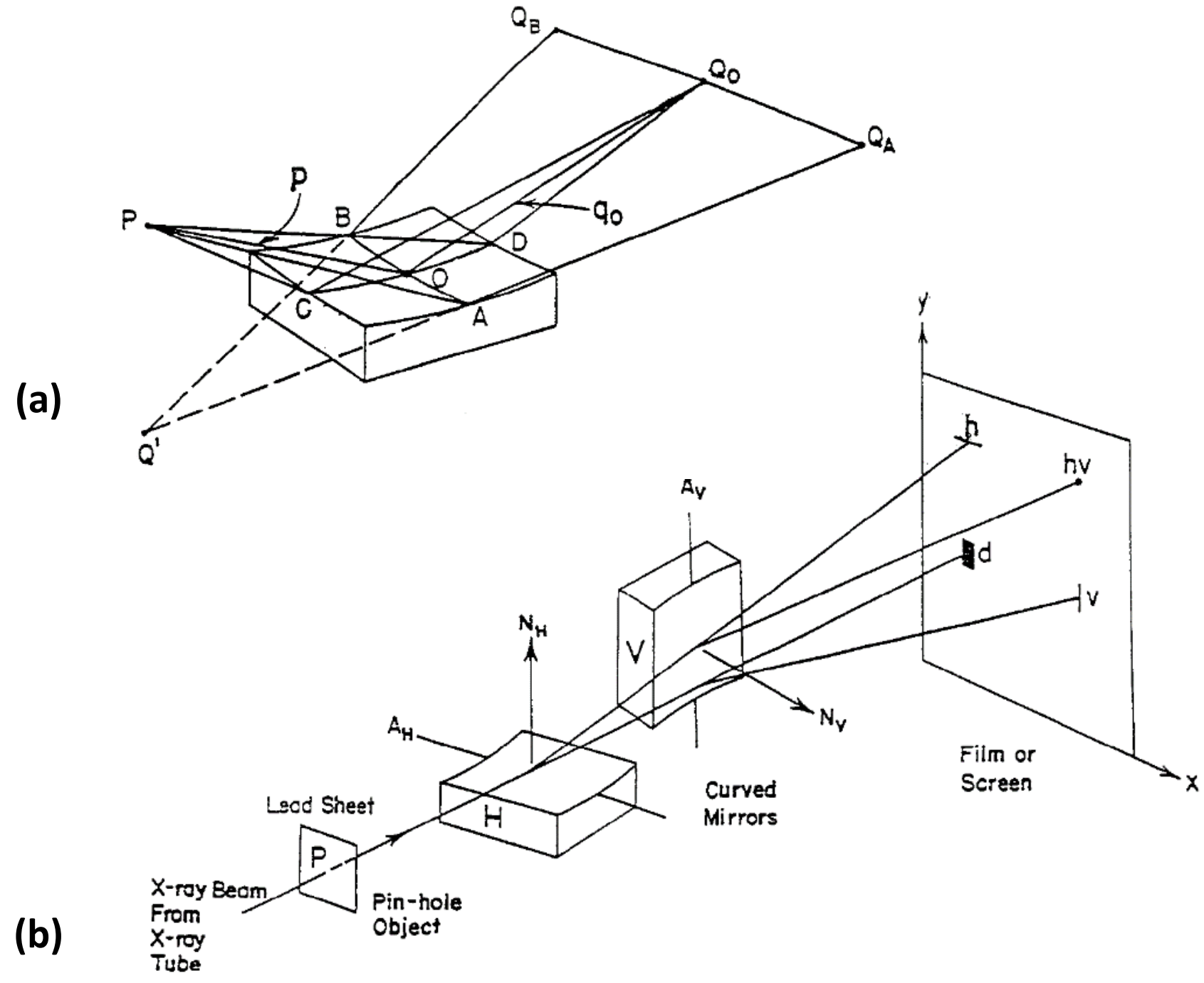

Figure 1.6: Reflection of X-ray beam using cylindrical mirrors. (a) X-rays diverging from a source are focused by a cylindrical surface to form an astigmatic image. (b) Arrangement of two cylindrical mirrors to focus the X-ray beam.

### 1.3.4 In 1960s-1970s

During this period, zone plates started to draw attention in X-ray microscopy. In principle, FZP is a circular diffraction grating, consisting of concentric rings with radially increasing ring density. Alternate rings should be opaque (or should reverse the phase) and transparent, The width of the outermost ring determines the numerical aperture and the limiting resolution. Baez was the first scientist to use Fresnel zone plates (FZPs) as lenses[34] to focus X-rays, see Fig. 1.7. His first free-standing metal zone plate, made by optical lithography, had just 19 zones with an outermost zone width of 20 $\mu$m.

In 1969, Gunter Schmahl and Dietbert Rudolph of the University of Göttingen came up with holographic approaches to zone plate fabrication[35]. The first microscope to use synchrotron radiation as the source was built by Horowitz and Howell[36] at the Cambridge Electron Accelerator in 1971. Unfortunately, this synchrotron light source closed down shortly. In 1972, David Sayre suggested an alternative approach for FZP fabrication with the electron-beam fabrication technology[37], which has been refined and implemented in several laboratories around the world since then. By 1974, the Göttingen group (including Bastian Niemann) had demonstrated 2 $\mu$m resolution zone plate imaging using a laboratory source[38], and in 1976 they produced the first zone plate images using synchrotron radiation[39], see Fig. 1.8.

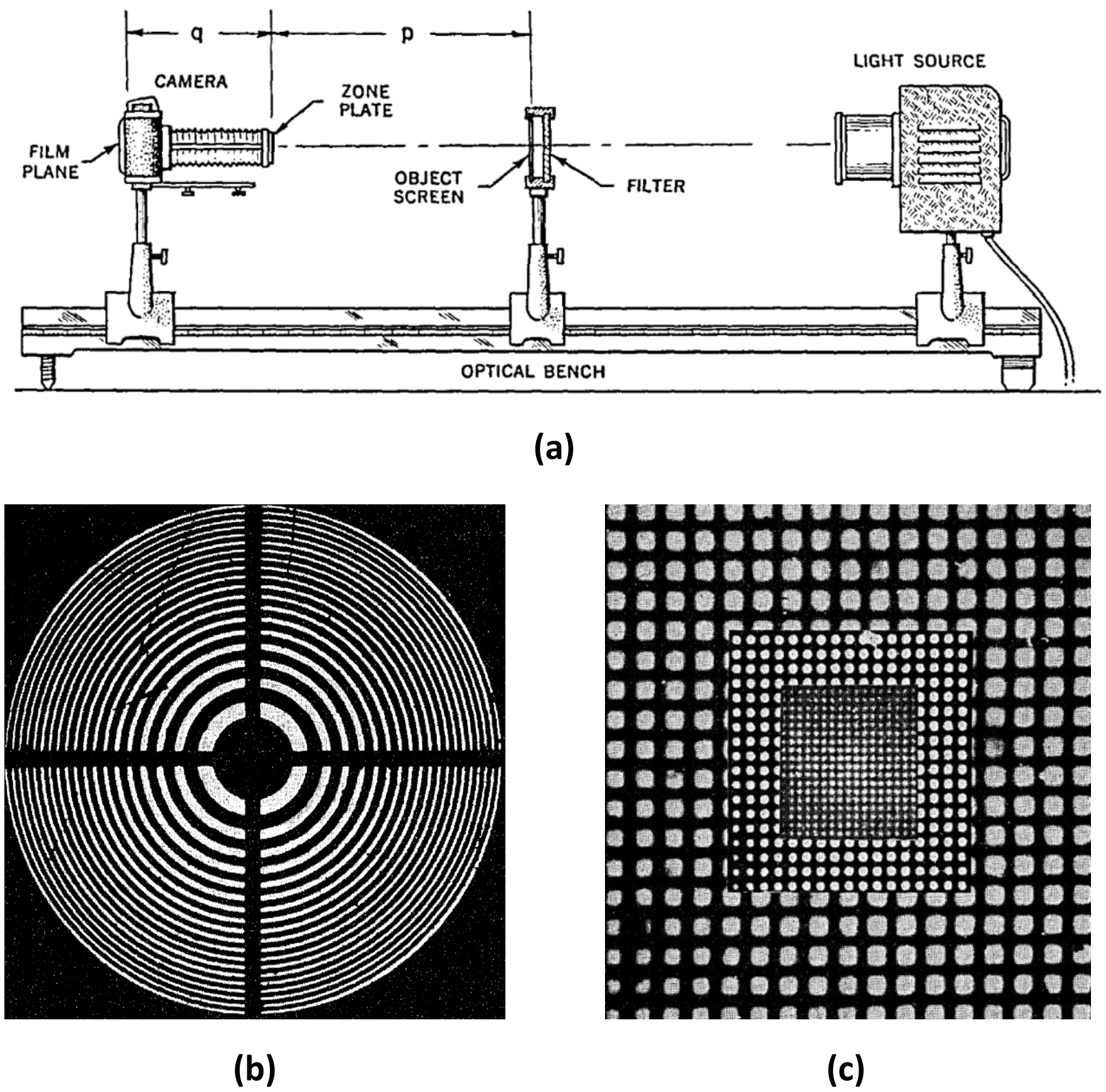

Figure 1.7: (a) Layout of the first X-ray microscope imaging apparatus with FZP on an optical bench. The light source consisted of a tungsten filament lamp and a condensing lens system. Object and image distances are labeled respectively. (b) The photograph of the self-supported gold zone plate used in the experiments. The diameter of the outer circle is 0.2596±0.0002 cm. The central circle has a diameter of 0.0426±0.0002 cm. The thickness of the gold is estimated as 10 $\mu$m. This zone plate was mounted at the open end of a bellows extension. (c) Pictures of a mesh with 4 lines/mm taken with the zone plate at three different wavelengths: 6700, 4358, and 2537 A. The shorter wavelength results not only in a longer focal length and hence a larger image size but also in improved resolution.

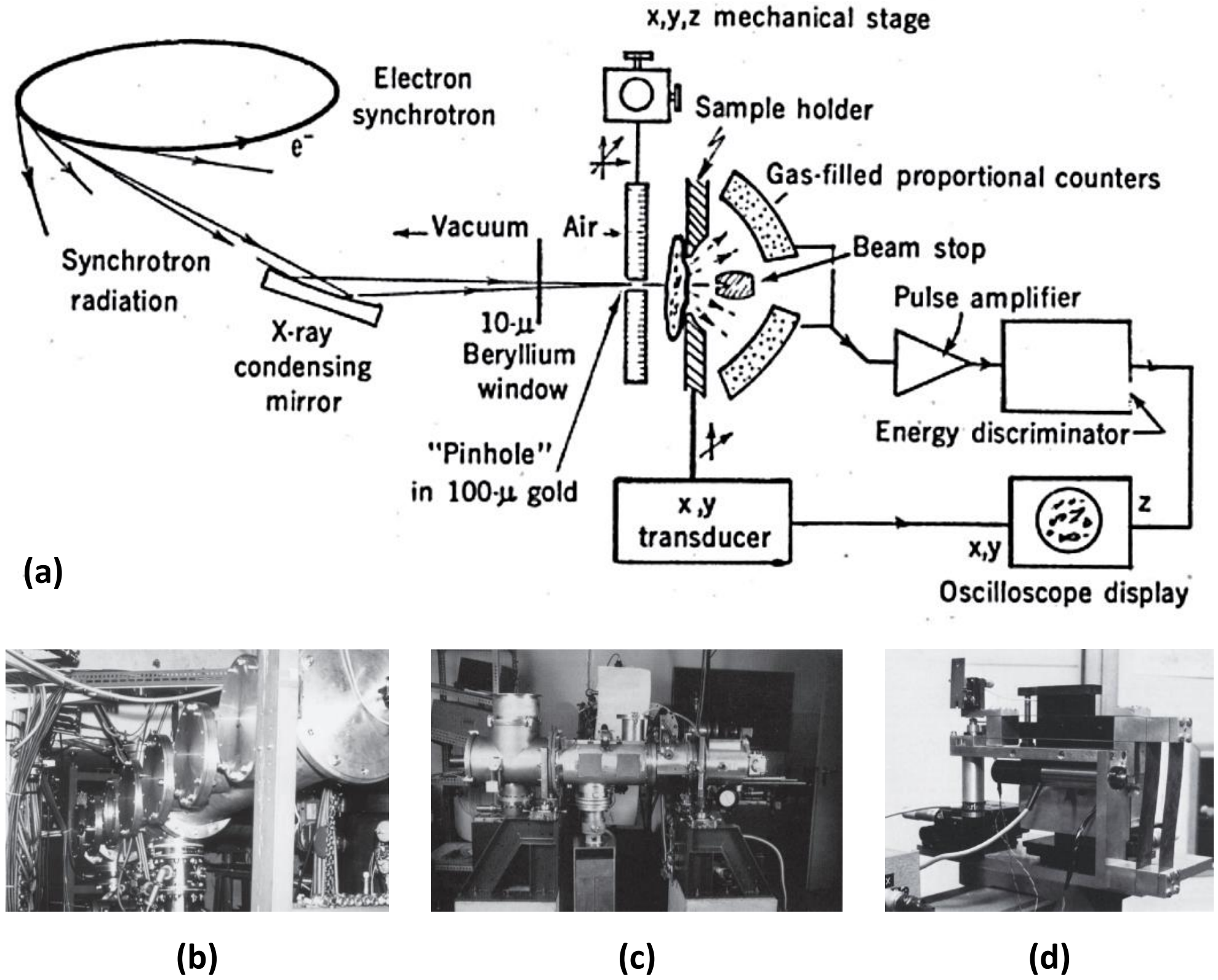

Figure 1.8: (a) The first synchrotron based x-ray microscope was developed at the Cambridge Electron Accelerator in 1972. The first zone plate transmission X-ray microscopes were developed (b) at DESY in Hamburg in 1976 and (c) at ACO in Orsay in 1983. (d) The first scanning transmission X-ray microscope using zone plate optics was constructed at U15 at the National Synchrotron Light Source at Brookhaven.

### 1.3.5 In 1980s-1990s

In brief, the developments of X-ray microscopy in the 1980s and 1990s were mainly impacted by the three-dimensional computed tomography (CT) technique and synchrotron light sources. The earliest X-ray micro-CT system, which produced micrometer-scale voxel sizes in reconstructed volume images, was conceived and built by Jim Elliott[40] in the early 1980s. The first published micro-CT images had voxel sizes of roughly 50 $\mu$m.

Early in 1980, Janos Kirz made several X-ray microscopy[41] tests using soft X-rays at the SPEAR ring at Stanford. In the mid-1990s, three large-scale synchrotron light sources began operation with higher electron beam energies: first 6 GeV at the European Synchrotron Radiation facility (ESRF) in Grenoble, France, then 7 GeV at the Advanced Photon Source (APS) at Argonne Lab near Chicago, USA; and finally 8 GeV at SPring-8 near Himeji, Japan. These facilities have led the way in advancing X-ray microscopy at multi-keV energies.

Table 1.2: The major X-ray microscopy (XRM) conferences held to date.

| **Year** | **Location** |
|---|---|
| 1956 | Cambridge, UK |
| 1959 | Stockholm, Sweden |
| 1962 | Stanford, USA |
| 1977 | New York, USA |
| 1980 | New York, USA |
| 1983 | Göttingen, Germany |
| 1987 | Brookhaven, USA |
| 1990 | London, England |
| 1993 | Chernogolovka, Russia |
| 1996 | Würzburg, Germany |
| 1999 | Berkeley, USA |
| 2002 | Grenoble, France |
| 2005 | Himeji, Japan |
| 2008 | Zürich, Switzerland |
| 2010 | Chicago, USA |
| 2012 | Shanghai, China |
| 2014 | Melbourne, Australia |
| 2016 | Oxford, England |
| 2018 | Saskatoon, Canada |
| 2022 | Hsinchu, Taiwan, China |
| 2024 | Lund, Sweden |
| 2026 | Campinas, Brazil |

Besides using the synchrotron light source, the Göttingen group demonstrated the feasibility of realizing X-ray microscopy in laboratories with plasma discharge sources[42] in 1990. Greater success was obtained by the group of Hans Hertz in Sweden by using lasers to excite plasma emission from in-vacuum liquid jets of ethanol[43] and, with improved emission, from liquid nitrogen[44]. In the late 1990s, Dr. Wenbing Yun founded the company Xradia (now Carl Zeiss X-ray Microscopy) and began delivering X-ray microscopes using characteristic line emission from microfocus X-ray sources, capillary reflectors as condensers, and FZP objectives for imaging and nanotomography at 5.4 keV[45]. No doubt this was a very important step for the wide applications of laboratory microscopes.

### 1.3.6 In 2000s-2020s

Since 2000s, the main research interests were shifted to lensless X-ray microscopy such as holography[46], coherent diffraction imaging (CDI)[47, 48], and ptychography[49–51]. The first experimental demonstration of CDI was perform in 1999 by Jianwei Miao[47]. As a powerful new variant of the CDI, ptychography was developed theoretically by Rodenburg and Faulkner[49], and implemented with X-rays first at the Swiss Light Source. In it, the resolution is limited only by the wavelength of the X-rays and the angular range over which the differential patterns can be recorded. Resolution as fine as 3 nm has already been demonstrated. To meet the requirement of coherent illumination, actually, X-ray free electron lasers (FEL)[52, 53] was introduced as a new dimension to X-ray microscopy with inherently coherent or nearly coherent beams. On the other hand, the availability and use of diamond substrates[54] under the X-ray tube anode target allowed more heat to be effectively dissipated at the focal spot thereby enabling increased X-ray flux for laboratory microscopes using FZPs in the future. Additionally, the novel liquid-metal anode[55, 56] based high brightness X-ray sources have also been demonstrated.

Over the past century, as discussed above, X-ray microscopy has been continuously developed, see Table 1.2. Many pioneering concepts and creative innovations were invented to significantly expand the capability of seeing ever smaller structures with X-rays. The main purpose of this book is to discuss the methods and practices for building a spatial domain laboratory X-ray microscopy from X-ray tube source and Fresnel zone plate.

# 2 Working environments

The X-ray microscope with FZP requires very critical working environments. Disturbances such as mechanical vibration, temperature instability, humidity variation, vacuum level (helium concentration) and power outage would severely impact the imaging outcome and long-term stability. In this chapter, the fundamental working environments, potential impacts, and optimization strategies are discussed in details.

## 2.1 Laboratory space

To build an X-ray microscope with FZP, first of all, one needs to find a certain laboratory space that meets the following special requirements.

The room cannot be too small. It is recommended that the main equipment room has at least an area of 4.0 m×5.5 m, and the operation room at least has an area of 1.5 m×5.5 m, see Fig. 2.1. Herein, it is assumed that the microscope itself occupies an area of 2.0 m×1.0 m, with a peripheral working space out-extended by 0.9 m. The necessary accessory facilities, for instance, vacuum pump, chiller, helium tank, dehumidifier, air compressor, uninterruptible power supply (UPS) and shelving unit, are put orderly surrounding the microscope. If possible, it is better to isolate the air compressor from other facilities with the purpose of reducing noise and heating. In the operation room, two tables are placed. One is used for sample preparation, and the other is used for data acquisition. Usually, the microscope housing made from iron/steel is thick enough to shield the hazard radiation, whose maximum beam energy is assumed to be 40 keV in this book. As a consequence, no additional shielding is needed in the room wall. The open width of the entrance door has to be larger than 1.2 m. Since the weight of the main body of the microscope may be more than 2.0 tons, therefore, the room must be placed directly above the ground.

Other main environment requirements are summarized in Table 2.1. They define the minimum acceptable requirements in ensuring stable and reproducible nano-resolution X-ray imaging. Please be aware that these requirements have to be fulfilled at the same time, and failure of any single parameter will degrade the overall imaging performance of the FZP-based X-ray microscope. During practical applications, these requirements have to be adjusted appropriately, especially with respect to the targeting spatial resolution. For instance, It was found that the environmental vibration requirements for systems with a spatial resolution of 50 nm are considerably less stringent than those for systems achieving a 30 nm spatial resolution.

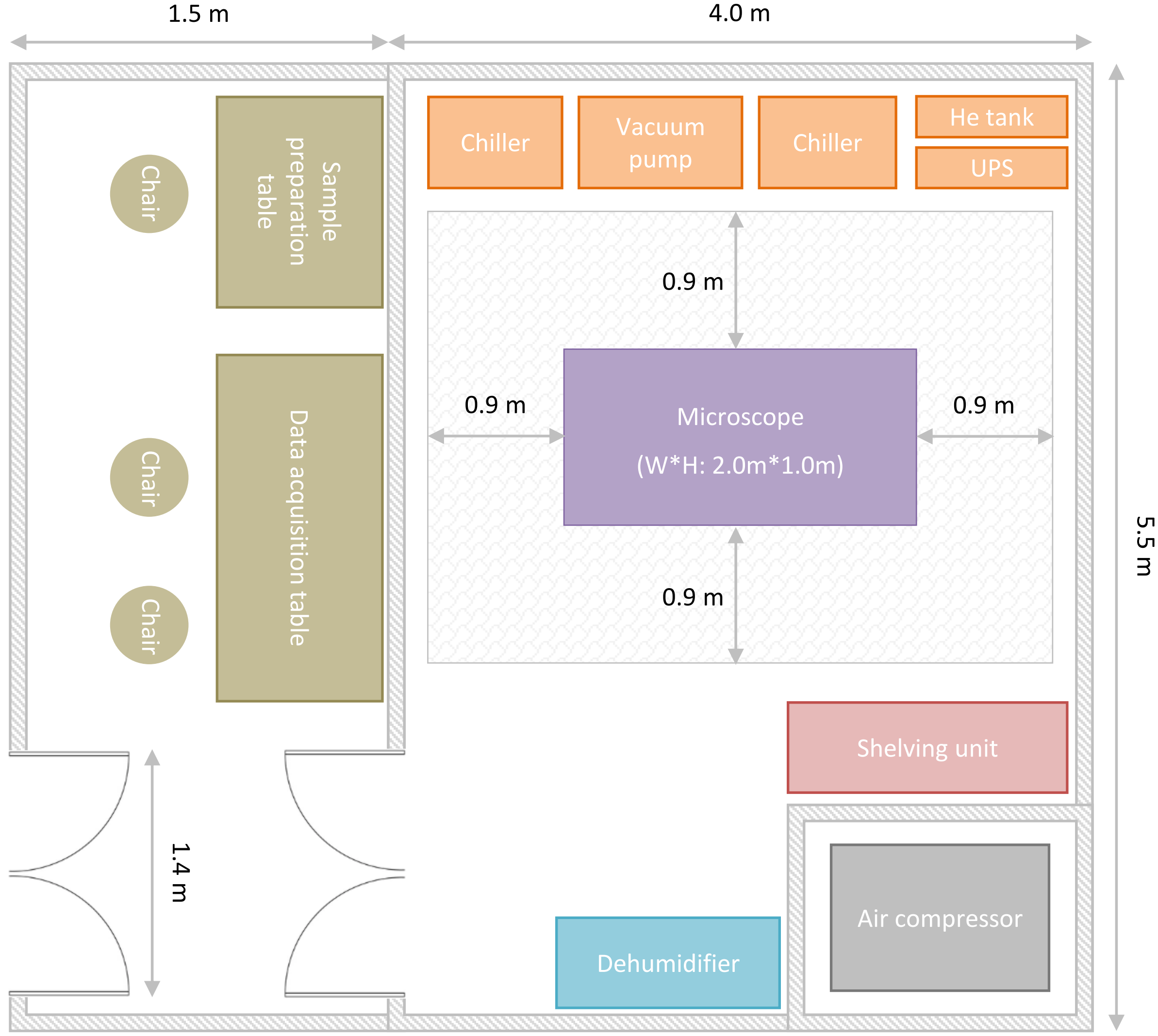

Figure 2.1: The recommended room layout for laboratory X-ray microscope with FZP.

Table 2.1: Key environment requirements for a laboratory FZP-based X-ray microscope.

| **Parameter** | **Value** |
|---|---|
| Vibration magnitude | $\leq 0.25 \times \Delta_s$ |
| Room temperature | 22°C ± 1 °C |
| Internal temperature fluctuation | ±0.15 °C/day |
| Power supply | 220 V + UPS |
| Relative humidity | 30%–50% |
| Airtightness holding time | ≥ 1 month |

Note: $\Delta_s$ denotes the highest targeting spatial resolution of the microscope.

## 2.2 Vibration

Vibration poses a major challenge to X-ray microscopy with FZP. In practice, numerous sources can contribute to environmental vibrations, including the operation of chillers and air compressors, passing vehicles and buses, subway trains, and nearby construction activities. Even walking in the laboratory can induce measurable ground vibrations. If the vibration results in rigid-body motion of the entire system, its impact on imaging may be limited and not necessarily be severe. However, if it induces relative motion between different components, the imaging performance of the X-ray microscopy with FZP would be significantly degraded. Because the vibration-induced relative motion is unavoidable, it is therefore essential to suppress environmental vibrations as much as possible.

In an X-ray microscope with FZP, different optical component has different vibration sensitivity. In general, they can be classified into two groups: critical components and less critical components. Critical components, including FZP, sample and gratings (important for phase contrast imaging), impose the most stringent vibration control requirements. For instance, the vibrations of zone plate and sample have direct impact to the targeting image resolution. As a consequence, vibration control with magnitude below ten nanometers has to be ensured. For gratings, the allowable vibration amplitude has to be smaller than a quarter of their periods. Excessive displacement would lead to degradation of interference fringe visibility. In contrast, the X-ray source and the detector are less sensitive to mechanical vibrations. Assuming the focal spot size of the X-ray source is 20 $\mu$m $\times$ 20 $\mu$m, obviously, micron level vibration will not produce significant impact. Similarly, micron level vibration at the detector plane corresponds to a much smaller, i.e., approximately two orders, effective displacement when projected onto the sample plane. Consequently, the maximum allowable vibration amplitude for each component must be determined independently based on its specific character rather than a common criterion.

Table 2.2: Properties of typical environmental vibrations.

| **Tpye** | **Frequency range (Hz)** |
|---|---|
| Walking | 1.8-2.5 |
| Jumping | 1.0-3.5 |
| Vacuum pump | 10-1000 |
| Vehicle/bus | 1.0-50 |
| Metro | 10-80 |
| Rotary drilling rig | 5-300 |

### 2.2.1 Vibration measurements

Vibrations have to be quantified. To do so, a dedicated full-field laser Doppler vibrometer (Model: Polytec PSV-500, KCF Technologies, USA) was employed to measure the vibration spectra at multiple critical locations, including the FZP, sample stage, air-bearing rotation stage, X-ray source assembly, detector, and vacuum chamber. The experimental setups for vibration measurements are shown in Fig. 2.2. To obtain precise results, the vibrometer and the object have to be placed on a rigid common floor, which of course should be well isolated from vibrations in prior.

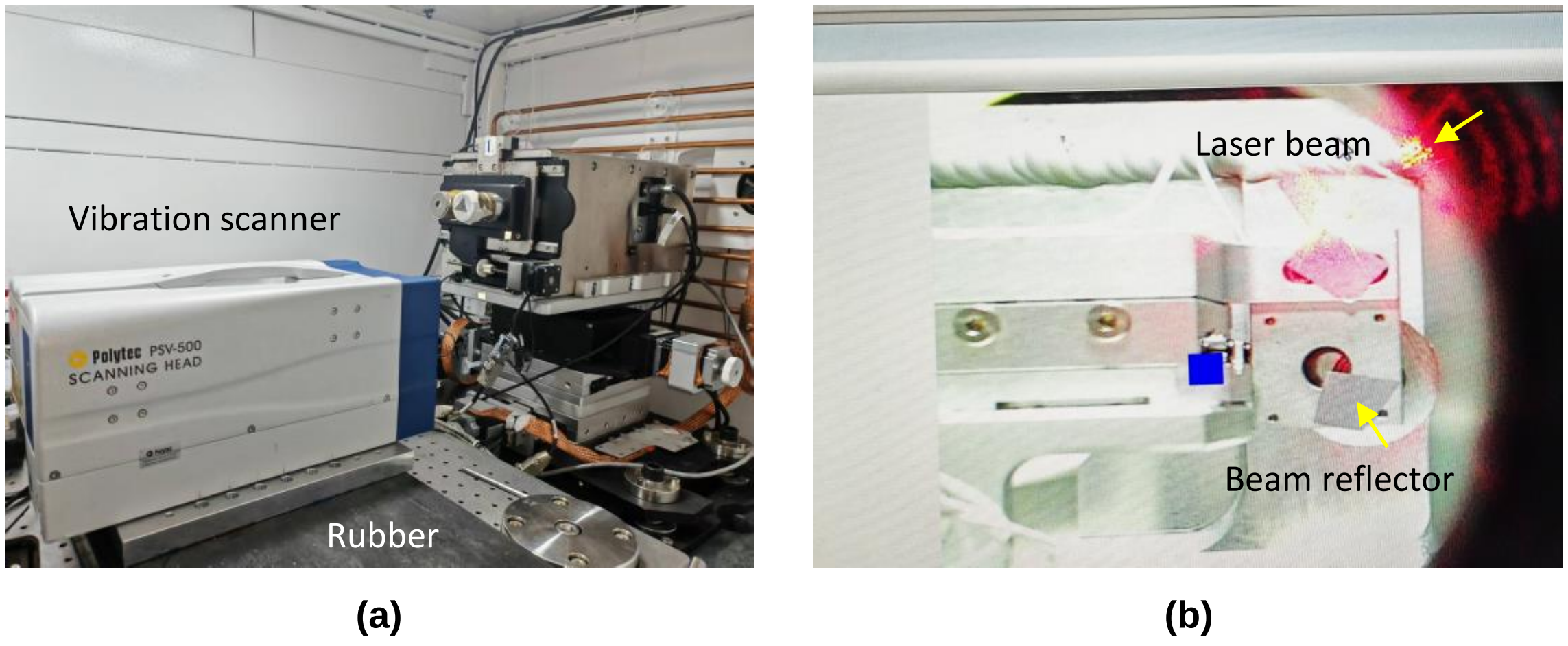

Figure 2.2: Vibration measurement setup at two representative locations of the FZP-based X-ray microscope: (a) detector, (b) FZP. A 5 cm thick rubber plate was placed beneath the vibrometer to isolate external vibrations.

The measured vibration spectra from six pre-selected locations are shown in Fig. 2.3(a). For our system, the vibrations at these locations are quite similar, with slightly higher values observed at the detector and X-ray source positions, see Fig. 2.3(b). We guess this is due to the fact that the vibrometer is not rigidly coupled to the detector and X-ray source, both are mounted on different independent mechanical stages. Due to the sequential scanning nature of the full-field laser Doppler vibrometry, slow system displacements and rigid-body motions will be mapped into low-frequency components, which lead to an overestimation of low-frequency vibrations. Furthermore, the numerical integration required for displacement reconstruction further amplifies such overestimation. Consequently, the measured low-frequency vibration results are more qualitative (less accurate for quantification). In contrast, high-frequency vibration measurements remain quite reliable. A pronounced vibration peak was observed around 80 Hz, particularly for the detector, sample stage, and zone plate, see Fig. 2.3(c). Careful and painful investigations ultimately traced such distinct vibration to the rotation of the detector cooling fan. When the fan was turned off, such distinct vibration signal disappeared completely, see Fig. 2.3(d). Unfortunately, the cooling fan cannot be turned off for this particular CCD detector (iKon-L 936, Andor, UK) from the software end, therefore, special modifications are performed to suppress the impact from this annoying vibration source.

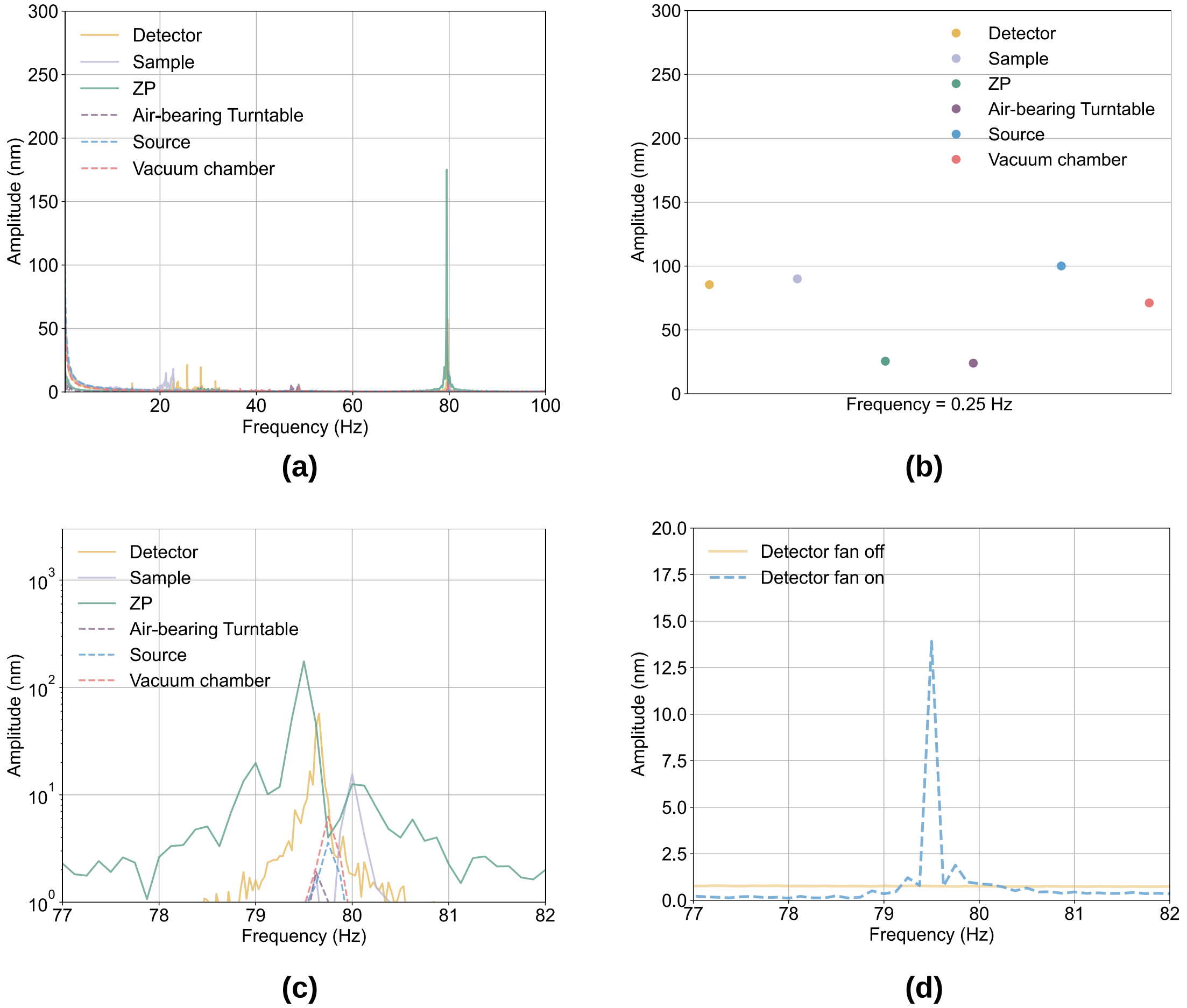

Figure 2.3: (a) The vibration spectra measured at six different representative locations. (b) Vibration amplitude distributions around 0.25 Hz. (c) Vibration amplitude distributions around 80 Hz. (d) Vibration amplitude distributions before and after turning on the detector cooling fan.

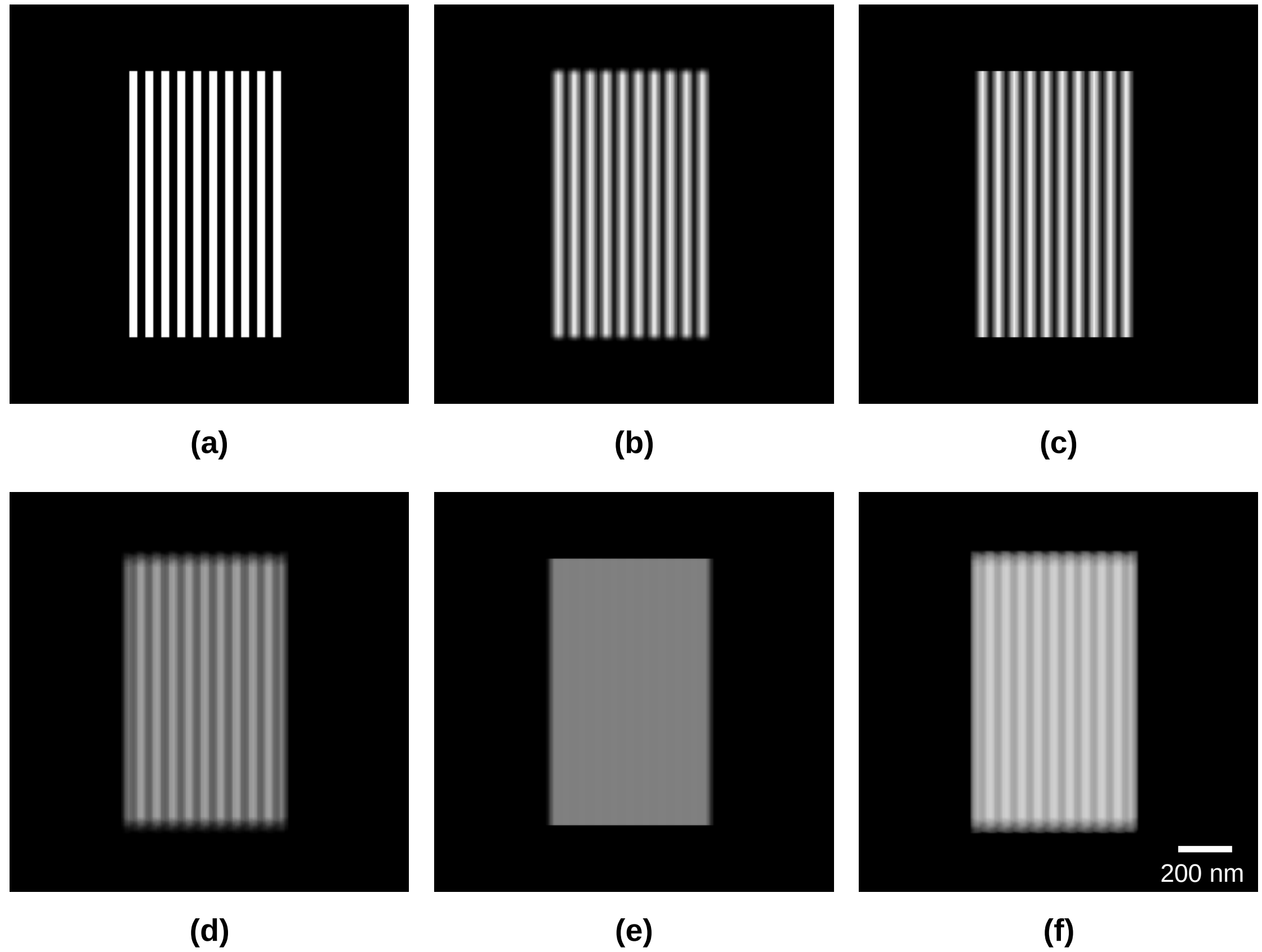

Figure 2.4: Simulation results of a 30 nm-period line-pair pattern under low- and high-frequency vibrations. (a) Imaging result without vibration. (b) Imaging result with a high-frequency vibration of 15 nm amplitude. (c) Imaging result with a low-frequency vibration of 15 nm amplitude. (d) Imaging result with a high-frequency vibration of 30 nm amplitude. (e) Imaging result with a low-frequency vibration of 30 nm amplitude. (f) Imaging result with a high-frequency vibration of 30 nm amplitude along both horizontal and vertical directions. Note that the above numerical simulation was performed as follows: a total of 120 frames were averaged. For the high-frequency vibration scenario, each frame was subjected to a random translation within the given amplitude. For the low-frequency vibration scenario, the total displacement over the 120 frames was two times of the given amplitude, with the displacement distributed uniformly across the frames.

### 2.2.2 Potential impacts

Mechanical vibrations, both high- and low-frequency components, can significantly degrade the high resolution imaging performance. Specifically, the high-frequency vibrations are particularly harmful and would significantly degrade the spatial resolution even for a short exposure period (several seconds). As a contrary, it was found that the low-frequency vibrations primarily cause image drift and long-term instability. If the exposure period is several tens or several hundreds seconds long, low-frequency vibrations can generate accumulating effect and thus leading to dramatic impact on the aiming spatial resolution.

To mimic the impact of mechanical vibrations under different frequencies, a 30 nm period line-pair pattern was simulated, see the results in Fig. 2.4. As shown in Figs. 2.4(b)-(c), the 30 nm structures remain resolvable if the vibration amplitude is less than half of the period, i.e., 15 nm. Under these conditions, the outcomes are less dependent on vibration frequency. Whereas, the structures can no longer be clearly distinguished when the vibration amplitude increases up to 30 nm, see Figs. 2.4(d)-(e). Additionally, it was demonstrated that the impact of low-frequency vibration is more pronounced than high-frequency vibration, see Fig. 2.4(e). Finally, when vibrations occur simultaneously along two orthogonal (horizontal and vertical) directions, the consequent outcome is likely to become much worse, see Fig. 2.4(f).

### 2.2.3 Vibration management

Vibration isolation and suppression are essential in the design and implementation of an FZP-based X-ray microscope. First, hardware that generates less vibration, for instance, detector without internal cooling fan or fan that can be turned off manually, should be considered. Second, anti-vibration marble table and honeycomb optical breadboards should be employed to suppress majority of vibrations. For instance, a 15-20 cm thick marble table with an area of 2.0 m (L) $\times$ 1.0 m (W) is needed as the main mounting base for all components. On top of the marble table, it is suggested to place three individual optical breadboards (10-15 cm thick) with proper sizes to mount the X-ray source, optical components and detector, respectively. By doing so, the cross-vibrating effect among different optical components can be minimized.

In fact, marble tables and honeycomb optical breadboards, which are more effective in suppressing high-frequency vibrations, alone are insufficient to provide the comprehensive vibration isolation required to achieve a spatial resolution of several tens of nanometers. As a result, other forms of vibration isolation are still required. During our experiments, the cost-effective anti-vibration rubber plates were added appropriately to suppress the relatively low-frequency vibrations or reduce the vibration magnitude. The imaging outcome before and after adding rubbers are depicted in

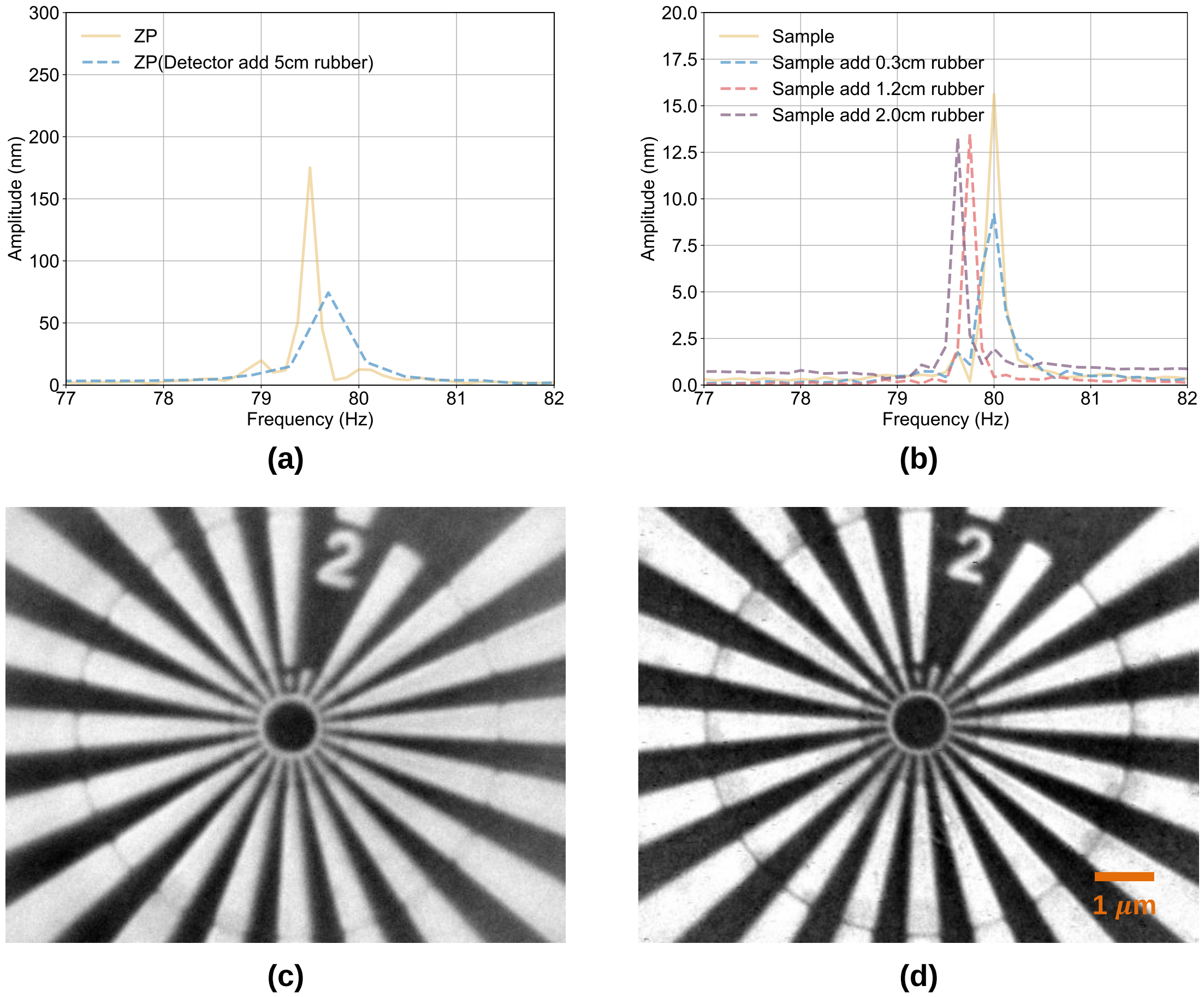

Figure 2.5: Vibration isolation and imaging outcome. (a) The measured vibration amplitude at the zone plate with and without adding a 5 cm rubber plate beneath the detector. (b) The measured vibration amplitude at the sample stage with different thicknesses of rubber plates. (c) The imaging outcome of the Siemens star pattern before vibration isolation. (d) The imaging outcome of the Siemens star pattern after vibration isolation.

Figs. 2.5(c)-(d). As demonstrated, the added 5 cm-thick rubber plates under the CCD detector was able to substantially reduce the impact of the internal 80 Hz vibration on FZP, see Fig. 2.5(a). In addition, rubber plates with optimal thickness are also designed and added under the sample stage, air-bearing rotation stage, honeycomb optical breadboards and the X-ray source assembly, see Fig. 2.5(b). With these rubber-based vibration isolation modifications, consequently, the image spatial resolution of our system was substantially improved from hundred nanometers down to thirty nanometers.

Table 2.3: Properties of marble, optical breadboard, and rubber.

| Tpye | Marble | Optical breadboard | Rubber |
|---|---|---|---|
| Primary mechanism | High mass | High rigidity | Viscoelastic dissipation |
| Target frequency | >50 Hz | 100 Hz-1000 Hz | 5 Hz-100 Hz |
| Damping ratio | 0.005-0.01 | 0.03-0.07 | 0.05-0.15 |
| Stiffness | High (50-80 GPa) | Extremely high | Very low (0.001-0.1 GPa) |
| Best used for | Heavy foundations | Lasers, optics | Foot pads, isolation joints |

## 2.3 Temperature and humidity

Temperature stability and humidity are also fundamental requirements for FZP-based X-ray microscopy. It was noticed that FZP is the most sensitive component to thermal variation among all the optics. The temperature-induced expansion or contraction can lead to direct image displacement and optical misalignment. Essentially, even very little temperature change can cause measurable displacement of FZP. For instance, experiments showed that the FZP may drift along one specific direction when the temperature increases (or decreases) gradually, see Fig. 2.6(c). For our system, a change of 1 °C in temperature can cause the FZP move by approximately 1.5 $\mu$m. Notice that this value may vary depending on specific instrument configurations and operating conditions. Such thermal induced FZP displacement would be directly transferred onto the image. For instance, a 10 nm shift of FZP would result in an approximately 10 nm shift of the image at the sample plane. As a consequence, the allowable FZP displacement during a single exposure must be kept well below the target spatial resolution, preferably less than half of it. If the thermal induced FZP displacement is more than 2 $\mu$m, significant deviations of the beam trajectory and severe disruption of the overall system alignment would happen. Our experience found that stringent temperature control is essential not only for preserving image quality during an individual exposure, but also for maintaining long-term stable system alignment. It should be noted that the allowable temperature fluctuation narrows as the highest achievable spatial resolution increases.

To meet such stringent temperature requirements, careful thermal managements are needed. At first, the temperature of the entire laboratory room has to be well controlled. Excessive external temperature fluctuations can propagate into the enclosure and thus degrade the internal thermal stability, even in the presence of insulation. In our laboratory, the room temperature is regulated by a central air-conditioning system, with fluctuations within $\pm 1\,^{\circ}$C. If possible, it is strongly suggested to install an independent air conditioning system to minimize the room temperature variation.

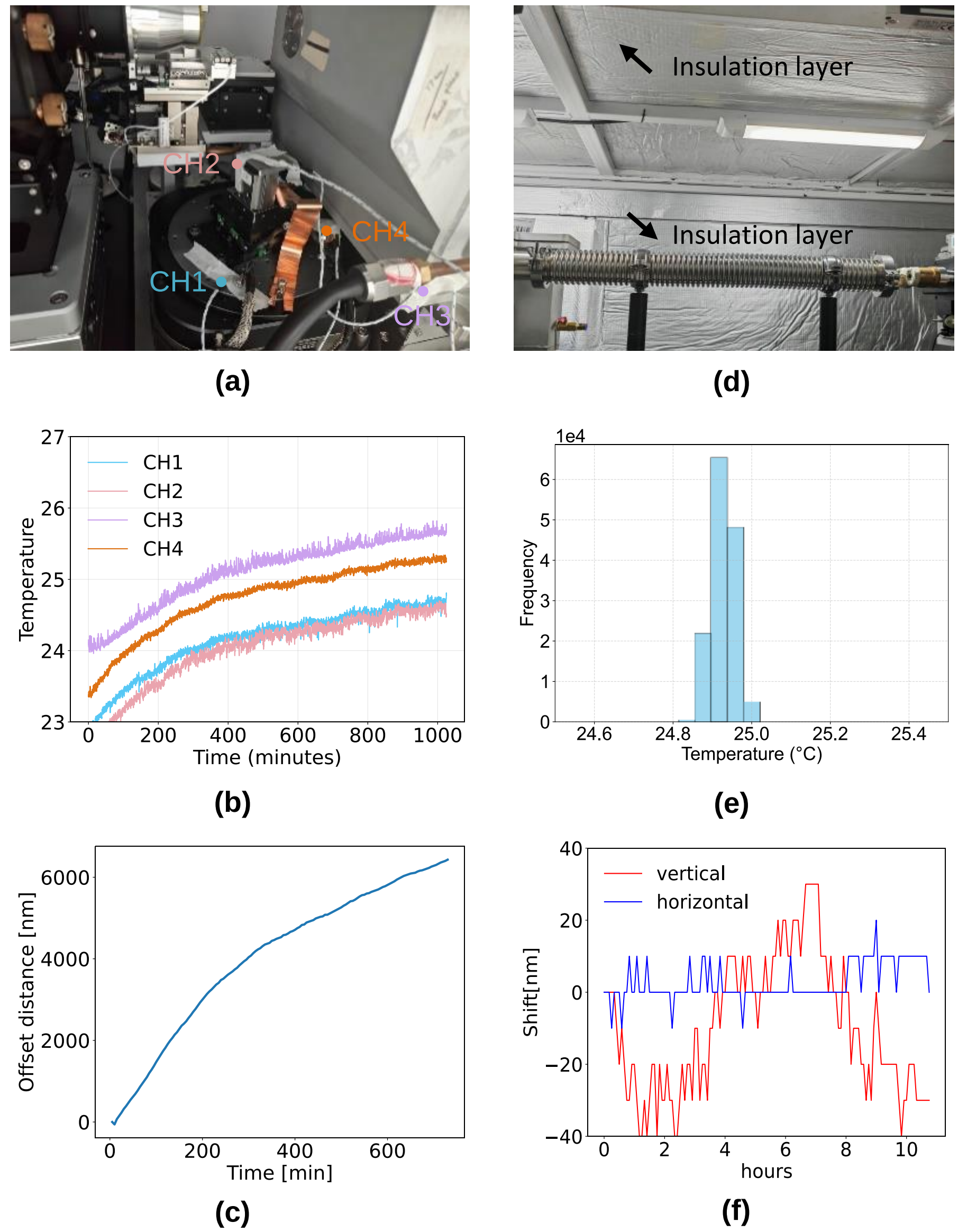

Figure 2.6: (a) Photograph of the temperature measurement setup. (b) Temporal evolution of temperature measured at different positions. (c) Corresponding FZP displacement during the temperature variation shown in (b). (d) Thermal insulation applied to the interior and exterior surfaces of the instrument enclosure. (e) Temperature distribution on the sample stage after thermal optimization. (f) FZP displacement under the optimized temperature conditions shown in (e).

The internal thermal insulation is more critical than the room temperature control. As shown in Fig. 2.6(d), all inner faces of the microscope were taped with 5 mm thick thermal insulation materials to minimize heat exchange.

Long-term temperature monitoring was conducted to test the effect of our thermal control at three different locations: the sample stage, the vacuum chamber wall, and a random place in the air. The measured temperature responses of the sample stage is illustrated in Fig. 2.6(e). As seen, the temperature variations at all locations remained within ±0.15 °C over 24-hours. Under these conditions, the measured FZP displacement during a single 300 seconds long exposure was less than 10 nm, which is far below the target spatial resolution of 30 nm, see Fig. 2.6(f). As long as the doors of the instrument remained closed, the temperature stability could be sustained within ±0.15 °C over a very long period, e.g., one week.

Humidity control, which is closely coupled with temperature management, must be carefully considered as well to prevent condensation on those low temperature components such as the chiller tubes connected to the X-ray source (approximately 10 °C) and CCD detector (approximately 18 °C), the air tube connected to the air-bearing rotation stage. Mathematically, the condensation can be estimated using the following dew-point relationship [57]:

$$\mathrm{T}_d = \frac{237.3\,\ln\left(\frac{\mathrm{RH}}{100}\exp\left(\frac{17.27\,\mathrm{T}}{\mathrm{T}+237.3}\right)\right)}{17.27-\ln\left(\frac{\mathrm{RH}}{100}\exp\left(\frac{17.27\,\mathrm{T}}{\mathrm{T}+237.3}\right)\right)}, \tag{2.3.1}$$

where $\mathrm{T}_d$ denotes the dew point temperature (in °C), T denotes the room air temperature (in °C), and RH is the relative humidity (in %). In our laboratory, the relative humidity was kept around 40%.

In addition, special manipulation is required for the air-bearing rotation stage, which relies on compressed air. We found that substantial amounts of water could still be generated by the air compressor in pipes even when the relative humidity is lower than 20%. Therefore, a refrigerated air dryer is strongly recommended to produce dry and cooled air. Sometimes, an additional three-stage filtration unit is also needed to remove water, oil, and particulate contaminants in the air to prevent damage of the air-bearing stage.

## 2.4 Vacuum system

To reduce the absorption loss of the 5.4 keV X-ray photons in the air, a vacuum chamber is needed to contain all the optical components such as FZP, gratings and so on. Essentially, low density helium gas was filled into the vacuum chamber to reduce the X-ray absorption in our experiments.

To maintain a long period of airtightness holding time, special sealing structures/units are required. For example, the entrance and exit windows of the vacuum chamber are made from 100 $\mu$m thick

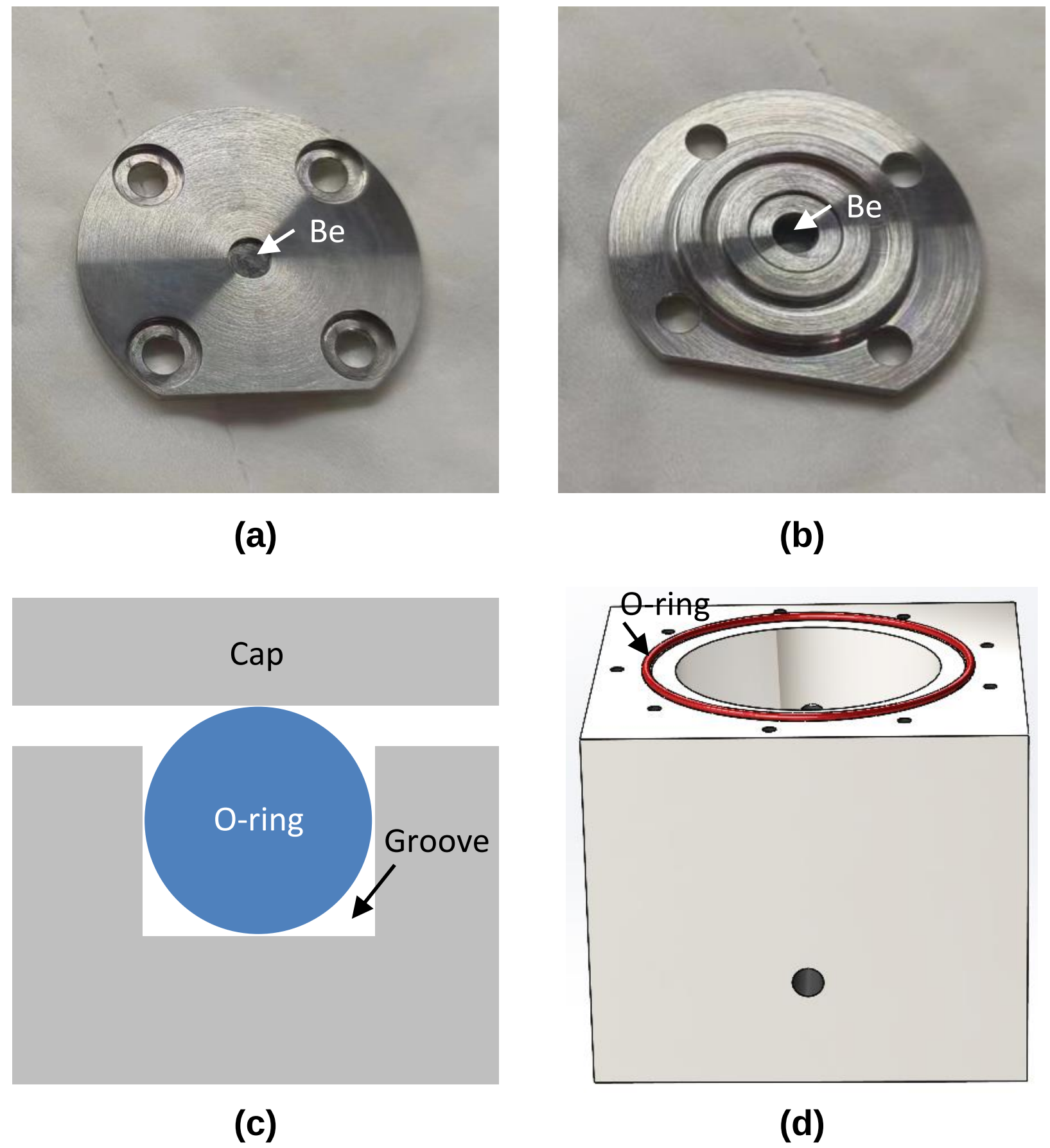

Figure 2.7: (a)–(b) The front side and back side of the beryllium window for the vacuum chamber, respectively. (c) The sealing architecture for O-ring, which sits inside the groove. (d) An example of vacuum chamber with the O-ring design.

Beryllium, see Figs. 2.7(a)-(b). In addition, the face seal joint structure with rubber O-ring is used to seal the cap of the aluminum vacuum chamber having 2.0 cm wall thickness. The depth and width of the groove have to be precisely designed, see Fig. 2.7(c). Moreover, a 50 cm long stainless steel tube with 12 cm diameter is used to extend the vacuum chamber space to the detector. The wing-nut type stainless steel KF clamp is utilized to seal and mount one end of the tube to the vacuum chamber. To allow signal communication between the desktop and the linear stages inside the vacuum chamber, a certain board containing multiple double-side male-male D-Sub (D-subminiature) connectors, whose pins are sealed with glue, is utilized. Herein, the signal communication board is connected to the vacuum chamber via an O-ring.

Based on our experimental observations, the vacuum chamber must maintain adequate airtightness for at least one month. Any degradation in the chamber sealing can result in rapid helium loss. As a result, the imaging performance, e.g., the exposure period and the image signal-to-noise-ratio (SNR), can be strongly affected. It is suggested to fill the chamber with helium gas to avoid any unexpected disasters due to the high pressure difference between the inside and outside of the chamber. To do so, a vacuum–helium exchange strategy needs to be adopted. Specifically, the chamber undergoes multiple cycles, e.g., three cycles, of evacuation followed by helium gas filling. The use of helium gas can also significantly relax the high requirements of expensive hardware, for instance, the vacuum compatible motorized stages that are equipped with vacuum compatible stepper motor to move the FZP, gratings and other optical components. In addition, another small vacuum chamber was introduced for the condenser assembly, see Fig. 2.7(d). A 15 days follow-up measurements indicate that the vacuum level inside the chamber can be well maintained. In fact, no significant intensity degradation was observed even over three months.

Lastly, we strongly recommend retaining the beryllium window and NOT replacing it with a Kapton polyimide film. For one aspect, this is because the Kapton film having the same thickness as of beryllium absorbs more X-ray photons, see Appendix A. Moreover, unexpected film rupture may happen during the vacuuming procedure. Such rupture is a deadly nightmare for the X-ray microscope, and should be absolutely avoided.

## 2.5 Emergency power supply

An uninterruptible power supply (UPS), which is a dedicated system that provides continuous and conditioned electrical power, is necessary to a laboratory X-ray microscope to protect the critical hardware from being damaged or displaced during unexpected power interruptions or outage. Essentially, the UPS has two primary functions. First, it ensures stable and clean power delivery to precision electronics and control systems, preventing transient voltage or current variations that could induce micron displacements in stages. Second, it protects sensitive hardware from sudden power outage, which may lead to mechanical shifts, misalignment of optical elements, or even permanent damages.

In an X-ray microscope, the protection of UPS is particularly crucial for both the X-ray source and the FZP. Our laboratory X-ray source was operated at 40 kV high voltage and was coupled with an ion pump used to maintain a vacuum level of about $10^{-10}$ Torr. A sudden power interruption could cause electrical breakdown and permanent damage to the X-ray tube. In addition, an unexpected loss of power may induce 10–20 $\mu$m or more displacement to the FZP along the optical axis, completely breaking the optimal imaging condition. Whenever this happens, a very time-consuming re-positioning of the FZP and the entire imaging system have to be performed. This whole procedure may take couple days, severely slowing the working efficiency and experiment throughput. As shown in Fig. 2.8, the accidental power loss caused obvious shift of the FZP focused beam spot, indicating significant axial displacement of the FZP along the optical path. Meanwhile, the corresponding imaging performance also gets degraded.

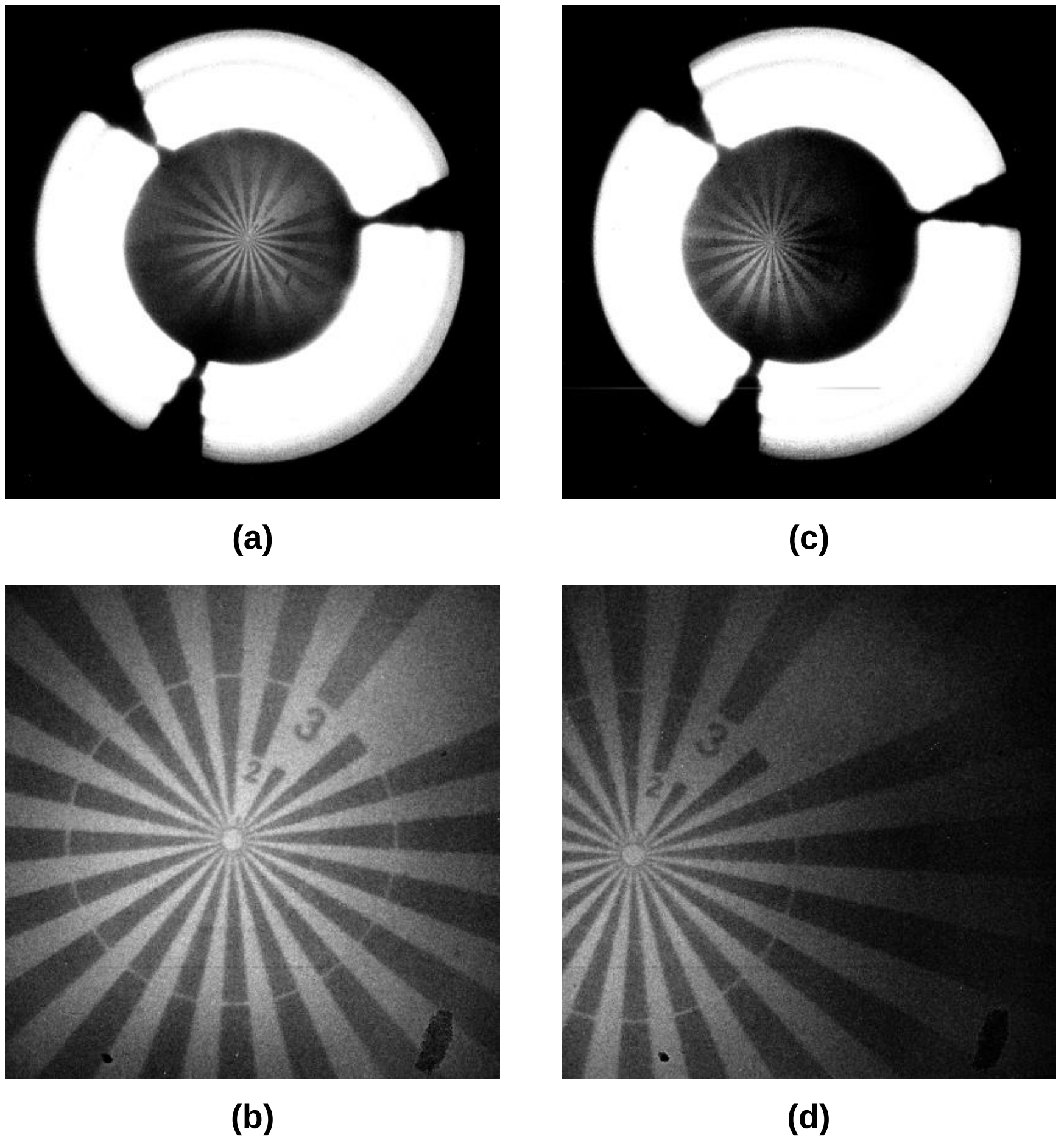

Figure 2.8: (a) The FZP focused beam spot before power interruption. (b) The imaging result of Siemens star pattern before power interruption. (c) The FZP focused beam spot after power interruption. (d) The imaging result of Siemens star pattern after power interruption.

If possible, the UPS device could also be integrated to the sample stage and air-bearing stage for comprehensive protection.

## 2.6 Transportation

For short-distance transportation, attentions are particularly needed to the major heavy components. For instance, the transferring of the anti-vibration marble table with weight of over 1000 kilograms and the steel frame base with weight of about 1000 kilograms from the loading deck to the laboratory. If the laboratory room is located in the basement, then one really has to double-check the elevator's key parameters such as the internal dimension and the maximum allowed weight capacity to make sure that the above two major heavy components can be safely transferred. If the laboratory room is located in the first/ground floor, then a forklift or a pallet jack are needed to safely transfer these heavy components.

For long-distance transportation of all the components from the manufacturers to the laboratory, special anti-vibration strategies are needed. Usually, vendors are assumed to be very experienced and responsible to provide reliable packing solutions to keep the components safe. If not, the buyer should make a detailed plan with the vendor to prevent any possible damages from any unexpected accidents.

# 3 Overall design

## 3.1 Optical microscope

Modern science believes that the universe can be divided into ever smaller units. Over the past two centuries, many probes with ultra-high spatial resolution have been invented to extend our vision boundary. Undoubtedly, microscopes are the most successful ones. The optical microscope is considered the oldest type of microscope, with the first two-lens designs appearing in Europe[58] around 1590–1620. Actually, the thin lens magnification mechanism[59] provides one of the most fundamental working principles of microscope, see the following formula,

$$\frac{1}{f} = \frac{1}{L_{obj}} + \frac{1}{L_{img}}, \tag{3.1.1}$$

where $f$ denotes the focal length of the thin lens, $L_{obj}$ denotes the distance from the imaging object to the thin lens, and $L_{img}$ denotes the distance from the thin lens to the image. The image is magnified by

$$\mathrm{M} = \frac{L_{img}}{L_{obj}} \tag{3.1.2}$$

times. Modern optical microscopes can achieve magnifications of up to several hundred-fold.

In practice, a microscope contains several primary components: the light source assembly (light source + condenser) that provides uniform illumination; the objective lens that produces the initial, highly magnified image; the focus knobs that adjust the distance between the sample and objective lens for clear imaging; the eyepiece lens that magnifies the image for the observer, see Fig. 3.1(a).

## 3.2 X-ray microscope with FZP

Similarly as in an optical microscope, the X-ray microscope with FZP discussed in this book was designed following the exact working principle as shown in Eq. (3.1.1). Moreover, the primary components of the X-ray microscope are also similar to the optical microscope, see Fig. 3.1(b). Specifically, the X-ray source assembly (X-ray tube + capillary condenser) is used to provide uniform sample illumination; the FZP plays the role of objective lens, which is responsible for producing the magnified image; the motorized stages that position the sample and the FZP function similarly to focus knobs, allowing precise alignment to form clear images; the optical lens mounted on the CCD detector can be treated as the eyepiece lens, enabling secondary image magnification.

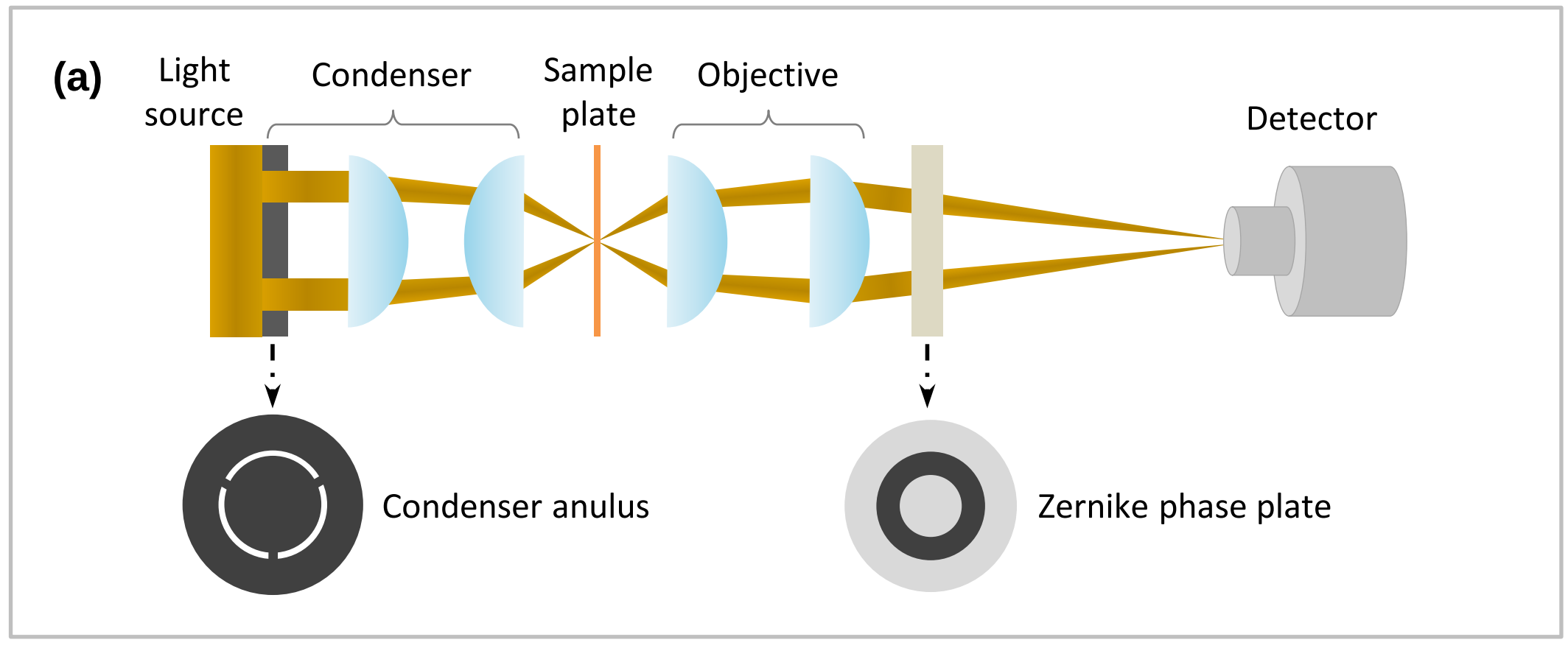

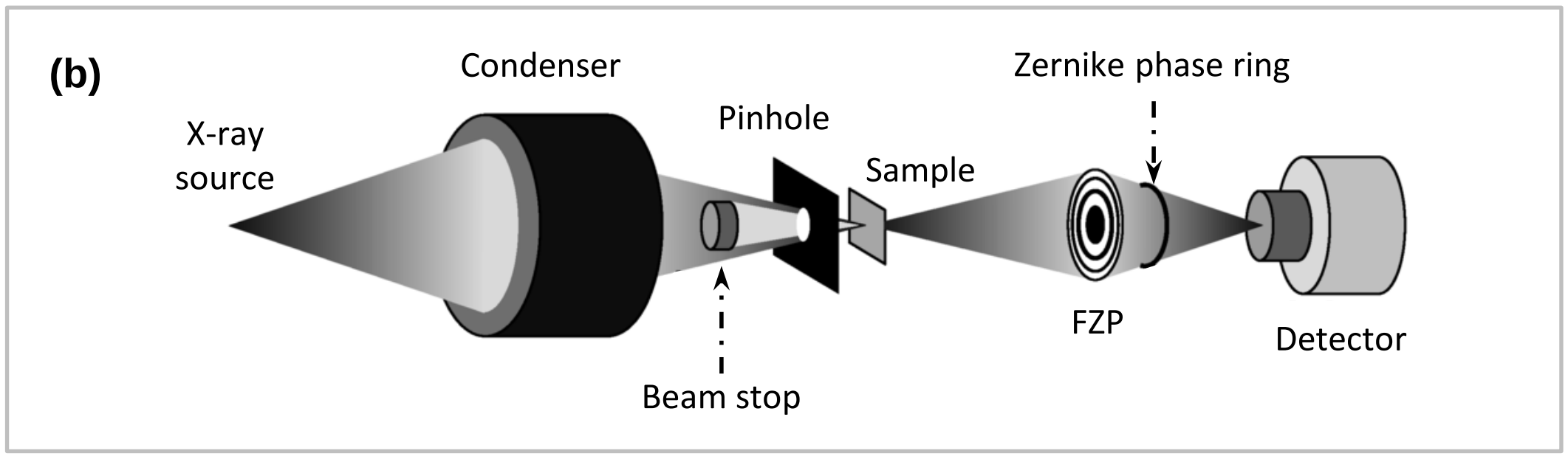

Figure 3.1: Illustrations of typical light path for (a) optical microscope, and (b) X-ray microscope with FZP.

### 3.2.1 Geometric design

To build an X-ray microscope, the imaging geometry has to be determined and designed in the first place. Because the imaging geometry is related to all the optical components, therefore, it must be designed with careful and comprehensive considerations. The key steps are schemed in Fig. 3.2.

Before starting the design, one needs to determine a targeting in-plane spatial resolution, denoted as $\Delta_s$ in this book. For instance, $\Delta_s$=50 nm or $\Delta_s$=30 nm. Usually, such spatial resolution depends on the special research needs. Moreover, the X-ray beam energy $E$ also needs to be determined in prior. For our X-ray microscope, $E$=5.4 keV.

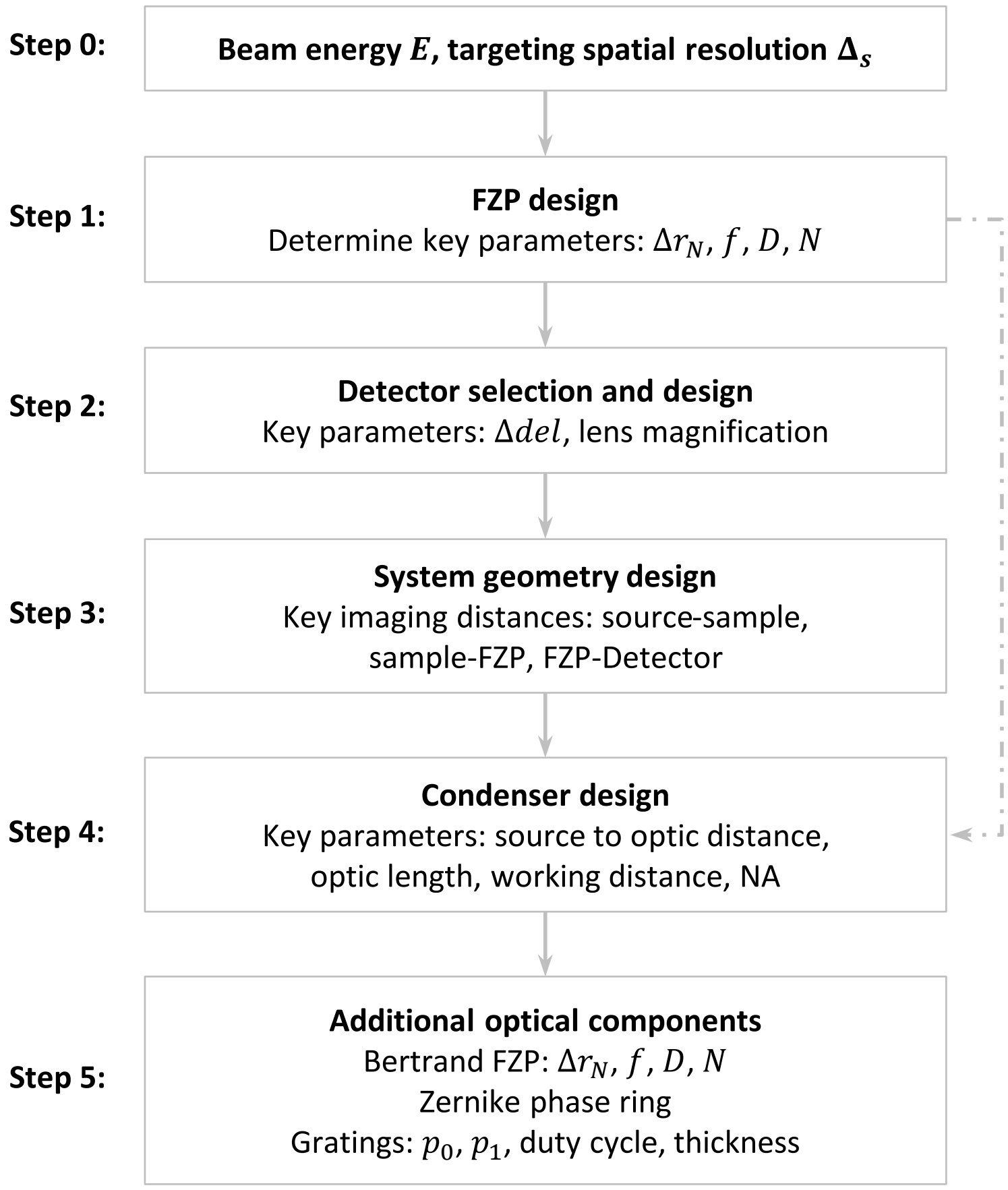

Figure 3.2: The main workflow used to design the FZP based X-ray microscope. In this particular case, it is assumed that the beam energy and targeting spatial resolution are determined in prior based on a specific user defined imaging task.

Under the diffraction limit, the relationship between the width of the outermost zone of the FZP and the spatial resolution $\Delta_s$ is expressed [7, 60] as follows,

$$\Delta_s = 1.22 \times \Delta r_N, \tag{3.2.1}$$

where $\Delta r_N$ denotes the width of the outermost zone of the FZP. As a result, $\Delta r_N$ can be determined immediately. Furthermore, the focal length $f$ of the FZP is found to be

$$f = \frac{D \times \Delta r_N}{n \times \lambda}, \tag{3.2.2}$$

where $\lambda$ denotes the X-ray wavelength, $n$ denotes the diffraction order, and $D$ denotes the diameter

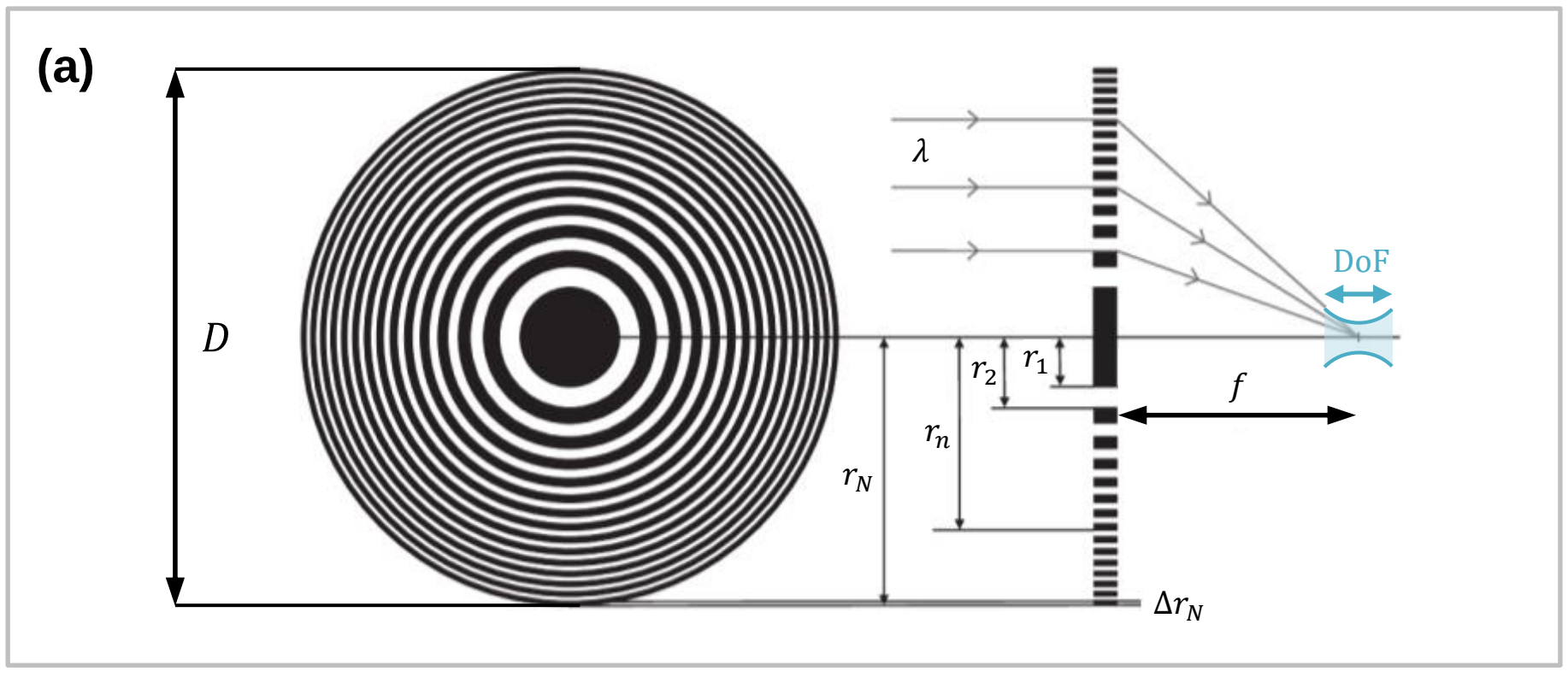

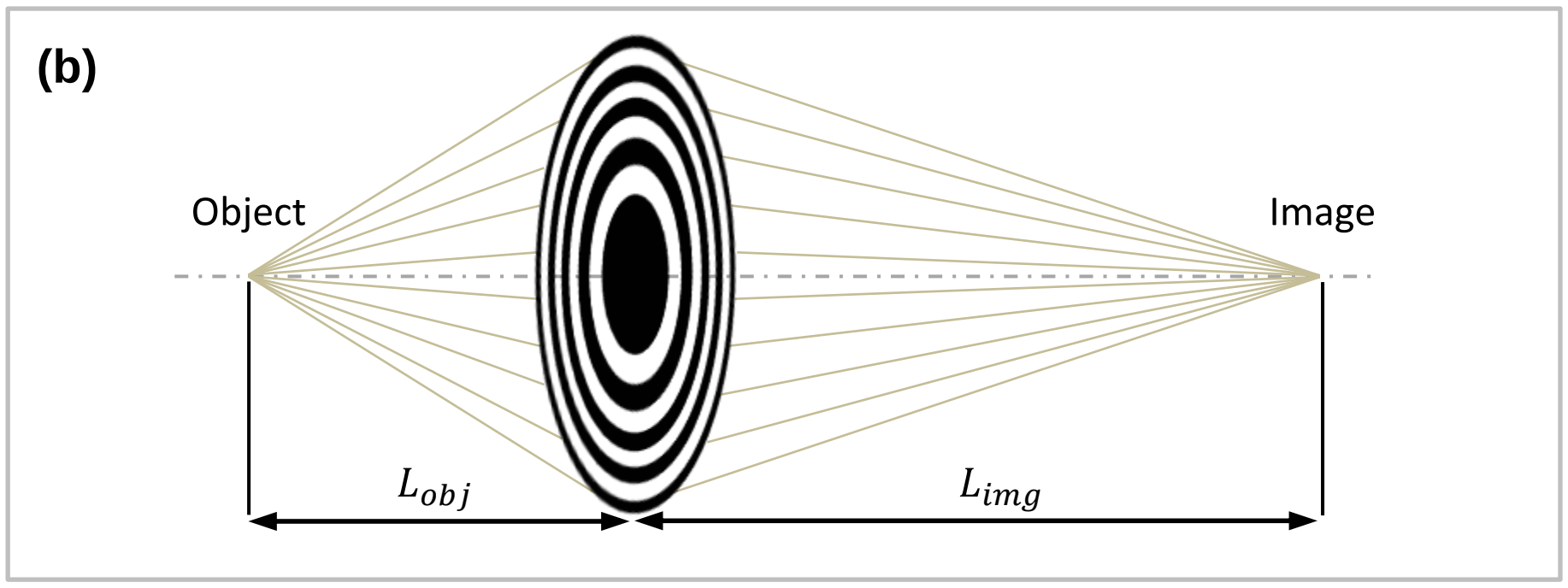

Figure 3.3: (a) Structure illustration of FZP, with diameter of $D$, focal length of $f$, and outermost zone width of $\Delta r_N$. The depth of field is highlighted in blue color. (b) The magnification imaging procedure of FZP. Herein, the object to FZP distance is denoted as $L_{obj}$, and the image to FZP distance is denoted as $L_{img}$.

of FZP, which is approximately expressed as follows:

$$D = 2r_N \approx 2\sqrt{N\lambda f}, \tag{3.2.3}$$

in which $N$ denotes the total number of zones, and $r_N$ denotes the radius of the boundary of the outermost zone. By default, the first diffraction order, namely, $n = 1$, is assumed in FZP based X-ray microscope. Based on Eq. (3.2.2), the numerical aperture of the FZP is derived as follows,

$$\text{NA} = \frac{D}{2 \times f} = \frac{\lambda}{2 \times \Delta r_N}. \tag{3.2.4}$$

Clearly, the numerical aperture depends on the focal length $f$ and the diameter $D$ at the same time. In particular, lower X-ray beam energy leads to larger numerical aperture, and higher spatial resolution corresponds to larger numerical aperture. The numerical aperture is so important that it not only impacts the design of the zone plate, but also adds constrains to the capillary condenser.

With the current lithography technology [61], usually, it is easy to fabricate zone plate with large diameter, especially when the width of the outermost zone is quite small.

Besides considering the in-plane spatial resolution of a FZP, defined in Eq. (3.2.1), which is perpendicular to the optical axis, the axial (or longitudinal) resolving power, which is measured parallel to the optical axis, also needs to be considered. Most often, such axial resolving power is referred to as depth of field (DoF), see Fig. 3.3(b). Mathematically, DoF is defined as:

$$\mathrm{DoF} = \frac{2\lambda}{\mathrm{NA}^2} = \frac{8 \times \Delta r_N^2}{\lambda}. \tag{3.2.5}$$

As seen, higher in-plane spatial resolution corresponds to shorter DoF. Usually, the focused X-ray beam within the DoF can be treated as parallel beam.

For laboratory microscopes working with 5.4 keV or 8.0 keV, it is suggested to keep the focal length somewhere between 8.0–13.0 mm. Too short focal length would add challenges to the sample placement, and too long focal length may cause the total length of the system exceeds 2.0 m, which is not favorable and acceptable for laboratory applications. Once the focal length is selected, the diameter of the zone plate could be determined immediately. Using Eq. (3.2.3), the total number of zones can also be readily estimated. In fact, both the focal length and diameter of the zone plate need to meet the lithography manufacture limitations. By far, the design of the most important optical component, FZP, in the X-ray microscope has completed.

In the next step, the design of the detector system is conducted. For ultra-high spatial resolution tasks, e.g., $\Delta_s$=50 nm or $\Delta_s$=30 nm, an optical lens coupled CCD/CMOS detector is often needed, see Fig. 3.4. In it, the X-ray sensitive scintillator is used to convert the zone plate magnified X-ray image into an optical image, which would be magnified by the coupled optical lens for another 10× or 20× times before eventually recorded by the two-dimensional (2D) pixel array of the CCD/CMOS detector sensor.

Nowadays, there are two major available candidates on the detector market. For the CCD type sensor, the typical pixel size is 13.5 $\mu$m×13.5 $\mu$m. For the CMOS type sensor, the typical pixel size is 6.5 $\mu$m×6.5 $\mu$m. As usual, the magnification ratio of the coupled optical lens is set to 20× and 10× for the CCD detector and CMOS detector, correspondingly. Compared to the CMOS detector, the CCD detector has lower dark current, particularly over a very long exposure period such as 10–30 minutes. Obviously, the scintillator based detector system is also an optical microscope system. To obtain clear images, usually, motorized linear stages are employed to fine tune the distances $L_{obj}$ and $L_{img}$ defined in Eq. (3.1.1) for the detector's internal optical microscope system, see Fig. 3.4.

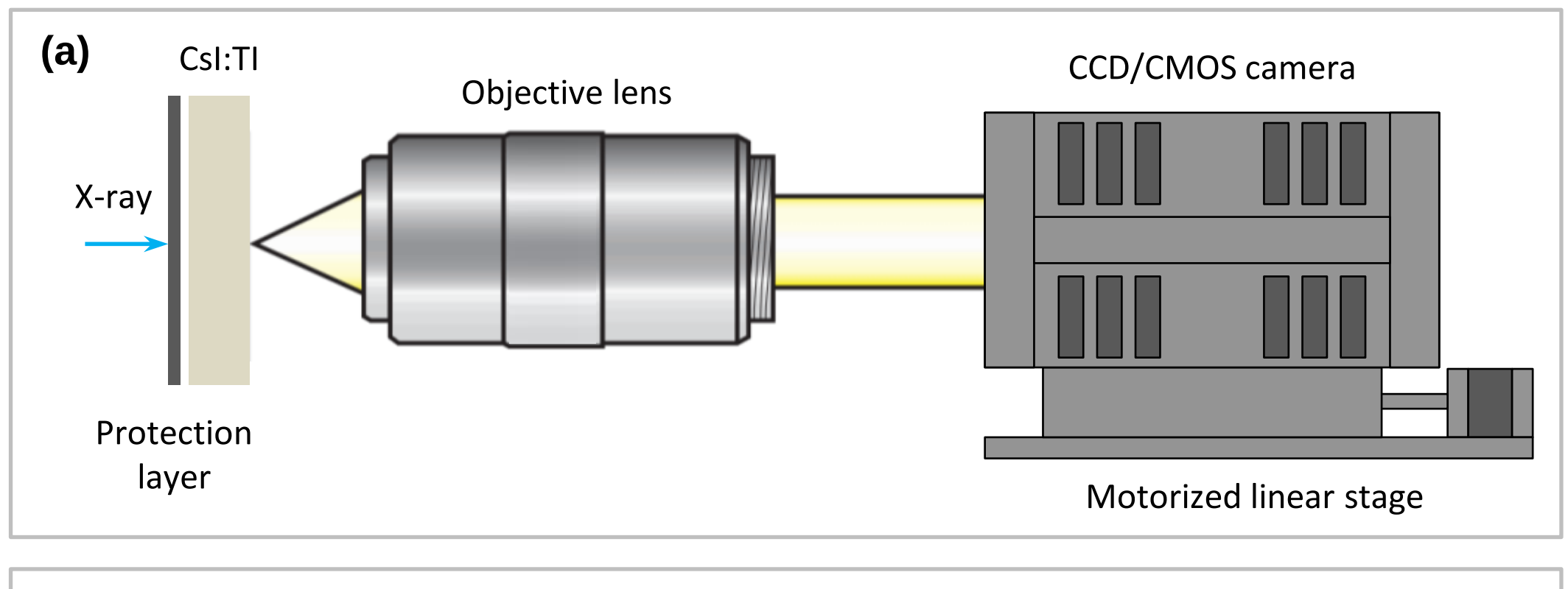

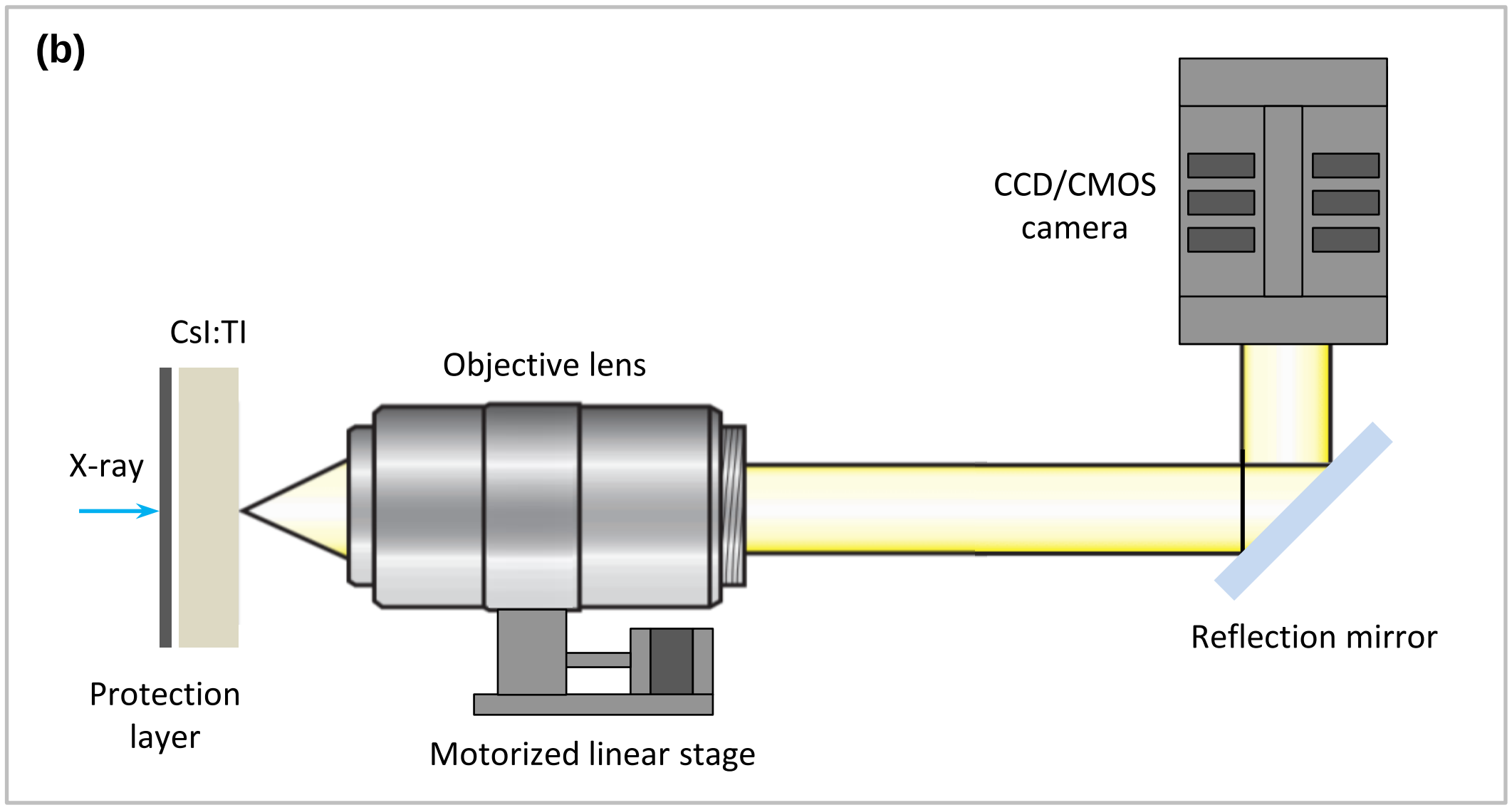

Figure 3.4: (a) Illustration of the (a) I-shaped and (b) L-shaped optical lens coupled CCD/CMOS detector systems. The motorized linear stage is used to fine tune the focal distance to generate clear projection images.

Design of the capillary condenser system are not discussed in this chapter. For more details, please see the discussions in the next chapter. Be aware that the X-ray beam numerical apertures formed by the condenser should match the ones that are allowed by the FZP in order to achieve the highest spatial resolution, see the illustration in Fig. 3.5.

A Bertrand lens, another FZP, is often added behind the main imaging FZP. Instead of generating ultra-high resolution X-ray images of the sample, the Bertrand FZP is used to assist the positioning of the optical components such as the Zernike phase ring or the grating. As a result, design of the Bertrand FZP is much less stringent than the main FZP. For example, the focal depth of Bertrand FZP is about 50 mm to provide enough free space to place the Zernike phase ring or the grating,

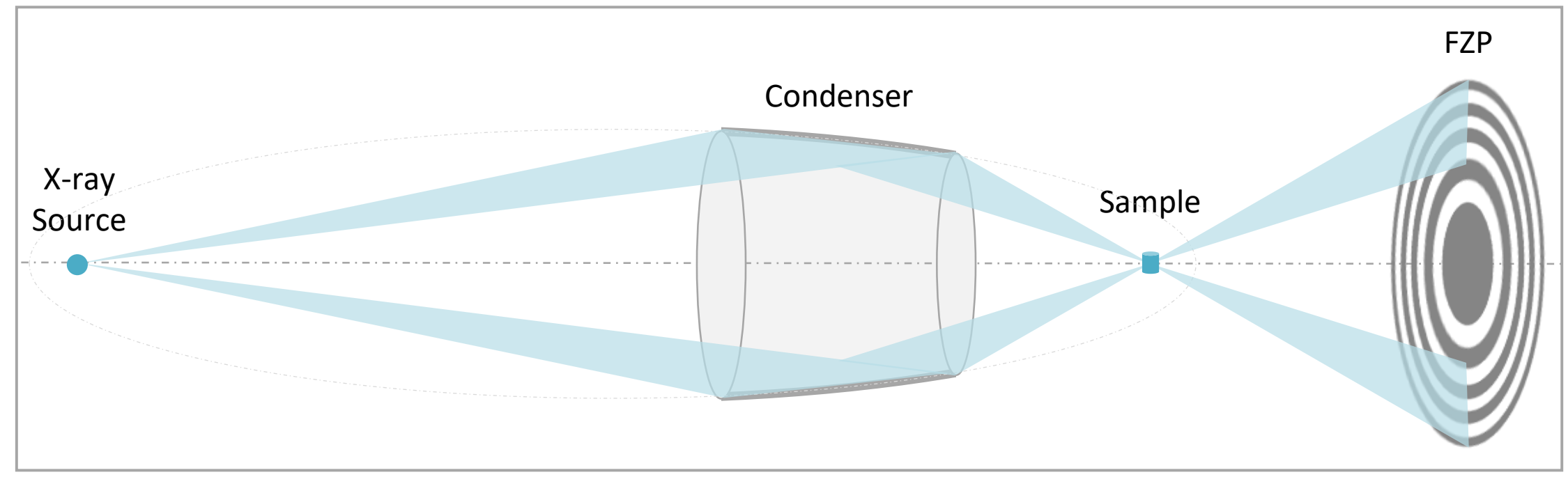

Figure 3.5: The beam path from the X-ray source to the FZP. The X-ray photons reflected by the condenser are focused on the sample. Afterwards, the X-ray beam is diverged and impinge on the FZP.

which is positioned around the rear focal plane of the main FZP.

### 3.2.2 Overall geometry

The entire geometry settings of our laboratory X-ray microscope with FZP is shown in Fig. 3.6. In brief, the total distance from the X-ray focal spot to the X-ray detector is 1203.69 mm, the distance between the sample and FZP is 13.19 mm, the source grating is positioned 38.81 mm in front of the sample, the phase grating is positioned 17.33 mm behind the FZP, and the Bertrand is positioned 36.41 mm downstream of the phase grating. In addition, the Vanadium (V) filter is taped on the outlet window of the condenser.

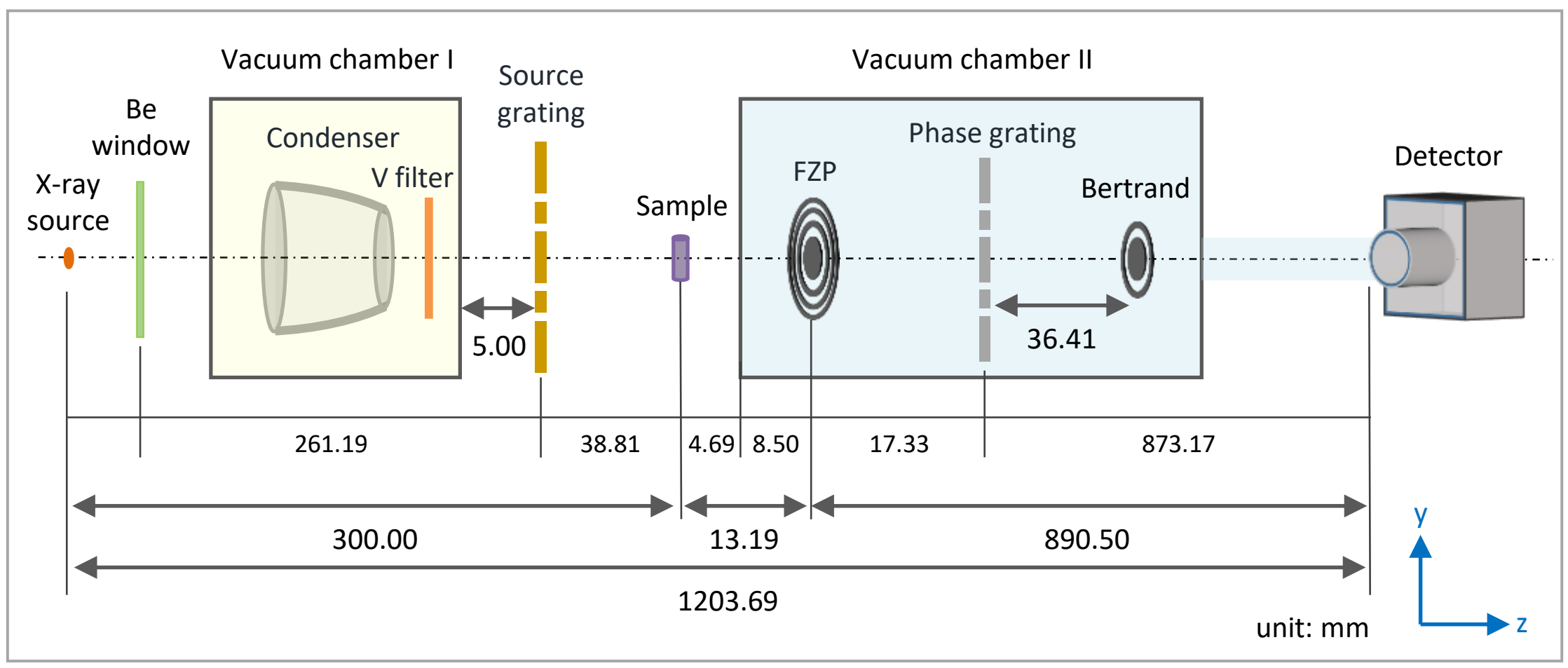

Figure 3.6: The entire geometry settings of our laboratory X-ray microscope with FZP.

# 4 Illumination system

When building an X-ray microscope, the illumination system (X-ray tube + capillary condenser) needs to be installed and adjusted in the first place. This is because nothing can be manipulated and proceeded without illumination.

## 4.1 X-ray tube source

As shown in Fig. 4.1, the X-ray tube, which illuminates the laboratory X-ray microscope with FZP, produces X-rays by heating a cathode filament (usually tungsten) to release electrons, which are then accelerated by high voltage (20–50 kV) across a glass or metal-ceramic high-vacuum housing to bombard the anode target metal. Eventually, only about 1% of this electron energy converts into bremsstrahlung and characteristic X-rays, while 99% becomes heat. Hence, water/oil cooling is necessary for the anode target.

### 4.1.1 Anode material

For laboratory X-ray microscope imaging, the most frequently utilized anode targeting materials are chromium (Cr) and copper (Cu), whose characteristic X-ray beam energies are 5.4 keV and 8.0 keV. In addition, anode targeting materials such as rhodium (Rh) and iron (Fe) are also available to generate 2.7 keV and 6.4 keV X-ray photons. For conventional X-ray absorption imaging, in general, the image contrast varies significantly with energy, impacting both data acquisition time and, more importantly, the visibility of interested object structures. Empirically, X-ray beam of 2.7 keV could significantly improve the image absorption contrast of polymers and cells, 5.4 keV X-ray beam is ideal for geological samples, and 8.0 keV X-ray photons are more sufficient for samples containing metal. In order to cover a broad range of samples, for instance, from polymers to metals, the most advanced X-ray tubes are able to seal up to multiple different target materials in a single illumination source [62].

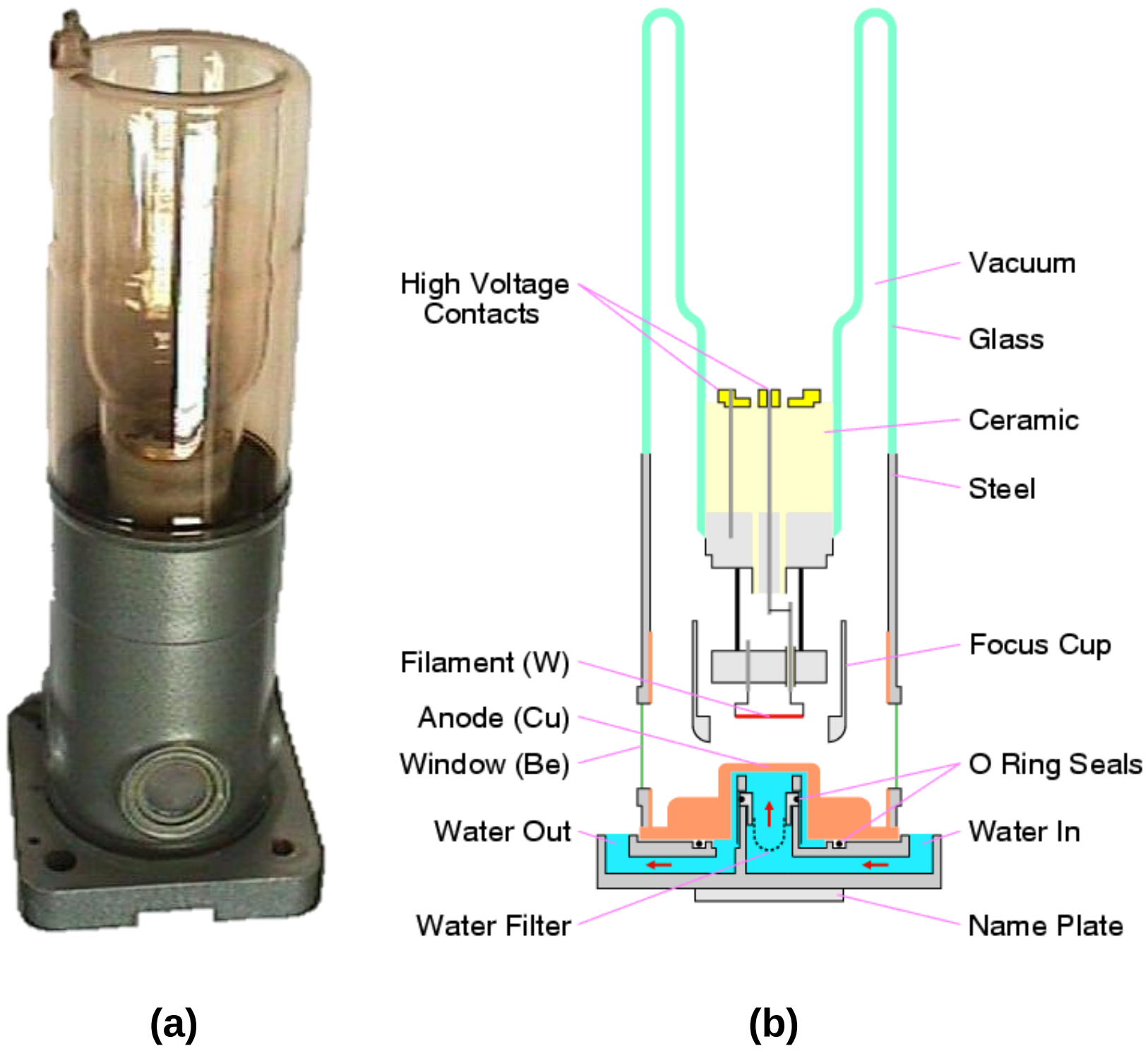

Figure 4.1: (a) Photo of an X-ray tube, (b) the internal structures of an X-ray tube. Image source: http://pd.chem.ucl.ac.uk/pdnn/inst1/xtube.htm.

### 4.1.2 Focal spot

The X-ray source focal spot is a specific area on the anode target struck by the electron beam, acting as the origin point for X-ray beam emission. Usually, the actual area on the anode surface struck by electron beam is denoted as the actual focal spot. Due to the anode angle, the finally obtained effective focal spot, the area projected downwards towards the object being imaged, is usually smaller than the actual focal spot. As shown in Fig. 1.4, a smaller focal spot decreases the penumbra (blurring) at the edges of structures, resulting in higher spatial resolution. However, smaller focal spot is more prone to melting the anode, limiting the maximum tube power.

In FZP based X-ray microscope, the effective focal spot determines the field of view (FOV), which corresponds to the maximum imaging area that the FZP can focus on. As a consequence, the effective focal spot with moderate sizes, for example, 20-70 $\mu$m, are preferred. There are many ways to measure the focal spot size. Herein, the full width half maximum (FWHM) based measurement approach of the focal spot size is employed, see the illustration in Fig. 4.2(a).

First, a high Z material edge plate, e.g., tungsten (W) plate, is positioned between the X-ray source

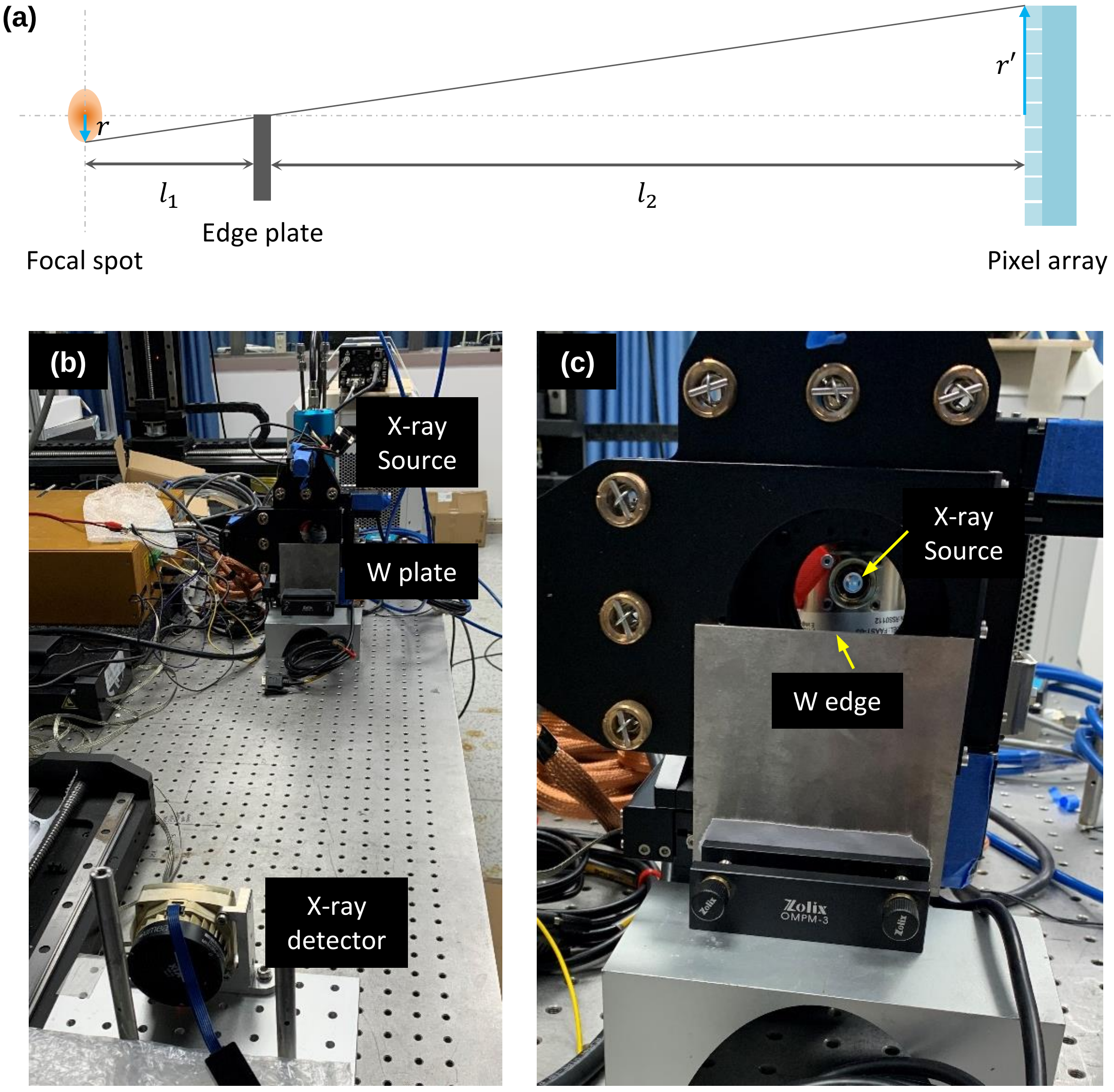

Figure 4.2: (a) Illustration of the full width half maximum (FWHM) based focal spot size measurement approach. By default, the vertical direction is parallel to the $y$-axis, and the horizontal direction is parallel to the $x$-axis. (b) Photograph of the entire experimental setup. (c) Photograph of the tungsten (W) wedge used in the experiment.

and the detector to block the bottom half of the focal spot, see the photographs in Figs. 4.2(b)-(c). Assuming that the distance between the X-ray focal spot and the edge plate is $l_1$, and the distance between the edge plate and the X-ray detector is $l_2$, as a result, the magnification of the imaging system is denoted as $\mathrm{M} = \frac{l_2}{l_1}$. Turn on the X-ray source and take an image of the half obstructed focal spot. Take a horizontal line profile of the image and compute the distance $r'$ it takes for the signal to drop from 80% to 20%. Eventually, the FWHM of the X-ray source along the horizontal

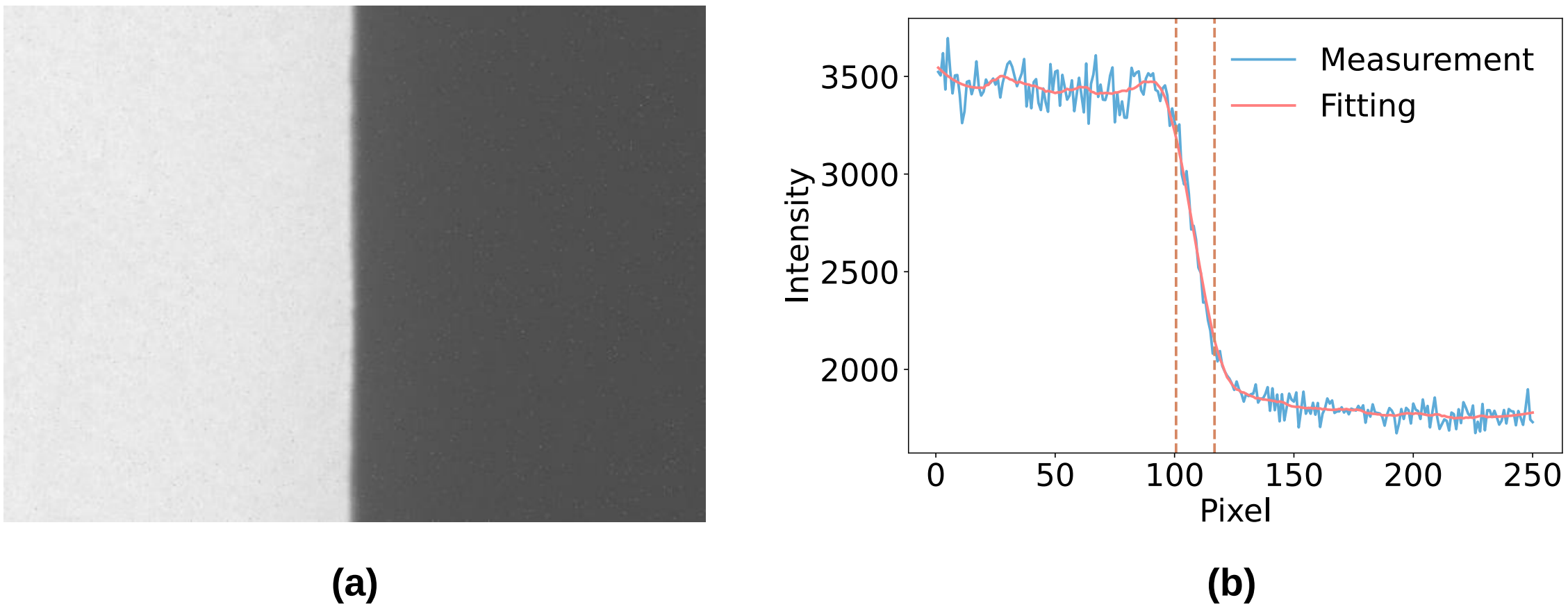

Figure 4.3: Measurement results of X-ray focal spot size. (a) The projection image of the tungsten wedge, (b) the measured edge profile.

direction is found to be

$$\mathrm{FWHM} = \frac{r'}{M} = \frac{r' l_1}{l_2}. \tag{4.1.1}$$

To determine the FWHM of the X-ray source along the vertical direction, the edge plate has to be placed such that the left or right half of the focal spot is blocked. By doing so, the focal spot size can be quantified. The measurement results are shown in Fig. 4.3. More mathematical details of the measurement can be found in Appendix B.

### 4.1.3 Beam brilliance

To quantify the intensity of an X-ray beam, usually, certain quantities are defined. Specifically, flux represents the integrated intensity of an X-ray beam and is defined as the number of X-ray photons emitted per unit time. The unit for flux is photon counts per second (cps). Moreover, flux density is defined as the flux passing through a unit area. The unit is cps/mm$^2$. Flux density is an appropriate parameter for measuring local counting rates. Brightness takes the beam divergence into account, and is defined as the flux per unit of solid angle of the radiation cone. The unit is cps/mrad$^2$. Brightness is an appropriate parameter to use when comparing two X-ray sources with identical focal spot size. In addition, brilliance takes the beam dimensions into account and is defined as brightness per mm$^2$. The unit is cps/mm$^2$/mrad$^2$. Note that the above four quantities are all within a 0.1% bandwidth (0.1% BW) represented by a wavelength range, $\Delta\lambda$, centered around a specific wavelength $\lambda$, i.e. $\Delta\lambda$

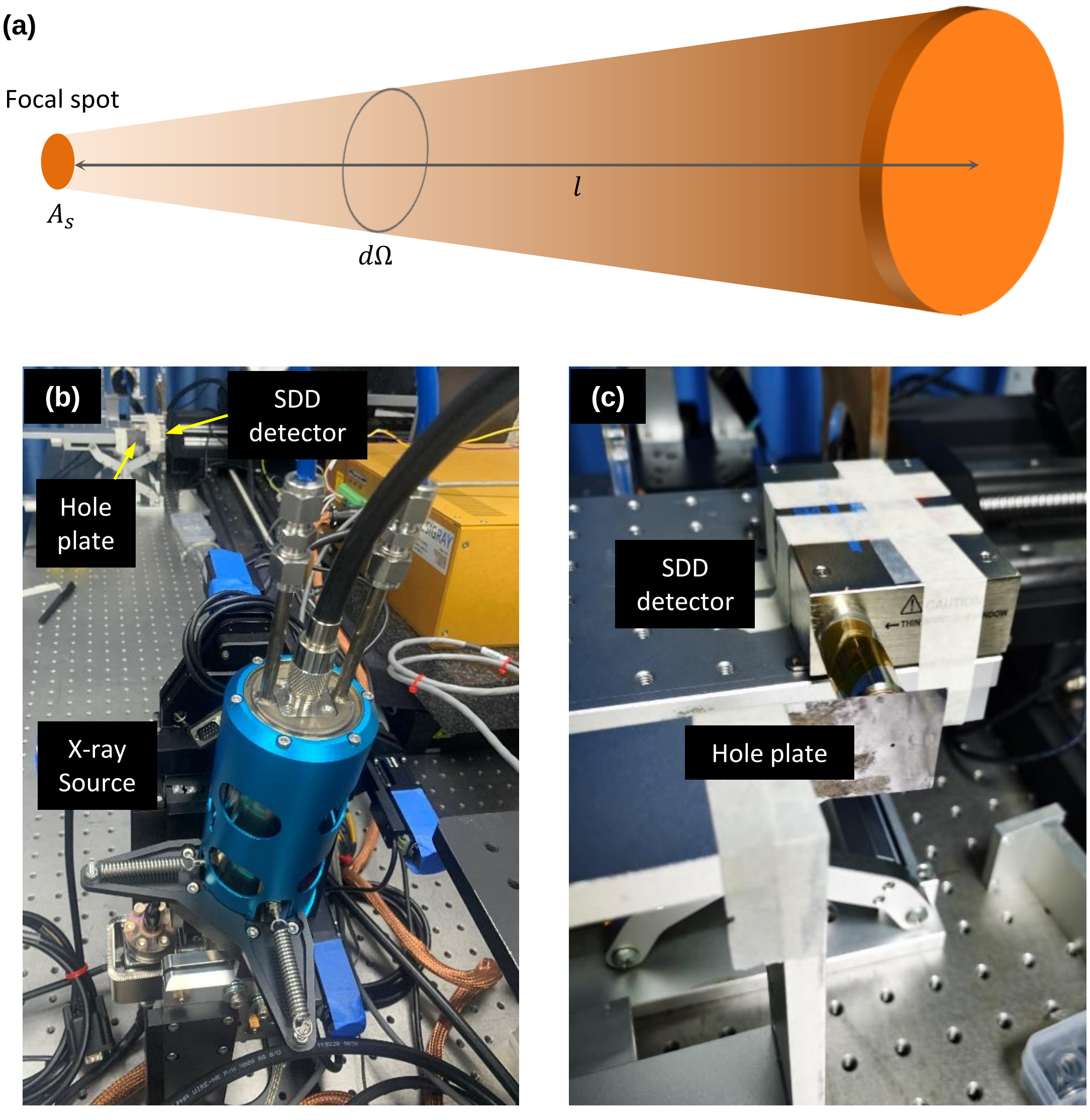

Figure 4.4: (a) Illustration of the source brilliance measurement approach. (b) Photograph of the entire experimental setup. (c) Photograph of the hole plate and the silicon drift detector (SDD) used in the experiment. In particular, the hole plate is attached on the entrance window of SDD.

is equal to 1/1000 of $\lambda$. In general, brightness is defined as follows:

$$\mathrm{B} = \frac{N_m}{\mathrm{A}_s \; d\Omega \; dt \; F_a \; 0.1\%\mathrm{BW}}, \tag{4.1.2}$$

in which $N_m$ denotes the measured 5.4 keV X-ray photons within $dt$ period, $\mathrm{A}_s$ denotes the focal spot size, $d\Omega$ denotes the solid angle of the radiation cone. In addition, $F_a$ denotes the attenuation

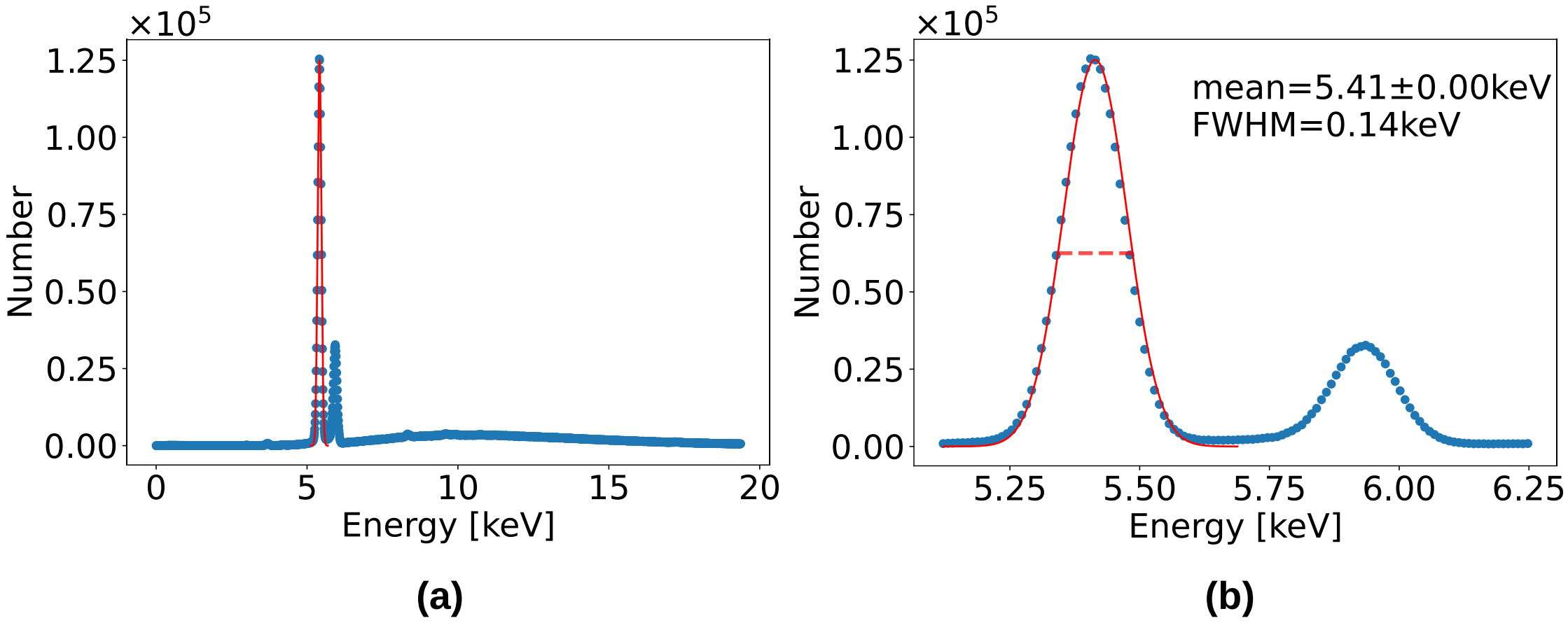

Figure 4.5: The measured X-ray beam spectra. (a) The full energy spectra, (b) The spectra containing the two characteristic peaks.

correction factor, namely,

$$F_a = e^{-\mu_{Be} t_{Be} - \mu_{Air} l_{Air}}. \tag{4.1.3}$$

Herein, $\mu_{Be}$ denotes the attenuation coefficient of the Be windows (including the X-ray source and the silicon drift detector (SDD)), $t_{Be}$ denotes the total thickness. Additionally, $\mu_{Air}$ denotes the attenuation coefficient of air, $l_{Air}$ denotes the total distance from the Be window to the SDD detector.

Essentially, the X-ray source brilliance is an invariant quantity, i.e. the brilliance at the specimen position cannot be improved by any optical techniques, but only by increasing the brilliance of the X-ray source. This is a consequence of Liouville's theorem, which states that phase space is conserved. Obviously, brilliance can be increased by making the focal spot size as small as possible, while making the tube power (proportional to the photon flux) as high as possible. Two X-ray beams may have the same flux density but different brilliance if they have different focal spot sizes. Brilliance is thus an appropriate parameter to use when comparing two X-ray sources with different focal spot sizes[63]. As a consequence, in the FZP based X-ray microscope, it is quite important to use a high brilliance X-ray source.

The general brilliance measurement setup is illustrated in Fig. 4.4. In particular, the silicon drift detector (SDD), which can measure X-ray photon energies with exceptional resolution and incredibly high count rates, is positioned $l$ distance away from the X-ray focal spot. In addition, a tungsten plate with a 0.5 mm hole is attached on the entrance window of SDD. The measured X-ray spectra are plotted in Fig. 4.5, from which the beam brilliance can be calculated using Eq. (4.1.2).

### 4.1.4 K-edge beam filter

In fact, the X-ray beam spectra, see Fig. 4.6, generated from the X-ray tube are polychromatic and composed of several X-ray characteristic peaks. To avoid chromatic aberration (maintaining high spatial resolution), it is important to make a quasi-monochromatic X-ray spectrum for the laboratory X-ray microscope with FZP. As a result, the undesired characteristic peaks in the X-ray spectra need to be suppressed. In practice, special K-edge metal foil filters can help to achieve this aim, see Table 4.1. Interestingly, it is found that the ideal choice of material for an X-ray filter is a metal whose atomic number, Z, is one less than that of the anode target metal. Essentially, the X-ray absorption edge of the particular filter material sits right between the $K_\alpha$ peak and the $K_\beta$ peak. As a result, the K-edge filters can significantly reduce the intensity of the $K_\beta$ peak in the X-ray spectrum compared to $K_\alpha$ peak, see Fig. 4.6. Note that the K-edge filters can also remove much of the high energy bremsstrahlung component, but it cannot remove the unwanted $K_{\alpha 2}$ component from the $K_\alpha$ peak.

The optimum thickness, $t_k$, of the K-edge filter can be determined from the Beer-Lambert law:

$$I(E) = I_0(E) \times e^{-\mu_k(E) t_k}, \tag{4.1.4}$$

where $I_0(E)$ denotes the incident X-ray beam spectra, $I(E)$ denotes the transmitted X-ray beam spectra, and $\mu_k(E)$ denotes the attenuation coefficient at beam energy $E$. It can be seen that in selecting the thickness of the filter, a compromise has to be reached between eliminating as much as possible of the undesired $K_\beta$ X-ray photons and maximizing the desired $K_\alpha$ X-ray photons. For instance, most commercial systems employing a Ni filter with a Cu anode target will choose the thickness of the foil so as to give a reduction ratio in the range 25:1 to 50:1, i.e., Ni foils between 15 and 20 $\mu$m thick. From the Table 4.2, it can be seen that this range of Ni foil thickness will diminish the desired X-ray photons of 8.0 keV by approximately a factor of 2.

Table 4.1: The choice of metal foil filter material depends upon the choice of anode material in the X-ray tube. The rule: choose for the filter an element whose K-edge is just to the high-energy side of the $K_\alpha$ peak of the target material.

| Anode material | Rh | Cr | Fe | Co | Cu | Mo |
|---|---|---|---|---|---|---|
| Filter material | LiCl | V | Mn | Fe | Ni | Zr |

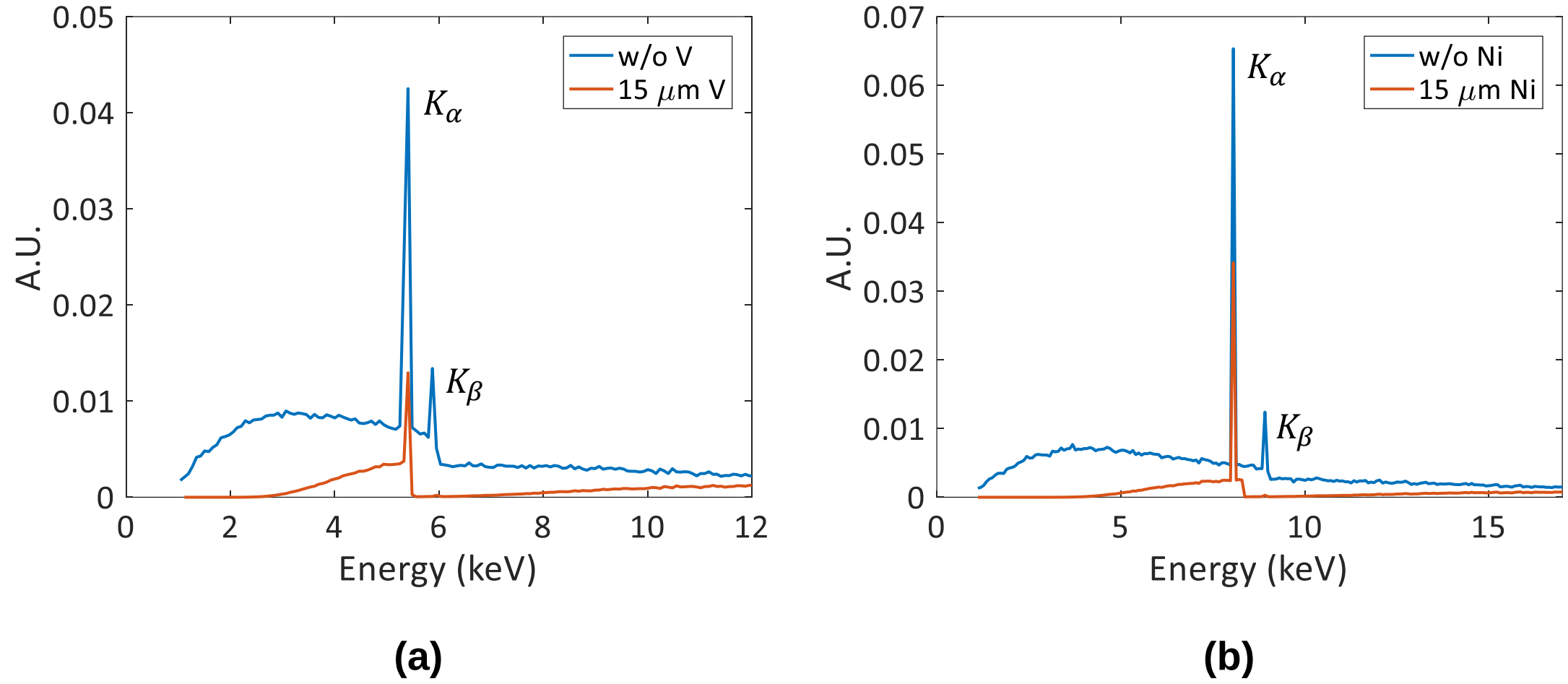

Figure 4.6: The normalized X-ray beam spectra obtained from (a) Cr target, (b) Cu target. The Cr spectrum filtered by 15 $\mu$m thick V and the Cu spectrum filtered by 15 $\mu$m thick Ni are also plotted, correspondingly.

Table 4.2: The percentage transmission for various thicknesses of K-edge foil materials. The top five rows are for the Cr target with V filter, and the bottom four rows are for the Cu target with Ni filter.

| Target | Filter | $t_k$ ($\mu$m) | $I/I_0$ (%) for $K_\alpha$ | $I/I_0$ (%) for $K_\beta$ | Reduction ratio |
|---|---|---|---|---|---|
| Cr | V | 10 | 64.7 | 6.2 | 10 |
| Cr | V | 15 | 52.1 | 1.5 | 33 |
| Cr | V | 20 | 42.0 | 0.4 | 108 |
| Cr | V | 25 | 33.7 | 0.1 | 350 |
| Cu | Ni | 10 | 64.5 | 7.8 | 8 |
| Cu | Ni | 15 | 51.8 | 2.2 | 24 |
| Cu | Ni | 20 | 41.6 | 0.6 | 68 |
| Cu | Ni | 25 | 33.4 | 0.2 | 197 |

## 4.2 Capillary

### 4.2.1 Total reflection

Capillaries are important in FZP based X-ray microscope. Usually, they are hollow glass tubes (single or in bundles) that focus, guide, or collimate X-ray beams using total external reflection. By guiding X-ray beams through curved or tapered channels, these devices concentrate X-rays into high-intensity, micron-sized illumination spots. In fact, X-ray capillaries can be viewed as special optical fibers that operating in the very short-wavelength regimes. Both are boundary-guided wave transport systems governed by the critical-angle physics. Whereas, the key differences arise from wavelength, refractive index, the number of reflections, and whether the core is solid (fiber) or hollow (capillary), see Fig. 4.7.

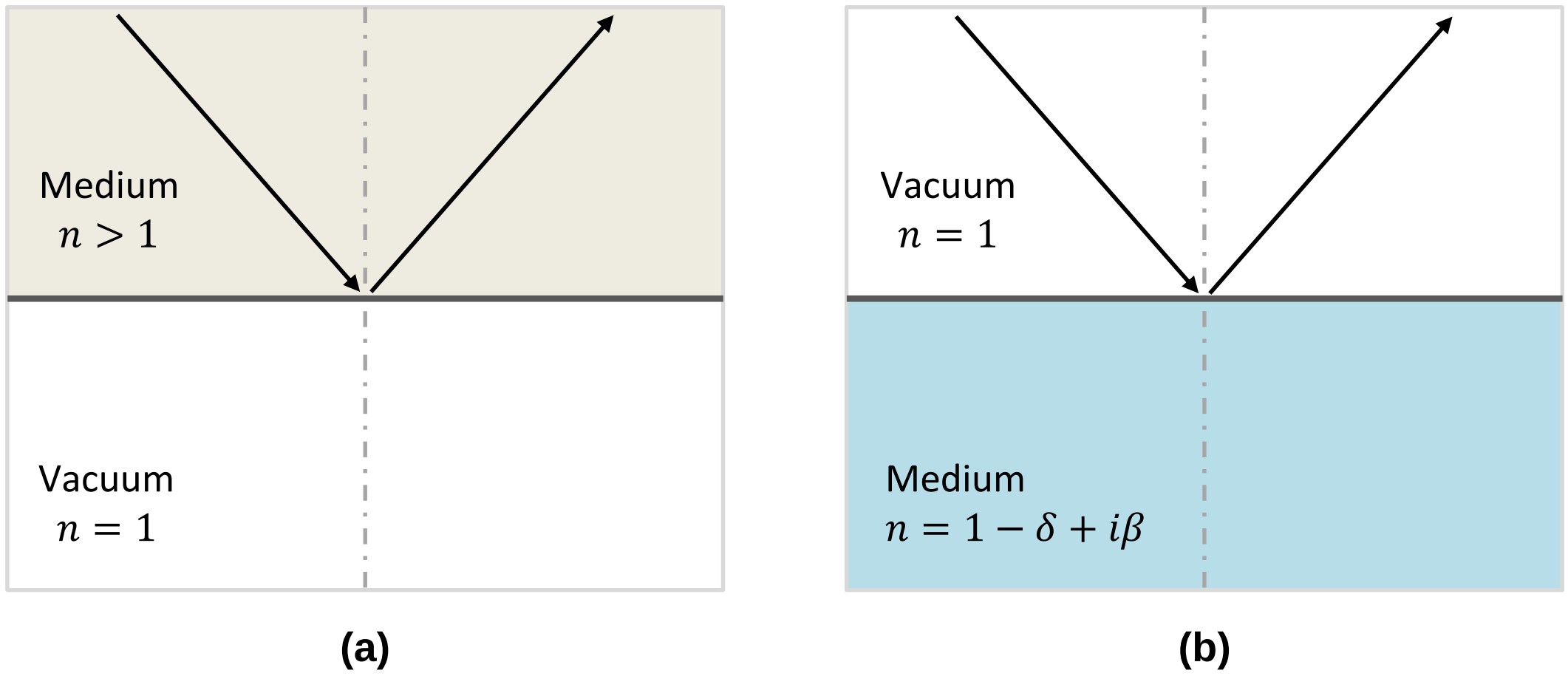

Figure 4.7: (a) Illustration of total internal reflection of light beam in medium. (b) Illustration of total external reflection of X-ray beam in vacuum. Note that the light beam is reflected and propagated inside the medium. Whereas, the X-ray beam is reflected and propagated in vacuum.

### 4.2.2 Types

In practice, the primary types include polycapillary optics (bundles of thousands of channels), single-capillary optics (tapered or curved), see Fig. 4.8. Specifically, the capillaries in the bundle are bend in a way that one side of each capillary is pointing to the X-ray focal spot and the other side is pointing to the focus. Polycapillary optics do not image the source point to a focus point in the sense of optical imaging: light entering the capillaries with a certain divergence angle is guided to the end of the capillary, where it leaves the capillary with an unchanged divergence angle. Consequently, the light spreads out on its way from the end of a capillary to the focus area, see Fig. 4.8(a). So these optics are illumination optics. The tapered capillary optics consist of a slightly conical tube with very smooth inner surface.The maximum cone angle is limited by the maximum angle of total external reflection for the material and photon energy used. The incoming light is reflected one or more times at the inner surface of the tube, mostly by total external reflection in a glass capillary, but other materials would be fine as well. There is no precise focus point, more like a line of increased intensity along the optical axis. These optics are simple, but have the disadvantage of providing very small working distances and do not meet the imaging needs, see Fig. 4.8(b). X-rays in monocapillary optics are guided in a single glass capillary by total external reflection. The inner surface of the monocapillary owns a rotational elliptic, or rotational parabolic geometry. For the rotational elliptic geometry, the source is often positioned in one of the two focal points of the ellipse, and the X-ray beam is redirected to the second focal point, see Fig. 4.8(c). For the single parabolic geometry, it focuses an incoming beam parallel to the optical axis to its focal point, see Fig. 4.8(d). Whereas, the source positioned in one of the two focal points is refocused to the second focal point in case of the double parabolic geometry, see Fig. 4.8(e).

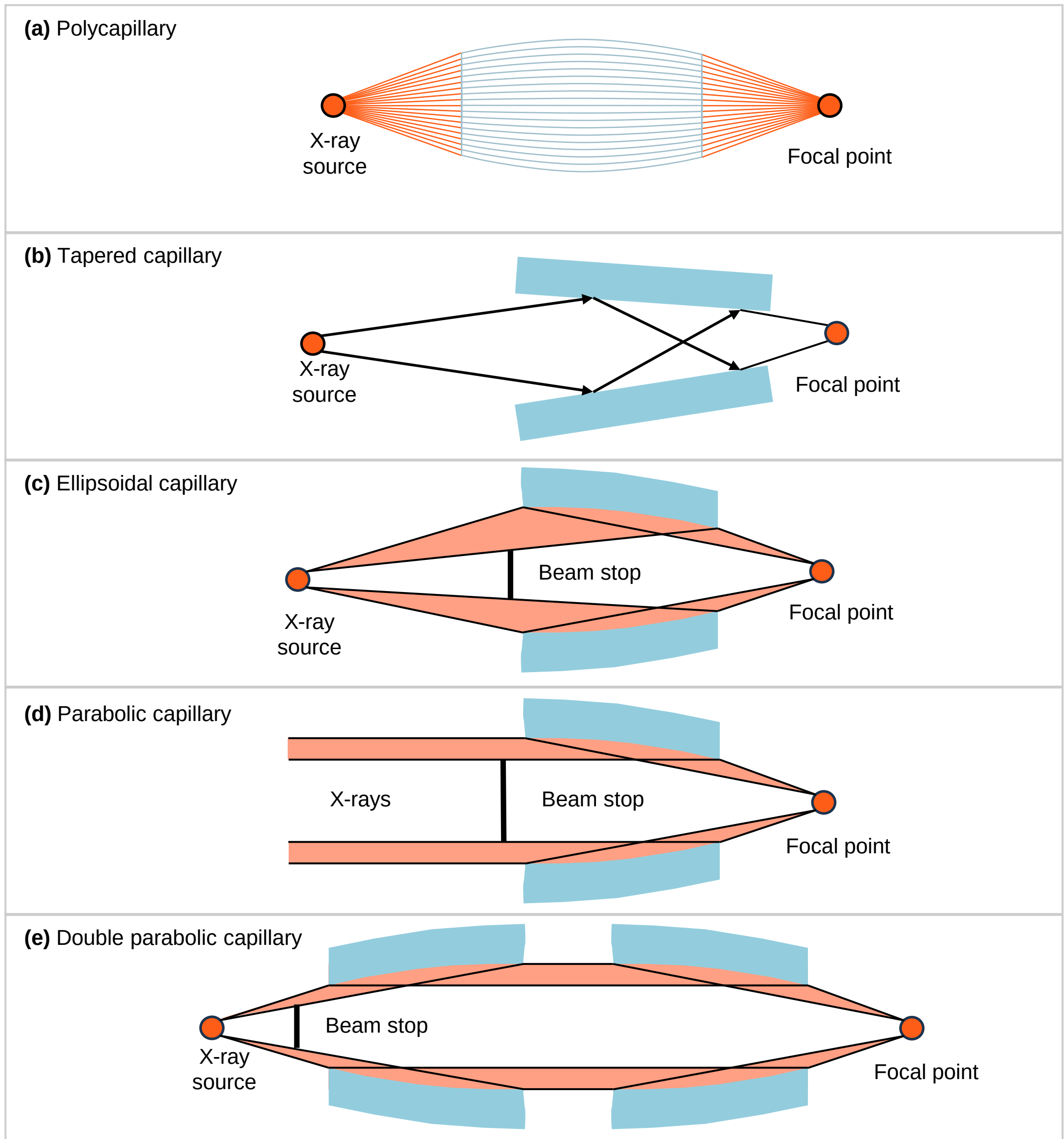

Figure 4.8: Typical capillaries used in X-ray imaging. (a) Polycapillary, (b) Tapered capillary, (c) Ellipsoidal monocapillary, (d) Parabolic capillary, (e) Double parabolic capillary. For laboratory X-ray microscopes with FZP, the ellipsoidal monocapillary is selected most frequently.

For X-ray microscope imaging, monocapillary optics are the most common options. The ellipsoidal monocapillary optics may have imaging limitations when the X-ray source has a finite size. Depending on where the X-ray beam is reflected on the inner surface, the magnification may vary significantly: higher magnification occurs near the beam entrance end and lower magnification (or even demagnification) near the beam exit end. As a result, the ellipsoidal monocapillary condenser produces a non-uniform illumination beam profile at the sample plane such that there is high intensity at the center of the focal spot and rapid fall-off (tailing aberrations) away from the center. The image aberration can also reduce the brightness of the focused X-ray beam.

Recently, the novel double paraboloidal monocapillary condenser, similar to the Wolter optic, is proposed[64] to overcome the above shortages in ellipsoidal monocapillary condenser. Essentially, the double paraboloidal monocapillary comprises two identical face-to-face paraboloidal monocapillaries. The first paraboloidal monocapillary is positioned and aligned to collect and collimate X-rays from the source, and the second paraboloidal monocapillary is used to focus the collimated X-rays to illuminate the sample.

### 4.2.3 Condenser design

In this discussion, only the design of ellipsoidal monocapillary condenser is provided. As illustrated in Fig. 4.9, the ellipsoidal monocapillary condenser strictly satisfies the following geometric constraints [65]:

$$
\begin{aligned}
&u_1 + v_1 = 2a, \\
&u_1 \sin\alpha_1 = v_1 \sin\beta_1, \\
&u_1 \cos\alpha_1 + v_1 \cos\beta_1 = 2c, \\
&u_2 + v_2 = 2a, \\
&u_2 \sin\alpha_2 = v_2 \sin\beta_2, \\
&u_2 \cos\alpha_2 + v_2 \cos\beta_2 = 2c, \\
&u_1 \cos\alpha_1 + v_2 \cos\beta_2 + L = 2c.
\end{aligned}
\tag{4.2.1}
$$

Given any four independent parameters, the other remaining geometric parameters can be uniquely

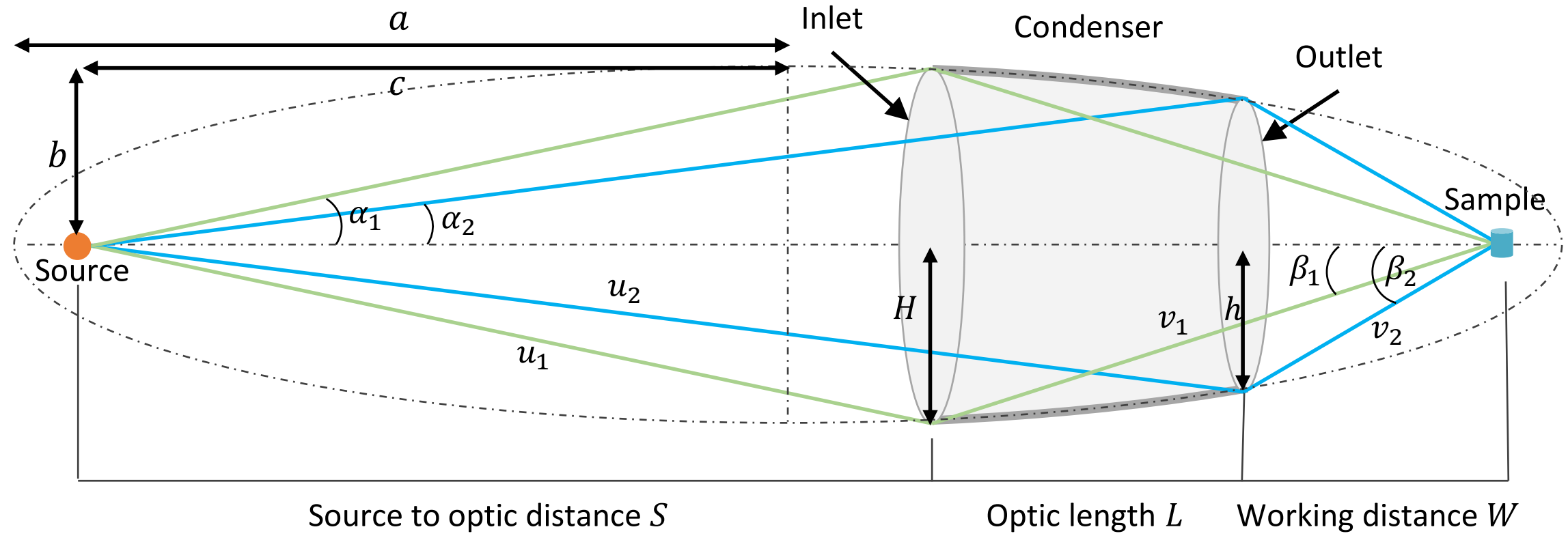

Figure 4.9: Geometric depiction of the ellipsoidal monocapillary condenser. The source to optic distance represents the distance from the X-ray source to the inlet plane of condenser. The optic length represents the distance from the inlet plane to the outlet plane of condenser. The working distance represents the distance from the outlet plane of condenser to the sample.

determined. Afterwards, the rest key parameters can be readily derived:

$$\begin{aligned} & b = \sqrt{a^2 - c^2}, \\ & S = v_1 \cos \beta_1, \quad W = v_2 \cos \beta_2, \\ & H = v_1 \sin \beta_1, \quad h = v_2 \sin \beta_2. \end{aligned} \tag{4.2.2}$$

Take our condenser as an example, it is designed that $c = 150$ mm, $L = 78$ mm, $\beta_1 = 1.91$ mrad, and $\beta_2 = 3.27$ mrad. With these known parameters, the rest parameters are solved and listed in Table 4.3. It can be seen that the measured parameters are highly consistent with the design.

Based on the above fundamental principles [66] of geometrical and wave optical propagation, a dedicated numerical simulation software was developed with Python 3.9 for the ellipsoidal monocapillary condenser, as illustrated in Fig. 4.10. This software can provide qualitative evaluations of condenser performance under both ideal and non-ideal conditions. For ideal settings, the software can generate the ray-traced optical path of the condenser, the corresponding intensity distribution on the detector plane, and the line profile along the central axis, as shown in Figs. 4.10(a)–(c). These outputs allow direct assessment of focusing behavior, beam uniformity, and spatial intensity modulation. For non-ideal conditions such as condenser misalignment, surface roughness, slope errors, and fabrication imperfections, the software treats them as independent or combined perturbation factors, and the final intensity distribution on the detector plane is presented in Figs. 4.10(d)–(e). These simulation results show strong agreement with experimental measurement, which is depicted in Fig. 4.10(f), validating the viability and robustness of the numerical simulation. In the future, this simulation framework can be utilized as a simple tool for parameter optimization and performance evaluation of condenser design prior to physical experiments.

Table 4.3: Geometric parameters of the ellipsoidal monocapillary condenser.

| Notation | $\beta_1$* | $\beta_2$* | $c$* | $L$* | $b$ | $u_1$ | $u_2$ | $v_1$ | $v_2$ | $S$ | $W$ | $H$ | $h$ |
|---|---|---|---|---|---|---|---|---|---|---|---|---|---|
| Unit | mrad | mrad | mm | mm | mm | mm | mm | mm | mm | mm | mm | mm | mm |
| Measured | 1.91 | 3.17 | 149.2 | 78.0 | 0.345 | 118.8 | 196.8 | 179.6 | 101.6 | 118.8 | 101.6 | 0.343 | 0.322 |
| Designed | 1.91 | 3.17 | 150 | 80 | 0.342 | 120 | 200 | 180 | 100 | 120 | 100 | 0.343 | 0.316 |

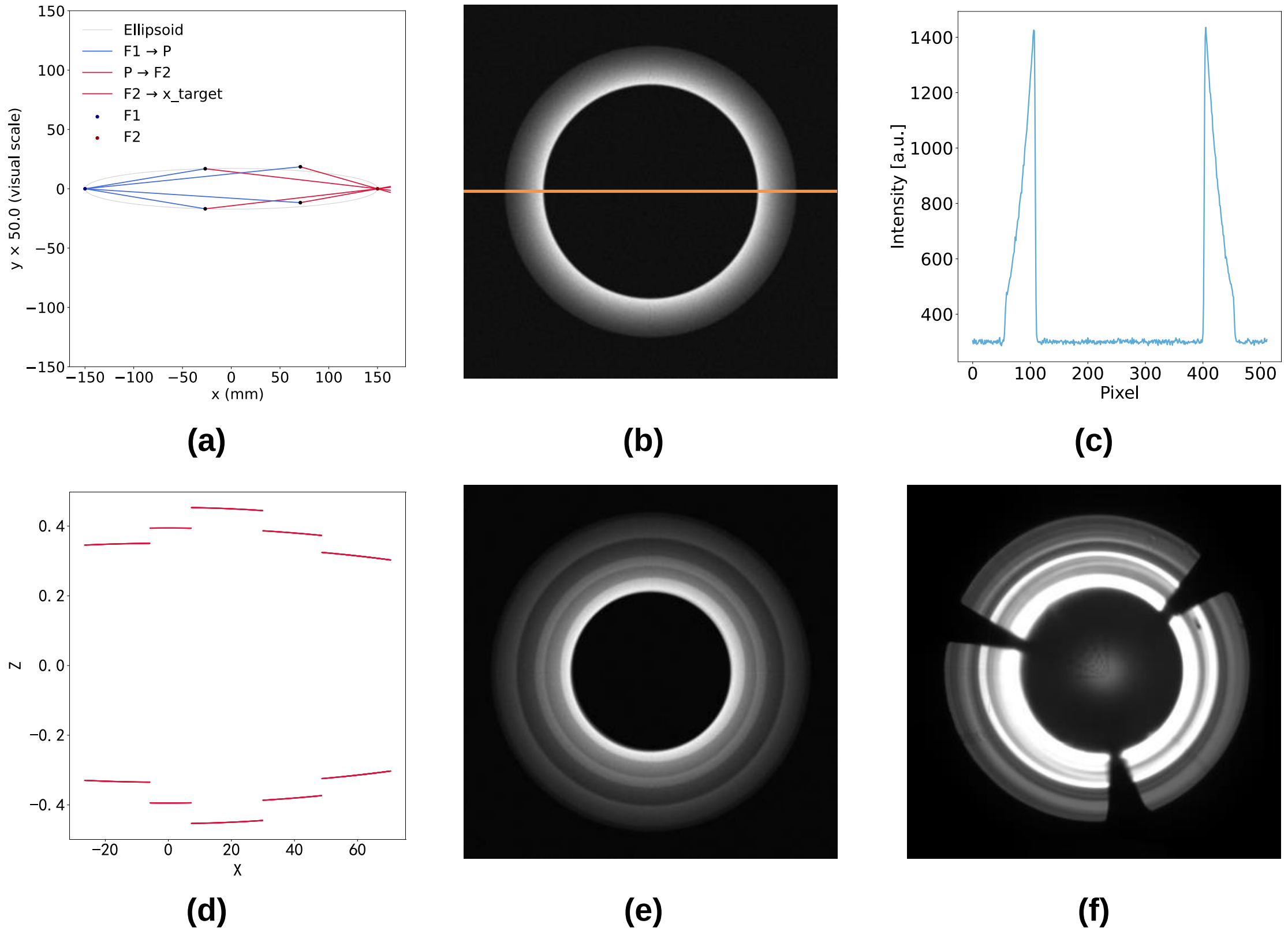

Figure 4.10: (a) Simulated optical path of the ideal condenser. The geometric dimension along the vertical direction is stretched by 10× for better visualization. (b) Simulated intensity distribution on the detector plane. (c) Intensity profile along the highlighted line in (b). (d) Cross-section of the non-ideal condenser. (e) Simulated intensity distribution on the detector plane for the condenser in (d). (f) Experimentally measured intensity distribution on the detector plane.

To achieve high spatial resolution, a high quality capillary condenser is needed to generate a narrow but uniformly distributed beam at the outlet plane of the condenser Most importantly, the beam angular coverage needs to match with that of the FZP, see Fig. 3.5, allowing for more efficient high resolution imaging.

### 4.2.4 Condenser adjustment

The alignment of condenser requires optimization in five degrees of freedom: horizontal translation, vertical translation, longitudinal translation, pitch adjustment and yaw adjustment. Therefore, a dedicated five-axis motion system is required, see Fig. 4.11. The alignment procedure is highly experience-dependent. However, the method illustrated in Fig. 4.12 provides a practical way to simplify the entire alignment procedure.

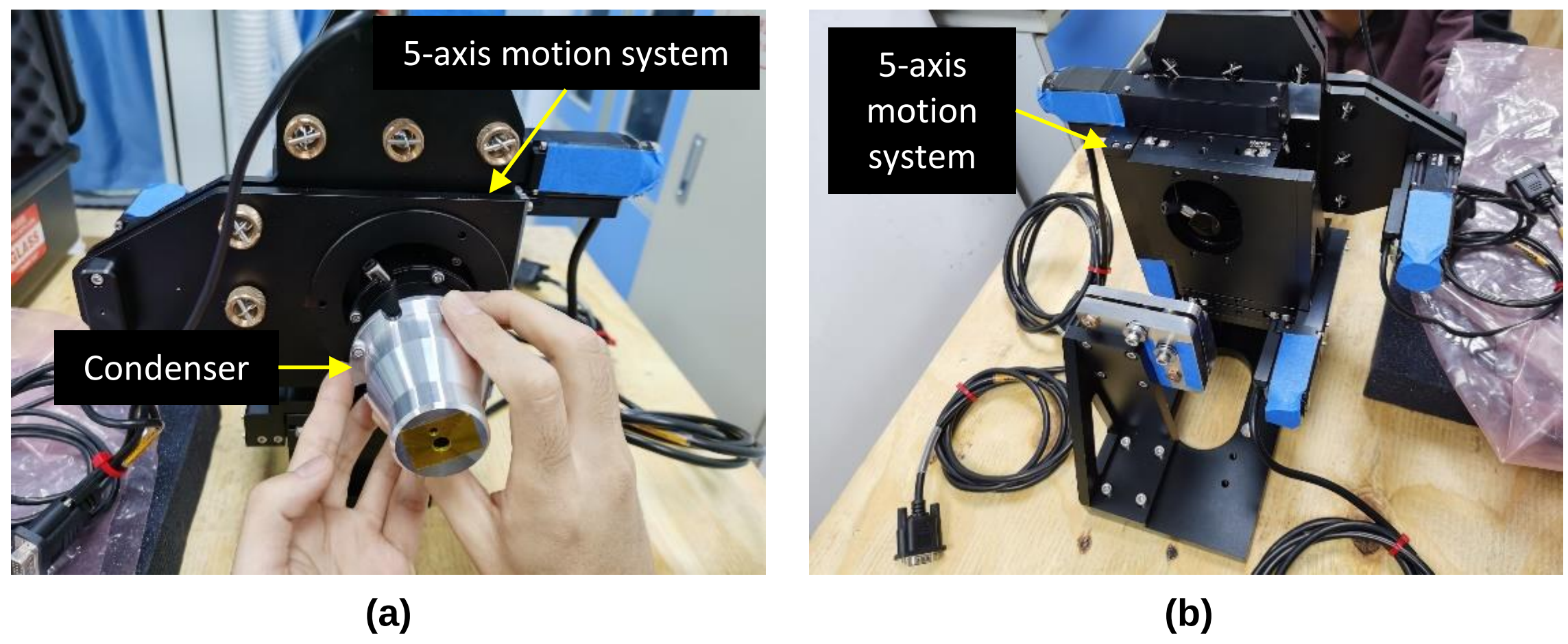

Figure 4.11: (a) The condenser installed on a five-axis motion system. (b) Backside photography of the five-axis motion system.

To proceed, it is assumed that the condenser is already aligned in a rough position with respect to the focal spot of the X-ray source. Namely, a clear annular beam profile is already obtained, see Fig. 4.12(a). Next, the zone plate is moved into the optical path. The detailed alignment procedure of zone plate can be found in Section 5.2.3. After completing the preliminary configuration, fine alignment of the condenser is performed following the procedures below. First, two yaw angles, denoted as $\mathrm{Y}_l$ and $\mathrm{Y}_r$, are adjusted accordingly to generate two identical distances: one is from the left outer annular edge to the left central semicircular pattern, as highlighted in Fig. 4.12(b), and the other is from the right outer annular edge to the right central semicircular pattern, as highlighted in Fig. 4.12(c). By adjusting the yaw angle to the average of the above two specific yaw angles, i.e., $(\mathrm{Y}_l+\mathrm{Y}_r)/2$, the condenser reaches to its most appropriate yaw orientation. Similarly, the pitch angle, denoted as $P_t$ and $P_b$, can be optimized using the same approach, as illustrated in Figs. 4.12(d)-(e). By jointly adjusting the yaw and pitch positions, the outer annular pattern becomes more symmetric, and the center will appear a uniform bright focal spot, see Fig. 4.12(f).

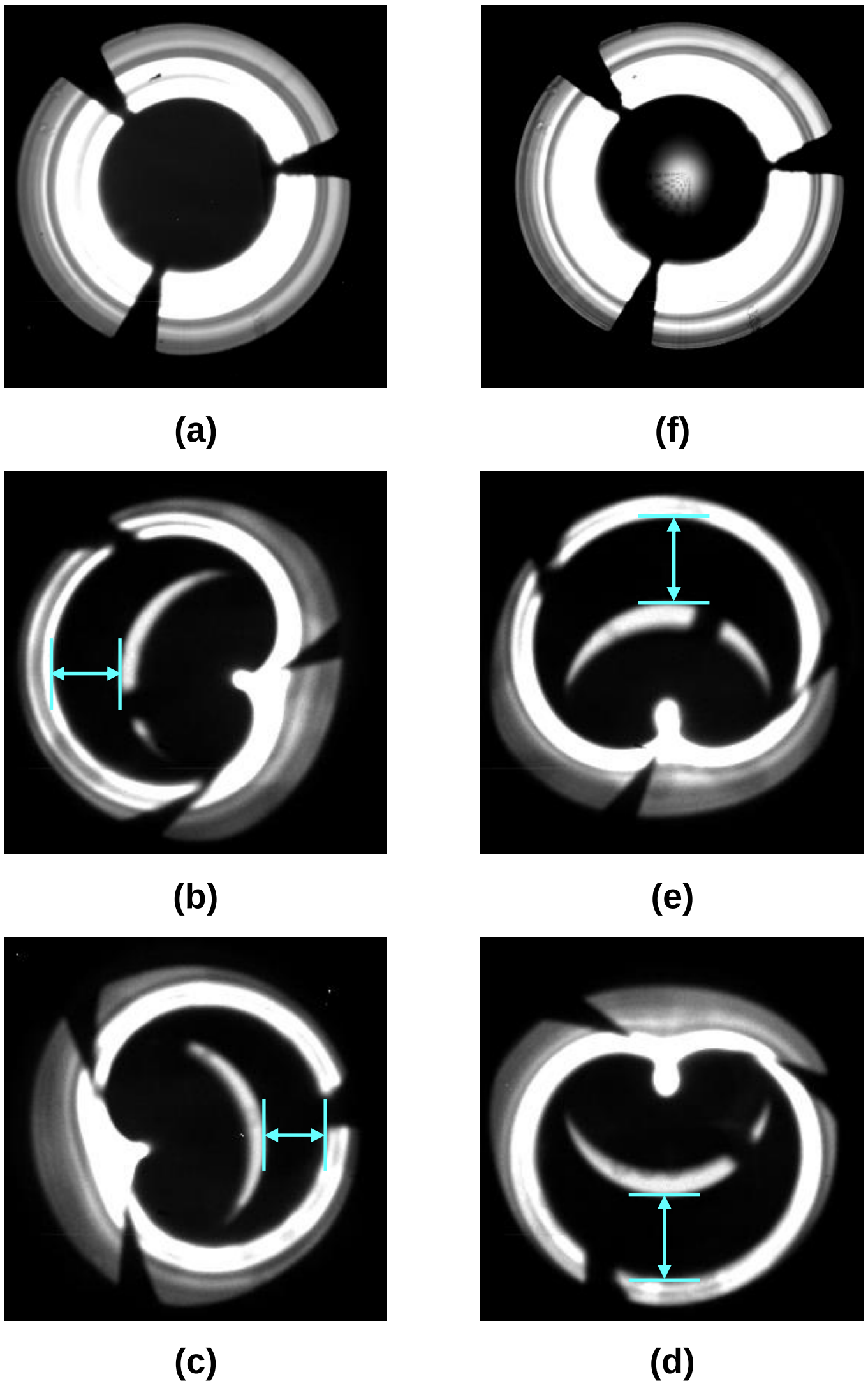

Figure 4.12: (a) The starting position of the condenser. (b) The $Y_l$ position along the horizontal direction. (c) The $Y_r$ position along the horizontal direction. (d) The $P_t$ position along the vertical direction. (e) The $P_b$ position along the vertical direction. (e) The focused imaging result after condenser alignment (the zone plate is added into the optical path).

Finally, the remaining degrees of freedom, such as horizontal translation, vertical translation and longitudinal translation, are iteratively adjusted to maximize the intensity of the focused spot at the center. Once the focus intensity reaches to the maximum, it is believed that the condenser alignment is accomplished.

It is worth noting that the alignment procedure presented here for condenser should be considered as a useful practical reference, but not the unique solution.

## 4.3 Radiation shielding

X-rays are hazard to human health. Therefore, radiation shielding, which uses high Z materials such as lead or steel to weaken the X-ray radiation to the environment background level, is very crucial during the development of FZP based X-ray microscope.

Typically, the maximum X-ray tube potential is under 50 kV. In addition, the maximum X-ray tube power is smaller than 1 kW. In general, the relatively low kV and low power settings make the radiation shielding much easier. For a rough estimation, it is assumed that the maximum X-ray photon energy is 50 keV. For 1 cm thick steel ($\mu = 15.38$/cm @ 50 keV) shielding, only 0.00002% X-ray photons can pass through. For 0.15 cm thick lead ($\mu = 91.68$/cm @ 50 keV) shielding, only 0.0001% X-ray photons can pass through.

# 5 FZP

Unlike visible light, it is not possible to focus X-ray beam, whose refractive index is approximately equal to 1, with a convex lens to achieve image magnification. Therefore, a Fresnel zone plate (FZP) is used. In fact, the FZP in X-ray microscope plays the same role as of the objective lens in an optical microscope, see Fig. 3.1. The FZPs used for X-ray microscope imaging can be either absorbing type or phase shifting type. In practice, the absorbing FZPs are more common than the phase shifting FZPs. Usually, FZPs are made from heavy X-ray absorbing material, e.g., gold, platinum or tantalum.

## 5.1 Design

In essence, X-ray FZPs are specialized circular, aperiodic diffractive gratings. Particularly, it consists of a set of concentric rings, which alternate between being opaque and transparent. The zones are spaced so that the diffracted X-ray beam constructively interferes at the focus. To achieve this, the boundaries of any two neighboring zones are chosen so that the optical path difference from adjacent zones to the first order focal spot differs by half a wavelength, namely, $\lambda/2$, see Fig. 5.1. By alternately blocking (or phase-shifting) these zones, all transmitted contributions arrive at the focus in phase and eventually form the constructive interference. Mathematically, the radius of the $N$-th zone satisfies the following equation:

$$r_N = \sqrt{N\lambda f + \frac{1}{4}N^2\lambda^2}, \tag{5.1.1}$$

where $N$ is a positive integer, $\lambda$ denotes the X-ray wavelength, and $f$ denotes the focal length. When the radius of the zone plate is small compared to the focal length, approximately, Eq. (5.1.1) becomes Eq. (3.2.3).

Be aware that a FZP usually has multiple focal spots, depending on the optical path difference from successive transmitted zones. As mentioned, for the first order focal spot, the optical path from successive transmitted zones differs by $\lambda/2$. Similarly, the optical path from successive transmitted zones differs by $3\lambda/2$ for the third order focal spot, and it differs by $5\lambda/2$ for the fifth order focal spot, see Fig. 5.1. By default, the first order focal spot is used in X-ray microscope.

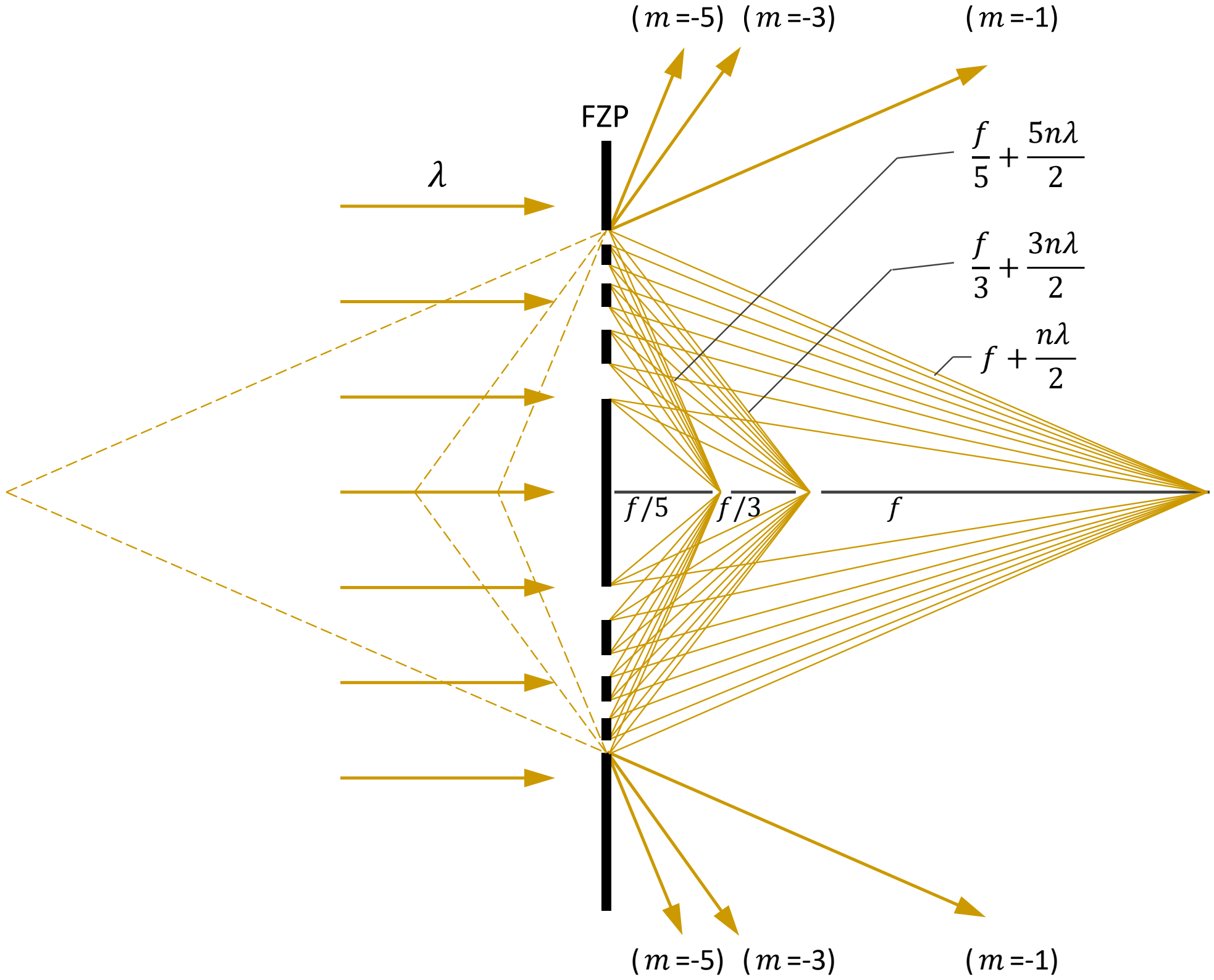

Figure 5.1: Illustration of the diffraction effect of FZP. Herein, the X-ray beam is assumed propagating from the left to the right. Three different diffraction orders, corresponding to three ordered focal spots, are highlighted on the right hand side of FZP. The optical paths from successive transmitted zones differ by $\lambda/2$, $3\lambda/2$ and $5\lambda/2$, respectively. Due to the principle of reversibility of light, the three ordered focal spots are also highlighted, see the dashed lines on the left hand side of FZP.

Nowadays, zone plates are mainly manufactured using lithography. As the lithography technology evolves, the fabricated width of the outermost zone had been significantly decreased. As a result, the minimum spatial resolution of zone plates gets increased dramatically. The loss of X-ray photons on FZP is quite high, and this remains the biggest challenge for X-ray microscopy. Stacking two identical zone plates back to back [67] can significantly increase the total material thickness and thus improve the overall diffraction efficiency of X-ray photons. In practice, it is always demanded to increase the efficiency of FZP, with the purpose of decreasing the total data acquisition time and enhancing the sample throughput efficiency.

## 5.2 Assembly and adjustments

### 5.2.1 FZP assembly

The fabricated FZP with frame size of 5 mm × 5 mm was glued to the aluminum holder by a thin layer of plain nail polish, see Fig. 5.2 for more details.

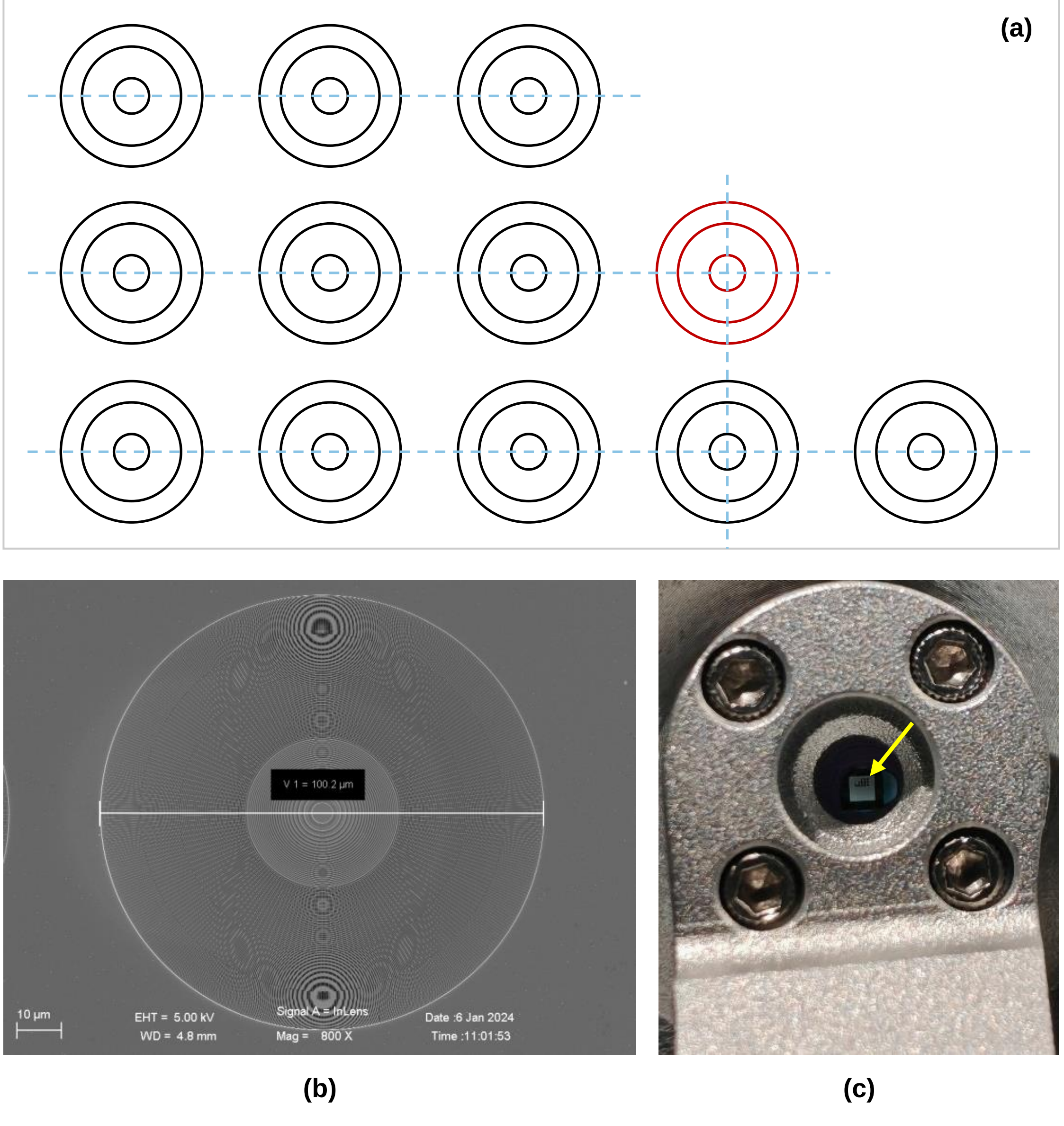

Figure 5.2: (a) Layout of 12 zone plates. The one highlighted in red has the best fabrication quality, whose outermost zone meets the design requirements according to SEM characterization in (b). (c) The backside photograph of the zone plate holder, where the layout of 12 zone plates is visible, highlighted by the arrow. This FZP was fabricated by Prof. Yifang Chen at Fudan University.

### 5.2.2 FZP selection

In our system, it needs to find out the best FZP from the twelve fabricated candidates that having varied morphologies and properties, as highlighted in Fig. 5.2(a). To identify, all of these twelve zone plates are scanned one after the other along both the horizontal and vertical directions, separately. The generated focal spot of each of the twelve zone plates is then individually evaluated and compared. The zone plates producing a more uniform and brighter focal spot are considered as good candidates. To find out the best zone plate from these above candidates, additional evaluations are needed by comparing their achievable imaging resolution. This requires additional optimization of the axial distance between the sample and the zone plate due to their slightly different focal lengths. Details are presented in the following discussions.

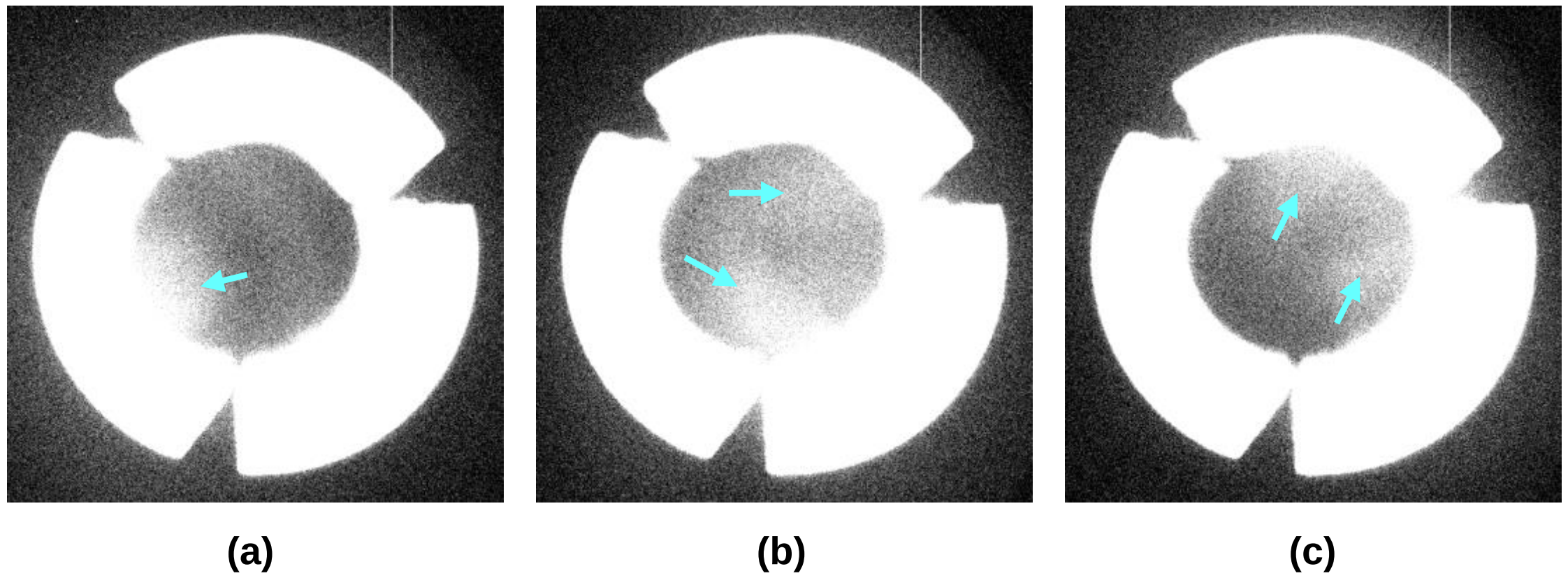

Figure 5.3: Different imaging effects and outcome are observed for the selected three zone plate candidates.

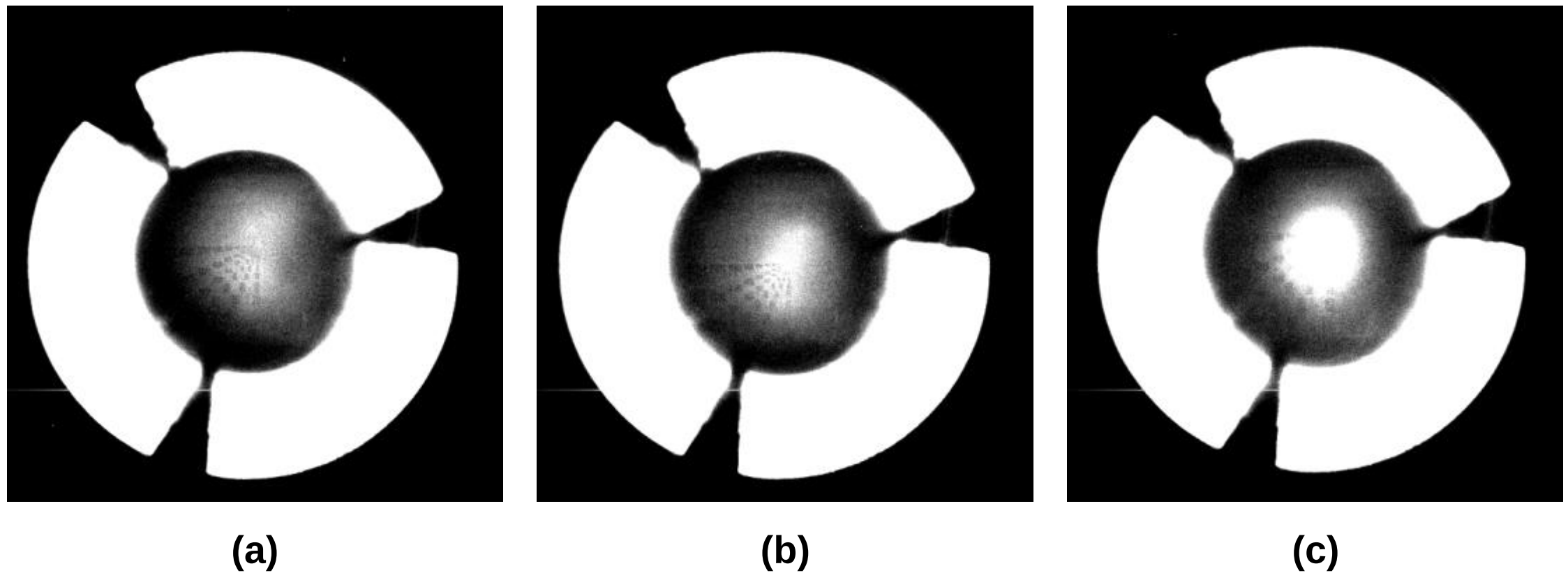

Figure 5.4: The beam focusing performance of different zone plates.

### 5.2.3 FZP alignment

In brief, alignment of the zone plate mainly consists of two steps. The first step is responsible for position and attitude alignment, aiming to obtain a uniform and bright focal spot at the center. The second step is responsible for axial alignment along the optical path, in which the relative distance between the zone plate and the sample is carefully adjusted to achieve the highest imaging resolution.

In particular, alignment in the first step mainly involves attitude adjustment of the condenser to match its orientation with that of the selected zone plate, see Fig. 5.5(a). Its importance lies in generating and ensuring the highest diffraction efficiency. The alignment details of the condenser have been discussed in Section 4.2.4. In fact, alignment of condenser must be repeated for multiple times before determining the most appropriate attitude matching with the selected zone plate. Meanwhile, the zone plate is carefully translated step by step along the horizontal and vertical directions until the focal spot image reaches the highest intensity. In this work, such position is regarded as the optimal working condition for the condenser and zone plate.

The second step mainly focused on axial (or longitudinal) alignment along the optical path. Its importance lies in generating and ensuring the highest in plane spatial resolution. In this step, it is assumed that the sample position is fixed. A strategy that combines coarse searching and fine scanning was proposed. Specifically, the whole procedure begins with coarse searching of the two starting axial positions, denoted as $\mathrm{A}_1^{(1)}$ and $\mathrm{A}_2^{(1)}$, of the zone plate over a large distance range [$\mathrm{A}_1^{(1)}$, $\mathrm{A}_2^{(1)}$], see Fig. 5.5(b). Then, a new image is acquired at the midpoint position of $(\mathrm{A}_1^{(1)}+\mathrm{A}_2^{(1)})/2$, denoted as $\mathrm{A}_1^{(2)}$. By comparing the spatial resolution of this current new image with the two starting images, a narrower position range can be determined. Herein, it is assumed that the image resolution at $\mathrm{A}_1^{(1)}$ is lower than the image resolution at $\mathrm{A}_2^{(1)}$, and the image resolution at $\mathrm{A}_1^{(2)}$ is higher than the image resolution at $\mathrm{A}_1^{(1)}$ but lower than the image resolution at $\mathrm{A}_2^{(1)}$. As a result, the new searching position range becomes [$\mathrm{A}_1^{(2)}$, $\mathrm{A}_2^{(1)}$]. Similarly, a new image is acquired at the midpoint position of $(\mathrm{A}_1^{(2)}+\mathrm{A}_2^{(1)})/2$, denoted as $\mathrm{A}_2^{(2)}$. As shown in Fig. 5.5(b), it is assumed that the image resolution at $\mathrm{A}_2^{(2)}$ is higher than the image resolution at $\mathrm{A}_1^{(2)}$. Therefore, the next new searching position range becomes [$\mathrm{A}_1^{(2)}$, $\mathrm{A}_2^{(2)}$]. The above coarse searching operations is repeated until the interval between the final $\mathrm{A}_1^{(k)}$ and $\mathrm{A}_2^{(k)}$ is reduced down to approximately 10–30 $\mu$m. Next, fine position scanning should be performed with a fixed step size, for instance, 2 $\mu$m. Images acquired at each position are compared to determine the most optimal axial position, denoted as $A_{\mathrm{best}}$, in generating the highest spatial resolution. Keep in mind that the fine scanning is the most time consuming procedure, special

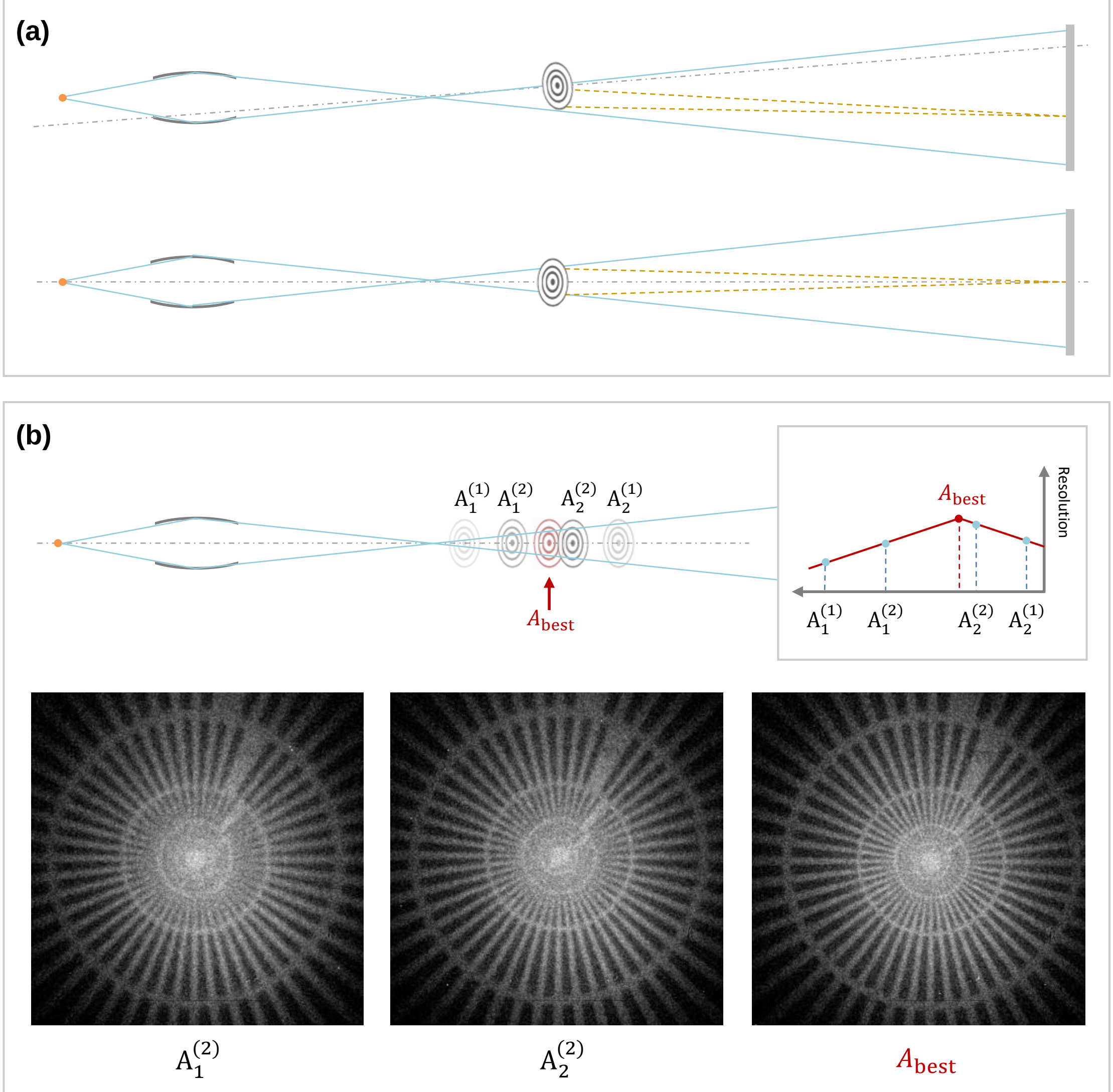

Figure 5.5: (a) Schematic illustrations of the relative axial adjustment of the capillary condenser and zone plate. Herein, the optical path is assumed parallel to the normal of zone plate plane. Without optical adjustment, the condenser significantly deviates from the optical path. After careful optical adjustment, the condenser well aligned with the optical path. (b) Procedure for longitudinal position adjustment of the zone plate. The representative images showing resolution improvement are depicted in the bottom.

attention and tremendous patience are particularly required.

Ideally, the depth of focus of our X-ray microscope is found to be 31.35 $\mu$m using Eq. (3.2.5), assuming $\Delta r_N = 30$ nm and beam energy of 5.4 keV. However, non-ideal factors such as imperfect condenser orientation and defects of the FZP may make it difficult to achieve the theoretical depth of focus. As a result, the effective range that provides optimal imaging performance may be reduced.

It should be noted that the relative positions of the X-ray source, condenser system and zone plate are strongly coupled and should be considered simultaneously during this alignment procedure. Therefore, a further optimization of the axial position of the X-ray source (together with the condenser assembly) maybe also needed using the same adjustment strategy as described above for FZP. After these careful axial adjustments, in principle, the designed image spatial resolution should be achieved. If NOT, one may need to double check the zone plate, detector, mechanical vibration, temperature variation, and other possible impacting factors.

# 6 X-ray detector system

## 6.1 General characteristics

X-ray detector, which is the second most critical component in microscope, is responsible for generating the final images by converting the incident X-ray photons into either visible light in scintillators or electrons and holes in semiconductors. The first type is known as indirect X-ray detector, and the second type is known as direct X-ray detector. No matter which type, they all share the following valuable characteristics[7]:

**Detective quantum efficiency (DQE):** the primary measure of an X-ray imaging system's ability to convert incident photons into a high-quality image. A higher DQE indicates better image quality at lower radiation dose levels.

**Dark signal/noise:** the signal that is recorded even though no X-ray photons are incident. Usually, the lower dark noise the better. In a photon-counting detector, the dark noise is zero.

**Dynamic range:** the ratio of the maximum to minimum signal value that can be recorded. Usually, this is related to the full-well capacity of the CCD/CMOS cameras.

**Solid angle:** for the detection of radiation that goes into a certain angular range. Mathematically, it is expressed as $\Omega = 2\pi(1 - \cos\theta)$, where $\theta$ is the half-angle of the maximum cone of X-ray beam.

**Pixel size:** the dimension of a pixel in pixelated CCD/CMOS detectors. Often, smaller pixel size is important for high spatial resolution imaging.

**Pixel array:** the active array of pixels. For a regular CCD/CMOS cameras, it is denoted as $N_x \times N_y$.

**Frame rate:** measured in frames per second (fps), corresponds to the frequency at which consecutive images can be recorded.

**Dead time:** the time after arrival of one photon before the detector is ready to record a subsequent photon. Usually, a smaller dead time is preferred.

## 6.2 Indirect conversion detector

### 6.2.1 Main structure

Until now, indirect conversion detectors are still the dominant driving horse with almost over 90% occupation in X-ray imaging applications. This type of detectors rely on the cascaded "X-ray photon–scintillation light–electric signal" stages to complete the X-ray detection, see Fig. 6.1. As depicted, a specific scintillator layer is placed in the front to convert the X-ray into scintillation light, which is then collected by a group of optical lens. Afterwards, the scintillation light is focused on the 2D CCD/CMOS array plane to generate the final digital image.

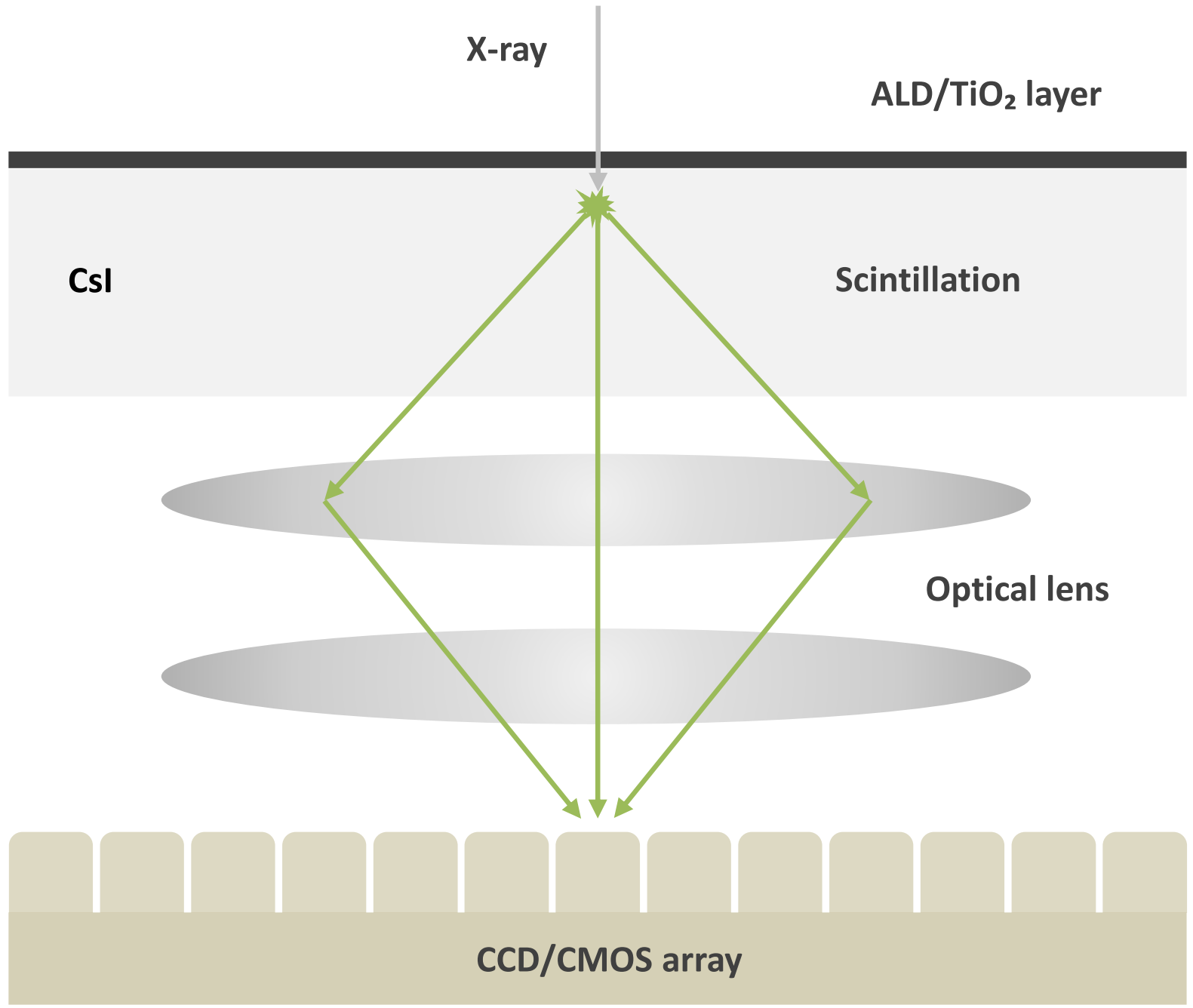

Figure 6.1: The internal structure of an indirect conversion detector. The incident X-ray photons are converted into optical light inside the scintillator layer. Afterwards, the scintillation light are collected by the optical lens and finally converted into electric signal in the CCD/CMOS pixel.

### 6.2.2 X-ray scintillator

By far, many scintillator materials have been applied in X-ray detection. For instance, CsI:TI (CsI), $Gd_3Al_2Ga_3O_{12}$:Ce (GAGG), $Lu_3Al_5O_{12}$:Ce (LuAG), $CdWO_4$ (CWO), $Y_3Al_5O_{12}$:Ce (YAG), $Bi_4Ge_3O_{12}$ (BGO), $Gd_2O_2S$:Tb (GOS) are some typical inorganic scintillator materials, with distinguished properties[68], see Fig. 6.2 and Table 6.2 for more details. Regarding to their light yield, spatial resolution, and image quality, experimental studies have demonstrated that the light yield of GOS was highest, followed in order by CsI, GAGG, CWO, YAG, LuAG, and BGO. Approximately, CsI produces 54–56 photons per keV of X-ray photon energy. Moreover, the spatial resolution in GAGG and LuAG was highest, followed by CsI, CWO, and YAG. The image qualities were almost the same, except for GOS (lowest signal-to-noise ratio (SNR)). Overall, results indicate that CsI is suitable for X-ray microscopy imaging with a low flux tube source in laboratory settings, and GAGG and LuAG are suitable for X-ray microscopy imaging with a high flux source in synchrotron settings.

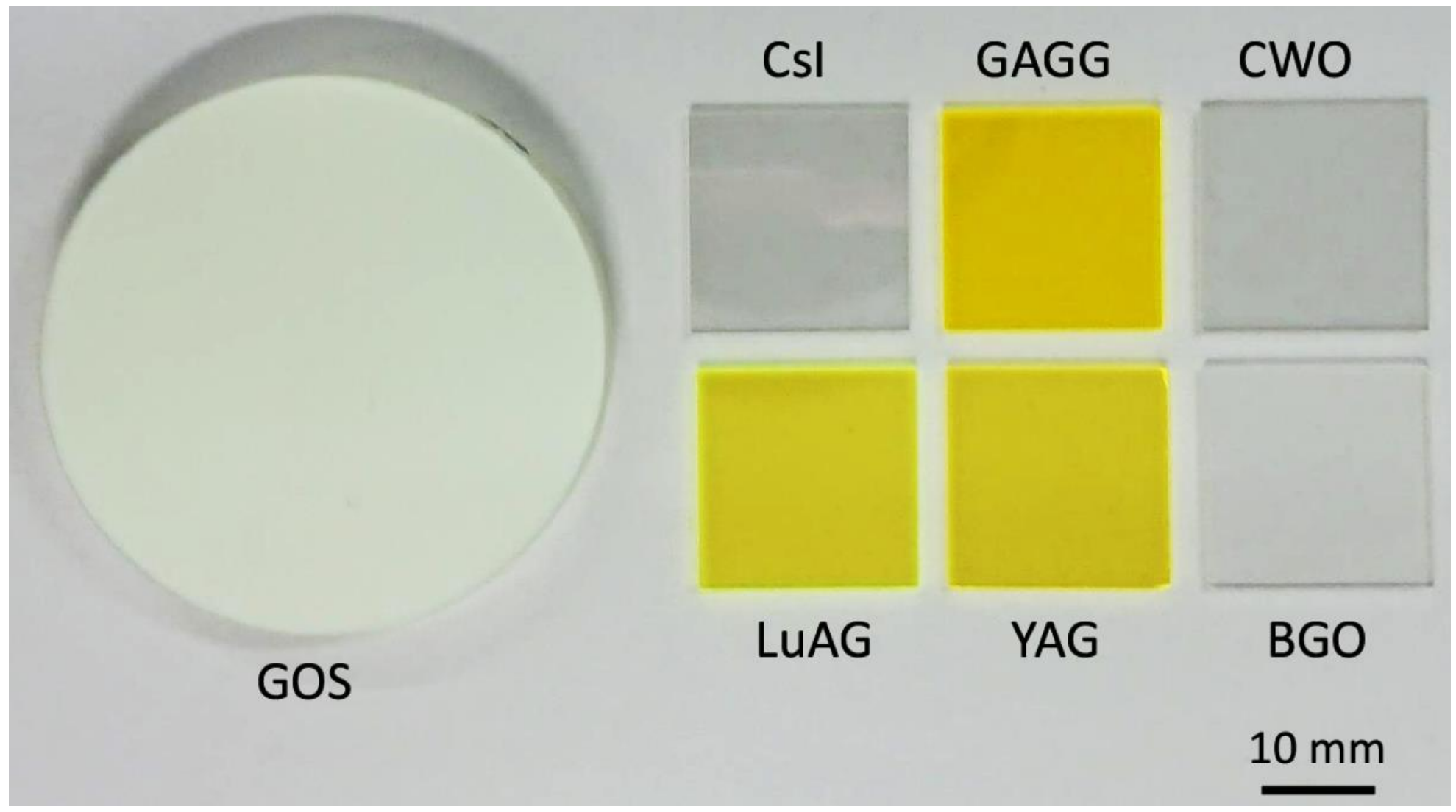

Figure 6.2: Photograph of the $Gd_2O_2S$:Tb (GOS), CsI:TI (CsI), $Gd_3Al_2Ga_3O_{12}$:Ce (GAGG), $CdWO_4$ (CWO), $Lu_3Al_5O_{12}$:Ce (LuAG), $Y_3Al_5O_{12}$:Ce (YAG), $Bi_4Ge_3O_{12}$ (BGO) scintillators.

Table 6.1: Main physical properties of each scintillator.

| Material | Density [g/cm$^3$] | Light yield per keV | $\mu$(5.4 keV) [1/cm] | $\lambda$ [nm] | Refractive index |
|---|---|---|---|---|---|
| CsI | 4.5 | 10 | 64.7 | 550 | 1.78 |
| GAGG | 6.6 | 15 | 52.1 | 520 | 1.93 |
| LuAG | 6.7 | 20 | 42.0 | 535 | 1.84 |
| CWO | 7.9 | 25 | 33.7 | 480 | 2.25 |
| YAG | 4.6 | 10 | 64.5 | 550 | 1.82 |
| BGO | 7.1 | 15 | 51.8 | 480 | 2.15 |
| GOS | 7.3 | 20 | 41.6 | 430 | 2.20 |

In practice, the image spatial resolution is strongly impacted by the CsI thickness, which primarily controls the lateral diffusion distance of scintillation light. As a consequence, increasing the CsI thickness would increase the optical diffusion length, which broadens the point-spreading function (PSF) and degrades the overall spatial resolution. For X-rays having less than 10 keV beam energy, empirically, 50 $\mu$m thick CsI film is enough to absorb 90%-99% X-ray photons. However, manufacturing a 50 $\mu$m thick CsI film is quite challenging because the film lies near the unstable nucleation regime where columnar structure is not fully developed, process tolerances become extremely tight, and mechanical/moisture vulnerabilities increase. In addition, it is most common to add an aluminum oxide ($Al_2O_3$, ALD) reflective layer or a high-reflectivity white dielectric coating layer such as $TiO_2$ on the top surface of CsI (away from the CCD/CMOS) to redirect the upward traveled scintillation photons (nearly 50% occupation) back toward the pixels. By doing so, the overall detective quantum efficiency (DQE) can be greatly improved.

Be aware that CsI material is strongly hygroscopic. Moisture will cause hydration, deliquescence to the CsI film. Hence, moisture protection is particularly important. Usually, a surface coating, e.g., ALD layer, can substantially slow the moisture degradation procedure of CsI, but long-term protection still requires a more comprehensive strategy combined with proper edge sealing and packaging. If the CsI film is hydrated[69], then "salt and pepper" noise (white spots/pixels) will emerge in the image. Additionally, a sharp decrease in scintillation light output will also happen and thus significantly degrades the overall image quality.

### 6.2.3 Objective lens

Objective lenses are arguably the most important element in an optical microscope. They are responsible for capturing light and producing a magnified, sharp, real image of the object. The objective itself is usually a cylinder containing one or more lenses that are typically made of glass, with very short focal length; its function is to collect light from the sample. One of the most important properties of microscope objectives is their magnification. The magnification typically ranges from 4× to 100×. In a microscope, the least powerful and shortest objective lens is called the scanning objective lens, and is typically a 4× objective. The second lens is referred to as the small objective lens and is typically a 10× lens. Modern microscopes are often designed to use infinity correction, in which the light coming out of the objective lens is focused at infinity. This is denoted on the objective with the infinity symbol ($\infty$), see Fig. 6.3.

In an X-ray microscope, similarly, objective lenses also play important roles in collecting light and magnifying the FZP generated image, which impinges on the scintillator film in front of the objective lenses. Similarly, two objective lenses are needed: a scanning objective lens with 4× magnification capability, and an imaging objective lens with 10× (or 20×) magnification capability. In particular, the scanning objective lens is used to search and determine the region-of-interests (ROIs) on the sample, see Fig. 6.3(a). In addition, the imaging objective lens is used to acquire high quality microscope images of the selected ROI, see Fig. 6.3(b).

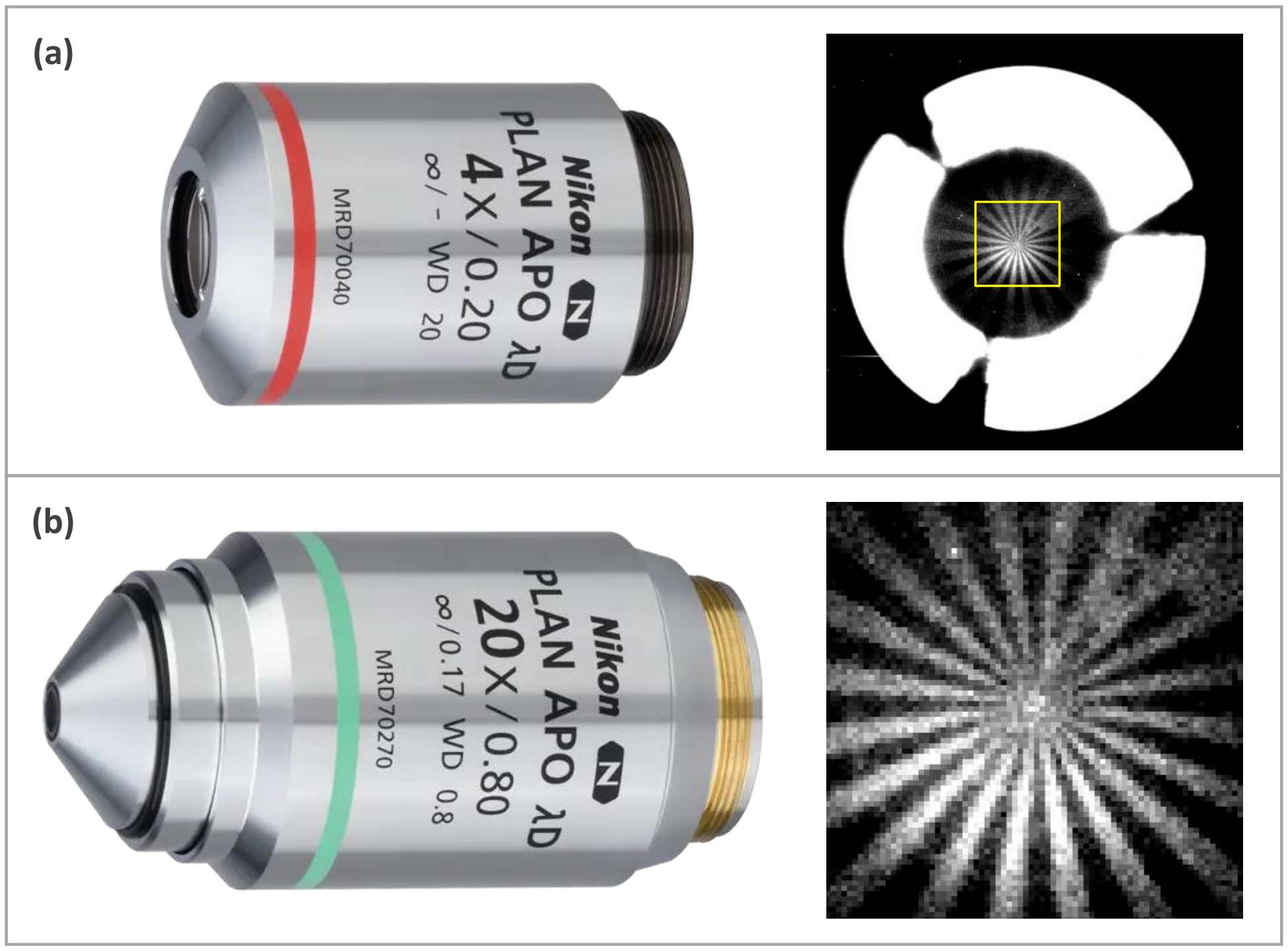

Figure 6.3: The newly developed Nikon CFI Plan Apochromat Lambda D is a high-performance objective series optimized for digital solutions. They can deliver high image quality across the large field of view of FN25 and chromatic aberration correction over a wide wavelength range. (a) the 4× scanning objective lens and the full-field low resolution image, (b) the 20× imaging objective lens and the high resolution image of the region-of-interest (ROI) highlighted in (a).

### 6.2.4 Digital camera

Since the early 1990s, digital cameras have become popular and dominant in imaging applications, largely replacing those that capture images on photographic film. Digital cameras use either CCD (Charge-Coupled Device) or CMOS (Complementary Metal Oxide Semiconductor) sensors to convert light into electrical signals. Both technologies possess inherent advantages. For instance, CCDs can have a 100% fill factor that captures all incoming light, whereas part of the CMOS sensor is occupied by transistors and metal wiring associated with each pixel. Historically, CCDs provided higher-quality images with lower noise at affordable prices. Whereas, CMOS chips can scale the number of read-out ports and achieve very high frame rates.

Electron-multiplying CCDs (EMCCDs) are a variant of silicon-based CCDs that use electron multiplication to amplify the electron signal greatly above the read noise floor to maximize sensitivity for low-light imaging. Due to this photoelectron amplification step, EMCCD cameras are able to achieve single-photon detection, with sub-electron read noise at high frame rates. As a result, EMCCD sensors are therefore used in a wide range of applications, from steady-state astronomical imaging to dynamic single-molecule tracing.

The sCMOS (scientific CMOS) is a high-performance version of CMOS technology tailored for scientific, low-noise, and high-speed applications, offering superior image quality compared to standard CMOS sensors. In short, sCMOS can be thought of as a higher grade, higher performance CMOS. To achieve higher quantum efficiency (QE), which describes how efficient the sensor is at converting the incoming signal in the form of light photons to an electrical signal, the back-illuminated CMOS sensors (also known as back-side illumination (BSI) sensors) are used to increase the amount of light captured and thereby improve low-light performance.

It is generally believed that there is no significant performance difference between the most popular CCD and CMOS cameras for scintillation light detection in X-ray microscope with FZP. Therefore, selection among commercial products may largely be restricted by the available budget.

### 6.2.5 Detector adjustment

The above three key components, X-ray scintillator, objective lens and digital camera, need to be carefully assembled according to the Eq. (3.1.1). As illustrated in Fig. 3.4(a), the scintillator layer and the objective lens are attached together and remain fixed. In particular, the scintillator layer is roughly positioned in front of the focal plane of the objective lens. In addition, the digital camera is independently mounted behind the lens on a motorized linear stage, which can move back and forth with micrometer precision until reaching to the best position to generate the most clear images. This is for the so called I-shaped optical lens coupled CCD/CMOS indirect conversion detector system. For the L-shaped optical lens coupled CCD/CMOS indirect conversion detector system, which is usually used for high energy X-ray detection, see Fig. 3.4(b), the objective lens is mounted on a motorized linear stage and moved back and forth until reaching to the highest image resolution, see

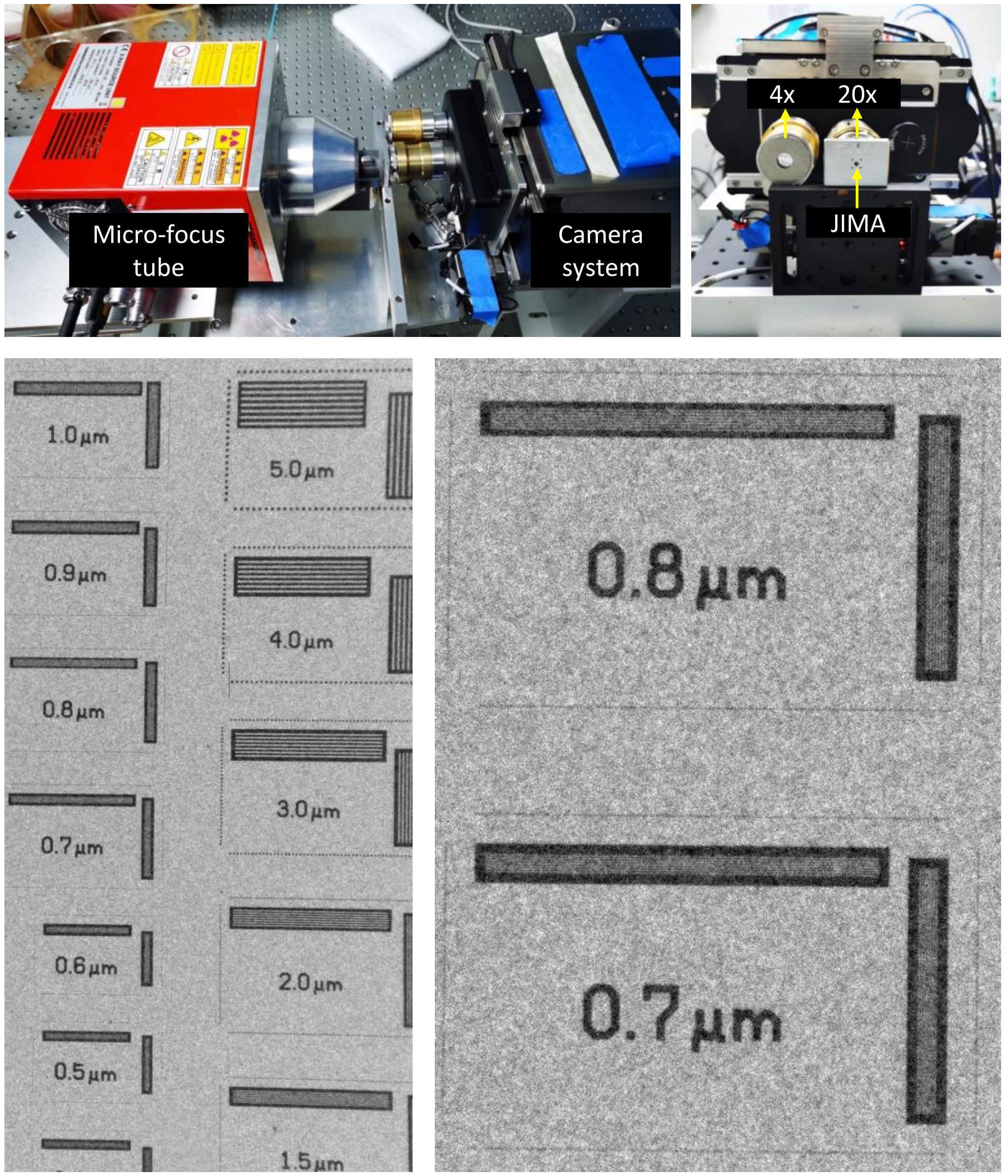

Figure 6.4: Setup for adjusting the indirect detector system in reaching to the highest imaging resolution. The micro-focus X-ray tube is positioned in front of the JIMA resolution chart, which is taped on top of the 20× camera lens. The acquired images in the bottom demonstrate that this indirect detector camera system is able to resolve 0.7 $\mu$m resolution pattern.

Fig. 6.4. During the position optimization procedure, the scintillator layer and the CCD/CMOS camera remain fixed.

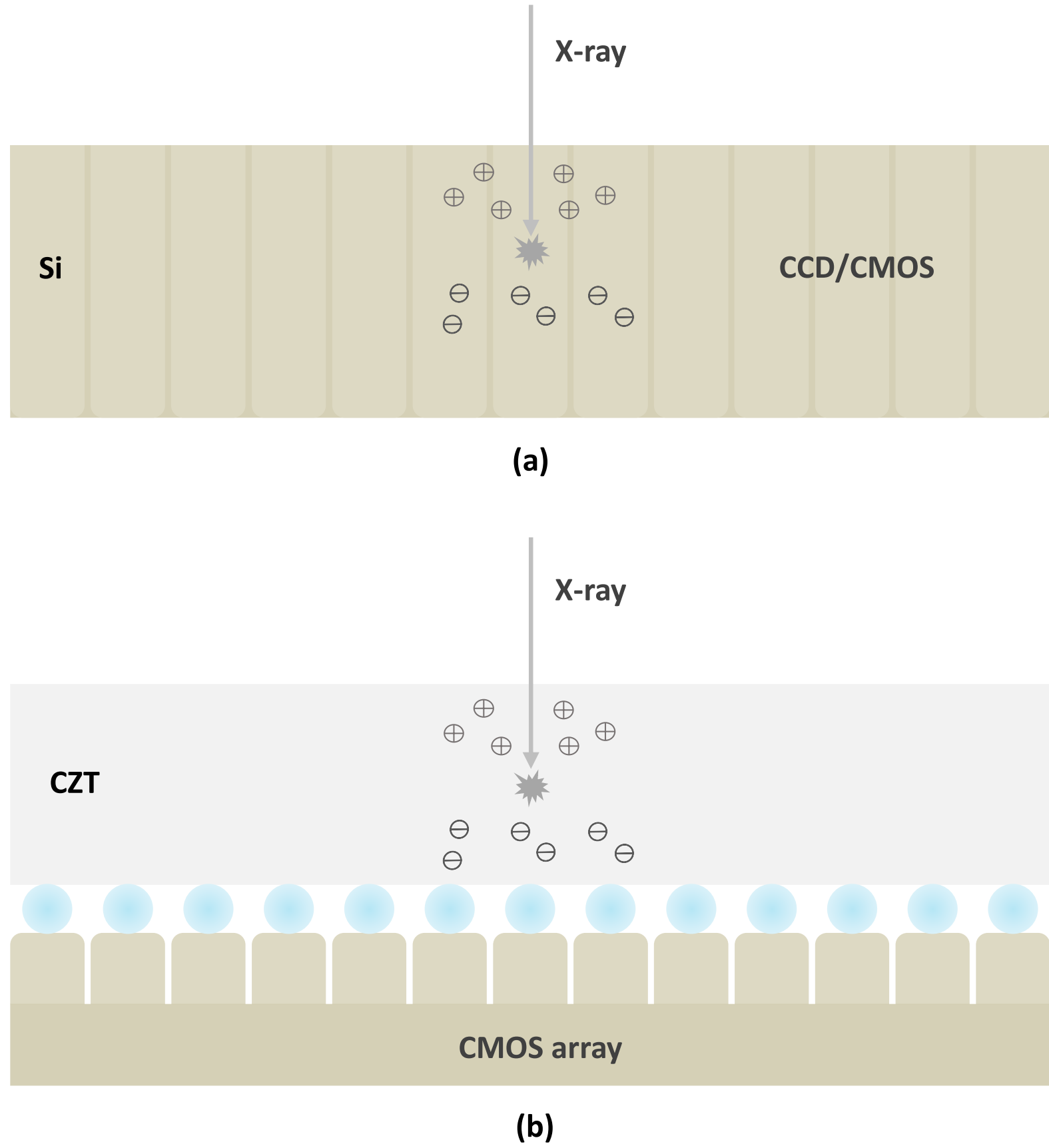

Figure 6.5: (a) The Si based direct conversion detector. (b) The cadmium zinc telluride (CZT) based direct conversion detector. For both semi-conductor materials, the incident X-ray photons are directly converted into electron-hole pairs, which are finally collected in the CCD/CMOS pixel array.

## 6.3 Direct conversion detector

### 6.3.1 Basic property

Direct conversion X-ray detector, as the next generation X-ray detector, has been applied in a lot of X-ray imaging tasks. Usually, semi-conductor materials, typically silicon (Si) and cadmium zinc telluride (CZT), are utilized to convert X-ray photons directly into electrical charge without an intermediate scintillation light-conversion step, as illustrated in Fig. 6.5. For low energy (<10 keV) X-ray photons, Si is a good candidate; for high energy (>20 keV) X-ray photons, CZT is more ideal due to its higher X-ray absorbing power.

In fact, the CCD and CMOS sensors made from Si can be utilized as direct conversion X-ray detectors very naturally. Specifically, X-ray detection occurs when an incident photon is absorbed within the Si layer, resulting in the production of the electron-hole (e-h) pairs. If this absorption occurs within the depletion region, the electrons and holes are separated by the applied internal electric field, generated by the sensor's electrode structure. The electrons are being trapped by the field and the holes undergo rapid recombination. The trapped electrons can subsequently be 'clocked' to the amplifier and readout. For this reason, only the electrons of the e-h pairs formed in the depletion zone will be read out as signal.

The average energy required to create an e-h pair in silicon is approximately 3.65 eV at room temperature (∼300 K). Although its fundamental band gap is 1.12 eV, additional energy is needed during the excitation process. When an incident X-ray photon is absorbed, the number of e-h pairs formed and the size of the cloud they form in the silicon is directly related to the energy of the incident photon[70], namely,

$$\bar{q} = \frac{E}{W}, \tag{6.3.1}$$

where $\bar{q}$ denotes the detected average number of quanta (e-h pairs), $W$ denotes the energy (∼3.65 eV) required to generate an e-h pair for silicon at root temperature, and $E$ denotes the energy of the incident X-ray photon. This equation holds for X-ray photon energies >10 eV, and assumes that the detected quanta $\bar{q}$ are produced directly by an X-ray photon. If there have some followed secondary processes, for instance, the electrons produced by an X-ray photon undergoing a number of inelastic scattering events inside the detector material, one might arrive at a statistical distribution that is described by the final processes rather than the very natural Poisson statistics. In order to account for these effects, Ugo Fano introduced an empirical correction parameter F (called as the Fano factor [71]) to quantify the variance of the detected quanta,

$$F = \frac{\text{Observed variance in } \bar{q}}{\text{Possion-predicted variance in } \bar{q}}. \tag{6.3.2}$$

Immediately, the variance of the detected signal becomes

$$\sigma_q^2 = F\bar{q} = F\frac{E}{W}. \tag{6.3.3}$$

For silicon, the Fano factor is found to be 0.118. The FWHM energy resolution of an X-ray detector can be estimated as follows:

$$\frac{\Delta E_{\text{FWHM}}}{E} = 2.35\frac{\sigma_q}{\bar{q}} = 2.35\sqrt{\frac{FW}{E}}, \tag{6.3.4}$$

Table 6.2: Relationship of electron-hole generation to X-ray photon energy.

| X-ray energy [keV] | Wavelength $\lambda$ [nm] | No. of e-h pairs | Size of electron cloud ($\mu$m) |
|---|---|---|---|
| 0.5 | 2.48 | 137 | 0.0051 |
| 2.7 | 0.45 | 739 | 0.0987 |
| 5.4 | 0.23 | 1479 | 0.3173 |
| 5.9 | 0.21 | 1616 | 0.3819 |
| 8.0 | 0.15 | 2191 | 0.6154 |

in which additional degradation in energy resolution due to the readout electronics noise (dark noise) is not accounted.

## 6.4 Performance comparison

To compare, the DQE performance was quantitatively evaluated for the indirect and direct X-ray detectors. After several earlier explorations of metrics for detector performance [72–75], detector DQE became accepted as the fundamental approach [76, 77]. For a number $\bar{q}_0$ of incident quanta on the detector, the number of these events that are actually recorded by the detector is called the noise-equivalent quanta (NEQ), which represents the effective number of X-ray quanta required to produce a specific image SNR. As a consequence, the zero frequency DQE is defined as:

$$\mathrm{DQE}(0) = \frac{\mathrm{NEQ}}{\bar{q}_0} = \frac{\mathrm{SNR}_{\mathrm{out}}^2}{\mathrm{SNR}_{\mathrm{in}}^2}, \tag{6.4.1}$$

where the second form written in terms of the SNR is true only for the case where the noise follows a Poisson distribution.

### 6.4.1 Cascaded model

In the next discussions, it is assumed that the intrinsic Poisson statistics of the incident X-ray photons, whose mean number is denoted as $\bar{q}_0$, is approximated into the Gaussian statistics. Due to the following conversions into other secondary quanta, e.g., electrons or lights, as a result, the statistics of the corresponding chain process needs to be investigated. In such a Markov chain [78], the expected final signal $q$ is expressed as:

$$q = g_{i+1} g_i, \tag{6.4.2}$$

in which the primary gain event $i$ causes a secondary gain event $i+1$ before detection. Statistically, the mean signal $\bar{q}$ is found to be

$$\bar{q} = \bar{g}_{i+1} \bar{g}_i, \tag{6.4.3}$$

and the variance of total signal $q$ is equal to

$$\sigma^2 = \bar{g}_{i+1}^2 \sigma_{g_i}^2 + \bar{g}_i \sigma_{g_{i+1}}^2. \tag{6.4.4}$$

Herein, the variance in both cascaded gain processes $g_i$ and $g_{i+1}$ are accounted[78, 79].

**Model-I:** In this fundamental model[7], the incident X-ray photons are absorbed by an X-ray sensitive detector, as depicted in Fig. 6.6. It is assumed that the X-ray detector has a detection efficiency of $\epsilon_{10}$, and generates no secondary quanta. Immediately,

$$\bar{q}_1 = \bar{g}_1 = \epsilon_{10} \bar{q}_0. \tag{6.4.5}$$

Under the Gaussian approximation to Poisson statistics, the variance is found to be

$$\sigma_1^2 = \bar{q}_1 = \epsilon_{10} \bar{q}_0. \tag{6.4.6}$$

The SNR of the input signal is

$$\text{SNR}_{\text{in}} = \frac{\bar{q}_0}{\sqrt{\bar{q}_0}} = \sqrt{\bar{q}_0}, \tag{6.4.7}$$

while the SNR of the output signal is

$$\text{SNR}_{\text{out}} = \frac{\bar{q}_1}{\sqrt{\bar{q}_1}} = \frac{\epsilon_{10} \bar{q}_0}{\sqrt{\epsilon_{10} \bar{q}_0}} = \sqrt{\epsilon_{10} \bar{q}_0}, \tag{6.4.8}$$

so the zero frequency DQE(0) is given by

$$\text{DQE}(0) = \frac{\text{SNR}_{\text{out}}^2}{\text{SNR}_{\text{in}}^2} = \frac{(\sqrt{\epsilon_{10} \bar{q}_0})^2}{(\sqrt{\bar{q}_0})^2} = \frac{\epsilon_{10} \bar{q}_0}{\bar{q}_0} = \epsilon_{10}. \tag{6.4.9}$$

As expected, the DQE(0) in model-I is merely determined by the detection efficiency $\epsilon_{10}$ of the X-ray

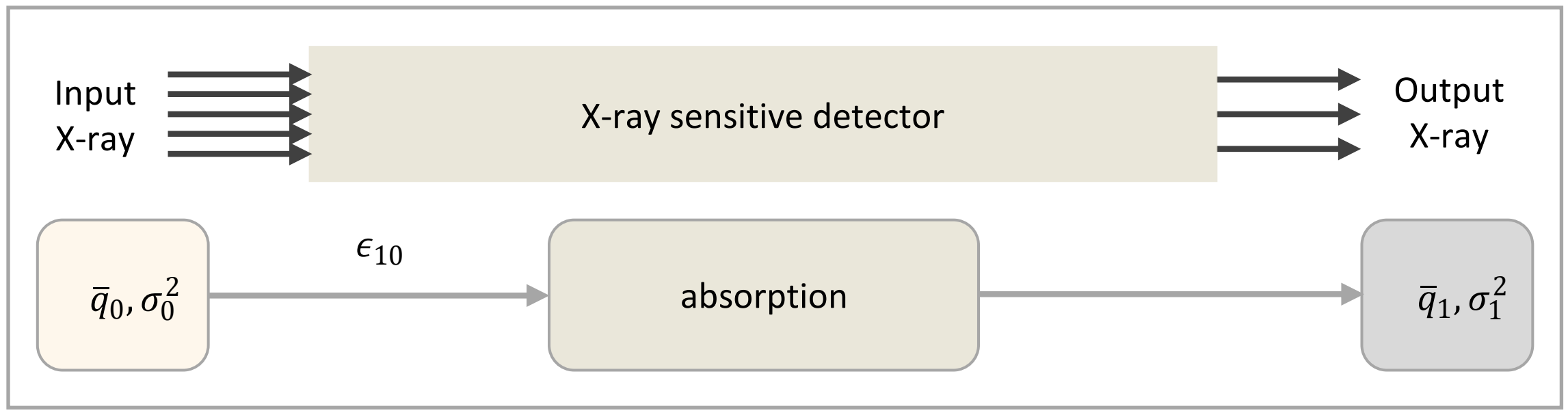

Figure 6.6: Illustration of the cascaded Model-I. In this model, the incident X-ray photons are absorbed by an X-ray sensitive detector. It is assumed that the X-ray detector has a detection efficiency of $\epsilon_{10}$, and generates no secondary quanta.

detector.

Be aware that the derived DQE(0) in Eq. (6.4.9) does not contain additional signal noise, such as dark current noise $\sigma^2_{\text{dark}}$, readout noise $\sigma^2_{\text{read}}$, and fixed pattern noise (FPN) $\sigma^2_{\text{FPN}}$. If denoting the variance of these additional signal noise as $\sigma^2_{add}$, then the DQE(0) in Eq. (6.4.9) can be rewritten as

$$\text{DQE}(0) = \frac{\text{SNR}^2_{\text{out}}}{\text{SNR}^2_{\text{in}}} = \frac{\epsilon_{10}}{1 + \sigma^2_{add}/(\epsilon_{10}\bar{q}_0)}. \tag{6.4.10}$$

where $\sigma^2_{\text{add}} = \sigma^2_{\text{dark}} + \sigma^2_{\text{read}} + \sigma^2_{\text{FPN}}$. The impact of $\sigma^2_{add}$ can be ignored when the detected signal $\epsilon_{10}\bar{q}_0$ is significantly higher than the additional signal noise $\sigma^2_{add}$. However, DQE(0) will be affected primarily at low detected signal value when $\epsilon_{10}\bar{q}_0$ is comparable or even lower than the additional signal noise $\sigma^2_{add}$.

**Model-II:** This model is for the indirect conversion detector[7]. As illustrated in Fig. 6.7, a cascading gain has to be considered. Let $\bar{q}_0$ be the number of incident X-ray photons, and $\epsilon_{10}$ be the fraction that are absorbed by the CsI:TI scintillator layer. Inside CsI:TI, the absorbed X-ray photons are converted into visible light photons. In this stage, it is assumed that one absorbed X-ray photon produces $\bar{\alpha}_{12}$ number of secondary light photon quanta, with a detection efficiency of $\epsilon_{21}$. In the following stage, these visible light photons are converted into electrons in the Si layer of the CCD or CMOS sensor. Similarly, it is assumed that one light photon produces $\bar{\alpha}_{23}$ number of secondary electron quanta, with a detection efficiency of $\epsilon_{32}$. For this particular Markov chain, the mean detected signal $\bar{q}_3$ is given by

$$\bar{q}_3 = \bar{g}_3\bar{g}_2\bar{g}_1 = (\epsilon_{32}\bar{\alpha}_{32})(\epsilon_{21}\bar{\alpha}_{21})(\epsilon_{10}\bar{q}_0). \tag{6.4.11}$$

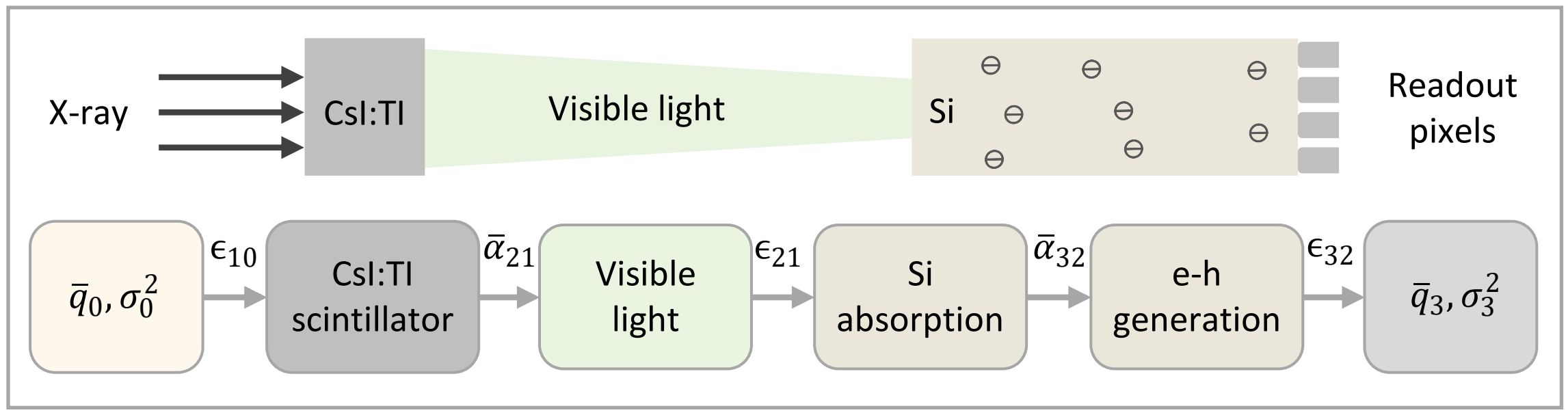

Figure 6.7: Illustration of the cascaded Model-II, which corresponds to the indirect conversion detector. In it, the incident X-ray photons are absorbed by a CsI:TI layer and converted into a certain amount of visible light photons. Afterwards, part of these visible light photons are collected and converted into a certain amount of electrons in the Si layer of the CCD or CMOS sensor.

If assuming that the individual variances are determined by the Gaussian approximation to Poisson statistics, one has

$$\sigma_{g_1}^2 = \epsilon_{10}\bar{q}_0, \tag{6.4.12}$$

$$\sigma_{g_2}^2 = \epsilon_{21}\bar{\alpha}_{21}, \tag{6.4.13}$$

$$\sigma_{g_3}^2 = \epsilon_{32}\bar{\alpha}_{32}. \tag{6.4.14}$$

Based on Eq. (6.4.4), the variance of signal $q_3$ is derived as following:

$$\sigma_3^2 = \bar{g}_3^2(\bar{g}_2^2\sigma_{g_1}^2 + \bar{g}_1\sigma_{g_2}^2) + \bar{g}_2\bar{g}_1\sigma_{g_3}^2. \tag{6.4.15}$$

The SNR of the input signal is

$$\text{SNR}_{\text{in}} = \frac{\bar{q}_0}{\sqrt{\bar{q}_0}} = \sqrt{\bar{q}_0}, \tag{6.4.16}$$

while the SNR of the output signal is

$$\text{SNR}_{\text{out}} = \frac{\bar{q}_3}{\sigma_3} = \sqrt{\frac{\epsilon_{32}\bar{\alpha}_{32}\epsilon_{21}\bar{\alpha}_{21}\epsilon_{10}\bar{q}_0}{1 + \epsilon_{32}\bar{\alpha}_{32} + \epsilon_{32}\bar{\alpha}_{32}\epsilon_{21}\bar{\alpha}_{21}}}, \tag{6.4.17}$$

so the zero frequency DQE(0) is given by

$$\text{DQE}(0) = \frac{\text{SNR}_{\text{out}}^2}{\text{SNR}_{\text{in}}^2} = \frac{\epsilon_{10}}{1 + 1/(\epsilon_{21}\bar{\alpha}_{21}) + 1/(\epsilon_{32}\bar{\alpha}_{32}\epsilon_{21}\bar{\alpha}_{21})}. \tag{6.4.18}$$

If considering the influence of the additional signal noise $\sigma_{add}^2$, the DQE(0) in Eq. (6.4.18) is

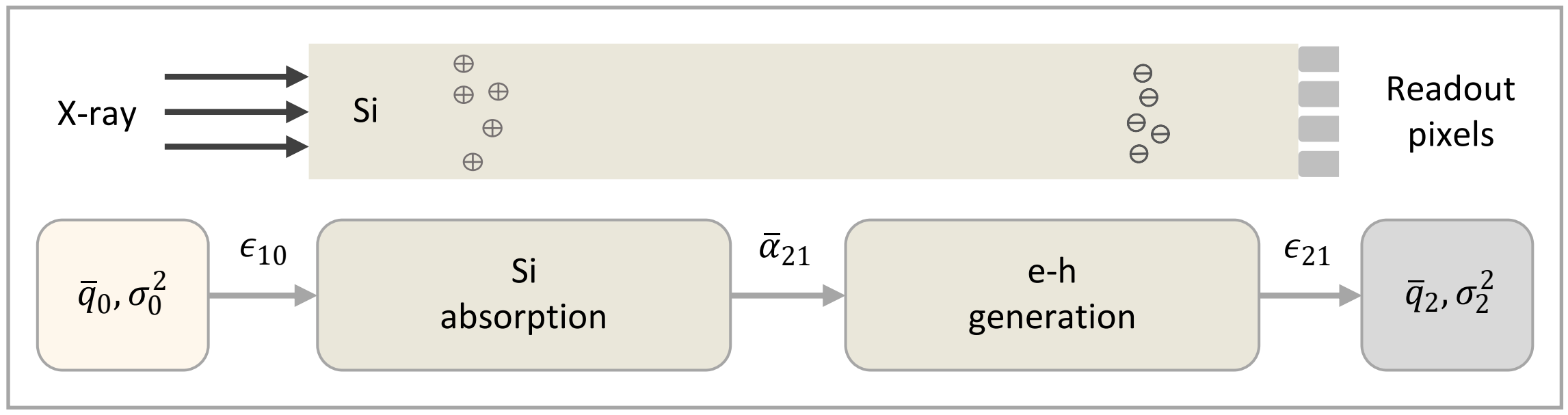

Figure 6.8: Illustration of the cascaded Model-III, which corresponds to the direct conversion detector. In this model, the incident X-ray photons are absorbed by the Si layer of the CCD or CMOS sensor, and converted into a certain amount of electrons.

re-expressed as

$$\mathrm{DQE}(0) = \frac{\mathrm{SNR}_{\mathrm{out}}^2}{\mathrm{SNR}_{\mathrm{in}}^2} = \frac{\epsilon_{10}}{1 + 1/(\epsilon_{21}\bar{\alpha}_{21}) + 1/(\epsilon_{32}\bar{\alpha}_{32}\epsilon_{21}\bar{\alpha}_{21}) + \sigma_{add}^2/(\epsilon_{32}^2\bar{\alpha}_{32}^2\epsilon_{21}^2\bar{\alpha}_{21}^2\epsilon_{10}\bar{q}_0)}. \tag{6.4.19}$$

**Model-III:** This model is for the direct conversion detector[7]. As illustrated in Fig. 6.8, similarly, a two-stage cascading gain procedure is considered. Let $\bar{q}_0$ be the number of incident X-ray photons, and $\epsilon_{10}$ be the fraction that are absorbed by the Si layer of the CCD or CMOS sensor. Inside Si, the absorbed X-ray photons are converted directly into electrons. In this stage, it is assumed that one absorbed X-ray photon produces $\bar{\alpha}_{12}$ number of secondary electron quanta, with a detection efficiency of $\epsilon_{21}$. For this particular Markov chain, the mean detected signal $\bar{q}_2$ is given by

$$\bar{q}_2 = \bar{g}_2\bar{g}_1 = (\epsilon_{21}\bar{\alpha}_{21})(\epsilon_{10}\bar{q}_0). \tag{6.4.20}$$

Similarly, assuming that the individual variances are determined by the Gaussian approximation to Poisson statistics, again, one obtains

$$\sigma_{g_1}^2 = \epsilon_{10}\bar{q}_0, \tag{6.4.21}$$

$$\sigma_{g_2}^2 = \epsilon_{21}\bar{\alpha}_{21}. \tag{6.4.22}$$

Based on Eq. (6.4.4), the variance of signal $q_2$ is derived as follows:

$$\sigma_2^2 = \bar{g}_2^2\sigma_{g_1}^2 + \bar{g}_1\sigma_{g_2}^2. \tag{6.4.23}$$

The SNR of the input signal is derived as

$$\mathrm{SNR_{in}} = \frac{\bar{q}_0}{\sqrt{\bar{q}_0}} = \sqrt{\bar{q}_0}, \tag{6.4.24}$$

while the SNR of the output signal is derived as

$$\mathrm{SNR_{out}} = \frac{\bar{q}_2}{\sigma_2} = \sqrt{\frac{\epsilon_{21}\bar{\alpha}_{21}\epsilon_{10}\bar{q}_0}{1+\epsilon_{21}\bar{\alpha}_{21}}}, \tag{6.4.25}$$

as a consequence, the zero frequency DQE(0) is given by

$$\mathrm{DQE}(0) = \frac{\mathrm{SNR_{out}^2}}{\mathrm{SNR_{in}^2}} = \frac{\epsilon_{10}}{1+1/(\epsilon_{21}\bar{\alpha}_{21})}. \tag{6.4.26}$$

If considering the influence of the additional signal noise $\sigma^2_{add}$, the DQE(0) in Eq. (6.4.26) is rewritten as

$$\mathrm{DQE}(0) = \frac{\mathrm{SNR_{out}^2}}{\mathrm{SNR_{in}^2}} = \frac{\epsilon_{10}}{1+1/(\epsilon_{21}\bar{\alpha}_{21}) + \sigma^2_{add}/(\epsilon^2_{21}\bar{\alpha}^2_{21}\epsilon_{10}\bar{q}_0)}. \tag{6.4.27}$$

### 6.4.2 Quantitative comparisons

Indirect and direct conversion X-ray detectors were two available detection systems in our laboratory, see Fig. 6.9. As a result, quantitative comparisons between these two types of detectors were performed. To do so, the aforementioned cascaded signal models were employed to analyze the signal generation, noise propagation, and DQE. Experiments were conducted on the our 5.4 keV laboratory X-ray microscope with FZP.

Specifically, the indirect conversion detector had 50.0 $\mu$m thick CsI:TI scintillator layer, and the direct detector had 8.0 $\mu$m thick Si absorption layer, respectively, see Table 6.3 for more detailed specifications. In addition, the indirect and direct detectors are mounted on the same linear translation stage and shared a common active imaging plane, see Fig. 6.9. Moreover, the conversion coefficient between the analog-to-digital unit (ADU) and the number of electrons is found to be 1.1 ADU/$e^-$ for the indirect detector, and 2.3 ADU/$e^-$ for the direct detector, respectively.

#### 6.4.2.1 Signal responses

During the data acquisitions, the exposure time was fixed at 60 s for both detectors. Dark signals were measured under the same exposure time, but without X-ray exposures. The measurements were repeated five times to minimize statistical errors, and the resulting signal distributions are shown in Fig. 6.10.

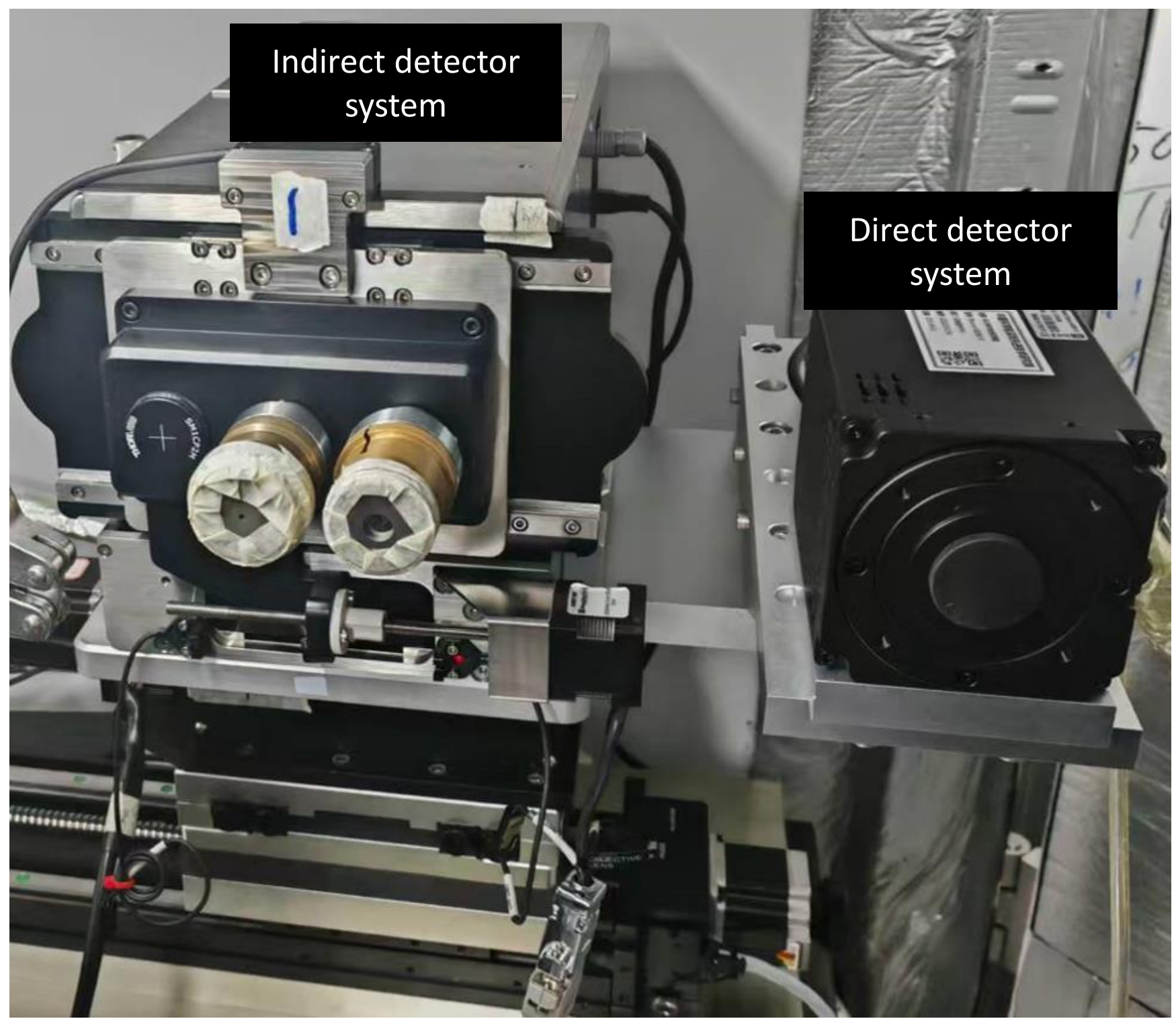

Figure 6.9: Photograph of the indirect detector and direct detector installed on our X-ray microscopy system. The two detectors are mounted on the same motorized mechanical stages. During the experiments, they can be easily switched via linear movement.

For the indirect conversion X-ray detector, it has: $\epsilon_{10}^{\text{ind}} \approx 1.0$, $\bar{\alpha}_{21}^{\text{ind}} = 54$ photons/keV, $\epsilon_{21}^{\text{ind}} = 0.7 \times 0.05$, where 0.7 denotes the visible light transmittance within the CsI:TI scintillator crystal [80] and 0.05 corresponds to the collection efficiency of the 20× objective lens. Additionally, the emission spectrum of CsI:TI is peaked at 550 nm [81], corresponding to a photon energy of 2.25 eV, yielding an electron-hole conversion ratio $\bar{\alpha}_{32}^{\text{ind}} = 2.25/3.65 = 0.62$ with a CCD detection efficiency $\epsilon_{32}^{\text{ind}} = 0.95$. From the measured mean output signal $\bar{q}_{\text{out}}^{\text{ind}} = 936\,\text{e}^-$, using Eq. (6.4.11), the incident X-ray photon number was estimated to be $\bar{q}_{\text{in}} = 155$. This corresponds to a beam flux of 2.58 X-ray photon counts per second (cps) per 6.5 $\mu$m × 6.5 $\mu$m area.

For the direct conversion X-ray detector, it has: $\epsilon_{10}^{\text{d}} = 0.3$, $\epsilon_{21}^{\text{d}} = 0.95$, and $\bar{\alpha}_{21}^{\text{d}} = 274\,\text{e}^-/\text{keV}$. The measured mean output signal $\bar{q}_{\text{out}}^{\text{d}} = 70060\,\text{e}^-$, corresponding to the incident X-ray photon number of $\bar{q}_{\text{in}} = 166$ using Eq. (6.4.20). This corresponds to a beam flux of 2.77 X-ray photon counts per second (cps) per 6.5 $\mu$m × 6.5 $\mu$m area.

Table 6.3: Specifications of the indirect conversion detector and direct conversion detector.

| Parameter | Indirect | Direct |
|---|---|---|
| X-ray energy | 5.4 keV | 5.4 keV |
| Conversion material | CsI:Tl | Silicon |
| Material thickness | 50 $\mu$m | 8 $\mu$m |
| $\varepsilon_{10}$ | 1.0 | 0.3 |
| Conversion ratio | 54 photons/keV | 274 $e^-$/keV |
| NA of lens | 0.5 | — |
| Lens magnification | 20× | — |
| Steradian ratio | 0.2 | 1.0 |
| Pixel size | 13.5 $\mu$m | 6.5 $\mu$m |
| Bit depth | 16 | 12 |
| Full well depth | $10^5$ | 1500 |
| Camera | Andor iKon-L 936 | Tucsen 400BSI |

As seen, the incident X-ray photons estimated independently from the indirect and direct detectors show high consistency, demonstrating the viability of the proposed cascaded gain signal model.

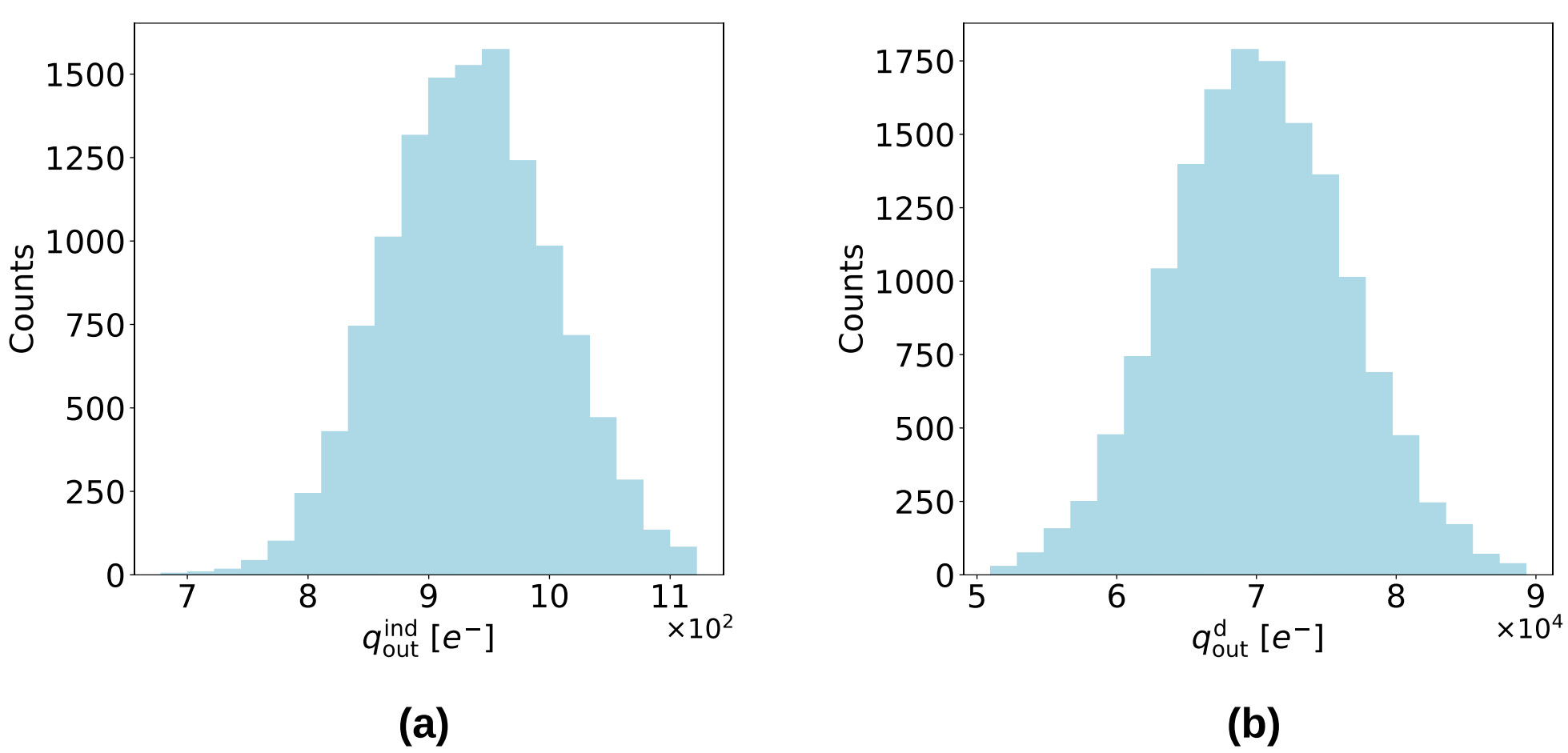

Figure 6.10: Distributions of the output signal acquired with 60 seconds exposure for the indirect (a) and direct (b) conversion detectors, respectively.

#### 6.4.2.2 Noise responses

The dark current responses were measured over a wide range of exposure times, see Fig. 6.11(a)-(b). Specifically, ten dark frames were acquired, baseline-subtracted, and linearly fitted to obtain the dark current coefficient $c_{\text{dark}}$. It is found that $c_{\text{dark}}^{\text{ind}} = 0.36\,e^{-}/\text{s}$ for the indirect detector, and $c_{\text{dark}}^{\text{d}} = 2.72\,e^{-}/\text{s}$ for the direct detector. In addition, the dark noise was approximated as $\sigma_{\text{dark}} = \sqrt{c_{\text{dark}}\,t}$ [82, 83].

In order to measure the readout noise $\sigma_{\text{read}}$, the exposure time was set at the shortest possible value to minimize the time-dependent impacts. The measured readout noise was $5.5\,e^{-}$ for the indirect detector (100 ms exposure time) and was $1.6\,e^{-}$ for the direct detector (1 ms exposure time). Clearly, the readout electronic noise was sufficiently small, therefore, it was neglected.

Results of $\bar{q}_{\text{out}}^2/\sigma_{\text{total}}^2$ with respect to exposure time were plotted in Fig. 6.11(c). As seen, the ratios of both detectors increased with exposure time, while the indirect detector exhibited higher values than the direct detector, indicating superior imaging performance. Moreover, results of $\bar{q}_{\text{out}}^2/(\sigma_{\text{total}}^2 - \sigma_{\text{FPN}}^2)$ were plotted in Fig. 6.11(d). Herein, the noise $\sigma_{\text{total}}^2 - \sigma_{\text{FPN}}^2$ was obtained from the variance of the difference between two adjacent averaged frames and divided by $\sqrt{2}$. In addition, results of $\bar{q}_{\text{out}}^2/\sigma_{\text{FPN}}^2$ shown in Fig. 6.11(e) are nearly independent to exposure time, indicating that $\sigma_{\text{FPN}}$ is proportional to signal level [84].

Based on the measured noise distributions, the ratio of DQE(0) between the indirect and direct conversion X-ray detectors were estimated, as shown in Fig. 6.11(f), with and without $\sigma_{\text{add}}$, respectively. Results indicated that $\sigma_{\text{add}}$ has negligible impact. Most importantly, the measured ratios agree well with the theoretically predicted ratios of about 2.1, meaning that the utilized indirect detector is superior to the utilized direct detector. Note that these ratios were obtained from a direct detector with 8.0 $\mu$m Si absorption layer. For the direct detector with 50.0 $\mu$m Si absorption layer, the ratio of DQE(0) between the indirect and direct conversion X-ray detectors would approximately become 0.7, meaning that the indirect detector gets inferior to the direct detector.

#### 6.4.2.3 CNR measurements

The CNR [85] is defined as:

$$\text{CNR} = \frac{\left|\bar{I}_{\text{obj}} - \bar{I}_{\text{bkg}}\right|}{\sqrt{\sigma_{\text{bkg}}^2 + \sigma_{\text{obj}}^2}}, \tag{6.4.28}$$

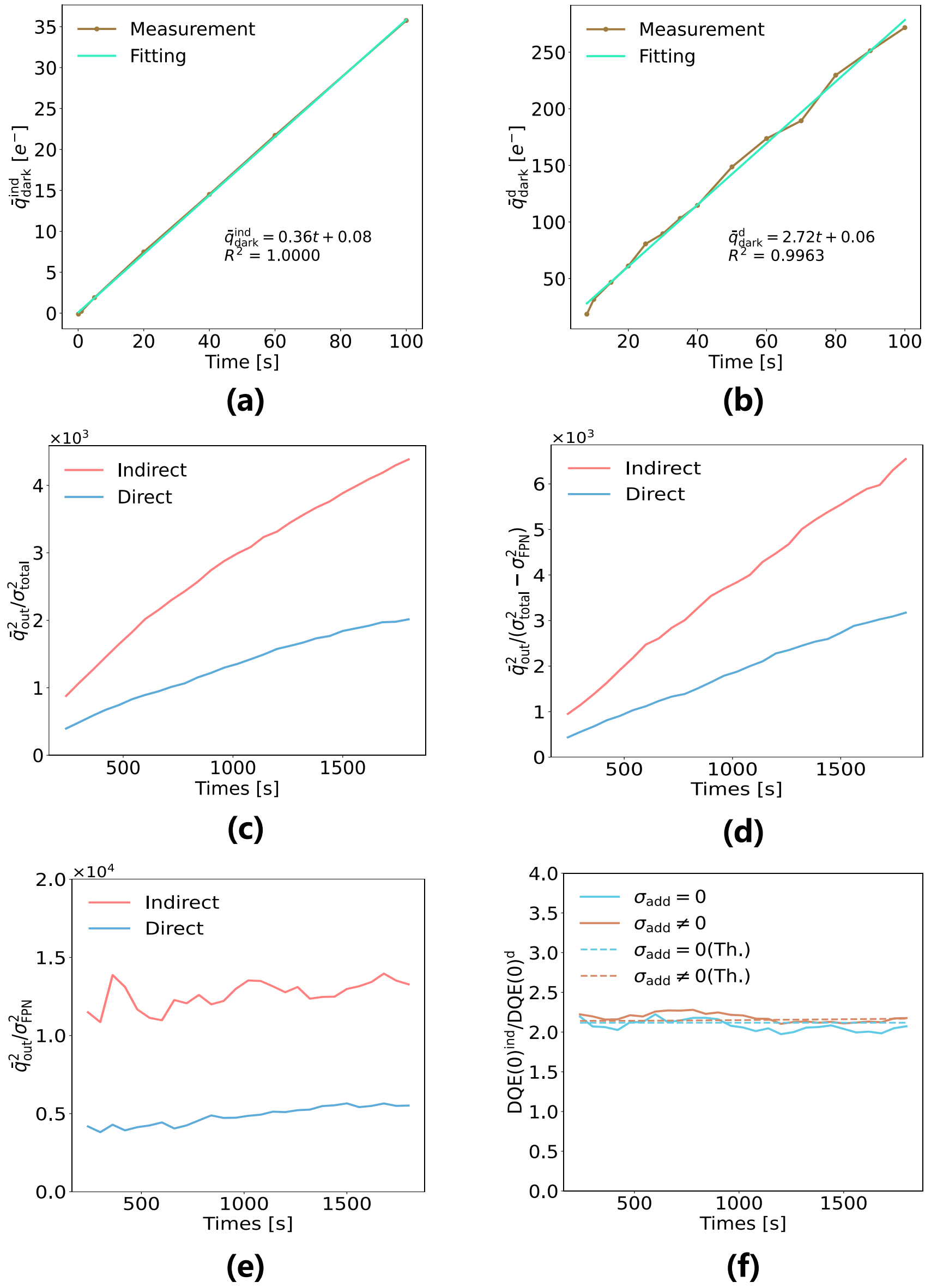

Figure 6.11: The measured dark current responses of the indirect (a) and direct (b) detectors. The ratio $\bar{q}^2_{\mathrm{out}}/\sigma^2$ with respect to exposure time for different noise conditions: $\sigma^2_{\mathrm{total}}$ (c), $\sigma^2_{\mathrm{add}}$ (d), and $\sigma^2_{\mathrm{FPN}}$ (e). The ratio $\mathrm{DQE}(0)^{\mathrm{ind}}/\mathrm{DQE}(0)^{\mathrm{d}}$ with respect to exposure time are plotted in (f) with or without considering $\sigma_{\mathrm{add}}$. The theoretically predicted values are plotted in dashed lines.

where $\bar{I}_{\mathrm{obj}}$ and $\sigma_{\mathrm{obj}}$ denote the mean attenuation value of the 0.4 $\mu$m thick Au and 0.2 $\mu$m thick silicon nitride ($Si_3N_4$) substrate and the standard deviation of the selected ROI, respectively. In addition, $\bar{I}_{\mathrm{bkg}}$ and $\sigma_{\mathrm{bkg}}$ denote the mean attenuation value of the $Si_3N_4$ substrate and the standard deviation of the selected ROI, respectively.

The experimental CNR comparison results are shown in Fig. 6.12. For 60 s exposure time, the indirect detector achieves $\mathrm{CNR}^{\mathrm{ind}} = 3.0$, outperforming the direct detector ($\mathrm{CNR}^{\mathrm{d}} = 2.7$) having 8.0 $\mu$m thick Si layer. We believe this is due to the limited X-ray absorption efficiency (about 30%) of the 8 $\mu$m thick Si layer. To mimic the CNR performance of a direct detector having 50.0 $\mu$m thick Si layer, whose X-ray absorption efficiency increases up to 90%, results corresponding to a threefold increased exposure time, namely, 180 s, was generated. Immediately, the predicted CNR becomes 4.5, substantially surpassing the indirect detector. Results confirmed that the direct conversion detector is superior to the indirect conversion detector, provided that the Si layer has high enough absorption efficiency ($\geq$90%).

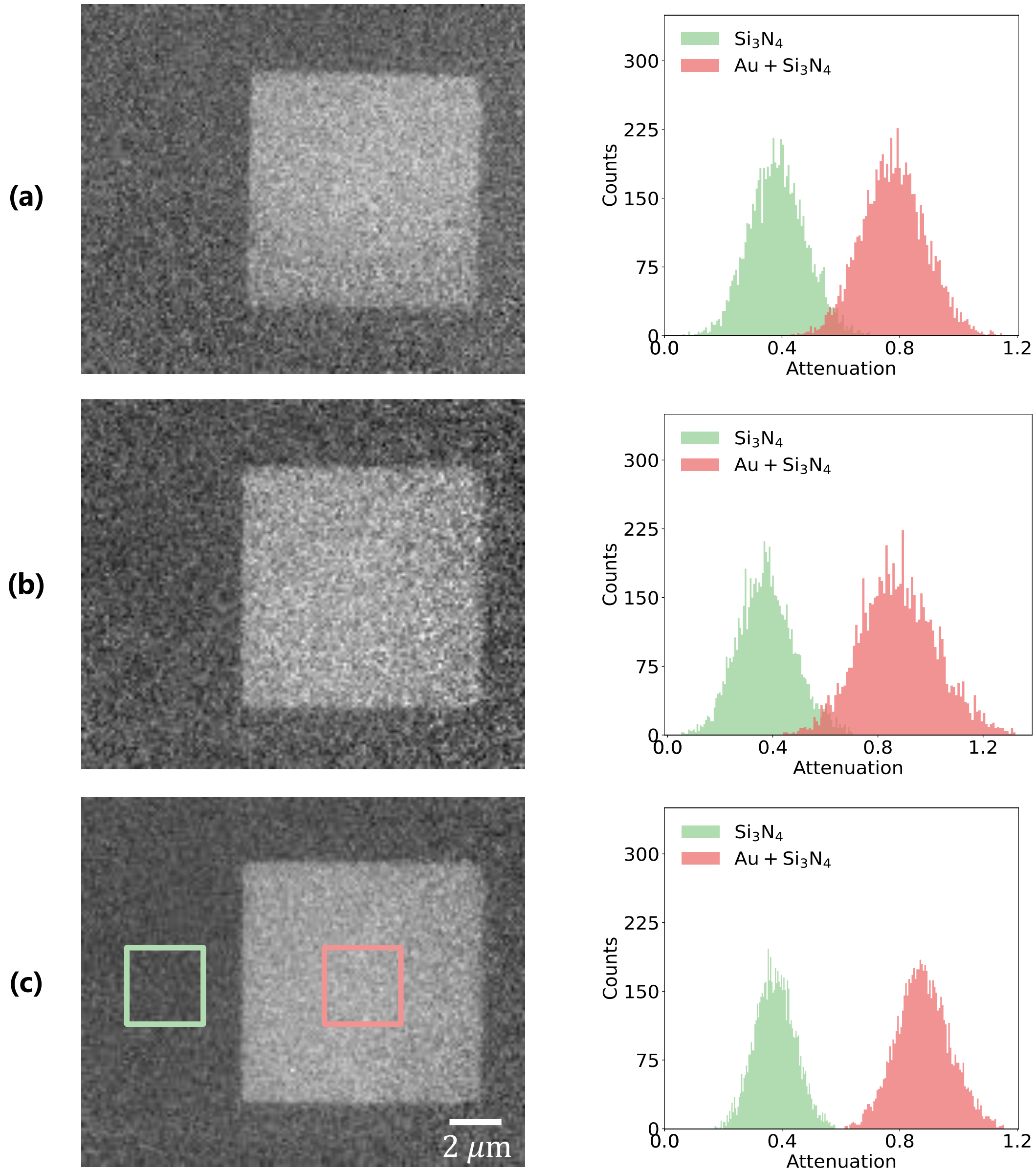

Figure 6.12: The CNR comparison results obtained from (a) indirect CsI detector with 60 s exposure time, (b) 8.0 $\mu$m Si direct detector with 60 s exposure time, and (c) 8.0 $\mu$m Si direct detector with 180 s exposure time (mimic the 50.0 $\mu$m Si direct detector). Distributions of the projection line integrals for the background 0.2 $\mu$m thick $Si_3N_4$ substrate and the 0.4 $\mu$m thick Au are plotted for the selected region-of-interests (ROIs). The scale bar denotes 2.0 $\mu$m.

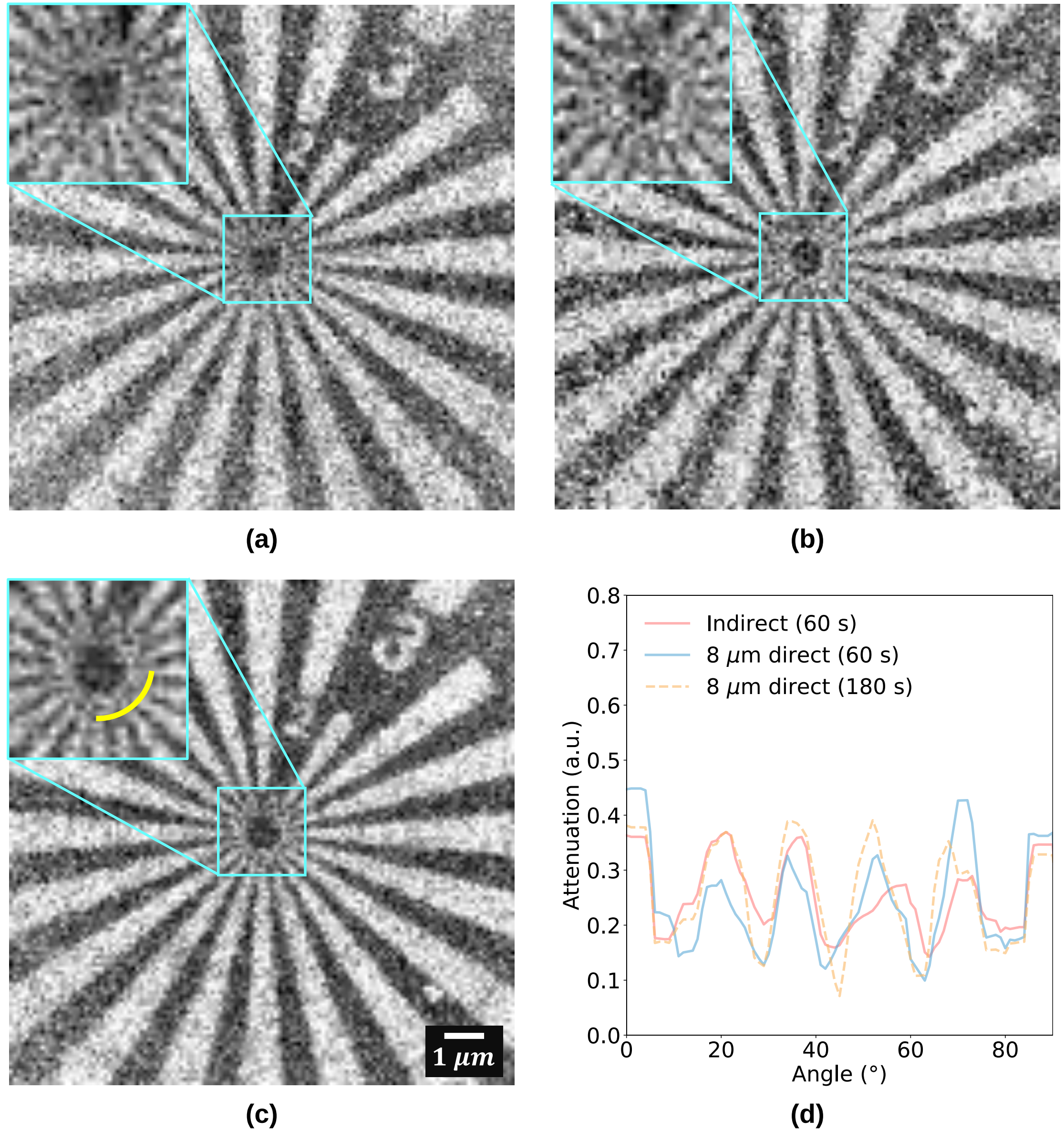

Figure 6.13: Imaging results obtained from (a) indirect CsI detector with 60 s exposure time, (b) 8.0 $\mu$m Si direct detector with 60 s exposure time, and (c) 8.0 $\mu$m Si direct detector with 180 s exposure time (mimic the 50.0 $\mu$m Si direct detector). The radial profiles along the highlighted yellow line are compared in (d). The scale bar denotes 1.0 $\mu$m.

#### 6.4.2.4 Imaging results

A standard Siemens star resolution pattern was imaged at 60 s exposure time. As shown in Fig. 6.13, the indirect detector outperforms the 8.0 $\mu$m Si direct detector in resolving the innermost 100 nm wide spikes. This is because the 8.0 $\mu$m Si only absorbs 30% of the total X-ray photons. However, increasing the Si layer thickness to 50.0 $\mu$m would significantly enhance the image quality, see Fig. 6.13(c). As a result, the direct detector with 50.0 $\mu$m Si absorption layer best delineated the innermost spikes having 100 nm period, see the corresponding sharp line profiles in Fig. 6.13(d). As a result, it is recommended to select direct conversion X-ray detector in the future to generate better image quality whenever it is possible.

### 6.4.3 Super resolution

With a direct conversion detector, it becomes possible to perform super resolution imaging, whose effective spatial resolution is smaller than the native physical pixel dimension. This is particularly viable when the detector is working under the photon counting mode [86], in which an individual X-ray photon can be identified at the time that they hit the detector. A strong requirement is that the generated electron cloud must be larger than the pixel size of the detection sensor. As a result, each incoming X-ray photon can deposit signal in a small cluster of pixels, see Fig. 6.14(b). Moreover, high-speed electronics are needed to recognize each X-ray photon event at fast enough speed, e.g., 1000 fps, to record all the responses of these interacted pixels. Approximately, the 2D coordinate of the response center, highlighted with orange dot in Fig. 6.14(c), is estimated by applying the standard Bi-linear interpolation algorithm, in which the center of the four pixels are used as the four vertices of the dashed orange square in Fig. 6.14(c). After doing two independent linear interpolations along the horizontal and vertical directions, the 2D coordinate of the X-ray hit can be estimated with sub-pixel precision, as shown in Fig. 6.14(d). Subsequently, the effective pixel dimension is reduced by half compared to the native pixel dimension. Namely, a 2$\times$ effective improvement of spatial resolution (far beyond the physical Nyquist limit) can be achieved along each image direction.

When the super resolution imaging technique is applied, the total system length of a laboratory X-ray microscope with FZP can be further shortened, assuming that the same FZP magnification and targeting spatial resolution are kept. In other words, the super resolution imaging technique is important for building compact laboratory X-ray microscope with FZP.

In the future, as the X-ray direct detection technology quickly develops and evolves, it is very prospective to implement the super resolution technique on a laboratory X-ray microscope with FZP. As a consequence, the achievable spatial resolution limit can be further reduced.

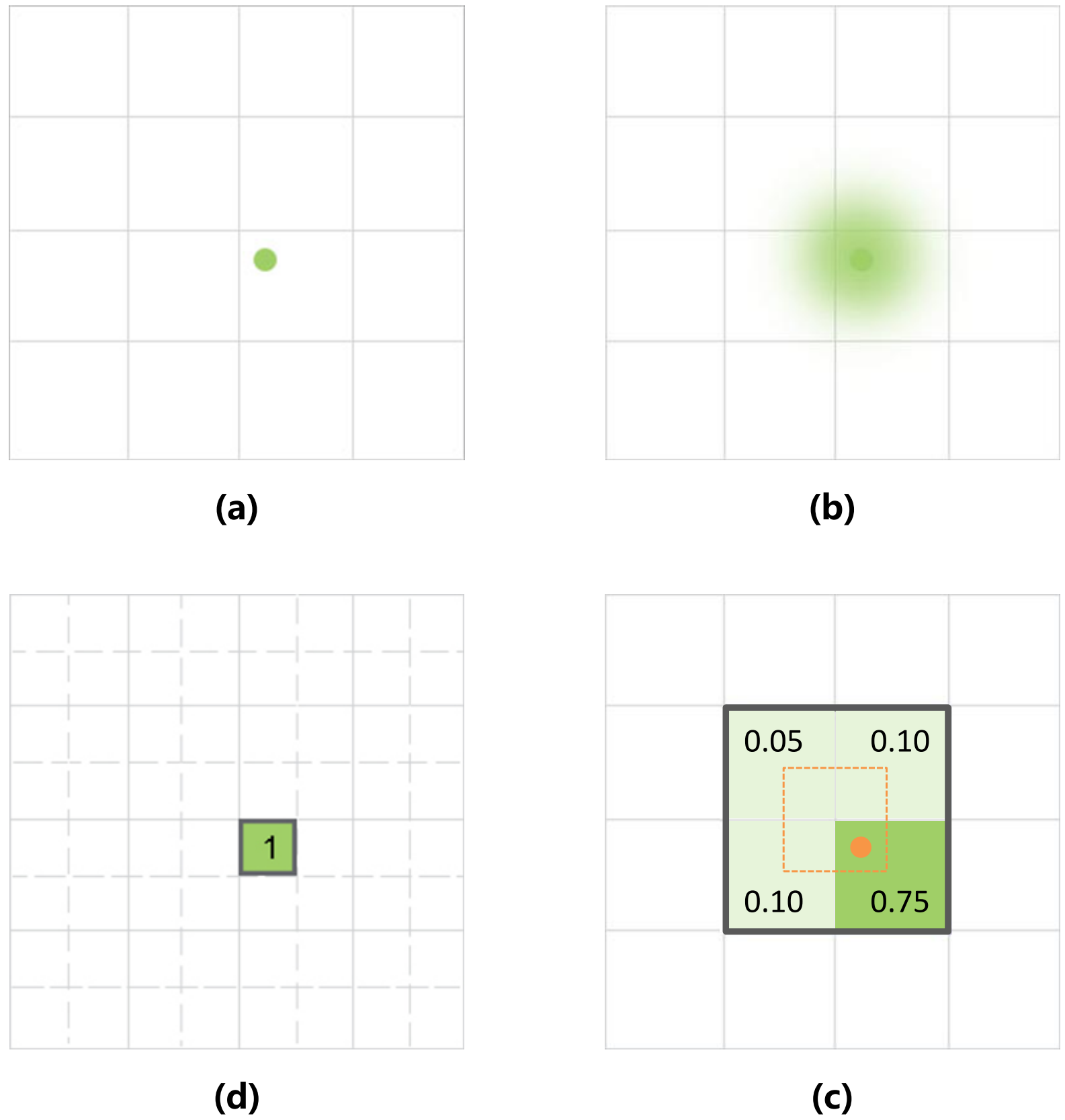

Figure 6.14: Illustration of the super resolution procedures with a direct conversion X-ray detector operating under the photon counting mode. (a) X-ray photon enters the silicon detector, (b) The X-ray stimulated electron cloud is diffused, (c) Charge signals are recorded in a cluster of neighboring pixels, (d) After estimation, the 2D position of the X-ray hit is localized to sub-pixel (half size of the native pixel dimension) accuracy.

# 7 Mechanical motion system

Like the skeletal system in our human body, the mechanical motion system in a zone-plate based X-ray microscope acts as a rigid, interconnected framework designed to support heavy components and guide precise, repeatable position adjustments through its various motion stages. In fact, the mechanical motion system plays a critical role in realizing the defined and designed imaging function of all the optical components by providing high-precision positioning, alignment, and adjustment. For instance, it enables accurate placement of X-ray source, condenser, sample, zone-plate, gratings, detector, and other important optical elements. In addition, it also supports repeated searching for the optimal imaging geometry, and facilitates fine alignment to maximize the imaging capability, e.g., achieving the highest spatial resolution.

For every single optical component in the X-ray microscope, at least a three dimensional mechanical motion system that translates along the x, y, and z axes with certain travel distance and precision is needed, see Table 7.1. The definitions of the x, y and z axes are depicted in Fig. 3.6. Specifically, the y axis is along the vertical direction, and the z axis is along the horizontal direction. In practice, the condenser may need mechanical rotation, i.e., pitch and yaw, motion system to adjust its orientation in space. Moreover, the grating may also need mechanical rotation, i.e., roll, motion system to accurately align the grating grooves along the vertical direction.

As mentioned, these mechanical motion stages should have the proper travel distance and precision to ensure the overall satisfactory imaging performance while saving the total expense. During the selection procedures, it is quite essential to choose reputable, widely accepted brands and conduct thorough on-site testings of their key specifications such as load capacity, repeated motion accuracy, anti-vibration capability, and so on. If possible, please consider to select all these motion stages from one supplier, or at least choose them from the least number of different suppliers. This can significantly reduce the efforts required for developing the motion control software and integrating them during the system assembling stages. In addition, it also simplifies troubleshooting, calibration, and maintenance when issues arise, leading to a more efficient and reliable prototype system.

NEVER to select mechanical motion stages from unproven and inexperienced start-up companies! Based on our unfortunate experiences, such decisions often leads to horrible integration challenges, reliability issues, and unexpected delivery delays. In the worst-case scenario, it can derail the entire project and force one to start over from the very beginning. Please do not waste valuable time, funding, and effort on these avoidable mistakes. Choosing well-established mechanical motion stage suppliers with validated products is always the safer and more cost-effective decision in the long run.

## 7.1 Motion system design

### 7.1.1 Motion axis classification

All motion axes are classified into two tiers:

**Ultra-precision axes**: they are critical for the source grating, sample, FZP and phase grating, which are important in achieving nano-resolution imaging and phase-contrast imaging. These motion axes need to have sub-micrometer resolution and excellent repeatability to avoid any positioning errors that could directly degrade the spatial resolution or fringe visibility. Usually, these high precision hardware are expensive.

**Precision axes**: they are required for the X-ray source, condenser, Bertrand and detector, which are important in adjusting the orientation and positions of the X-ray beam. Usually, cost-effective stepper-driven motion stages are enough.

### 7.1.2 Ultra-precision axes

**Piezo motor.** Enables nanometer linear resolution or micro-degree angular resolution with minimal self-vibration. For instance, they can be used for the ultra-precision translation axis and the roll axis of both gratings, where fine adjustment is essential in generating high fringe visibility.

**High-precision stepper motor.** Balances moderate travel range with sub-micrometer resolution. They are employed for the sample, FZP and the remaining translation axes of the gratings.

**Air-bearing rotary stage.** Achieves angular accuracy better than $0.001°$ in a frictionless condition. Configured for the sample rotation (roll) axis to meet the stability demands of CT acquisition. Note that this stage requires a dedicated supply of dry compressed air.

Examples of ultra-precision motion stages used in the system are shown in Fig. 7.1. The stage configuration for each ultra-precision component is described below.

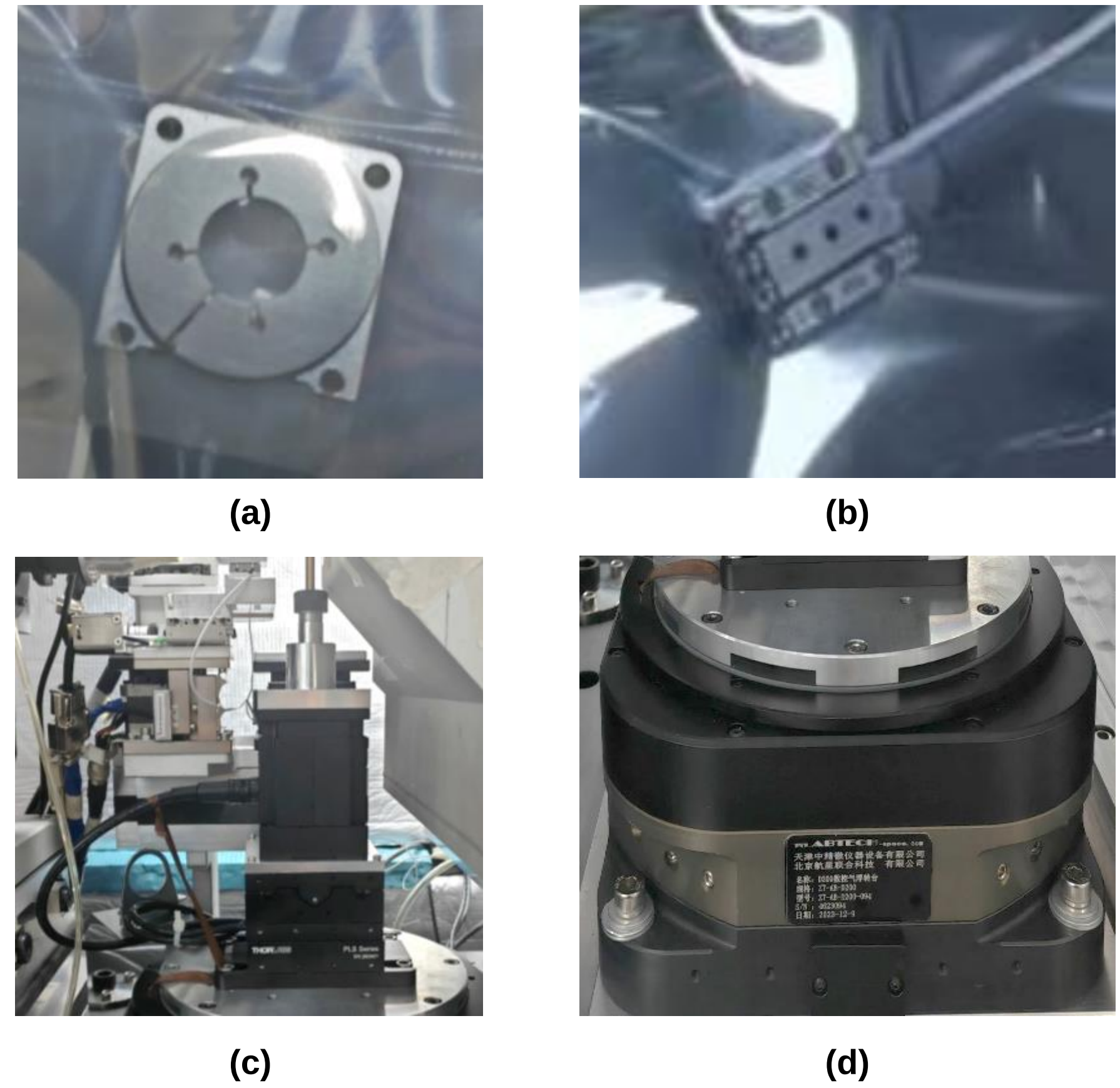

Figure 7.1: Photographs of the main ultra-precision motion stages used in our X-ray microscope: (a) grating rotation axis (Model: SR-1908, Smaract GmbH, Germany), (b) grating translation axis (Model: SLC-1720-L, Smaract GmbH, Germany), (c) sample $x - z$ translation (horizontal plane) stages (Model: PLSXY, Thorlabs, Inc., USA) in the bottom and $y$ translation (vertical direction) stage (Model: OSMS60-10ZF, OptoSigma, Japan) on the top, (d) air-bearing rotary stage (Model: ZTAB-D200, ZJW Instruments, China).

**Source grating:** The source grating is one of the key component of the Talbot-Lau interferometer. Both of its $x$-axis and roll axis are driven by Piezo motors with linear-scale feedback, while the $y$ and $z$ axes are driven by high-precision stepper motors. Results found that even a nanometer level displacement or a microradian rotation error may degrade the visibility of the interference fringe.

**Sample stage:** The sample is carried by a three-axis stepper-driven translation stage plus one air-bearing rotary stage. It is recommended that the $z$ axis stage be equipped with linear scales to enhance the repeatability. The air-bearing rotary stage ensures smooth, wobble-free rotation during CT data acquisition.

**Zone plate:** As the objective lens of the microscope, the FZP has the most restrict positioning requirement among all optical components. Since the depth of field is only about 20 $\mu$m, therefore, all three translation axes have to be driven by ultra-precision stepper motors with linear-scale feedback to avoid positioning errors. The entire alignment procedure is discussed in details in Section 5.2.3.

**Phase grating:** The phase grating shares the similar configuration as of the aforementioned source grating. Namely, stepper-driven translation axes and a Piezo motor roll axis. The source grating and phase grating together form the Talbot-Lau interferometer to allow multi-contrast imaging (absorption contrast, phase contrast, and dark-field contrast) in this microscope.

### 7.1.3 Precision axes

Precision axes, whose positional tolerances are much relaxed by 1∼2 of orders compared with the ultra-precision axes, are mainly utilized among the following components:

**X-ray Source:** The X-ray source is attached on a three-axis stepper motion stage, which is used for coarse alignment of the system optical path, for instance, alignment of the source–condenser.

**Condenser:** The condenser is mounted on a five-axis stage comprising three stepper-driven translation axes and two stepper-driven rotation axes (yaw and pitch) for orientation adjustment in space.

**Bertrand Lens:** The Bertrand lens is used to image the rear focal plane of the main FZP when optimizing the position of the phase ring or the phase grating. Since its depth of field is much deeper than that of the main FZP, so stepper-driven translation axes are enough.

**Detector:** The detector module is equipped with three stepper driven translation axes, who are used to adjust the scintillator imaging plane onto the designed imaging plane and the field-of-view.

Table 7.1: The recommended specifications of the mechanical motion stages with minimum configuration requirements.

| Component | Motion | Range (mm) | Precision ($\mu$m) | Notes |
|---|---|---|---|---|
| X-ray source | x | $\geq$50 | $\leq$5 | Stepper, closed-loop |
| | y | $\geq$50 | $\leq$5 | Stepper, closed-loop |
| | z | $\geq$50 | $\leq$5 | Stepper, closed-loop |
| Condenser | x | $\geq$5 | $\leq$3 | Stepper, closed-loop |
| | y | $\geq$5 | $\leq$3 | Stepper, closed-loop |
| | z | $\geq$10 | $\leq$5 | Stepper, closed-loop |
| | yaw | $\geq$5° | $\leq$0.001° | Stepper, closed-loop |
| | pitch | $\geq$5° | $\leq$0.001° | Stepper, closed-loop |
| Source grating | x | $\geq$10 | $\leq$0.01 | Piezo, closed-loop, linear scales |
| | y | $\geq$20 | $\leq$1 | Stepper, closed-loop |
| | z | $\geq$20 | $\leq$1 | Stepper, closed-loop |
| | roll | $\geq$ 360° | $\leq$ 0.00001° | Piezo, open-loop |
| Sample | x | $\geq$20 | $\leq$0.3 | Stepper, closed-loop |
| | y | $\geq$20 | $\leq$0.5 | Stepper, closed-loop |
| | z | $\geq$20 | $\leq$0.3 | Stepper, closed-loop, linear scales |
| | roll | $\geq$ 360° | $\leq$ 0.001° | Stepper/Air bearing, closed-loop |
| Zone-plate | x | $\geq$20 | $\leq$0.1 | Stepper, closed-loop, linear scales |
| | y | $\geq$20 | $\leq$0.1 | Stepper, closed-loop, linear scales |
| | z | $\geq$20 | $\leq$0.1 | Stepper, closed-loop, linear scales |
| Phase grating | x | $\geq$10 | $\leq$0.01 | Piezo, closed-loop, linear scales |
| | y | $\geq$20 | $\leq$1 | Stepper, closed-loop |
| | z | $\geq$20 | $\leq$1 | Stepper, closed-loop |
| | roll | $\geq$ 360° | $\leq$ 0.00001° | Piezo, open-loop |
| Bertrand | x | $\geq$50 | $\leq$5 | Stepper, closed-loop |
| | y | $\geq$50 | $\leq$5 | Stepper, closed-loop |
| | z | $\geq$50 | $\leq$5 | Stepper, closed-loop |
| Detector | x | $\geq$150 | $\leq$5 | Stepper, closed-loop |
| | y | $\geq$50 | $\leq$5 | Stepper, closed-loop |
| | z | $\geq$50 | $\leq$5 | Stepper, closed-loop |

# 8 X-ray phase contrast imaging

X-ray phase contrast imaging is particularly important for imaging low-density and low-atomic number objects. It relies on detecting variations in the phase shift of X-rays induced by their interaction with the imaged object, which arises from the wave nature of X-rays and their wave-particle duality. Unlike conventional X-ray attenuation imaging, in which the imaging signal is directly encoded into the X-ray intensity, the phase shift induced by the object usually cannot be measured directly. Instead, the phase information must first be converted into measurable intensity variations, from which the phase shift can subsequently be retrieved. To do so, specialized X-ray optical components and imaging methods are required to encode and modulate the phase information.

For FZP based X-ray microscopy, there have two major approaches to encode the X-ray phase information: the Zernike phase contrast imaging method [87, 88], and the grating interferometer based phase contrast imaging method [89–91].

## 8.1 Zernike phase contrast imaging

### 8.1.1 Combing Zernike phase ring with FZP

By far, the most widely applied phase contrast imaging approach for X-ray microscopy with FZP is based on Zernike phase contrast imaging, which has been well established in visible light microscopy by Frits Zernike[87, 88]. The Zernike phase contrast imaging in optical microscopy has two fundamental requirements: First, it requires coherent illumination; Second, it requires additional phase filtration of the image of the illumination source. As illustrated in Fig. 8.1(a), the point source provides the coherent illumination, and the phase plate performs phase filtering on the image of the point source. Note that the objective lens focuses the coherent illumination light and forms a demagnified image of the point source at its back focal plane (BFP). Interestingly, the objective lens forms a magnified image of the sample at the same time on the detector plane.

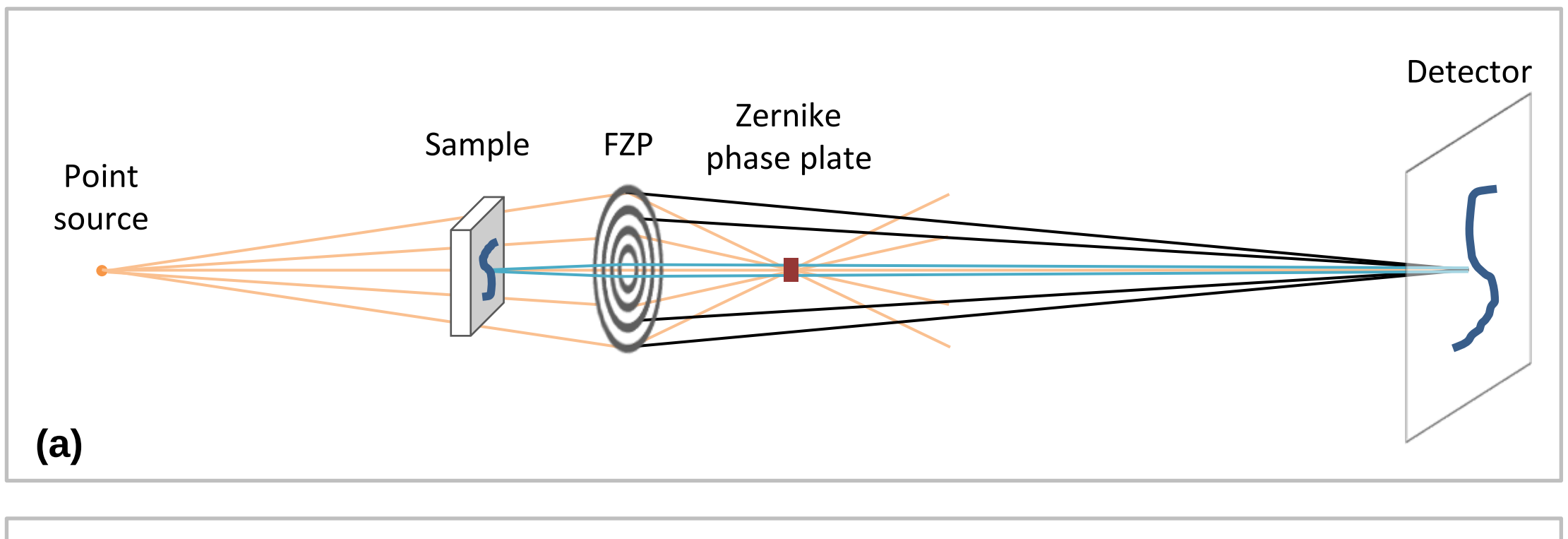

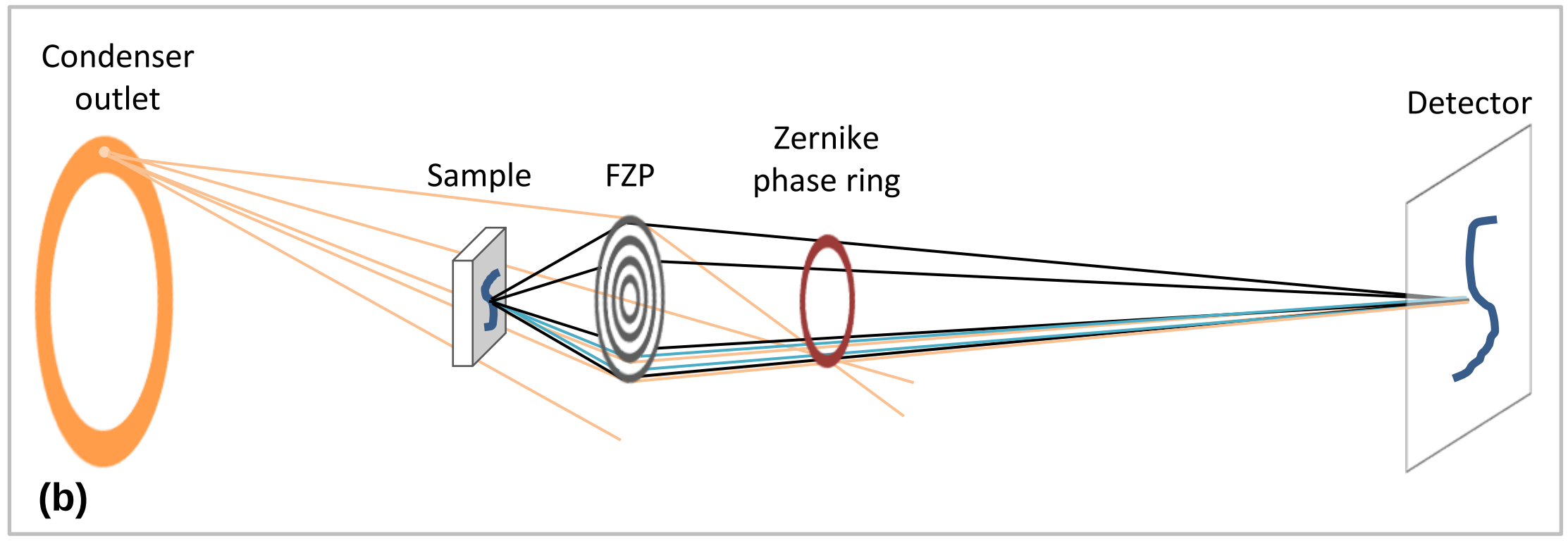

Figure 8.1: Two schemes of Zernike phase contrast imaging. (a) Point-source illumination, where the Zernike phase plate is located at the back focal plane of the FZP and is aligned with the optical axis. (b) Ring-shaped source illumination on the condenser outlet plane. The ring-shaped source can be considered as an infinite number of point sources, each corresponding to a point-line Zernike phase plate. For each point source, the source point, the center of the FZP, and the corresponding Zernike phase plate are collinear, as in (a). Due to the off-axis geometry of the ring-shaped source, these lines are tilted relative to the optical axis. Consequently, a Zernike phase ring is formed at the back focal plane.

The Zernike phase contrast imaging method was firstly introduced into X-ray microscopy with FZP by G. Schmahl[92, 93]. Instead of employing a phase plate, a Zernike phase ring is placed near the BFP of the FZP to shift the phase of the undiffracted X-ray beam by $\pm\frac{\pi}{2}$. The shape of the Zernike phase ring was basically determined by the annular shape of the outlet illumination of the capillary condenser. Essentially, the illumination to the sample in a FZP based X-ray microscopy is tilted to prevent excessively strong undiffracted X-ray photons from reaching the detector. In this configuration, the extended ring-shaped source can be considered as an infinite number of point sources, with each point source corresponding to a Zernike phase plate. For each point source, the source point, the center of the FZP, and the corresponding Zernike phase plate are collinear, as illustrated in Fig. 8.1(a). However, owing to the off-axis geometry associated with the ring-shaped source, this collinear line is tilted with respect to the optical axis. Consequently, the infinite number of Zernike phase plates collectively form a Zernike phase ring at the back focal plane of FZP, as illustrated in Fig. 8.1(b). As a result, the image contrast for weakly absorbing, low-density samples, such as cells or polymers, can be significantly enhanced[94, 95], see the results in Fig. 8.2.

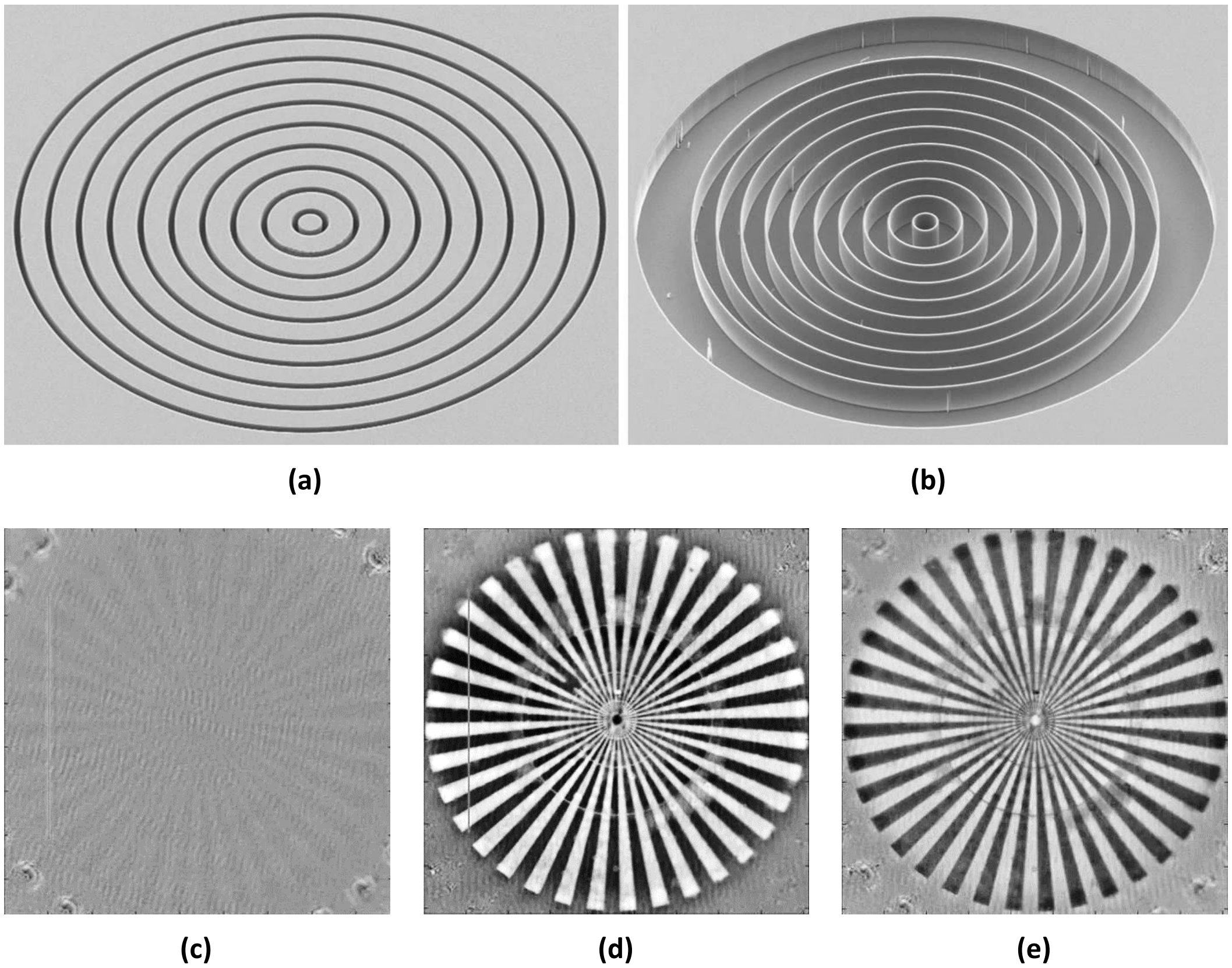

Figure 8.2: SEM images of two different Zernike phase plates: (a) positive phase shifting plate, (b) negative phase shifting plate. The outer diameter is 100 microns. X-ray microscopy images of the Siemens star pattern made of 200 nm thick Nickel correspond to (c) absorption contrast, (d) positive Zernike phase contrast, and (e) negative Zernike phase contrast, respectively.

Mathematically, the complex amplitude of the wavefront arriving upon the object plane can be expressed as

$$\mathrm{U}_1(x_1, y_1) = \frac{1}{i\lambda d_1} e^{ik\left[d_1 + \frac{(x_1 - x_s)^2 + (y_1 - y_s)^2}{2d_1}\right]}, \tag{8.1.1}$$

where coordinates $(x_s, y_s)$ represent the location of a point source, whose amplitude is set to unity here for convenience. Herein, the parabolic approximation of the spherical wavefront is utilized (derivation details can be found in Appendix D).

When shining upon the object, the X-ray wavefront starts to interact with the electrons and atomic nucleus that composite the object in a very complicated way. However, with appropriate approximations, the exit wave can be estimated using the projection approximation, in which the primary direction of the X-ray beam is assumed remain unchanged. Due to the presence of the object, the exit beam will change its amplitude and phase accordingly. As a result, the amplitude of the exiting wave field after the object is expressed as follows:

$$\begin{aligned} \mathrm{U}_1'(x_1, y_1) &= \frac{1}{i\lambda d_1} e^{ik\left[d_1 + \frac{(x_1-x_s)^2+(y_1-y_s)^2}{2d_1} - \Psi(x_1,y_1)\right]}, \\ \Psi(x_1, y_1) &= \phi(x_1, y_1) - i\alpha(x_1, y_1) = \int \delta(x_1, y_1, z)\, dz - i \int \beta(x_1, y_1, z)\, dz, \end{aligned} \tag{8.1.2}$$

where the line integrals are performed along the $z$-axis direction. The decrement of the real part of the refractive index, $\delta$, is related with the electron density distribution $\rho_e$ of the imaging object via

$$\delta(x, y, z, E) = \frac{h^2 c^2 r_e}{2\pi E^2} \rho_e(x, y, z), \tag{8.1.3}$$

in which $r_e$ denotes the classical electron radius, $h$ is the Planck's constant, and $c$ is the light speed in vacuum. Hence, the detection of phase information reflects the internal electron density distribution. In addition,

$$\beta(x, y, z) = \frac{\lambda}{4\pi} \mu(x, y, z). \tag{8.1.4}$$

Clearly, the imaginary part, $\beta$, of the refractive index is linearly proportional to the material attenuation coefficient $\mu$ and the beam wavelength $\lambda$.

Be aware that Eq. (G.2) holds true only if the thickness of the object is relatively thin compared to the total length of the imaging system.

Once leaving the object, the X-ray wave continues traveling towards the FZP in free space. Right upon the surface of FZP, the wave field becomes

$$\mathrm{U}_2(x_2, y_2) = \frac{e^{ikd_2}}{i\lambda d_2} \int\int \mathrm{U}_1'(x_1, y_1) e^{ik\left[\frac{(x_2-x_1)^2}{2d_2} + \frac{(y_2-y_1)^2}{2d_2}\right]} dx_1 dy_1, \tag{8.1.5}$$

Assuming that the FZP is mathematically expressed as

$$\mathrm{T}_{\mathrm{FZP}}(x_2, y_2) = \frac{1}{\pi} e^{-i\frac{k(x_2^2+y_2^2)}{2f}}, \tag{8.1.6}$$

where $f$ denotes the first order focal distance of the FZP. Immediately, the above $\mathrm{U}_2(x_2, y_2)$ wave front meets the FZP and is modified accordingly by the transmission function $\mathrm{T}_{\mathrm{FZP}}(x_2, y_2)$, namely,

$$\mathrm{U}_2^{'}(x_2, y_2) = \mathrm{U}_2(x_2, y_2)\mathrm{T}_{\mathrm{FZP}}(x_2, y_2). \tag{8.1.7}$$

Equation (8.1.7) represents the wave field right behind the FZP. Based on the Fresnel diffraction formula (see Appendix H for more details), the wave field distribution that locates $d_3$ distance behind the FZP can be expressed as following

$$\mathrm{U}_3(x_3, y_3) = \frac{f}{i\lambda\pi q}\, e^{ik(d_1+d_2+d_3+\frac{S_3(x_3,y_3)}{2q})} e^{-ik\Psi\left(\theta_x(x_3),\theta_y(y_3)\right)}, \tag{8.1.8}$$

where

$$\begin{aligned}
&q = f(d_1 + d_2 + d_3) - (d_1 + d_2)d_3, \\
&S_3(x_3, y_3) = f\left[(x_3 - x_s)^2 + (y_3 - y_s)^2\right] - d_3(x_s^2 + y_s^2) - (d_1 + d_2)(x_3^2 + y_3^2), \\
&\theta_x(x_3) = \frac{d_1 f x_3 + \left[(d_2 + d_3)f - d_2 d_3\right]x_s}{q}, \\
&\theta_y(y_3) = \frac{d_1 f y_3 + \left[(d_2 + d_3)f - d_2 d_3\right]y_s}{q}.
\end{aligned} \tag{8.1.9}$$

According to the Zernike phase contrast principle, the Zernike phase ring introduces additional amplitude attenuation and phase shift only to the background zero-frequency direct wave. As a result, the constant term 1 is replaced by $\gamma e^{i\phi_r}$, and $\gamma$ denotes the slight X-ray absorption of the Zernike phase ring, and $\phi_r$ denotes the phase shift induced by the Zernike phase ring. Namely

$$e^{-ik\Psi(\theta_x(x_3),\theta_y(y_3))} = 1 + \left[e^{-ik\Psi(\theta_x(x_3),\theta_y(y_3))} - 1\right] = \gamma e^{i\phi_r} - 1 + e^{-ik\Psi(\theta_x(x_3),\theta_y(y_3))}. \tag{8.1.10}$$

Consequently, Eq. (8.1.8) becomes:

$$\mathrm{U}_3^{\prime}(x_3, y_3) = \frac{f}{i\lambda\pi q}\, e^{ik\left(d_1+d_2+d_3+\frac{S_3(x_3,y_3)}{2q}\right)} \left[\gamma e^{i\phi_r} - 1 + e^{-ik\Psi(\theta_x(x_3),\theta_y(y_3))}\right]. \tag{8.1.11}$$

Be aware that the Zernike phase ring must be placed at the BFP of the FZP. This is because the spatial separation between the undiffracted (background) and diffracted light reaches its maximum at the BFP, allowing the phase plate to shift only the background light phase. Eventually, the complex amplitude of the X-ray beam arriving at the detector plane is expressed as

$$\mathrm{U}_4(x_4, y_4) = A_z(x_4, y_4)\left[\gamma e^{i\phi_r} - 1 + e^{-ik\Psi(\theta_x(x_4),\theta_y(y_4))}\right], \tag{8.1.12}$$

where

$$\begin{aligned}
&A_z(x_4, y_4) = \frac{f}{i\lambda\pi\tau}\, e^{ik\left(d_1+d_2+d_3+d_4+\frac{(f-d_1-d_2)(x_4^2+y_4^2)-2f(x_4x_s+y_4y_s)-(d_3+d_4-f)(x_s^2+y_s^2)}{2\tau}\right)},\\
&\tau = (d_1+d_2+d_3+d_4)f - (d_1+d_2)(d_3+d_4),\\
&\theta_x(x_4) = \frac{d_1 f x_4 + \big[(d_2+d_3+d_4)f - d_2(d_3+d_4)\big]x_s}{\tau} = -\frac{x_4}{M},\\
&\theta_y(y_4) = \frac{d_1 f y_4 + \big[(d_2+d_3+d_4)f - d_2(d_3+d_4)\big]y_s}{\tau} = -\frac{y_4}{M}.
\end{aligned} \tag{8.1.13}$$

In the derivations of $\theta_x(x_4)$ and $\theta_y(y_4)$, it is assumed that the Zernike phase ring is placed at the BFP of the FZP, under which the thin-lens magnification formula Eq. (3.1.1) holds. Additionally, $M = (d_3+d_4)/d_2$ denotes the geometric magnification.

Assuming that the phase shift induced by the Zernike phase ring is $\phi_0 = \frac{\pi}{2}$, consequently, the beam intensity recorded by the X-ray detector can be derived as below

$$\begin{aligned}
I_z(x_4, y_4) = |A_z(x_4, y_4)|^2\Big[&\gamma^2 + 1 + e^{-2k\alpha(-\frac{x_4}{M},-\frac{y_4}{M})}\\
&- 2\gamma e^{-k\alpha(-\frac{x_4}{M},-\frac{y_4}{M})}\sin\big(k\phi(-\frac{x_4}{M},-\frac{y_4}{M})\big)\\
&- 2e^{-k\alpha(-\frac{x_4}{M},-\frac{y_4}{M})}\cos\big(k\phi(-\frac{x_4}{M},-\frac{y_4}{M})\big)\Big],
\end{aligned} \tag{8.1.14}$$

If assuming that the scanned object weakly absorbs and shifts the X-ray wave, namely,

$$\begin{aligned}
&\alpha(-\frac{x_4}{M},-\frac{y_4}{M}) \approx 0,\\
&\cos\big(k\phi(-\frac{x_4}{M},-\frac{y_4}{M})\big) \approx 1,\\
&\sin\big(k\phi(-\frac{x_4}{M},-\frac{y_4}{M})\big) \approx k\phi(-\frac{x_4}{M},-\frac{y_4}{M}),
\end{aligned} \tag{8.1.15}$$

Immediately, Eq. (8.1.14) becomes

$$I_z(x_4, y_4) \approx |A_z(x_4, y_4)|^2\left[\gamma^2 - 2\gamma k\phi(-\frac{x_4}{M},-\frac{y_4}{M})\right], \tag{8.1.16}$$

Denoting the background intensity as:

$$I_{\text{bkg}}(x_4, y_4) = |A_z(x_4, y_4)|^2\gamma^2, \tag{8.1.17}$$

Immediately, Eq. (8.1.16) can be rewritten as

$$I_z(x_4, y_4) \approx I_{\text{bkg}}(x_4, y_4)\left[1 - \frac{2k}{\gamma}\phi\left(-\frac{x_4}{M}, -\frac{y_4}{M}\right)\right]. \tag{8.1.18}$$

The above equation indicates that the phase information of the object can be directly encoded into the beam intensity with the help of the Zernike phase ring, and finally detected by the detector.

### 8.1.2 Limitations

However, the FZP based X-ray microscopy does not fully satisfy the aforementioned requirements of the Zernike phase contrast imaging. First, the rotating-anode X-ray source is not a point source and therefore does not meet the requirement for coherent illumination. Second, the Zernike phase ring with a finite width (usually 4–6 $\mu$m) would block the diffracted low-frequency and high-frequency X-ray photons and therefore cause distortion of X-ray beam intensity. As a result, halo image artifacts would be produced[96]. To eliminate such halo artifacts, in principle, one has to reduce the width of the annular condenser aperture, the size of the condenser image and the phase plate, so that the phase plate blocks as little low-frequency diffracted X-ray beam as possible. In fact, the width of the annular condenser aperture is directly related to that of the corresponding phase ring, resulting in a trade-off between X-ray photon flux and phase contrast performance. To shorten the exposure time, the annular aperture must be widened to increase the number of X-ray photons transmitted from the source. However, a wider condenser aperture requires a correspondingly wider phase ring, allowing a broader range of low-spatial-frequency X-ray components to pass through. Consequently, the desired 90° phase shift cannot be uniformly maintained between the X-ray components passing through the object and the corresponding illumination. One potential solution[94] is to replace the annular source aperture with an aperture containing a large number of small holes and to place a corresponding array of matched phase-shifting dots at the back focal plane of the objective.

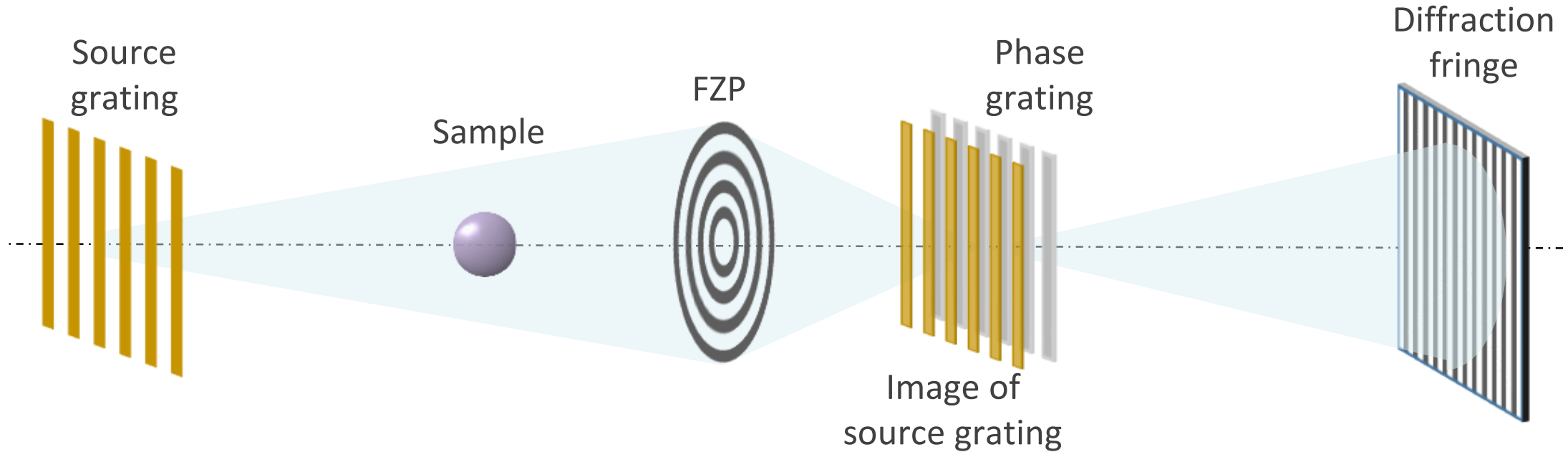

Figure 8.3: Illustration of integrating the grating interferometer into the X-ray microscope with FZP. The source grating is used to generate spatially coherent X-ray beam, and the phase grating is used to encode the phase information and generate large period diffraction fringe via the self-imaging effect. Usually, the phase grating is positioned near the back focal plane of FZP.

## 8.2 Grating interferometer phase contrast imaging

During 1960s–2000s, X-ray phase contrast imaging developed very quickly as many novel imaging techniques were emerged. For example, the analyzer-based phase imaging method (also known as the diffraction-enhanced imaging (DEI) method) [97–110], the propagation-based imaging (also known as the in-line holography) [111–125], the grating-based imaging technique [126–138], and the edge-illumination technique (also known as the coded-aperture method) [139–142].

Particularly, the so-called grating-based X-ray phase contrast imaging method, which makes use of the Talbot effect, discovered by Henry Fox Talbot in 1836, was pioneered in 1965 by Bonse and Hart[143] with crystal interferometer that was made from a large and highly perfect single silicon crystal. In early 2000s, this technique was further developed by Christian David[144] and Atsushi Momose[128]. In essence, the self-imaging effect generates an interference pattern downstream of a diffraction grating, usually a $\pi$ or $0.5\pi$ phase grating having period of several microns. At a particular distance this pattern resembles a similar periodic structure as of the phase grating. For a $\pi$ phase grating, the period of the formed self-image is half size of the phase grating; for a $0.5\pi$ phase grating, the period of the formed self-image is exactly the same size as of the phase grating. Such interference pattern can be displaced, usually by several microns, when adding an object into the beam. To measure the slight displacement of the interference pattern, an analyzer grating is added to generate the Moiré fringes with large period, whose movements can be easily detected. Finally, the phase information is obtained with certain signal retrieval method.

### 8.2.1 Combing grating interferometer with FZP

The pioneer work of integrating the grating interferometer into the X-ray microscope with FZP, as illustrated in Fig. 8.3, was proposed by Momose and Yashiro[89–91]. Experiments have been performed on both the synchrotron and the laboratory system (Model: Xradia 800, Zeiss, Germany).

#### 8.2.1.1 Beam coherence

In physics, two wave sources are perfectly coherent if they have a constant phase difference and the same frequency. It is an ideal property of waves that enables stationary interference, i.e., temporally or spatially, or both. The coherent property describes the correlations between physical quantities of a single wave, or between several waves[145]. The higher coherence, the tighter correlation. When considering coherent waves, two waves can add together to create a wave of greater amplitude than either one (constructive interference) or subtract from each other to create a wave of lesser amplitude than either one (destructive interference), depending on their relative phase. It is more convenient to define the level of coherence by the interference visibility.

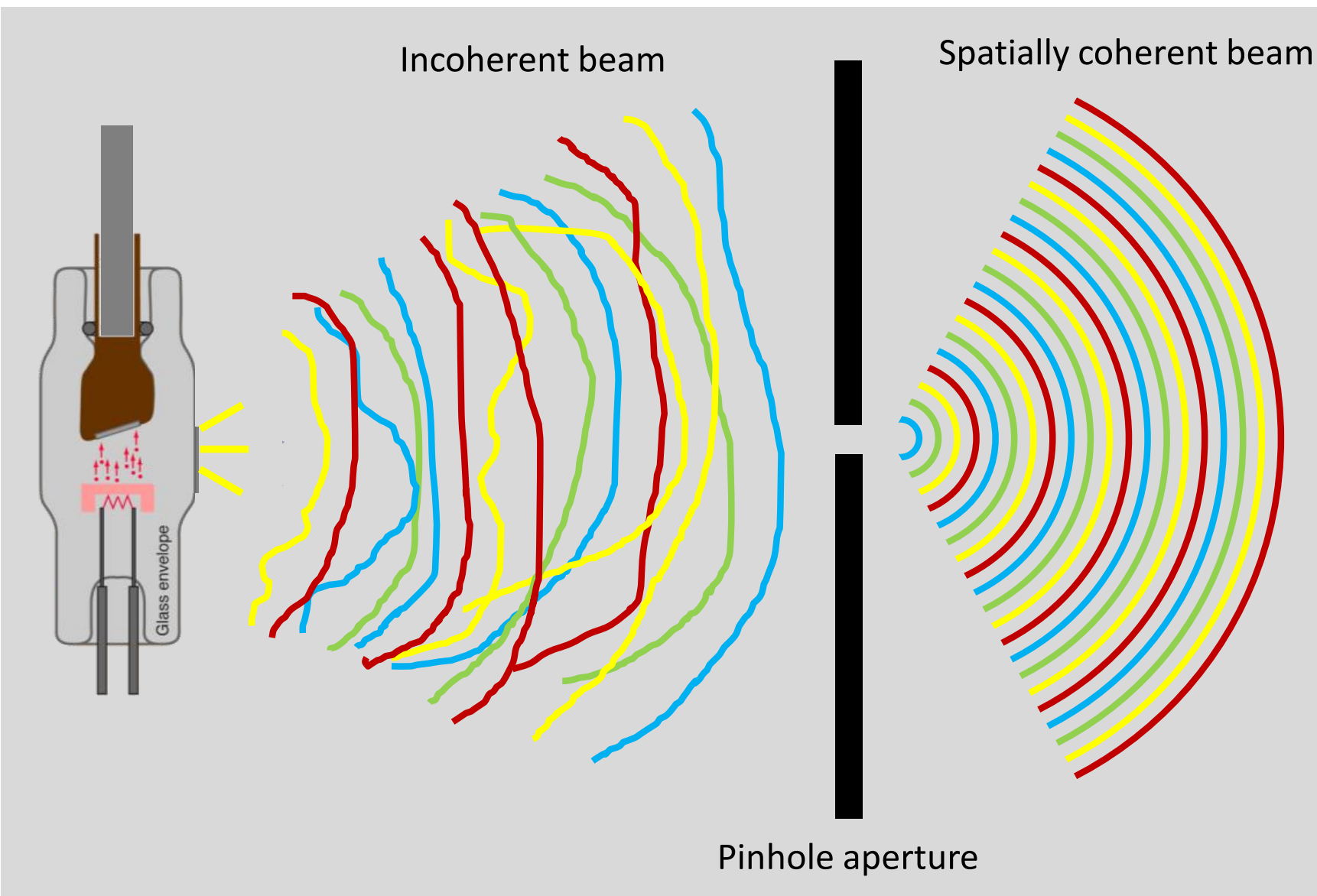

Figure 8.4: Generating spatially coherent X-ray wave from an incoherent X-ray source.

**Spatial coherence**

Spatial coherence describes the correlation between waves at different points in space. If two or more waves are spatially coherent, the phase difference between them will be constant, no matter where the observation is taken in the space. The goodness of spatial coherence is usually quantified by the transverse coherence length $\zeta$. For a source of size $s$, and a propagation distance of $L$, the spatial coherence length $\zeta$ is defined as following:

$$\zeta = \frac{L}{s}\lambda, \tag{8.2.1}$$

where $\lambda$ is the beam wavelength. Clearly, the farther the propagation distance $L$, the smaller the source size $s$, the longer the transverse spatial coherence length $\zeta$.

**Temporal coherence**

Temporal coherence describes the correlation between waves observed at different moments in time. As contrary to the transverse coherence length, the longitudinal coherence length is usually used to characterize the temporal coherence.

Be aware that in X-ray phase contrast imaging, the spatial coherence is more important than the temporal coherence.

#### 8.2.1.2 Diffraction grating

In optics, a diffraction grating is an optical component with a periodic structure, which splits and diffracts light into several beams traveling in different directions. After interacting with the diffraction grating, the initial wavefront will be modulated accordingly. Depending on the optical property of the diffraction grating, they can be categorized into two main types: the amplitude grating and the phase grating. In this book, the absorption portion of the amplitude grating is assumed to completely absorb the beam, while the grating opening portion allows the beam to pass through without introducing any beam change. For the phase grating, the beam amplitude maintains the same. However, the phase of the beam wavefront will be modified corresponding to the used phase grating. Two different types of phase grating will be discussed: the $\frac{\pi}{2}$ grating and the $\pi$ grating.

As depicted in Fig. 8.5, the incident beam is diffracted into different orders after interacting with the grating. Herein, only the first two orders, i.e., $|n| \leq 2$, are illustrated. Moreover, the period of the grating is denoted as $p$, and the height of the grating groove is equal to $h$. Usually, the duty cycle of the grating is defined as the ratio of $\frac{p-t}{p}$. In this book, the duty cycle of the gratings is assumed to 50% by default. In reality, the absolute phase shift of a phase grating with certain material thickness $h$ heavily relies on the used x-ray beam energy $E$. Due to this reason, the material thickness $h$ of a fabricated phase grating is always designed according to the mean energy of the polychromatic X-ray beam.

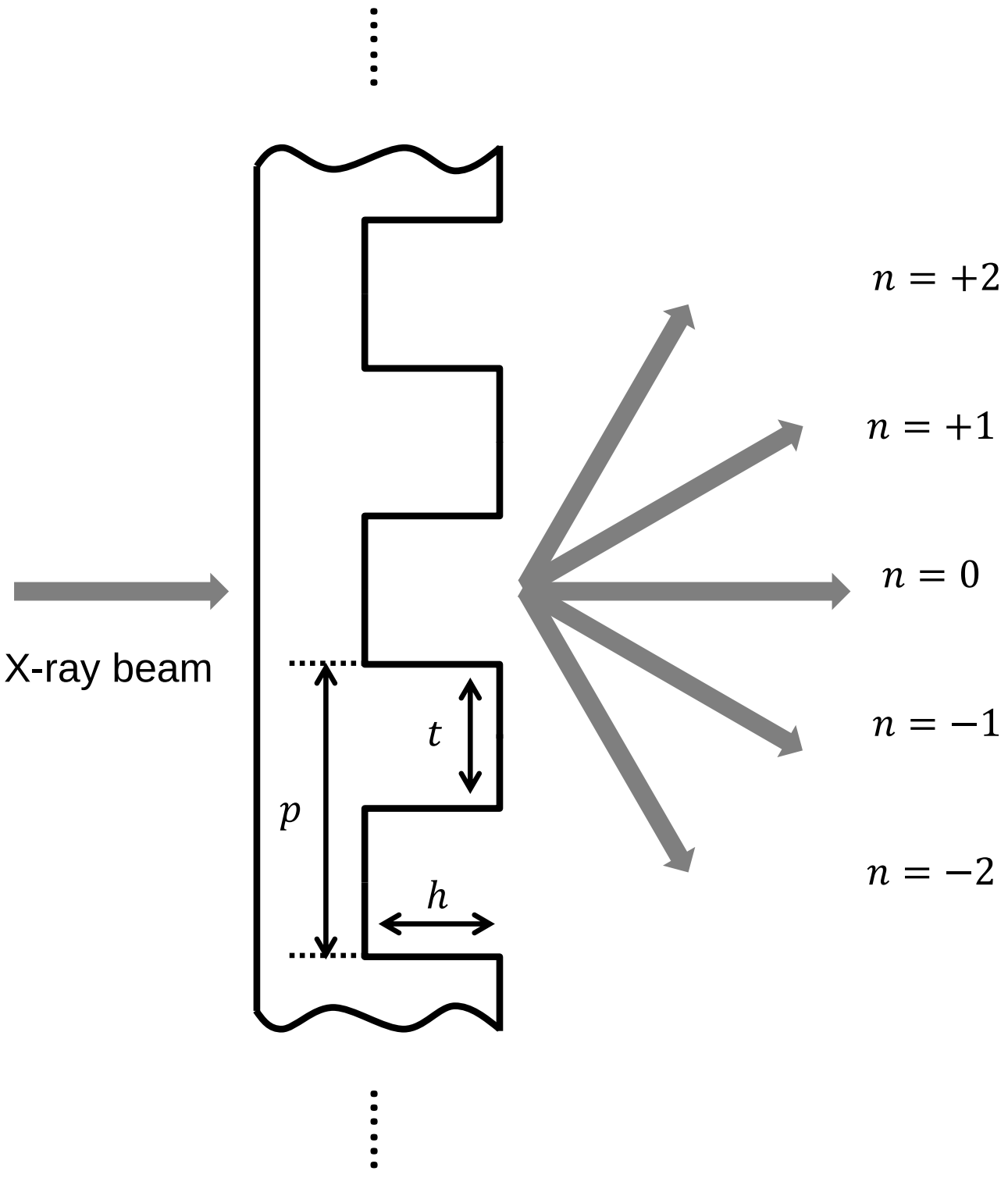

Figure 8.5: The incident x-ray beam is diffracted into different directions after interacting with the diffraction grating. Only the first two diffraction orders, i.e., $|n| \leq 2$, are illustrated.

Usually, the grating is described through a periodic transmission function, denoted as $\mathrm{T}(x)$. In optics, it is also convenient to use its Fourier series representation to describe the transmission function,

$$\mathrm{T}(x) = \sum_{n=-\infty}^{n=\infty} a_n e^{\frac{i2\pi nx}{p}}. \tag{8.2.2}$$

Depending on the diffraction order of $n$, the coefficient $a_n$ will have different values. In the following parts, the coefficient $a_n$ will be calculated for three different types of 1D diffraction gratings: the pure amplitude grating, pure $\frac{\pi}{2}$ phase grating, and the pure $\pi$ phase grating.

**Pure amplitude grating:** Mathematically, the transmission function of a pure amplitude grating can be described as:

$$\mathrm{T}(x) = \begin{cases} 1 & \text{if } \mathrm{mod}(x,p) \in [0 \;\; p/2) \\ 0 & \text{if } \mathrm{mod}(x,p) \in [p/2 \;\; p) \end{cases} \tag{8.2.3}$$

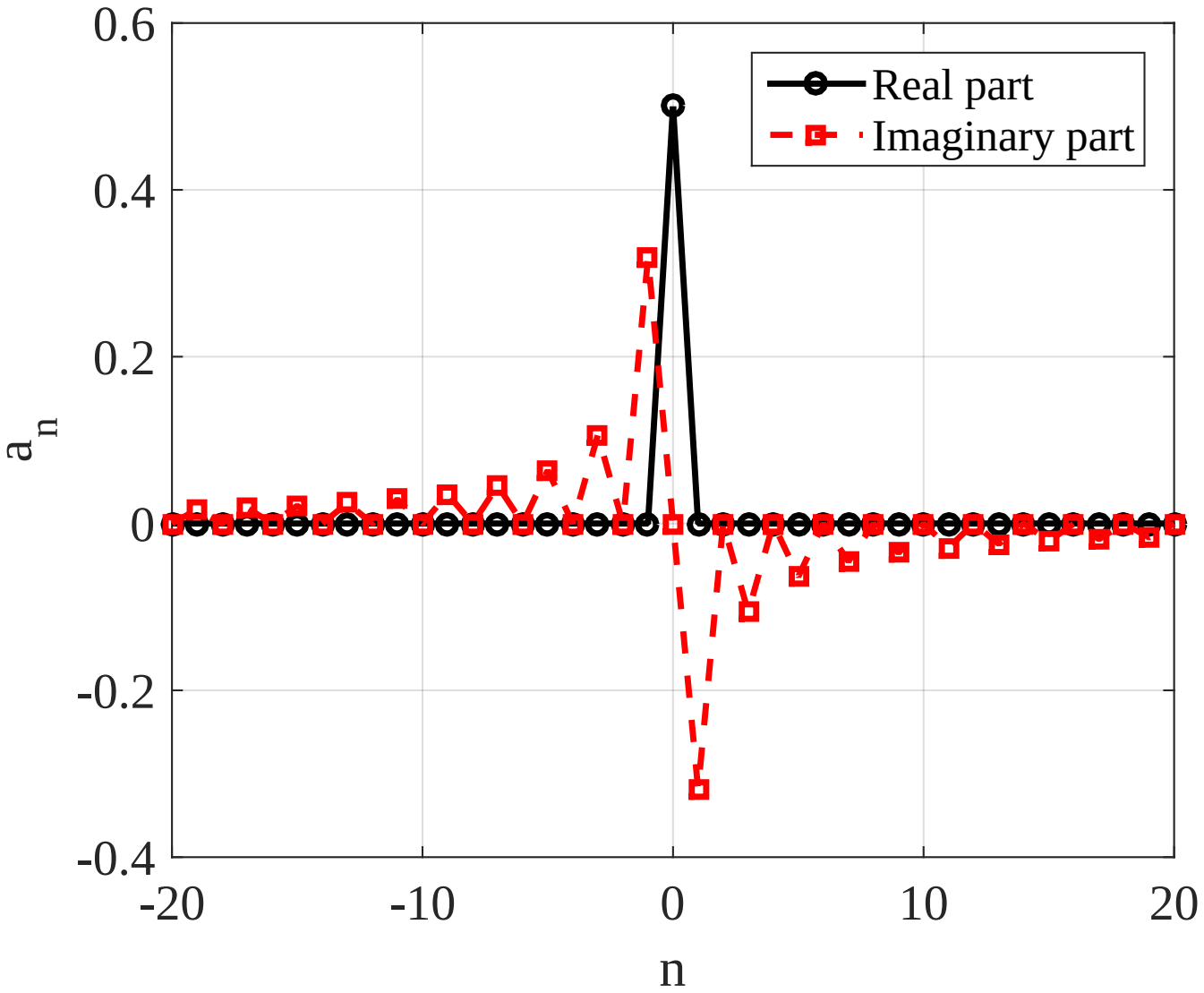

Figure 8.6: The theoretical values of the complex coefficient $a_n$ for a pure amplitude grating when $n \in [-20,\ 20]$. Both the real and imaginary part of $a_n$ were plotted.

Correspondingly, the coefficient $a_n$ are found to be

$$a_n = \begin{cases} \frac{1}{2} & \text{if n=0} \\ 0 & \text{if } \mathrm{mod}(n,2) = 0, \text{ but } \ n \neq 0 \\ -\frac{i}{n\pi} & \text{if } \mathrm{mod}(n,2) = 1 \end{cases} \tag{8.2.4}$$

The plot in Fig. 8.6 numerically shows the complex coefficients $a_n$ for different diffraction order $n$. As can be seen, the value of $a_n$ drops to zero quickly as $|n|$ increases.

**Pure $\frac{\pi}{2}$ phase grating:** For a given pure $\frac{\pi}{2}$ phase grating, the transmission function is written as:

$$\mathrm{T}(x) = \begin{cases} 1 & \text{if } \mathrm{mod}(x,p) \in [0 \ \ p/2) \\ e^{i\frac{\pi}{2}} & \text{if } \mathrm{mod}(x,p) \in [p/2 \ \ p) \end{cases} \tag{8.2.5}$$

Immediately, the complex coefficient $a_n$ are found to be

$$a_n = \begin{cases} \frac{1+i}{2} & \text{if } n = 0 \\ 0 & \text{if } \mathrm{mod}(n,2) = 0, \text{ but } \ n \neq 0 \\ -\frac{1+i}{n\pi} & \text{if } \mathrm{mod}(n,2) = 1 \end{cases} \tag{8.2.6}$$

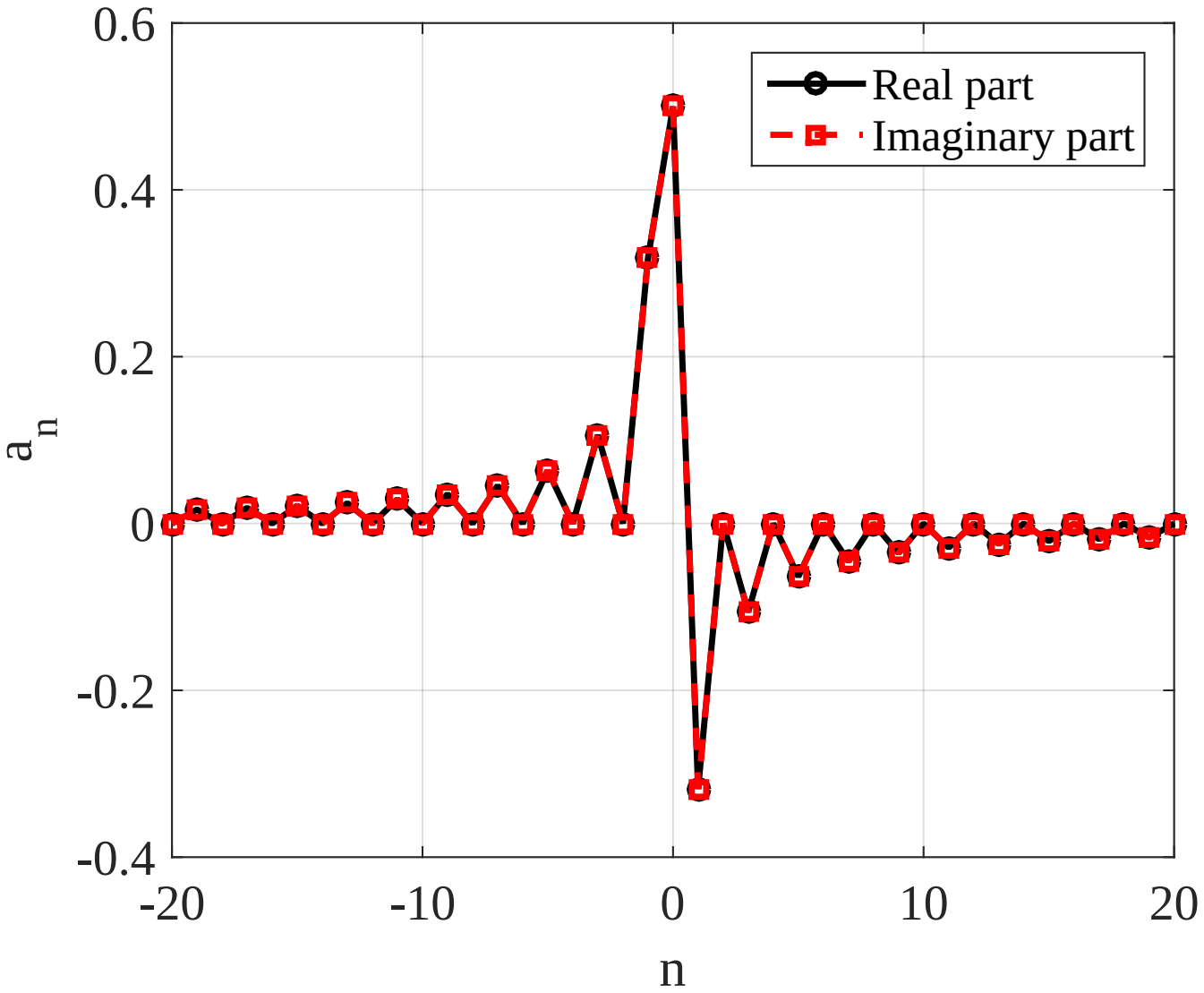

Figure 8.7: The theoretical values of the complex coefficient $a_n$ for a pure $\frac{\pi}{2}$ phase grating when $n \in [-20,\ 20]$. Both the real and imaginary part of $a_n$ were plotted.

The plot in Fig. 8.7 numerically shows the coefficients for different diffraction order $n$. Again, the value of $a_n$ approaches to zero quickly as $|n|$ increases.

**Pure $\pi$ phase grating:** In this case, the transmission function of the $\pi$ phase grating can be described as:

$$\mathrm{T}(x) = \begin{cases} 1 & \text{if } \operatorname{mod}(x,p) \in [0 \quad p/2) \\ e^{i\pi} & \text{if } \operatorname{mod}(x,p) \in [p/2 \quad p) \end{cases} \tag{8.2.7}$$

Similarly, the coefficient $a_n$ is equal to

$$a_n = \begin{cases} 0 & \text{if } \operatorname{mod}(n,2) = 0 \\ -\frac{2i}{n\pi} & \text{if } \operatorname{mod}(n,2) = 1 \end{cases} \tag{8.2.8}$$

The plot in Fig. 8.8 numerically shows the coefficients for different diffraction order $n$. As can be seen, the most dominant diffraction term is $|n| = 1$.

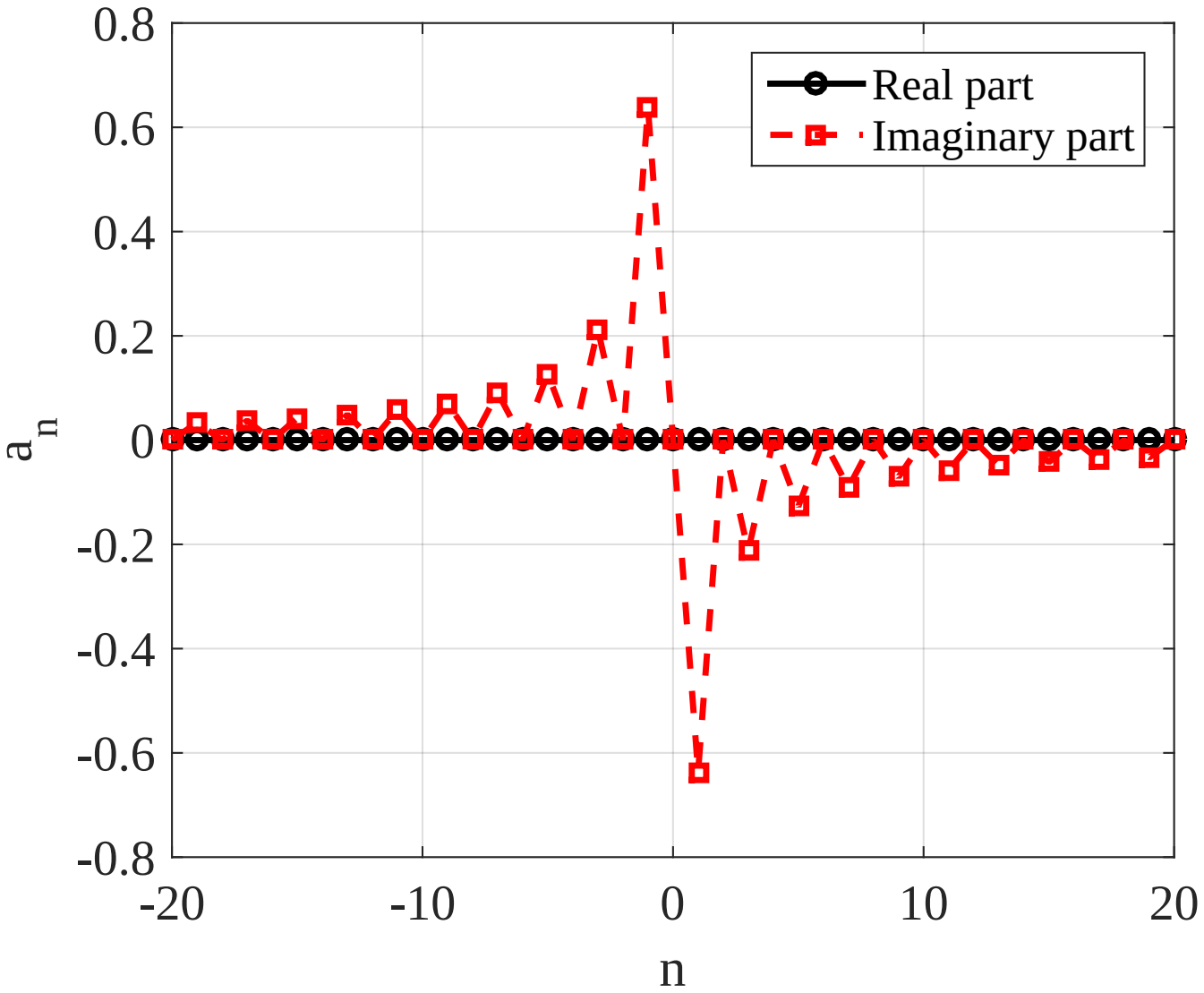

Figure 8.8: The theoretical values of the complex coefficient $a_n$ for a pure $\pi$ phase grating when $n \in [-20,\ 20]$. Both the real and imaginary part of $a_n$ were plotted.

#### 8.2.1.3 Grating interferometer

Similar to the Zernike phase ring, the phase grating $\mathrm{G}_1$ is also placed around the BFP of FZP. Based on Eq. (8.1.8) and Eq. (8.2.2), the wave front right after $\mathrm{G}_1$ becomes

$$\mathrm{U}_3^{'}(x_2, y_2) = \mathrm{U}_3(x_2, y_2)\mathrm{T}_{\mathrm{G}_1}(x_2, y_2). \tag{8.2.9}$$

Eventually, the wave front arriving at the detector plane becomes

$$\begin{aligned}\mathrm{U}_4(x_4, y_4, \omega) &= A_4 \sum_{n=-\infty}^{\infty} a_n\, e^{ik[D_n + \frac{2n\pi\omega}{kp} - \Psi(\theta_x(n), \theta_y)]} \\ &= A_4 \sum_{n=-\infty}^{\infty} a_n\, e^{-k\alpha(\theta_x(n), \theta_y)} e^{ik[D_n + \frac{2n\pi\omega}{kp} - \phi(\theta_x(n), \theta_y)]},\end{aligned} \tag{8.2.10}$$

where $\omega$ denotes the coefficient of phase variation induced by the displacement of the phase grating, and

$$
\begin{aligned}
A_4 &= \frac{f}{i\lambda\pi\tau}\, e^{ik\left(d_1+d_2+d_3+d_4+\frac{S_4(x_4,y_4)}{2\tau}\right)}, \\
S_4(x_4,y_4) &= (f-d_1-d_2)\left(x_4^2+y_4^2\right) - 2f(x_4x_s+y_4y_s) - (d_3+d_4-f)\left(x_s^2+y_s^2\right), \\
D_n &= \frac{2n\pi\,\{[(d_1+d_2)d_3-(d_1+d_2+d_3)f](d_4n\pi-kpx_4)+d_4fkp\,x_s\}}{k^2p^2\tau}, \\
\theta_x(n) &= \frac{d_1f(kpx_4-2d_4n\pi)+k\left[(d_2+d_3+d_4)f-d_2(d_3+d_4)\right]px_s}{k\tau p} = -\frac{x_4}{M}+\frac{nd_4\lambda}{Mp}, \\
\theta_y &= \frac{d_1fy_4+\left[(d_2+d_3+d_4)f-d_2(d_3+d_4)\right]y_s}{\tau} = -\frac{y_4}{M}.
\end{aligned}
\tag{8.2.11}
$$

For a $\frac{\pi}{2}$ phase grating, the finally detected beam intensity is found equal to [146]

$$
\begin{aligned}
I(x_4,y_4,\omega) = |A_4|^2 \Bigg\{ & \frac{1}{2}e^{-2k\alpha_0} + \frac{2}{\pi^2}\left(e^{-2k\alpha_1}+e^{-2k\alpha_{-1}}\right) \\
& - \frac{2}{\pi}e^{-k(\alpha_0+\alpha_1)}\cos\left[k(-D_1-\frac{2\pi\omega}{kp}-\phi_0+\phi_1)\right] \\
& + \frac{2}{\pi}e^{-k(\alpha_0+\alpha_{-1})}\cos\left[k(-D_{-1}+\frac{2\pi\omega}{kp}-\phi_0+\phi_{-1})\right]\Bigg\},
\end{aligned}
\tag{8.2.12}
$$

where

$$
\begin{aligned}
\alpha_n &= \alpha\left(-\frac{x_4}{M}+\frac{nd_4\lambda}{Mp}, -\frac{y_4}{M}\right), \qquad n=0,\pm1. \\
\phi_n &= \phi\left(-\frac{x_4}{M}+\frac{nd_4\lambda}{Mp}, -\frac{y_4}{M}\right), \qquad n=0,\pm1.
\end{aligned}
\tag{8.2.13}
$$

As seen, the signal model in Eq. (8.2.12) contains three different components: the zero order signal, and the $\pm1$ order signals. These three signals were mixed and overlapped. With Eq. (8.2.12), the absorption signal is found to be

$$
I_0(x_4,y_4) = |A_4|^2\left[\frac{1}{2}e^{-2k\alpha\left(-\frac{x_4}{M},-\frac{y_4}{M}\right)} + \frac{2}{\pi^2}\left(e^{-2k\alpha\left(\frac{d_4\lambda}{Mp}-\frac{x_4}{M},-\frac{y_4}{M}\right)} + e^{-2k\alpha\left(-\frac{d_4\lambda}{Mp}-\frac{x_4}{M},-\frac{y_4}{M}\right)}\right)\right],
\tag{8.2.14}
$$

When the sample's absorption is weak, i.e., $\alpha_n = 0$, the detected signal can be approximated as below

$$
\begin{aligned}
I(x_4,y_4,\omega) \approx |A_4|^2\Bigg[ & \frac{1}{2}+\frac{4}{\pi^2}+\frac{4}{\pi}\sin\left\{\frac{k}{2}\left[-D_1-D_{-1}-2\phi_0+\phi_1+\phi_{-1}\right]\right\} \\
& \cdot\cos\left\{\frac{k}{2}\left[D_1-D_{-1}+\phi_{-1}-\phi_1\right]+\frac{\pi}{2}+\frac{2\pi\omega}{p}\right\}\Bigg],
\end{aligned}
\tag{8.2.15}
$$

Since $D_1$ and $D_{-1}$ terms are not related to the phase stepping factor $\omega$, therefore, it can be further approximated as:

$$I(x_4, y_4, \omega) \approx |A_4|^2 \left[ I_0 + I_1(x_4, y_4) \cos \left\{ \varphi_g(x_4, y_4) + \frac{2\pi\omega}{p} \right\} \right], \tag{8.2.16}$$

where

$$\begin{aligned} I_0(x_4, y_4) &= \frac{1}{2} + \frac{4}{\pi^2}, \\ I_1(x_4, y_4) &= \frac{4}{\pi} \sin \left\{ \frac{k}{2} \left[ -D_1 - D_{-1} - 2\phi_0 + \phi_1 + \phi_{-1} \right] \right\}, \\ \varphi_g(x_4, y_4) &= \frac{\pi}{2} + \frac{k}{2} \left[ D_1 - D_{-1} + \phi_{-1} - \phi_1 \right]. \end{aligned} \tag{8.2.17}$$

It is easy to derive that the detected phase image contains two splitting phase signals:

$$\Delta\varphi_g(x_4, y_4) = \frac{k}{2} \left[ \phi \left( -\frac{d_4\lambda}{Mp} - \frac{x_4}{M}, -\frac{y_4}{M} \right) - \phi \left( \frac{d_4\lambda}{Mp} - \frac{x_4}{M}, -\frac{y_4}{M} \right) \right], \tag{8.2.18}$$

#### 8.2.1.4 Differential phase contrast imaging

If the splitting distance $\Delta s = \frac{2d_4\lambda}{Mp}$ is small enough, then having

$$\Delta\varphi_g(x_4, y_4) \approx -\frac{k\Delta s}{2} \phi' \left( -\frac{x_4}{M}, -\frac{y_4}{M} \right). \tag{8.2.19}$$

Obviously, this is nothing but the most familiar differential phase contrast (DPC) image acquired from a standard X-ray grating interferometer.

For DPC-CT images, we would like to mention that the finite splitting distance $\frac{d_4\lambda}{Mp}$ could cause minor spatial resolution degradation compared to the conventional absorption contrast CT images. Study[146] has demonstrated that such degradation might be related to the object composition, splitting distance, detector sampling interval, and so on.

### 8.2.2 Limitations

Based on those above grating coefficient plots, it is found that a $\frac{\pi}{2}$ phase grating behaves much similar like an amplitude grating. Therefore, it is possible to replace the $\frac{\pi}{2}$ phase grating by an amplitude grating. However, one drawback of using the amplitude grating is that the radiation dose efficiency is reduced, leading to prolonged total exposure period. Note that the $\pi$ phase grating behaves quite differently from the $\frac{\pi}{2}$ phase grating and the amplitude grating.

In an X-ray microscope with FZP, the use of grating interferometer can easily cause signal splitting in both absorption image and phase contrast image. To avoid, one may choose gratings having large periods to reduce the signal splitting distance. However, such selections will significantly degrade the detection sensitivity[146, 147] of phase signal. In practice, it is always needed to reduce the grating period to enhance the detection sensitivity of phase signal. As the grating period becomes smaller, splitting of the two phase signals will be significantly increased, leading to sophisticated signal retrieving. One particular case is for small sized imaging object (smaller than the splitting distance), whose phase signal recovery might be easy because the overlapping between the two splitting signals can be ignored.

An interesting possible alternative to the Zernike phase ring is the absorption ring. Since the absorption ring is able to completely absorb the X-ray beam, therefore, it can be considered as a variant of a large period amplitude grating having a two-dimensional circular shape and a small duty cycle. With the absorption ring, as a consequence, the differential phase contrast image of the object would be acquired. During the data acquisition, the absorption ring has to be laterally shifted step by step. Again, the detection sensitivity of the DPC signal acquired from absorption ring might be limited.

# 9 System assembling

After spending so much efforts on discussing every individual subsystem such as illumination (Chapter 4), FZP (Chapter 5), detector (Chapter 6) and mechanical (Chapter 7), eventually, we come to the most exciting and critical step to assemble them into a fully functional X-ray microscope. Before getting started, it is assumed that everything we need for system assembling are ready in hand.

## 9.1 Commissioning procedure

The following commissioning procedure assumes that the entire mechanical system has been designed and manufactured in accordance with the content outlined in Chapter 7. Once all mechanical components and motion stages have been delivered to the laboratory, system commissioning proceeds along the optical axis in a downstream sequence. The overall commissioning procedures are depicted in Fig. 9.1. The major stages include:

1. Install and level the marble base on the laboratory floor, and verify its stability with vibration measurements. Mount the segmented optical breadboards onto the marble base and confirm that the vibration isolation between adjacent boards meets the specifications.

2. Install each component stage one by one from the X-ray source to the detector along the optical axis, verifying both mechanical compliance and optical functionality at each step before proceeding to the next step, as detailed in Section 9.2.

3. Performing fine alignment to achieve the target ultra-high spatial resolution after all components are mounted and the basic imaging capability is confirmed, following the procedures described in Sections 4.2.4, 5.2.3, and 6.2.5.

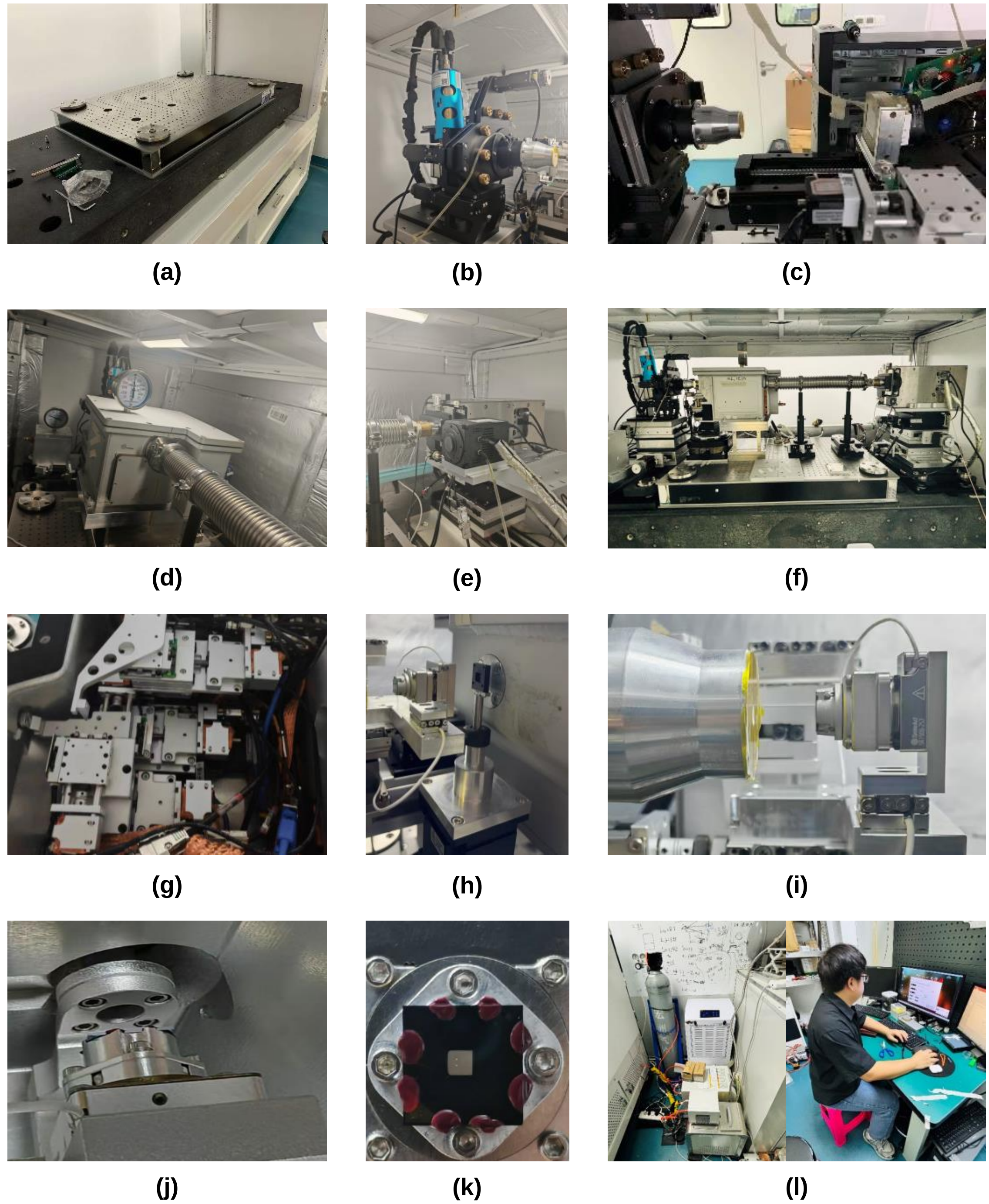

Figure 9.1: Major commissioning procedures: (a) optical breadboard installation, (b) X-ray source installation, (c) condenser adjustment, (d) vacuum chamber installation, (e) detector system installation, (f) overall system, (g) FZP integration, (h) sample stage installation, (i) source grating installation, (j) phase grating installation, (k) phase grating, (l) He gas tank, cooling system and operation table.

## 9.2 Optical path alignment

Optical path alignment is the most time-consuming and experience-dependent step in the entire integration process. Based on our experience, we believe the most effective strategy is to build up the optical path incrementally, verifying each component via measurable physical signatures before moving onto the next optical component. Do NOT try to align all the components at one time. This is because the whole X-ray microscope involves many cascaded optical components along a single axis, and troubleshooting the entire system is extremely challenging, i.e., it is nearly impossible to find out which component is misaligned immediately when the system fails to produce a favourable image.

In general, the rough alignments of all components begin by positioning them along a reference laser beam. Specifically, the system is assembled one component after the other from the source end toward the detector end. Whenever a new component is added, the downstream output needs to be verified by the detector to check that the preceding optical path and the newly added component are functioning correctly. Three representative verification checkpoints are summarized below.

**Condenser output:** After installing the X-ray tube and capillary condenser, one needs to verify that the condenser is correctly aligned and the illumination geometry meets the overall design. At this step, one should verify that the focused X-ray beam by the condenser is exactly located at the designed sample position, see Fig. 9.2(b), and indeed has the smallest size along the optical axis. To ease such longitudinal scanning along the $z$-axis, a portable X-ray detector with micron grade pixel dimension, e.g., 2-5 $\mu$m, is preferred and recommended.

**Beam profile at the FZP plane:** After installing the sample stage (without sample), one should verify that the beam profile at the FZP plane forms a hollow annular ring, whose diameter matches the diameter of the designed FZP, see Fig. 9.2(c). The X-ray beam downstream of the sample plane diverges and propagates toward the FZP. This verification helps confirming that the beam geometry and the condenser–FZP distance are consistent with the design, and that the illumination numerical aperture falls within the acceptance angular range of the zone plate.

**Image formation at the detector plane:** Once the FZP is inserted into the optical path, a bright beam spot is expected at the detector plane, see Fig. 9.2(d). This confirms that the FZP is correctly positioned along the optical axis and is receiving the full annular beam. Note that the bright hollow annular ring on Fig. 9.2(d) corresponds to the transmitted X-ray beam on FZP without being diffracted. When a test sample, e.g., a line pair resolution pattern, is placed at the sample plane, a magnified image of the sample should appear immediately at the center of the bright ring, see Fig. 9.2(d). This confirms that the complete optical path — from X-ray source to condenser, sample, FZP, and detector objective lens — is functioning properly as designed.

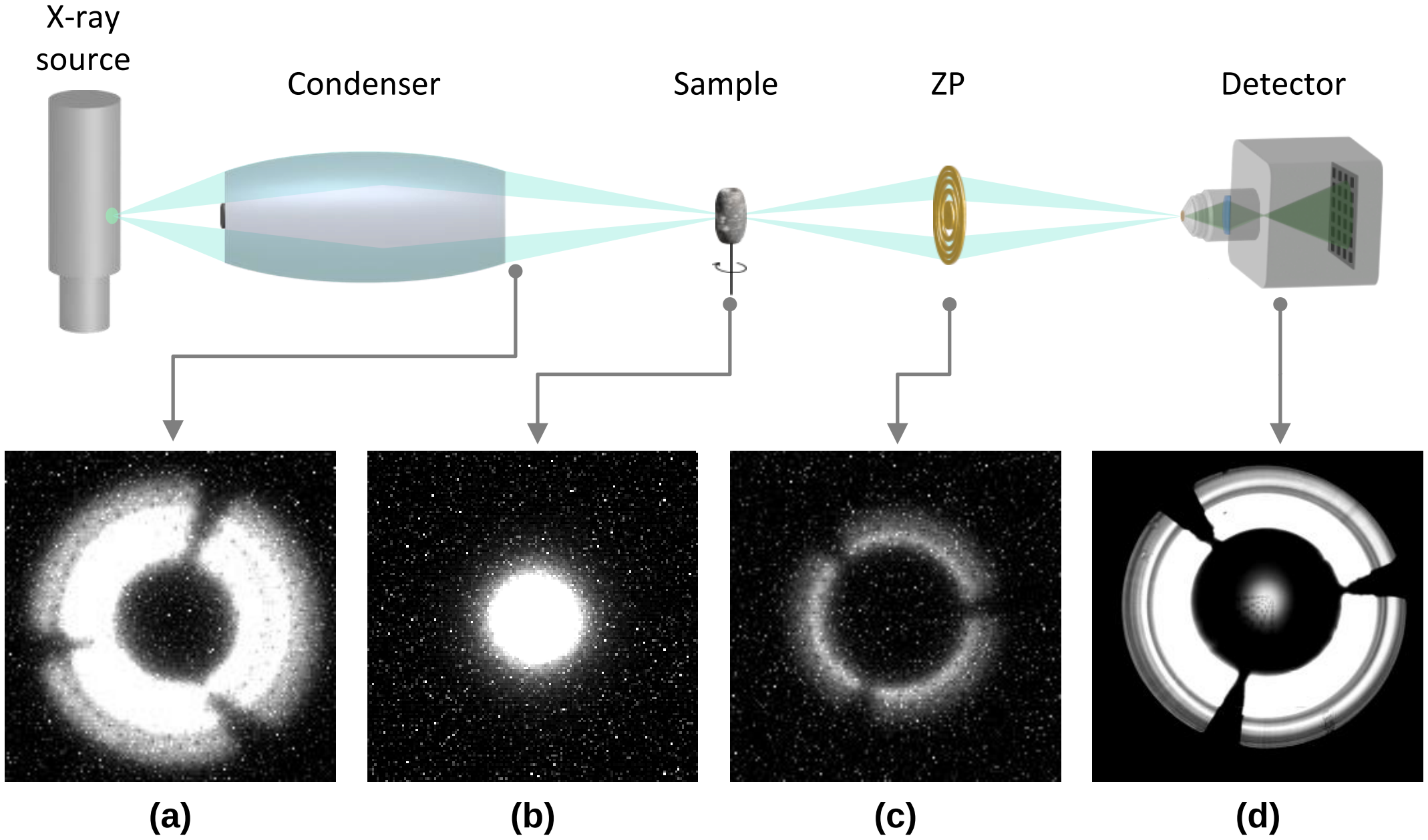

Figure 9.2: The condenser reflected X-ray beam intensity distributions at different positions along the optical path: (a) close to the condenser exit window, (b) at the sample plane, (c) at the FZP plane, (d) at the detector plane (with a sample).

After these three checkpoints, the system should be able to generate some imaging results, though typically at a resolution still below the designed value. Further improvement requires fine alignment of the condenser orientation, the axial positions of the sample and FZP, and the detector focal distance, following the procedures in Sections 4.2.4, 5.2.3, and 6.2.5. If the resolution still falls below, mechanical vibration (Section 2.2) and temperature stability (Section 2.3) need to be re-examined.

## 9.3 Sample mounting and preparation

There are two ranges limit the generation of desired high resolution images in a FZP-based X-ray microscope: 1. The lateral range, which is limited by the annular illumination geometry of the capillary condenser or FZP, is approximately 20 $\mu$m in diameter. Objects that are larger than this size cannot be fully illuminated and imaged. 2. The longitudinal range, namely, the depth of field of the FZP, defined as $\mathrm{DoF} = 4\Delta r_N^2/\lambda$, is approximately 30 $\mu$m for the 30 nm zone plate at 5.4 keV.

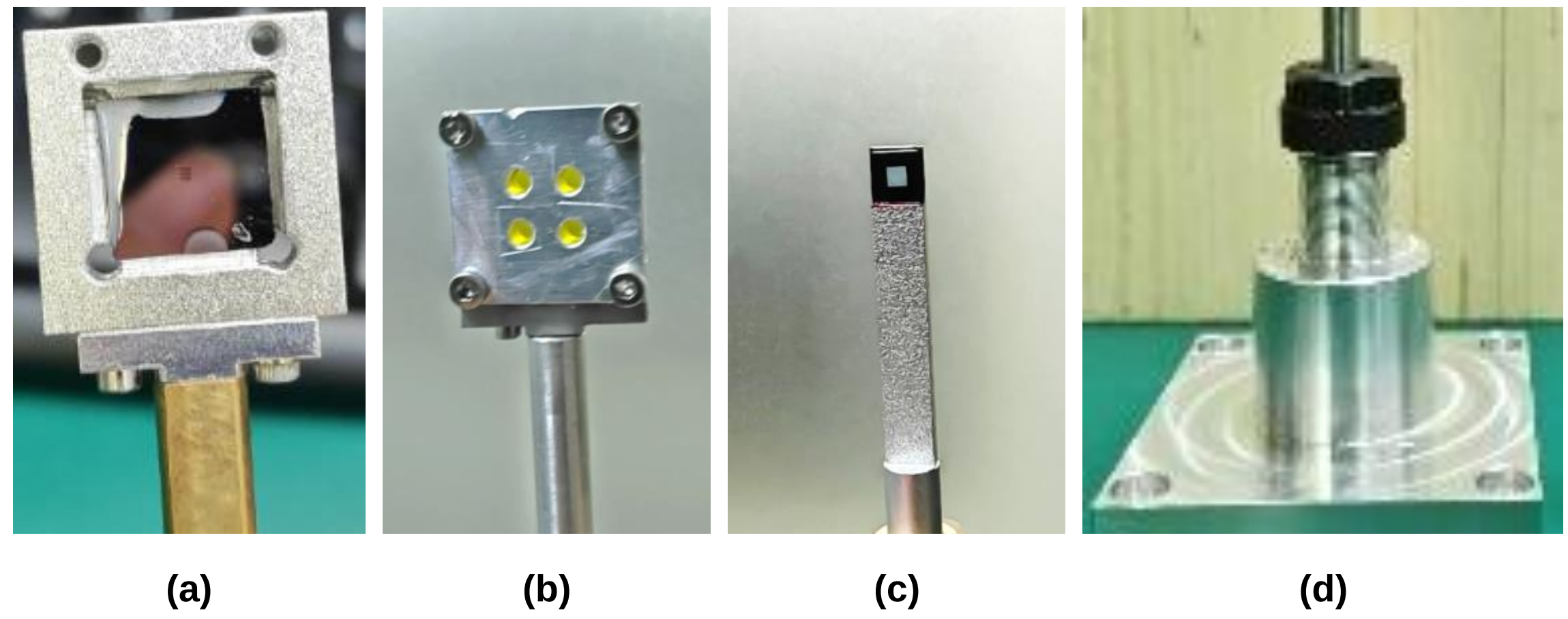

Figure 9.3: Photographs of the self-made sample holders: (a) single-sample holder, (b) multi-sample holder, (c) rotation holder. (d) The collet chuck base used to fix all holders.

Samples thicker than 30 $\mu$m may stay outside the depth of field and thus get defocused, resulting in resolution degradation. In practice, the sample thickness is typically constrained to $\sim 20\,\mu$m to stay well within the depth of field.

In fact, the nanometer spatial resolution imaging requires that the sample vibration should be within a fraction of the target resolution during the entire exposure period. Any random drift or vibration would cause image blurring. Consequently, both the sample preparation and the mounting strategy must ensure mechanical rigidity at the nanometer level.

### 9.3.1 Sample mounting

A widely adopted mounting approach is to support the sample on an ultra-thin silicon nitride ($Si_3N_4$) membrane window fabricated on a silicon frame. The $Si_3N_4$ membrane is almost transparent to 5.4 keV X-rays, minimizing the absorption loss on the support. In addition, the silicon frame provides mechanical rigidity and is compatible with standard laboratory sample handling.

Depending on the imaging task, different sample holders can be selected, as shown in Fig. 9.3. Typically, they include:

**Single-sample holder:** Mounts one single $Si_3N_4$ membrane window at a fixed position. Designed for projection only imaging such as resolution testing.

**Multi-sample holder:** Integrates multiple membrane windows in a single holder, enabling rapid sequential projection imaging of different specimens.

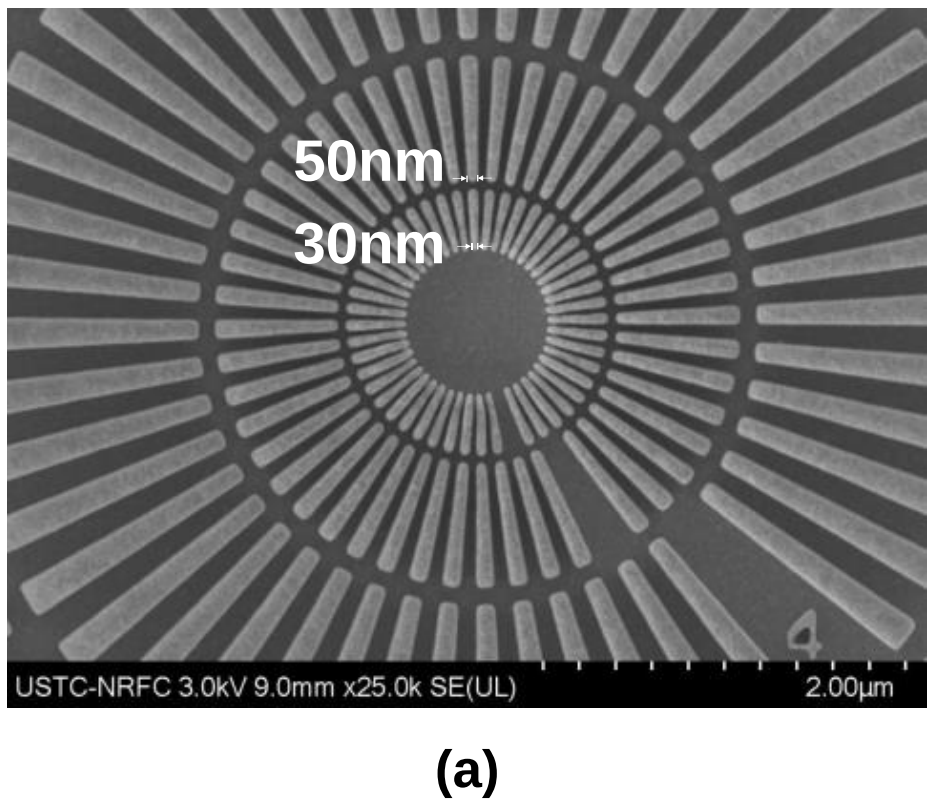

(a)

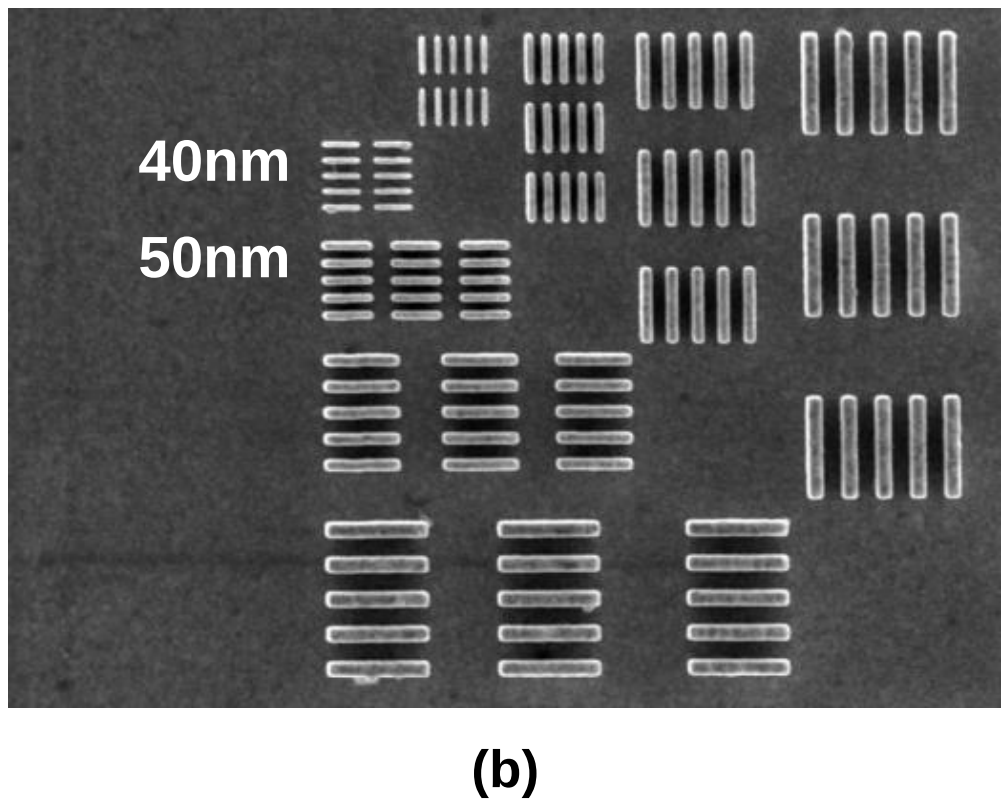

(b)

Figure 9.4: Standard samples used for resolution testing: (a) Siemens star pattern, (b) line-pair chart. Both resolution samples were fabricated by Prof. Yangchao Tian at the University of Science and Technology of China.

**Rotation holder:** Used for large-angle ($\geq 120°$) CT data acquisition. The sample is centered on the rotation axis with minimal wobble.

Besides, needle-like sample holders [148] can also be utilized to increase the total rotation angle up to 180 degrees.

All of these holders share a common base equipped with an elastic collet chuck, see Fig. 9.3, which allows convenient and rapid sample exchange while making negligible positional deviation and rotation center shift between different holders.

### 9.3.2 Sample preparation

A variety of samples can be imaged using the FZP-based X-ray microscope. Among them, usually, standard phantom samples are selected for imaging performance characterizations, such as spatial resolution, geometric alignment and quantitative evaluations. Usually, the Siemens star pattern and the line-pair chart are utilized, see Fig. 9.4(a). The Siemens star pattern consists of a group of radial spokes whose line width decreases continuously from the periphery toward the center. Unlike the periodic bar-pattern test chart, see Fig. 9.4(b), which typically offers discrete spatial frequencies along certain orientations (horizontal or vertical), the Siemens star pattern probes resolution isotropically in all directions and provides a continuous frequency gradient. This makes it particularly advantageous for detecting resolution anisotropy caused by optical aberrations or mechanical vibrations along a specific axis. The line-pair chart, in contrast, is highly useful for quantitative measurements of the contrast transfer function (CTF) at specific frequencies, making it a complementary tool in system evaluation. In practice, the Siemens star pattern is more convenient to utilize than the line-pair chart when searching for the highest spatial resolution.

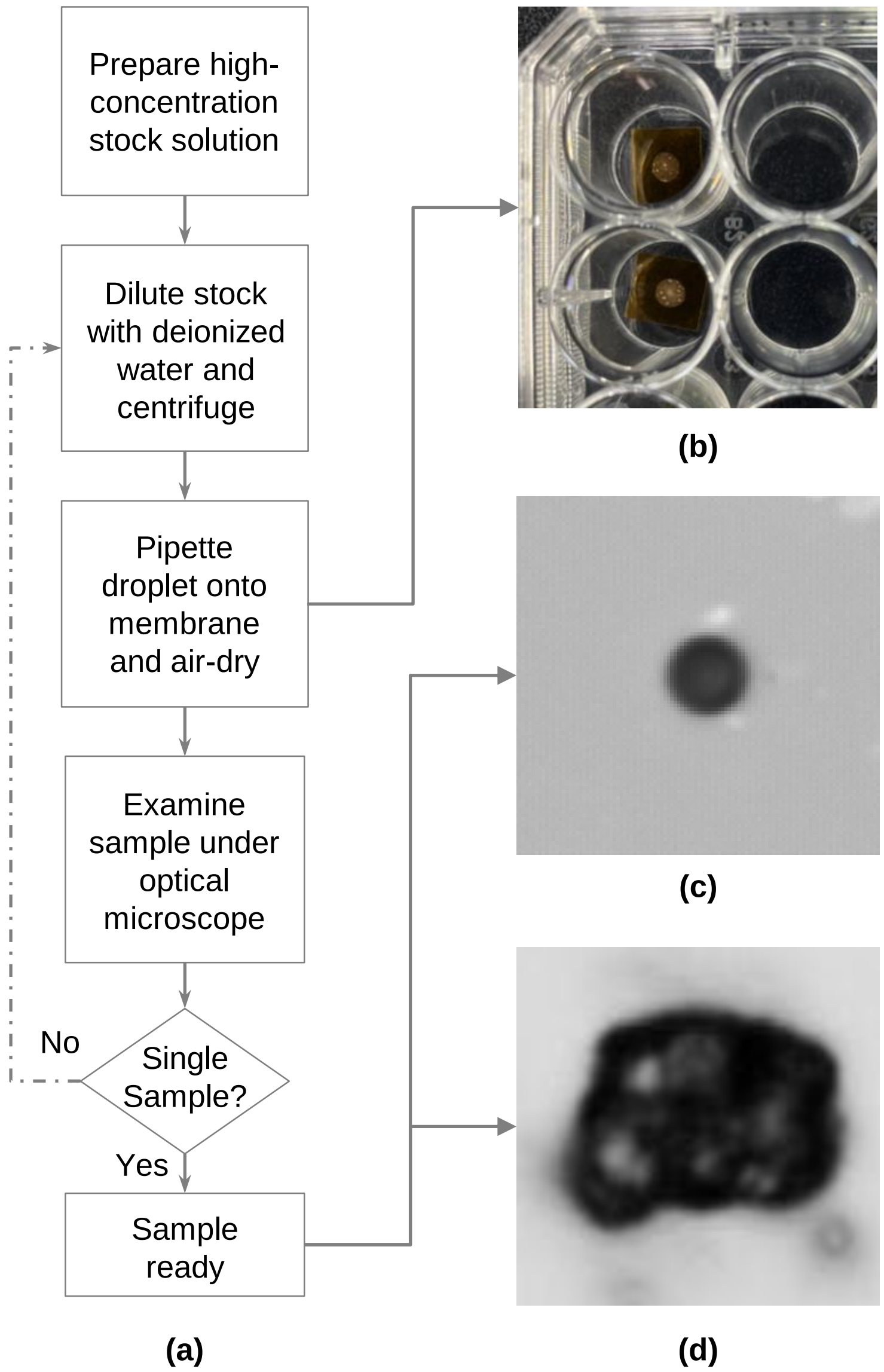

Figure 9.5: (a) The entire sample preparation workflow. (b) drop-cast particles on a Kapton membrane after air-drying, (c) optical microscope image of an individual 5 $\mu$m polystyrene (PS) microsphere, (d) optical microscope image of a biological cell with a diameter of approximately 10 $\mu$m.

Other samples may include particles and colloids, and the main sample preparation procedure is schemed in Fig. 9.5(a). First, a stock solution with known high concentration of particles or cells is prepared. Second, the stock solution is diluted with deionized water to achieve an appropriate concentration, and the sample is thoroughly dispersed by centrifugation. Third, a single small droplet of the diluted suspension is deposited onto the $Si_3N_4$ membrane via a micropipette and left to air-dry, which typically takes 5–10 minutes. Fourth, the membrane is inspected under an optical microscope to verify that individual particles or cells are well separated. If the concentration is too high (overlapping particles) or too low (no particles in the field of view), the dilution ratio has to be re-adjusted and the deposition step is repeated until a suitable distribution of isolated individual specimens is obtained. For biological cell samples, the cells are cultured, fixed, and if necessary stained[149, 150] before the dilution and deposition operations, with the final thickness kept below $\sim 20\,\mu$m to stay within the depth of field of our FZP.

### 9.3.3 Coordination mapping

It is essential to establish a coordination mapping from the mechanical stage coordinates to the X-ray optical coordinates. In practice, this is achieved by imaging a reference structure, e.g., a high-contrast grid or a Siemens star, at several positions within the field of view. By mapping the known physical dimensions of the reference structure to the corresponding pixel intervals on the recorded image, a linear transformation between the stage coordinate system and the X-ray image coordinate system can be determined and established. Essentially, such mapping procedure calibrates the system's real magnification response. By doing so, reliable navigation can be achieved. In particular, when a region of interest is identified on the image plane, the corresponding stage coordinates on the sample plane can be estimated such that the sample can be positioned accurately for subsequent high-resolution imaging or tomographic acquisition.

### 9.3.4 Sample localization via montage imaging

Since the field of view is limited to approximately $20\,\mu$m in diameter, therefore, locating the target from a much larger sample plane (e.g., $500\,\mu\text{m} \times 500\,\mu\text{m}$ on a $Si_3N_4$ membrane) would be extremely difficult and time-consuming. To address, a montage-based sample localization approach is employed. Specifically, the sample stage is scanned along the $x$ and $y$ directions (horizontal and vertical) with a step size of $20\,\mu$m, and a single projection is acquired at each position. These individual projections are then stitched together in sequence to form a large-area montage image. By comparing the montage image with the corresponding optical microscopy image, the target object can be identified, as illustrated in Fig. 9.6.

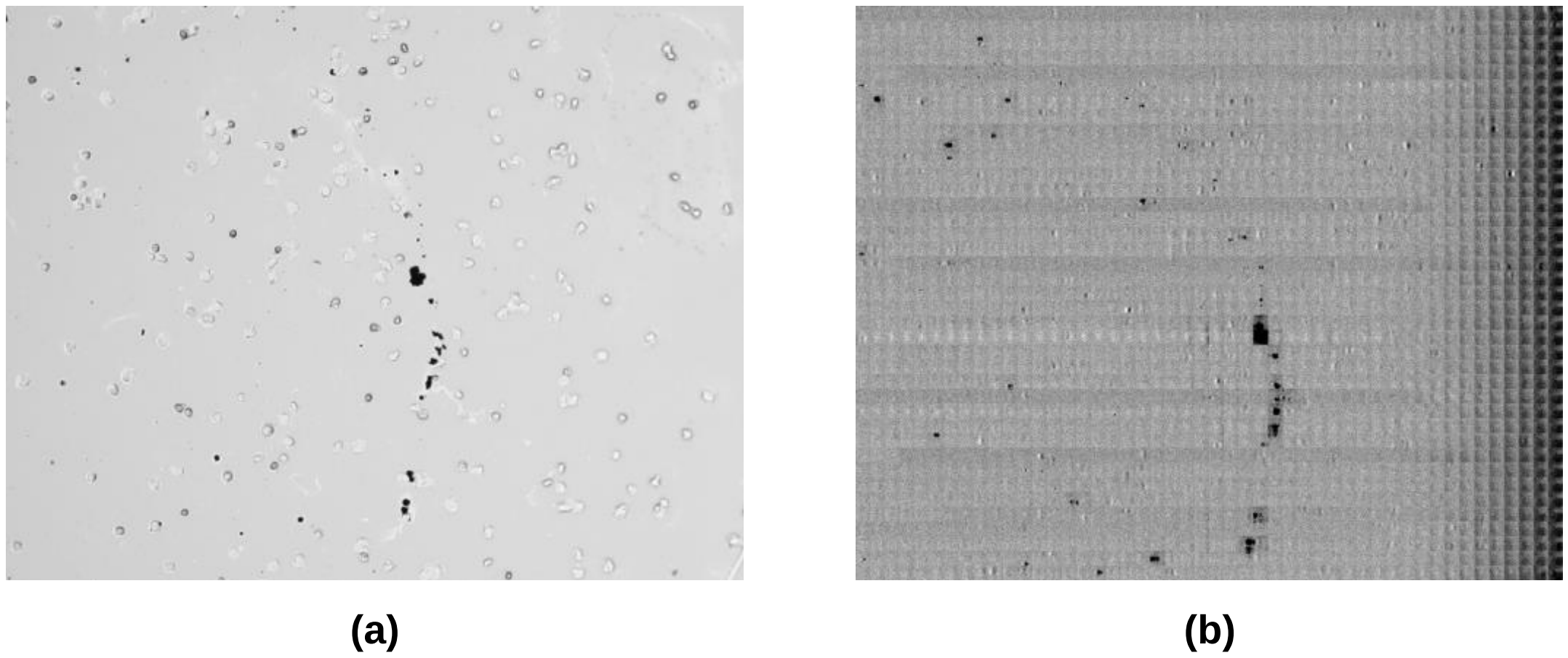

Figure 9.6: Sample localization procedure. (a) The optical microscopy image of the sample. (b) The montage image composed of a series of X-ray projections acquired at sequential stage positions. Each square in (b) corresponds to a 20 $\mu$m $\times$ 20 $\mu$m field of view of the X-ray microscope.

Additionally, sample localization can also be achieved using a special optical system that is pre-calibrated between the sample holding coordinates and the sample preparation coordinates[151].

## 9.4 Integrating the grating interferometer

The X-ray grating interferometer based phase contrast imaging system is added after the FZP based absorption contrast imaging system has been fully aligned and verified. Specifically, two additional gratings need to be integrated: the source grating $G_0$ is placed between the condenser and the sample to generate periodic coherent illumination, and the phase grating $G_1$ is placed near the back focal plane of FZP to form the Talbot self-imaging diffraction fringes on the detector plane, see the optical path layout shown in Fig. 9.7(a). The underlying working principles are discussed in details in Chapter 8.

The main challenge of integrating the grating interferometer lies in the grating alignments. Because neither grating is positioned at the imaging plane of the microscope, therefore, simply adding a grating into the beam path does not produce any periodic fringes on the detector plane except for a reduction of overall beam intensity. This makes the interferometer alignment and adjustment quite challenging. In fact, adjusting the phase grating $G_1$ is more important than the source grating $G_0$. To overcome, the Bertrand lens is added behind the FZP to focus the diffraction fringes on the detector, enabling visual feedback for the positioning and orientation alignment of both $G_0$ and $G_1$ gratings. The imaging results corresponding to the complete step by step alignment procedure are depicted in Fig. 9.7.

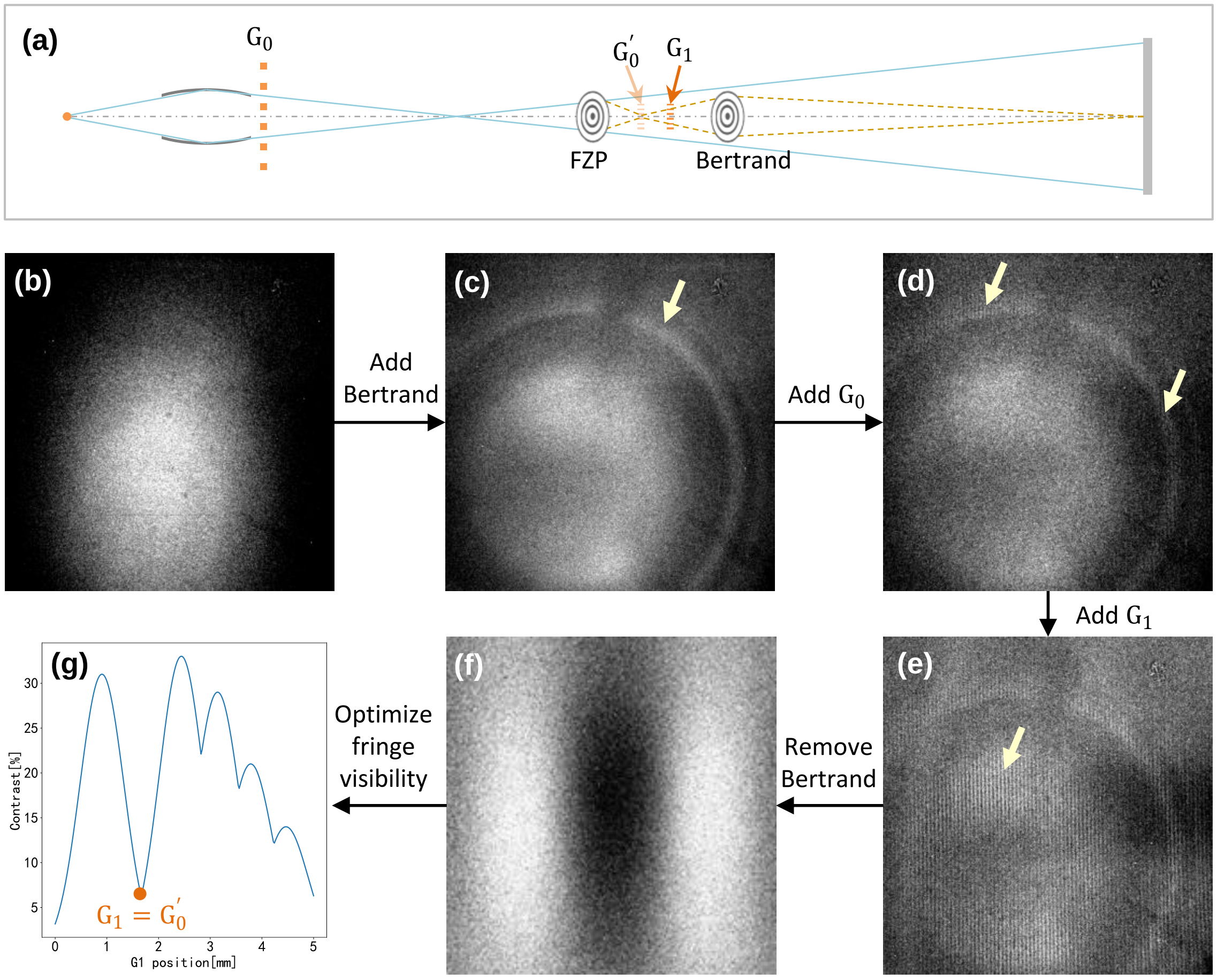

Figure 9.7: (a) Schematic of the optical path adjustment for the grating interferometer. $G_0'$ denotes the demagnified image of the absorption grating $G_0$. (b) Imaging result of the absorption-contrast system without samples. (c) Result after adding the Bertrand lens. (d) Result after adding the absorption grating $G_0$. (e) Result after adding the phase grating $G_1$. (f) Result after removing the Bertrand lens. (g) The fringe visibility responses at different $G_1$ grating positions.

**Establishing the reference image.** With the well aligned absorption contrast microscope (without sample), making sure that the FZP is able to produce a focused bright spot on the detector, see Fig. 9.7(b). This serves as the starting reference for aligning the grating interferometer system.

**Imaging the condenser outlet window with Bertrand.** Placing the Bertrand lens into the beam path. With it, the Bertrand lens is able to image the rear focal plane of the main FZP, which corresponds to the demagnified image of the condenser outlet window, see the annular ring in Fig. 9.7(c).

**Imaging the source grating $G_0$ with Bertrand.** Now adding the source grating $G_0$ into the beam path and positioning it close to the condenser outlet window. Immediately, the demagnified image of the periodic structure of $G_0'$ falls onto the object plane of the Bertrand lens and is captured by the detector. Rotating $G_0$ can align the grating structures along the vertical direction, see Fig. 9.7(d).

**Imaging the phase grating $G_1$ with Bertrand.** Next, adding the phase grating $G_1$ and position it near the plane of $G_0'$, i.e., the back focal plate of FZP. When the phase grating is roughly aligned with the image of the source grating, and is positioned at the right focusing position, similarly, the diffraction pattern formed on the phase grating should also become visible on the detector, see Fig. 9.7(e). By far, the Bertrand lens can be removed out of the beam path.

**Optimizing the fringe visibility.** Introducing a small angular tilt between the two gratings by rotating one of them with respect to the optical $z$ axis to generate large period Moiré fringes on the detector. By scanning within a certain angular range, the variation curve of the fringe visibility can be measured, see Fig. 9.7(f). It is found that the longitudinal position of $G_0$ is kind of relaxed compared to that of $G_1$, which is far more sensitive. Notably, the fringe visibility reaches to a local minimum rather than a maximum if $G_1$ is placed exactly at the $G_0'$ plane. As a consequence, moving $G_1$ along the optical axis away from this plane in either direction would yield higher fringe visibility, which is always expected for high quality phase contrast imaging. The main procedure to find the optimal $G_1$ position is as follows:

(i) Scanning $G_1$ grating along the optical axis within a certain range and identify the position where the minimum visibility appears. Usually, the visibility responses are symmetric with respect to this particular position on both sides;

(ii) Moving $G_1$ grating toward the detector to get the highest visibility response, which tends to be slightly higher than the highest visibility response on the opposite side in our experiment. Additionally, it also contains multiple local visibility maxima;

(iii) Verifying it by moving $G_1$ grating to its symmetric position with respect to the minimum point. At this symmetric position, the diffraction fringe should have a larger period. By far, the entire alignment of the grating interferometer has been completed.

Upon completion of the above alignments, now the grating interferometer is fully integrated into the FZP based X-ray microscope. Thus, it becomes a FZP based X-ray phase contrast microscope, and is ready to acquire the multi-contrast (absorption, phase and dark-field) imaging data.

## 9.5 Testing and verification

Rigorous testings and verifications are needed to confirm the designed specifications. The main verifications contain the following three key aspects:

**Spatial resolution.** A Siemens star resolution pattern is imaged at first with certain beam energy, exposure time and detector settings. The finest resolvable feature is measured. If the obtained image spatial resolution is worse than expected due to some unknown reasons, then the FZP axial alignment, detector focusing, vibration, and temperature stability should be re-examined in sequence, as described in Section 9.2.

**Stability and repeatability.** To do so, a fixed sample is imaged repeatedly over a prolonged period, e.g., several hours. Ideally, the spatial drift between two acquired consecutive images should remain well below the target resolution, i.e., 50 nm. Based on our experience, any systematic drift may indicate thermal instability (Section 2.3) or mechanical motion stage relaxation (Section 7.1). For CT imaging, the stability requirement is even more stringent. Namely, the sample drift must remain far smaller than the depth of field to avoid resolution degradation, and the sample must stay in the field of view over the entire acquisition duration, which may last for several hours or several days.

## 9.6 Maintenance

In fact, regular maintenance is needed to support and enable long-term system service. For instance, periodic re-alignment of key optical components such as FZP, sample, and gratings to correct their slight drifts from the optimal working positions. In addition, the pressure status of the vacuum chamber needs to be monitored and checked constantly to avoid strong X-ray attenuation loss along the beam path. Finally, the system has to be operated with proper frequency and intensity. Do NOT overuse it or underuse it, for instance, turn off the power for two weeks or ever longer time. All of these extreme situations would be risky (may make damage) to the hardware.

# 10 Data acquisition and post-processing

This chapter describes the details of data acquisition for both absorption contrast and phase contrast imaging applications, as well as the key post-processing procedures that converts the original detector measurements into the final outcome.

## 10.1 Data acquisition

Due to the extremely low source brilliance, as a consequence, the laboratory X-ray microscope needs a very long data acquisition period ranges from seconds/minutes up to hours/days to ensure adequate image signal-to-noise ratio. However, unavoidable mechanical drifts of the sample stage or the FZP may cause motion blurring artifacts. Therefore, to mitigate, a multi-frame acquisition scheme is adopted. Specifically, the single-exposure time is chosen such that the drift within one frame remains well below the targeting spatial resolution. Afterwards, a series of such short-exposure frames are acquired at the same projection angle. Finally, these frame sequences are post-processed (registration) and averaged to reduce the quantum noise.

Depending on the certain imaging mode, the aforementioned overall acquisition workflow may differ accordingly:

**Absorption contrast imaging mode.** For this particular imaging mode, the grating interferometer is assumed stay out of the beam path. For each projection view, the multi-frame registration, see Section 10.2.3, and averaging are performed to produce a good enough absorption projection. After gathering all the projections from a wide angular range, inter-angle rotational alignment is performed to correct for any sample drift or rotation-axis wobble, see Section 10.2.3, in order to finally reconstruct sharp CT images with minimum motion artifacts.

**Phase contrast imaging mode.** The phase-stepping [152] based data acquisition scheme is employed for this particular imaging mode. Specifically, the source grating $G_0$ (or the phase grating $G_1$) is stepped $M_{ps}$ times within one grating period. At any given projection angle, such phase-stepping procedure is repeated $M_{ps}$ times. The absorption, phase and dark-field signals are extracted from these phase-stepping data, see Section 10.2.1. Due to the strong magnification effect of the zone plate, the extracted phase signal exhibit a characteristic signal splitting, which must be recovered with certain algorithm, see Section 10.2.2. Finally, image registration and alignment are performed to reconstruct sharp CT images, see Section 10.2.4.

## 10.2 Signal post-processing

As mentioned, dedicated signal post-processing algorithms are needed to extract, correct, denoise and reconstruct the final outcome.

### 10.2.1 Signal extraction

For absorption-contrast imaging mode, the line integral signal is extracted via the following logarithmic calculation with respect to the ratio of the measured beam intensity $I_{bkg}$ without object and beam intensity $I_{obj}$ with object:

$$\mu L = \ln\left(\frac{I_{bkg}}{I_{obj}}\right), \tag{10.2.1}$$

For phase contrast imaging mode, the following formulas can be utilized to extracted three different signals from the measured phase stepping dataset $\{I^{(m)}\}$ (see Eq. (8.2.12)):

$$I_0 = \frac{1}{M_{ps}} \sum_{m=1}^{M_{ps}} I^{(m)}, \tag{10.2.2}$$

$$\varphi = \tan^{-1}\left[-\frac{\sum_{m=1}^{M_{ps}} I^{(m)} \sin(\frac{2\pi m}{M_{ps}})}{\sum_{m=1}^{M_{ps}} I^{(m)} \cos(\frac{2\pi m}{M_{ps}})}\right], \tag{10.2.3}$$

$$I_1 = \frac{2}{M_{ps}} \sqrt{\left[\sum_{m=1}^{M_{ps}} I^{(m)} \sin(\frac{2\pi m}{M_{ps}})\right]^2 + \left[\sum_{m=1}^{M_{ps}} I^{(k)} \cos(\frac{2\pi m}{M_{ps}})\right]^2}. \tag{10.2.4}$$

Consequently, the absorption contrast image, the phase contrast image and the dark-field contrast image are obtained as below:

$$\mu L = \ln\left(\frac{I_{0,bkg}}{I_{0,obj}}\right), \tag{10.2.5}$$

$$\Delta\varphi = \varphi_{obj} - \varphi_{bkg}, \tag{10.2.6}$$

$$\mu L_d = \ln\left(\frac{\varepsilon_{bkg}}{\varepsilon_{obj}}\right) = \ln\left(\frac{I_{1,bkg} I_{0,obj}}{I_{0,bkg} I_{1,obj}}\right), \tag{10.2.7}$$

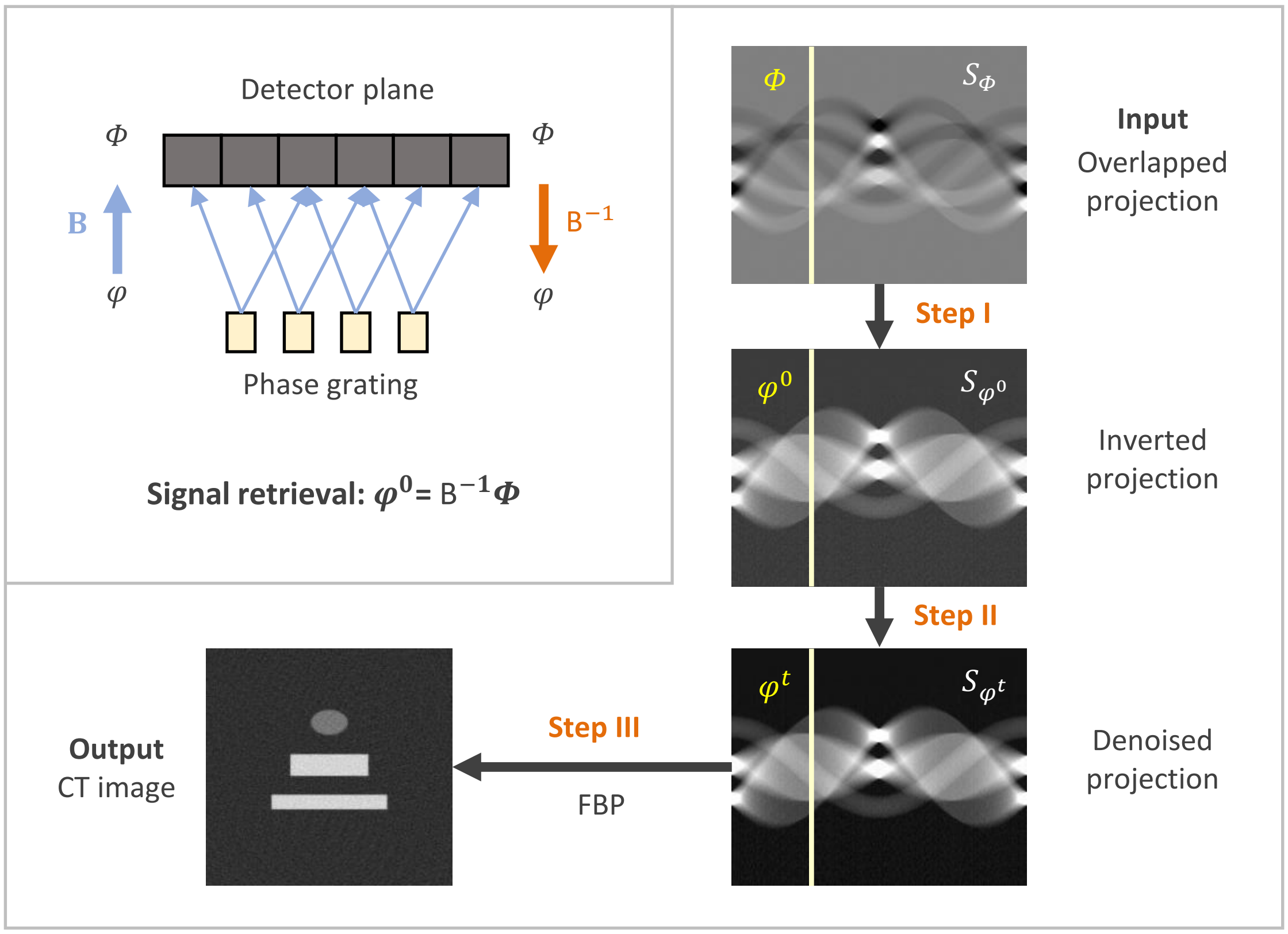

Figure 10.1: Workflow of the splitted phase recovery algorithm. In step one, the split projection $S_\Phi$ is converted to the unsplit projection $S_{\varphi^{(0)}}$ using $\mathbf{B}^{-1}$. In step two, $S_{\varphi^{(0)}}$ is iteratively denoised according to Eq. (10.2.14) for all projection views. In the final step, the phase contrast CT image is reconstructed from the recovered sinogram $S_{\varphi^{(t)}}$ using the FBP algorithm. The highlighted line denotes the phase projection acquired at one representative view.

### 10.2.2 Phase signal recovery

As have been demonstrated [146, 147, 153], the phase contrast signal extracted from a zone-plate based X-ray microscope is splitted. According to Eq. (8.2.18), the extracted phase signal $\Phi$ is rewritten as

$$\Phi(x) = \varphi\left(x + \frac{\Delta s}{2}\right) - \varphi\left(x - \frac{\Delta s}{2}\right), \tag{10.2.8}$$

where $\varphi$ denotes the phase signal before splitting, and $\Delta s$ denotes the splitting distance.

To recover the unsplitted signal $\varphi$, Eq. (10.2.8) is re-expressed into the following matrix form

$$\Phi = \mathbf{B}\varphi + n_\varphi, \tag{10.2.9}$$

where $n_\varphi$ represents measurement noise, and matrix $\mathbf{B}$ is defined as

$$\mathbf{B} = \mathbf{b}_{m\times m} + \boldsymbol{O}, \tag{10.2.10}$$

where $\mathbf{b}_{m\times m}$ denotes a sparse banded matrix with entries $+1$ and $-1$ located at offsets corresponding to $\pm\Delta$ from the diagonal, encoding the shifting and subtraction in Eq. (10.2.8). Herein, $m$ denotes the number of detector elements, $\Delta = \Delta s/(2w)$ denotes the half-splitting distance expressed in detector element units of width $w$, and $\boldsymbol{O} = \mathrm{diag}(\gamma, \gamma, \ldots, \gamma)$ with $\gamma = 10^{-12}$ denotes a small regularization term that ensures the invertibility of $\mathbf{B}$. When $\Delta$ corresponds to a non-integer number of detector elements, fractional weights are used to interpolate between adjacent elements.

If noise $n_\varphi$ is ignored, in principle, the unsplitted phase signal can be recovered via $\varphi = \mathbf{B}^{-1}\Phi$, see Fig. 10.1. In practice, however, such analytical inversion of Eq. (10.2.9) would dramatically amplify noise. To suppress noise while preserving signal fidelity, a penalized weighted least-squares model with total variation (PWLS-TV) regularization [154] is applied

$$\varphi^* = \arg\min_{\varphi\geq 0} \left\{ (\Phi - \mathbf{B}\varphi)^T \Lambda^{-1} (\Phi - \mathbf{B}\varphi) + \alpha\, \mathbf{R}_{\mathrm{TV}}(\varphi) \right\}, \tag{10.2.11}$$

where $\alpha$ denotes a smoothing parameter controlling the trade-off between data fidelity and noise suppression, $\Lambda$ denotes a diagonal weight matrix whose $q$-th element is [155]

$$\sigma^2_{\varphi_q} = \frac{2}{\varepsilon^2 \sum_{m=1}^{M_{ps}} I_q^{(m)}}, \tag{10.2.12}$$

with $\varepsilon$ denotes the fringe visibility, $M_{ps}$ denotes the total number of phase steps, and $I_q^{(m)}$ denotes the detected photon count at the $q$-th detector element during the $m$-th step. The total variation prior $\mathbf{R}_{\mathrm{TV}}(\varphi)$ is defined as

$$\mathbf{R}_{\mathrm{TV}}(\varphi) = \sum_q \sqrt{(\varphi_q - \varphi_{q-1})^2 + \upsilon}, \tag{10.2.13}$$

where $\upsilon$ denotes a small constant that keeps the prior term differentiable. The optimization in Eq. (10.2.11) is solved iteratively:

$$\varphi^{(t)} = \varphi^{(t-1)} - \eta^{(t-1)} \mathbf{B}^T \left( \Lambda^{-1} (\mathbf{B}\varphi^{(t-1)} - \Phi) \right) - \alpha\tau\, \frac{\nabla \mathbf{R}_{\mathrm{TV}}(\varphi^{(t-1)})}{\|\nabla \mathbf{R}_{\mathrm{TV}}(\varphi^{(t-1)})\|}, \tag{10.2.14}$$

where $\varphi^{(t)}$ denotes the recovered phase image at iteration $t$, $\tau$ denotes a scalar factor, and $\eta^{(t-1)}$

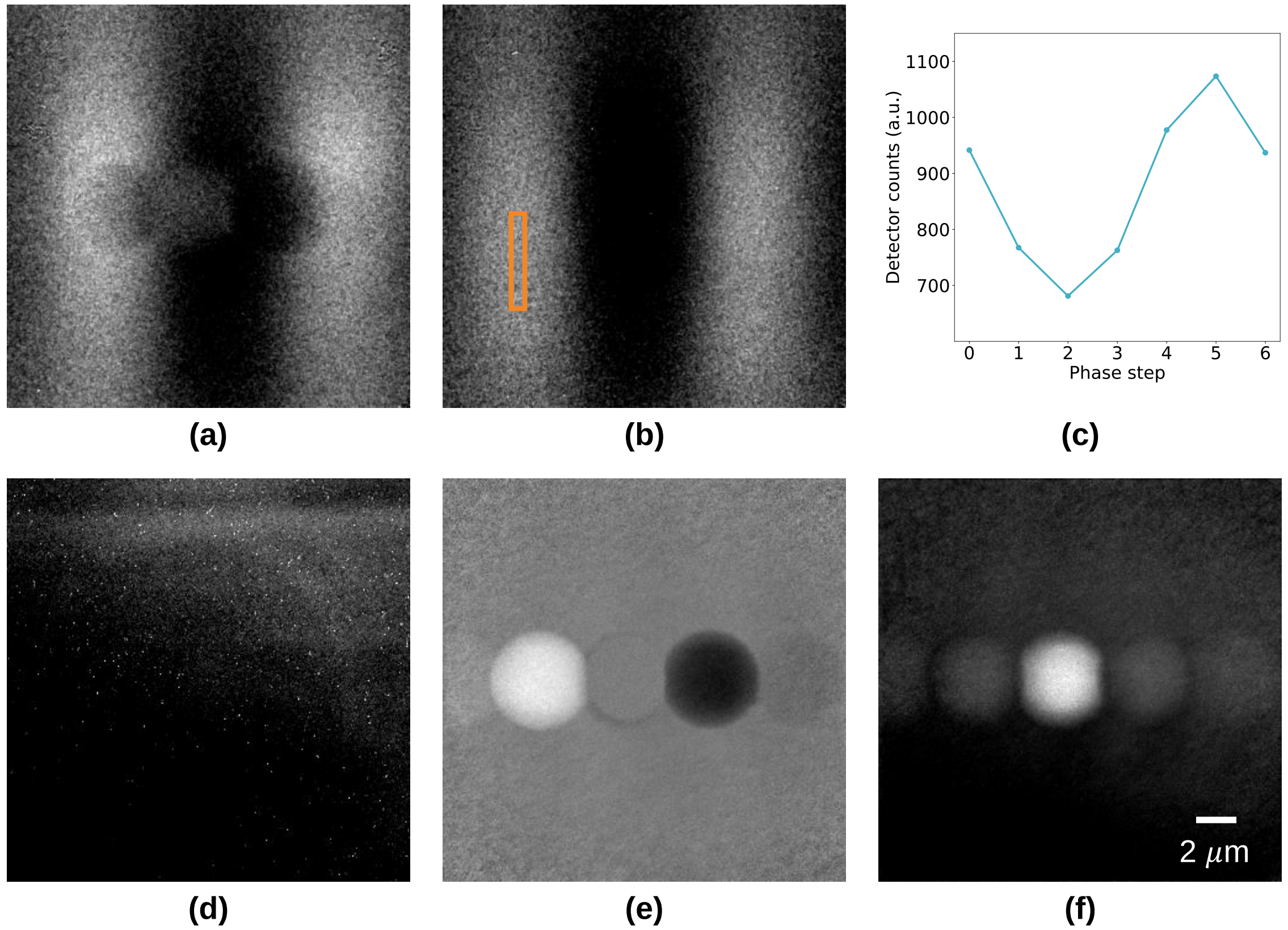

Figure 10.2: Multi-contrast imaging results of a single 5 $\mu$m diameter PS microsphere. (a) and (b) show the phase-stepping projection with and without the sample, respectively. (c) The obtained phase-stepping curve. (d)–(f) The extracted absorption, phase, and dark-field signals, respectively.

denotes the gradient step size [156]

$$\eta^{(t-1)} = \frac{G^T G}{(\mathbf{B}G)^T(\Lambda^{-1}(\mathbf{B}G))}, \quad G \triangleq \mathbf{B}^T\left(\Lambda^{-1}(\mathbf{B}\varphi^{(t-1)} - \Phi)\right). \tag{10.2.15}$$

As illustrated in Fig. 10.1, the split sinogram $S_\Phi$ is first converted to an unsplit sinogram $S_{\varphi^{(0)}}$ via $\mathbf{B}^{-1}$, then is iteratively denoised according to Eq. (10.2.14) for all projection views to yield the recovered sinogram $S_{\varphi^{(t)}}$.

To validate the splitting model and recovery algorithm, a single 5 $\mu$m diameter polystyrene (PS) microsphere was imaged. As shown in Fig. 10.2, the absorption, phase, and dark-field projections of the PS microsphere are simultaneously extracted from a single phase-stepping dataset, demonstrating the multi-contrast capability of this system. As shown in Fig. 10.2 and Fig. 10.3, the measured splitting distance of the split phase contrast projection agrees well with the theoretical prediction. After applying the recovery algorithm, a clear single phase contrast projection is obtained. Moreover, the measured phase value of approximately 1.2 rad is consistent with the theoretically expected value for a 5 $\mu$m microsphere, demonstrating the quantitative imaging capability of this system.

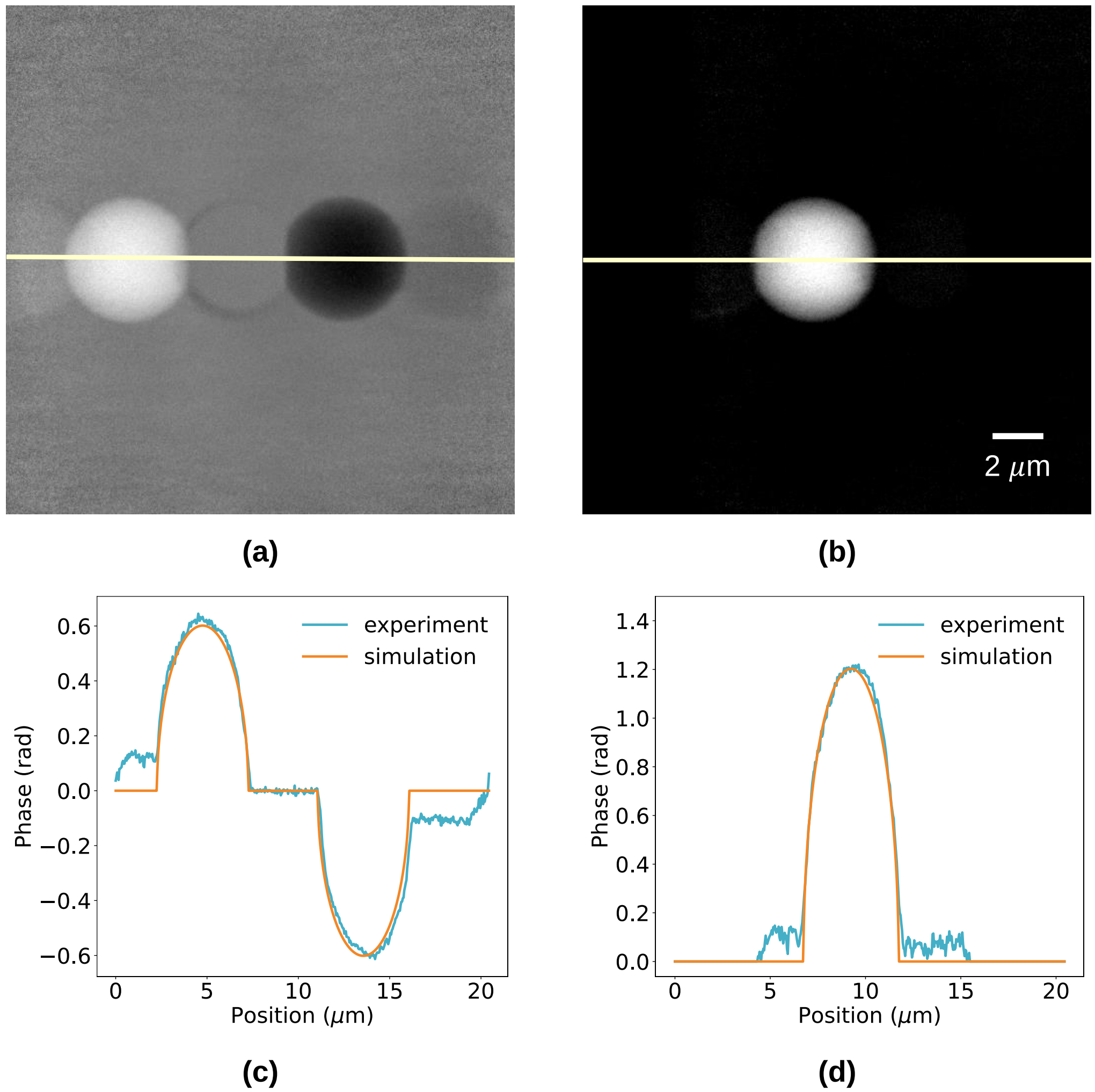

Figure 10.3: Phase-contrast projections of a single 5 $\mu$m diameter PS microsphere: (a) split phase contrast projection, (b) recovered single phase contrast projection after applying the splitting recovery algorithm. (c) and (d) show the line profiles along the highlighted yellow lines in (a) and (b), respectively, compared with the corresponding simulated profiles.

### 10.2.3 Image registration

**Intra-angle frame registration.** This post-processing procedures help to mitigate the motion artifacts between adjacent frames at a specific projection angle. In particular, a sum-of-squared-differences metric is minimized between the feature profiles of the correction frame and the reference frame. In particular, the reference frame could be selected as the preceding adjacent frame, or the first frame, or any suitable frame. In our research, the feature profiles were obtained by averaging the image along both the horizontal and vertical directions, separately, as plotted in Fig. 10.4. As shown, these structural features vary from frame to frame before registration. After registration, the structural features are well aligned across all frames. As a result, the spatial resolution of the Siemens star pattern gets substantially improved.

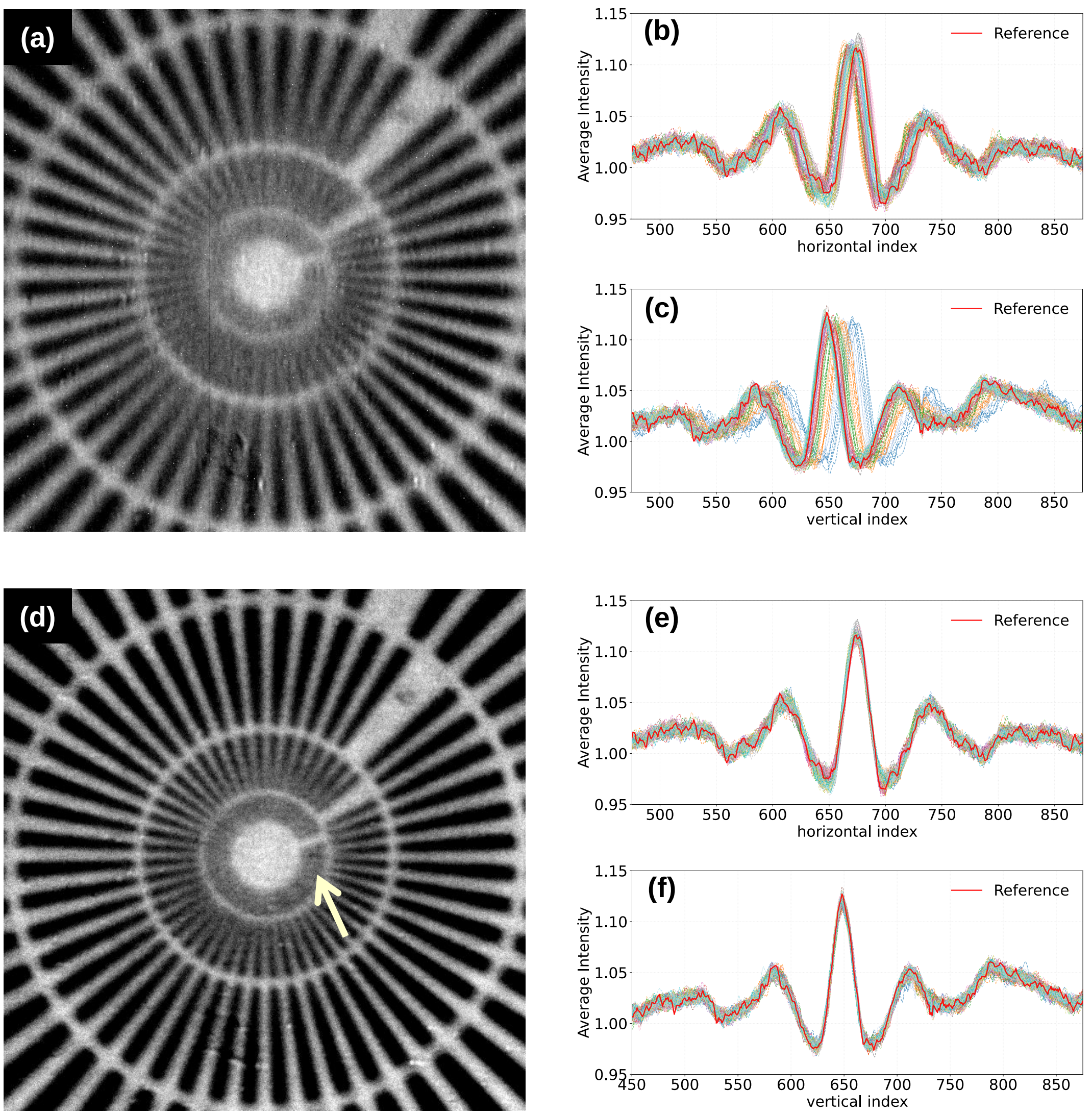

Figure 10.4: (a) The averaged projection images before registration. (b) The horizontal profiles of every single projection image without registration. (c) The vertical profiles of every single projection image without registration. (d) The averaged projection images after registration. (e) The horizontal profiles of every single projection image with registration. (f) The vertical profiles of every single projection image with registration.

**Inter-angle projection registration.** During the entire CT data acquisition period, the strength and direction of mechanical/thermal drifts vary significantly from one projection view to the other. Such motion artifacts may lead to data inconsistency for CT image reconstruction. Therefore, these inter-angle motions need to be corrected before reconstruction. Such registrations contain two major steps: First, computing the 2D center of mass of the intra-angle registered projection for every view angle. Second, translating each projection horizontally or vertically (independently along the two directions) until the deviation of its center of mass from the mean center of mass falls within one pixel. By doing so, the vertical and horizontal center of mass should remain constant across all projection angles. As a consequence, the rotation axis is properly corrected to align with the center of the sinogram. The reconstructed CT images before and after inter-angle projection registration are compared in Fig. 10.5.

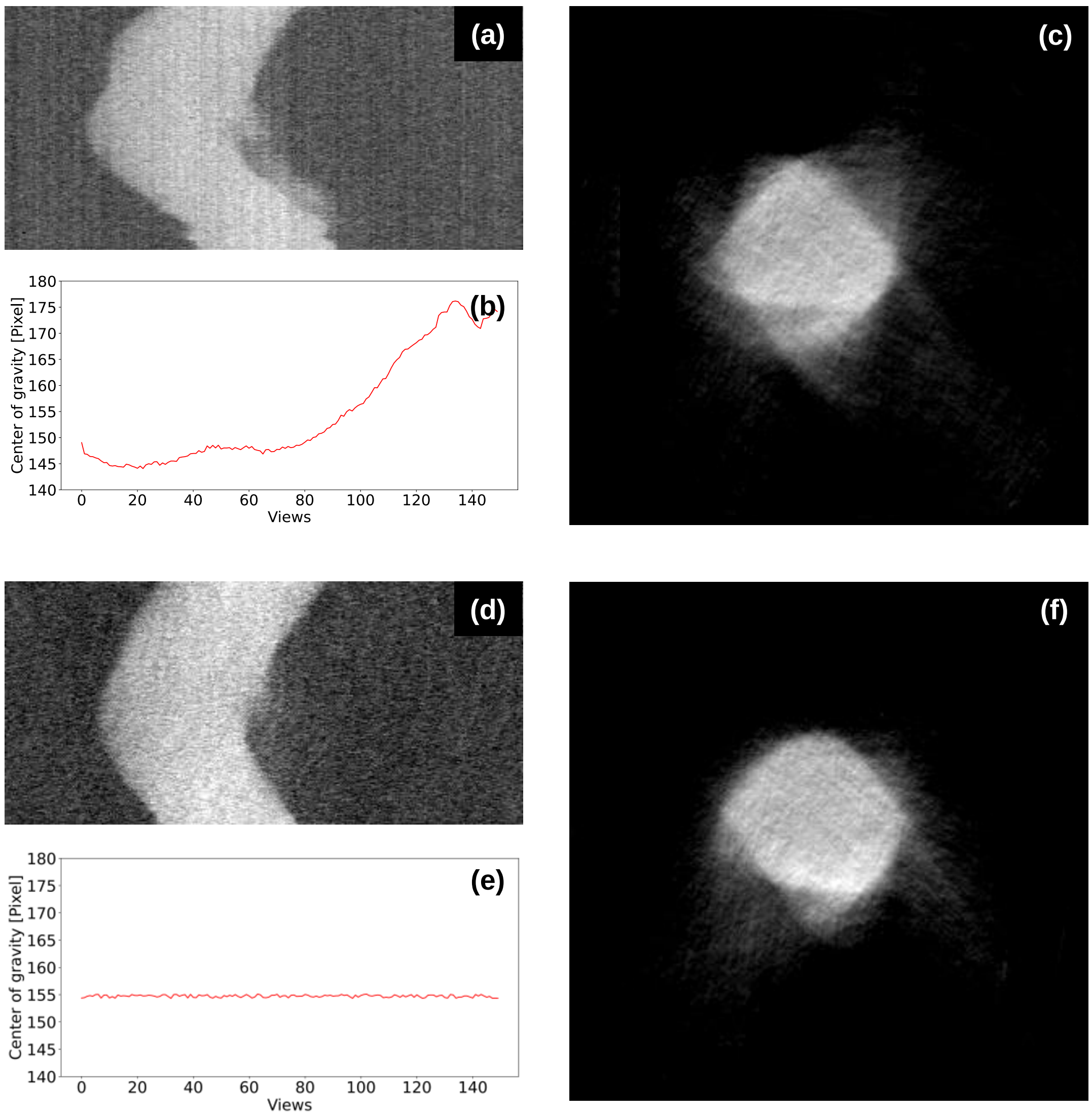

Figure 10.5: (a) The sinogram before correction, (b) The horizontal center of mass versus projection view angle, (c) The axial slice of the reconstructed CT image. (d) The sinogram after correction, (e) The horizontal center of mass versus projection view angle, (f) The axial slice of the reconstructed CT image.

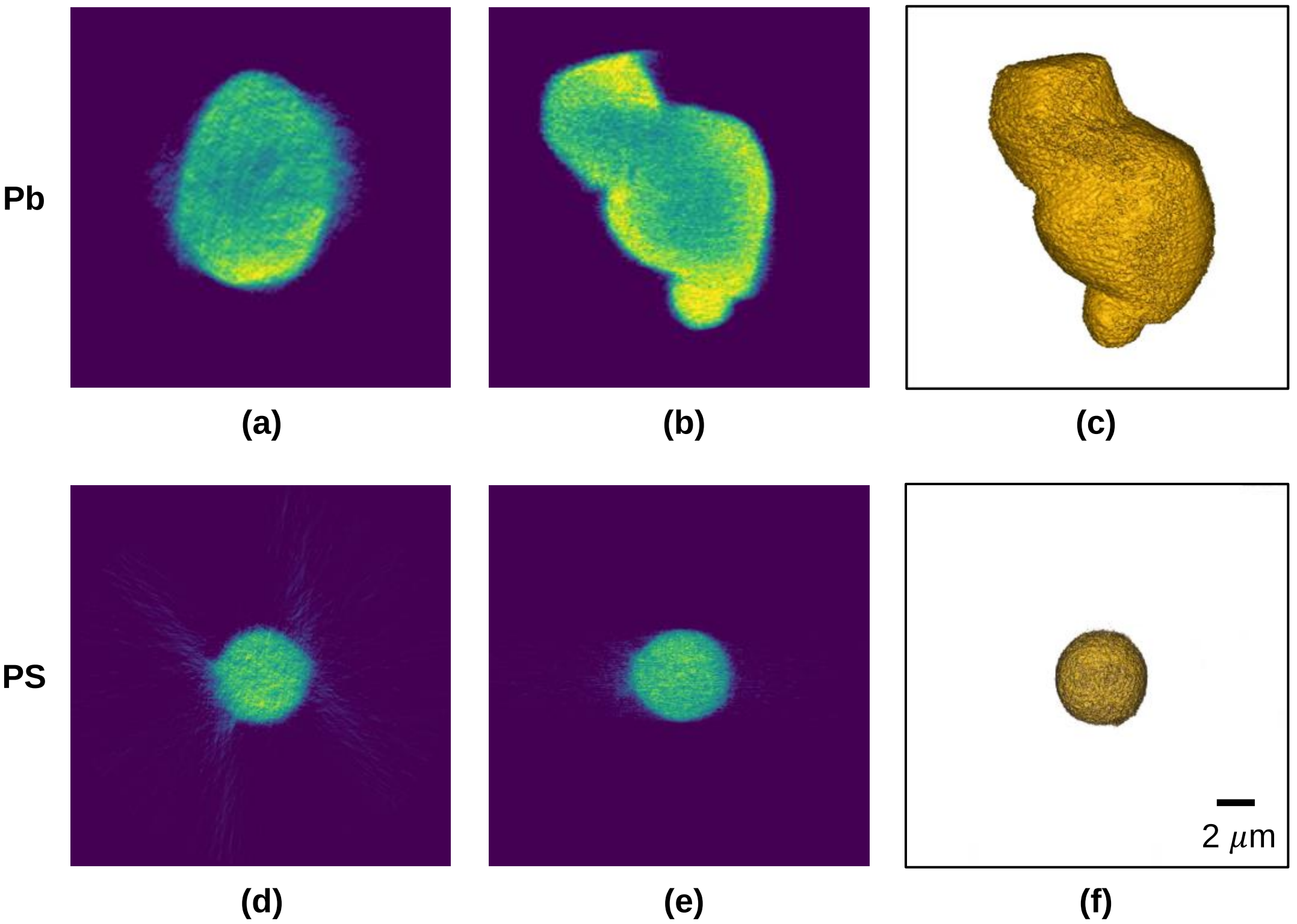

Figure 10.6: The CT reconstruction results. (a)–(b) are the absorption contrast CT slices of tungsten particles along the $x-y$ cross-section and $y-z$ cross-section, respectively, (c) denotes the three-dimensional volume rendering result. (d)–(e) are the phase contrast CT slices of a single $5\,\mu$m diameter PS microsphere along the $x-y$ cross-section and $y-z$ cross-section, respectively, (f) denotes the three-dimensional volume rendering result.

Be aware that the above registration schemes can be utilized for both absorption-contrast imaging mode and the phase contrast imaging mode.

### 10.2.4 CT Reconstruction

The final step is the reconstruction of the three-dimensional CT image volume. In our implementation, the reconstruction is performed using the LEAPCT[157, 158] library, which supports both analytical and iterative reconstructions. The system geometry is configured as a fan-beam model with finite source-to-object distance ($L_{\mathrm{SOD}}$) and source-to-detector distance ($L_{\mathrm{SDD}}$), which provides a more accurate description of the actual beam divergence than the parallel-beam approximation. Specifically, the filtered back-projection (FBP) algorithm with a standard Ramp filter is suggested if the projections have adequate signal-to-noise ratio. Otherwise, the Simultaneous Algebraic Reconstruction Technique (SART) algorithm is recommended as the default method due to its robustness against noise and its ability to produce satisfactory results with a moderate number of iterations (typically 3 iterations with 10 subsets). In addition, a median-based ring artifact removal filter is applied to suppress detector-specific stripe artifacts in the sinogram before CT image reconstruction.

The reconstructed CT image volumes are shown in Fig. 10.6. Results in the first row are the absorption-contrast CT images of a tungsten particle sample. As seen, the microstructure can be clearly resolved. Results in the second row are the phase contrast CT images of a single PS microsphere.

## 10.3 Quantitative analyses and non-ideal impacts

### 10.3.1 Spatial resolution characterization

The high resolution imaging results of a Siemens star pattern are shown in Fig. 10.7, and the innermost spike structures can be clearly resolved, see the highlighted arrows in Fig. 10.7(b). Two different approaches, the modulation transfer function (MTF) analysis and the power spectral density (PSD) analysis, are employed to quantify the image spatial resolution.

Specifically, the MTF approach provides a quantitative measurement of the image resolution according to the Rayleigh criterion. It is measured by scanning line profiles across features with different spike widths, denoted as $\Delta w$, and computing the intensity modulation at each spatial frequency. The normalized intensity modulation is defined as

$$\mathrm{MTF}(\Delta w) = \frac{I_{\mathrm{Au}}(\Delta w) - I_{\mathrm{bkg}}(\Delta w)}{I_{\mathrm{Au}}(\Delta w)}, \tag{10.3.1}$$

where $I_{\mathrm{Au}}$ denotes the signal intensity in the gold region, $I_{\mathrm{bkg}}$ denotes the signal intensity at the center of the background region, and $\Delta w$ denotes the spike width of the selected line structures. The measured intensity modulations are plotted as a function of the spatial frequency (lines per micron), see the plot in Fig. 10.7(c). The spatial resolution is determined by the Rayleigh criterion: the feature size corresponding to a 26.5% modulation level [159, 160]. Based on this analysis, a spatial resolution of 35.0 nm is obtained for this FZP based X-ray microscope.

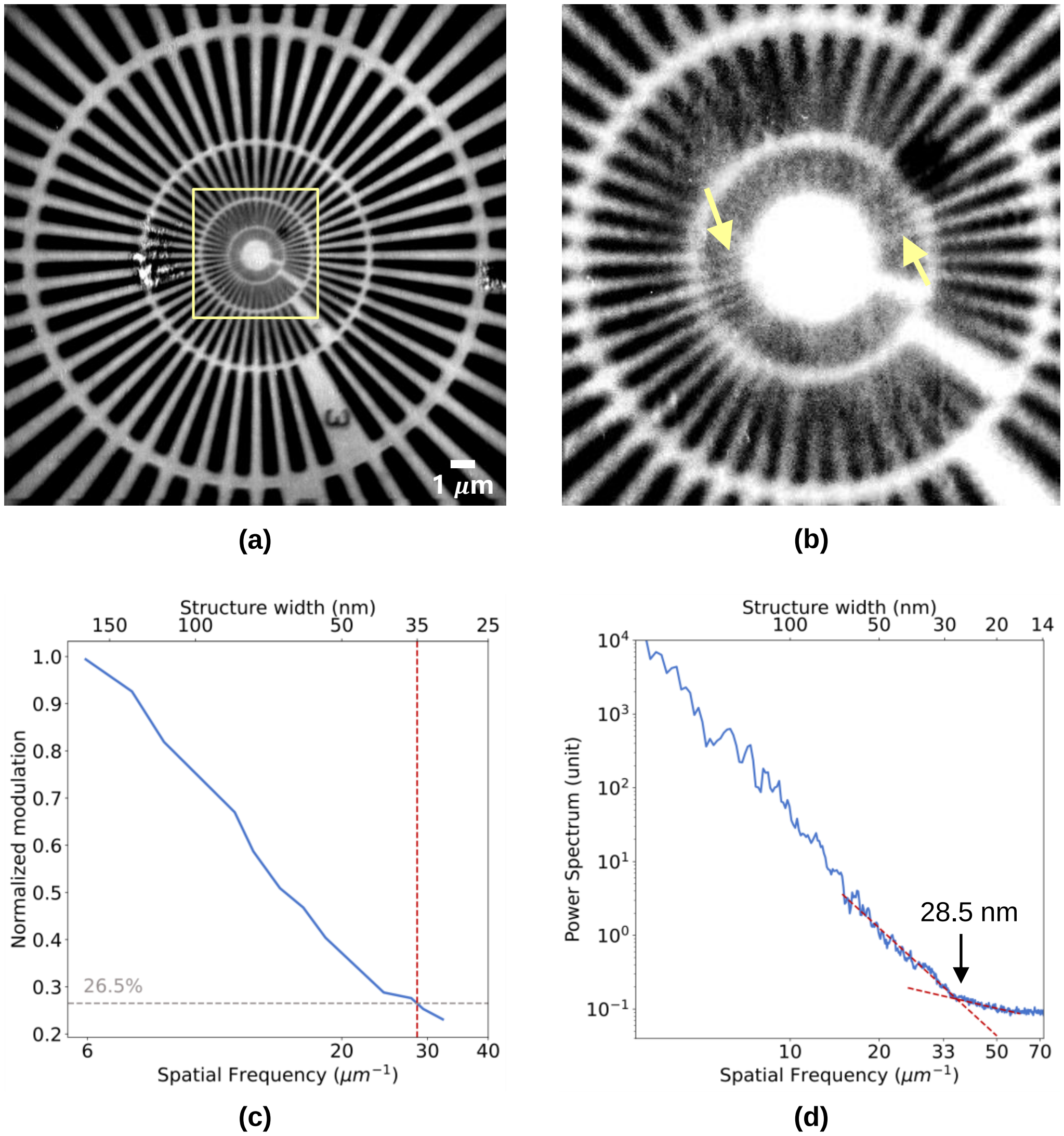

Figure 10.7: Results of the Siemens star resolution pattern. (a) the overall X-ray projection, (b) the zoomed-in result, and the arrows highlight the 35.0 nm line structures. (c) the MTF curve (the gray dashed line corresponds to the 26.5% modulation level). (d) the PSD curve, the arrow points the cutoff frequency $f_c$, corresponding to the system's limiting spatial resolution capability of 28.5 nm.

The PSD analysis [160, 161] results are shown in Fig. 10.7(d). The physical meaning of the PSD curve is interpreted as follows: At low spatial frequencies, the object spectrum is dominated by strong structural signal contributions from features of varying sizes. At high spatial frequencies, the object spectrum decreases monotonically and converges to the background noise level, which contributes a constant offset across all frequencies. Therefore, the normalized ratio of the two spectra clearly reveals the transition between these two regimes: below the cutoff frequency $f_c$ the ratio reflects the object morphology, while above $f_c$ the ratio becomes structureless, indicating that only noise is present. In brief, the PSD curve determines the minimum detectable structure size by identifying the cutoff frequency between the signal-dominated and noise-dominated regimes. The entire PSD calculations involve the following steps: First, a two-dimensional Fourier transformation is applied onto the acquired X-ray image of the Siemens star pattern. Second, the resulting two-dimensional power spectrum is averaged azimuthally, namely, along concentric circles in the frequency space, at each fixed radial frequency to yield a one-dimensional power spectrum. Third, the above power spectrum of the Siemens star pattern is normalized by that of a reference image acquired without any object to mitigate the impact of background noise. Finally, a linear fitting is applied to identify the transition point, corresponding to the cutoff frequency $f_c$, in the normalized spectrum. By default, the spatial resolution is determined as one half of the inverse cutoff frequency, i.e., $1/(2f_c)$. Consequently, the extracted minimum detectable structure size was found to be 28.5 nm, which should be interpreted as the limiting spatial resolution capability of this FZP based X-ray microscope.

### 10.3.2 Impact of Au thickness and duty ratio

It is worth mentioning that both the Au thickness and duty ratio of the Siemens star pattern can affect the signal contrast. To demonstrate, a numerical simulation study was performed, see the results shown in Fig. 10.8. Specifically, three Au thicknesses were studied: 2000 nm, 700 nm, 250 nm, and the duty ratio varies from 1:1 to 2:1. Results indicate that as the duty ratio increases, the signal contrast diminishes progressively. Moreover, the image contrast drops much more dramatically as the Au thickness decreases. To compensate the reduction of image contrast, much longer exposure time will be needed to maintain the high contrast-to-noise ratio. For example, if the image contrast is reduced by a factor of 2 due to the thinner Au material, as a consequence, the exposure time would be prolonged by a factor of 4. Thereby, the Au in Siemens star pattern sample should be sufficiently thick to minimize the entire imaging time. In addition, the fabrication accuracy should be high enough to keep the duty ratio as close to 1:1 as possible.

### 10.3.3 Impact of quantum noise

For high spatial resolution imaging, the image quality may be significantly limited by quantum noise. To investigate, a numerical simulation study with varying dose levels was performed, and results are shown in Fig. 10.9. At the similar dose level of our system, only mitochondrial clusters can be barely distinguished, see Figs. 10.9(c)-(d). Whereas, the mitochondrial distribution becomes clearly resolved if the exposure time is increased by 16-folds, i.e., image noise reduced by 4 times, see Figs. 10.9(e)-(f).

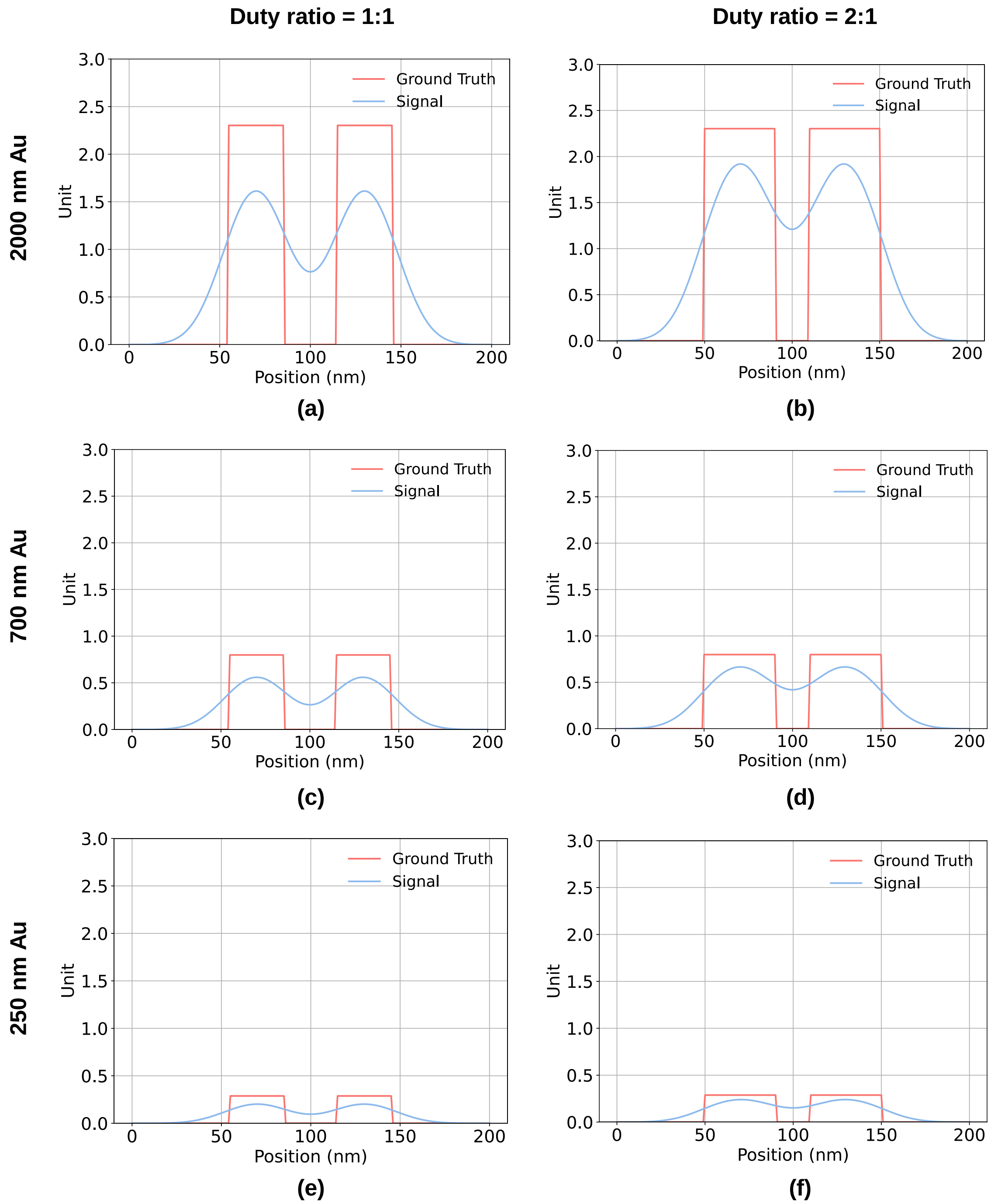

Figure 10.8: Simulation study of the impact of duty ratio and gold thickness on signal contrast for the Siemens star pattern. Plots in the first column are with 1:1 duty ratio, and plots in the second column are with 2:1 duty ratio. Plots in the first row are obtained with 2000 nm thick Au, plots in the second row are obtained with 700 nm thick Au, and plots in the third row are obtained with 250 nm thick Au.

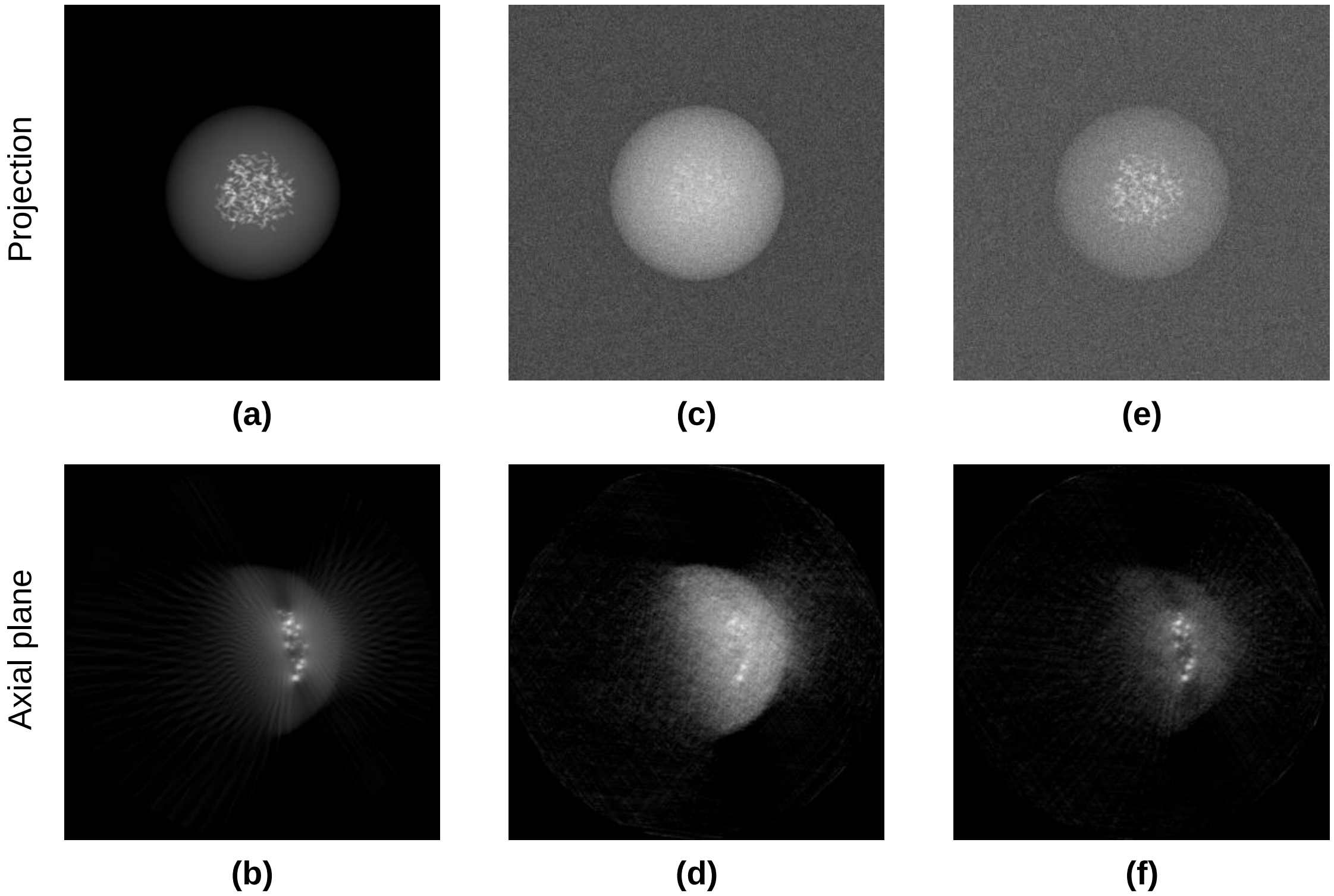

Figure 10.9: Simulation results obtained at different dose levels. (a)–(b) Ground truth, (c)–(d) Results obtained at the similar dose level as of our system (with an exposure time of 300 s per projection), (e)–(f) Results obtained with a 16-fold higher dose level.

These numerical simulations clearly demonstrate that it is quite necessary to increase the X-ray source brightness for laboratory FZP based X-ray microscopy when imaging low density objects. Otherwise, the ultra-high spatial resolution capability would be hindered by the degraded image quality due to the increased image noise.

### 10.3.4 Impact of phase grating size

Another design detail that may be easily overlooked is that the size of the phase grating $G_1$ must be no greater than the condenser outlet image at the $G_1$ plane. Otherwise, the direct X-ray beam would be splitted, producing overlapped ring artifacts in the central region, see Fig. 10.10. The splitting mechanism is consistent with the theoretical model described in Section 10.2.2. To better explain this issue, both numerical simulations and experimental validations were performed. When a large phase grating $G_1$ ($700\,\mu$m $\times$ $700\,\mu$m) is used, both the numerical simulations and measurement results exhibit the overlapped ring artifacts, see Fig. 10.10(a) and (c). Herein, the simulation was carried out by applying the splitting model to a projection acquired without $\mathrm{G}_1$ grating. In contrast, when a smaller $\mathrm{G}_1$ grating ($85\,\mu\mathrm{m} \times 85\,\mu\mathrm{m}$) is employed, such splitting artifacts are mitigated, see Fig. 10.10(b) and (d). Note that the condenser outlet image on the $\mathrm{G}_1$ plane in our system has a diameter of approximately $85\,\mu\mathrm{m}$.

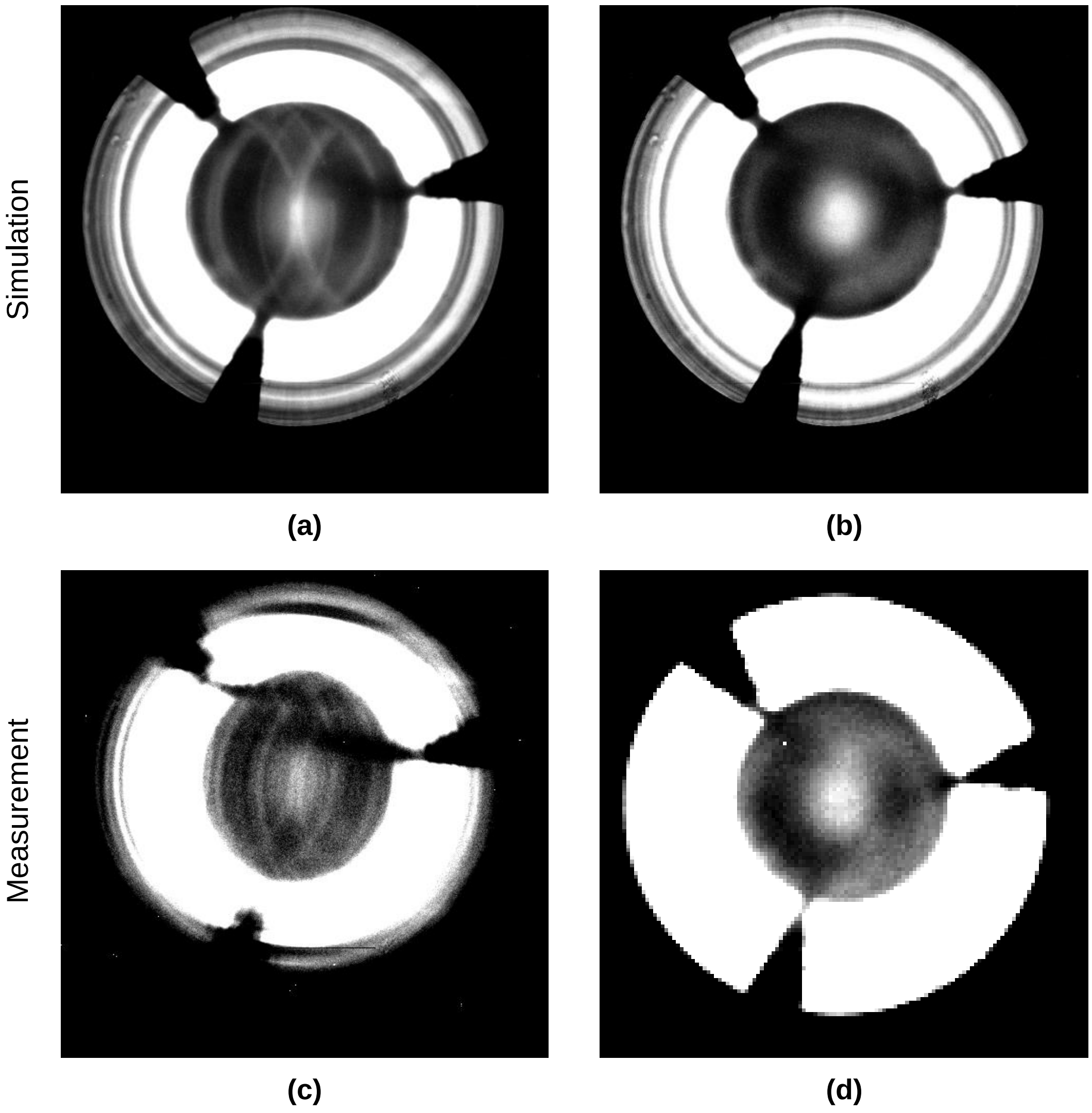

Figure 10.10: Impact of the phase grating size. Images in the top row are obtained from numerical simulations, and images in the bottom row are obtained from experimental measurements. Images in the left column are obtained with $700\,\mu\mathrm{m} \times 700\,\mu\mathrm{m}$ sized $\mathrm{G}_1$ grating, and images in the right column are obtained with $85\,\mu\mathrm{m} \times 85\,\mu\mathrm{m}$ sized $\mathrm{G}_1$ grating.

# 11 Outlook

The major steps involved in building a FZP based X-ray microscope are discussed in depth throughout this book. Even if readers do not fully master every technical detail after reading it, we believe that developing an overall understanding of the system will provide them with the confidence and knowledge needed to independently explore and develop a FZP based X-ray microscope from scratch in their own laboratories.

It is our belief that the zone-plate based laboratory X-ray microscope, like any other microscopy, is a kind of important high end scientific instrument for advanced research findings. As a result, it needs special attention and will evolve as time goes by. We strongly believe that X-ray microscope with FZP will play much more important and wider roles in future scientific research activities. It is not a backup tool to any technique, it will never disappear. On the contrary, it is indeed a complimentary tool to all the existing microscopes, and in fact it has been developing very quickly in recent years.

In the coming 10-30 years, we are expecting to see many exciting innovative FZP based laboratory X-ray microscope with unexpected features. It should become more user-friendly, more convenient to access, and more efficient for imaging. For instance, the microscope may become an AI powered instrument. As a consequence, alignments of all the optical components will be fully automatic and can be completed with one button or one command. Moreover, new monochromatic X-ray sources having much higher loading power could be integrated into the X-ray microscope. With it, the entire data acquisition time would be greatly shortened, and the image quality would be significantly improved. Depending on the application tasks, the beam energy and focal spot size could be switched automatically to achieve multi-resolution, multi-scale and multi-contrast X-ray microscope imaging. Additionally, the multi-scale information obtained from different microscopes such as optical, X-ray, electron and so on, will eventually be unified. Possibly, these microscopes could be integrated together as a multi-model microscope, together with X-ray spectroscopy.

# A Attenuation chart

The following table summarizes the attenuation properties of different materials at 5.4 keV and 8.0 keV.

Table A.1: Attenuation chart for He (gas), Be, Air (gas), Ar (gas), Kapton, SiN, Si, CsI, Au materials for 5.4 keV and 8.0 keV, correspondingly. In addition, the thickness for 10% attenuation of X-rays photons are also listed, respectively.

| Material | $\mu$ (cm$^{-1}$)@5.4 keV | 10% thickness (cm) | $\mu$ (cm$^{-1}$)@8.0 keV | 10% thickness (cm) |
|---|---|---|---|---|
| He (gas) | 0.00008 | 1317.01 | 0.00005 | 2107.21 |
| Be | 6.3689 | 0.0165 | 2.0521 | 0.0513 |
| Air (gas) | 0.0384 | 2.7437 | 0.0118 | 8.9288 |
| Ar (gas) | 0.7378 | 0.1428 | 0.2076 | 0.5075 |
| Kapton | 28.78 | 0.0036 | 8.68 | 0.0121 |
| $Si_3N_4$ | 440.39 | 0.0002 | 142.01 | 0.0007 |
| Si | 458.25 | 0.0002 | 148.61 | 0.0007 |
| CsI | 3489.98 | 0.00005 | 1367.38 | 0.00008 |
| Au | 10586.30 | 0.00001 | 3949.20 | 0.00003 |

# B Focal spot size

Assuming that the electron beam distribution on the target can be modeled by the following two dimensional (2D) Gaussian function (normalized to one):

$$\mathrm{G}(x,y) = \frac{1}{\pi\sigma_x\sigma_y} e^{-\left(\frac{x}{\sigma_x}\right)^2} e^{-\left(\frac{y}{\sigma_y}\right)^2}. \tag{B.1}$$

Note that the origin of the $x$-$y$ coordinate system is positioned at the maximum intensity part of the focal spot. See Fig. 3.6 for the coordinate definition.

Then, inserting the tungsten plate into the X-ray beam path and making sure that it blocks half area of the X-ray focal spot (either along the x-axis or along the y-axis). In the following discussion, it is assumed that the tungsten plate is placed parallel to the y-axis. Namely, the right half area of the X-ray focal spot is blocked (the $x$-axis coordinate of the tungsten plate is approximately equal to zero). On the detector plane, in general, the X-ray intensity distribution can be calculated via the following integration

$$\begin{aligned} I &= \frac{1}{\pi\sigma_x\sigma_y} \int_{-\infty}^{\infty} \int_{-r}^{\infty} e^{-\left(\frac{x}{\sigma_x}\right)^2} e^{-\left(\frac{y}{\sigma_y}\right)^2} dxdy & \text{(B.2)} \\ &= \frac{1}{\pi\sigma_x\sigma_y} \int_{-\infty}^{\infty} \int_{-l_1/l_2 r'}^{\infty} e^{-\left(\frac{x}{\sigma_x}\right)^2} e^{-\left(\frac{y}{\sigma_y}\right)^2} dxdy & \text{(B.3)} \\ &= \frac{1}{\sqrt{\pi}\sigma_x} \int_{-l_1/l_2 r'}^{\infty} e^{-\left(\frac{x}{\sigma_x}\right)^2} dx & \text{(B.4)} \\ &= \frac{1}{2} + \frac{1}{2} erf\left(\frac{l_1}{l_2}\frac{r'}{\sigma_x}\right). & \text{(B.5)} \end{aligned}$$

Note that the actually measured X-ray beam intensity is proportional to the above derived $I$. To determine $\sigma_x$, the distance $\Delta r'$ on the detector surface, which changes the function value from 0.8 to 0.2 (corresponding change of x in $erf(x)$ function from -0.6 to 0.6) is measured on the acquired

X-ray beam intensity distribution profile. Mathematically, having

$$\frac{l_1}{l_2}\frac{\Delta r'}{\sigma_x} = 1.2. \tag{B.6}$$

As a consequence,

$$\sigma_x = \frac{l_1}{l_2}\frac{\Delta r'}{1.2}, \tag{B.7}$$

and

$$\mathrm{FWHM}_x = 2.35\sigma_x. \tag{B.8}$$

To determine the $\sigma_y$ and $\mathrm{FWHM}_y$, the tungsten plate has to be placed parallel to the x-axis. By doing so, the bottom half of the X-ray focal spot is blocked. Finally, $\sigma_y$ and $\mathrm{FWHM}_y$ can be calculated by repeating the above measurements and calculations.

# C Maxwell equations

Because X-ray photons can also be treated as waves, therefore, the Maxwell electromagnetic wave equations are introduced and used as the theoretical foundations in describing the X-ray wave propagation and interactions between X-ray and medium. The solutions of these Maxwell electromagnetic wave equations [162] determine how the electric fields distribute after penetrating through the object.

**Maxwell electromagnetic wave equations in vacuum**

When electromagnetic fields propagate through the vacuum, the substantial equations that govern the spatial and temporal evolutions are known as the Maxwell electromagnetic wave equations. Using the International System (SI) units, they are given by [162]

$$\nabla \cdot \mathbf{E}(x, y, z, t) = 0, \tag{C.1}$$

$$\nabla \cdot \mathbf{B}(x, y, z, t) = 0, \tag{C.2}$$

$$\nabla \times \mathbf{E}(x, y, z, t) + \frac{\partial}{\partial t}\mathbf{B}(x, y, z, t) = \mathbf{0}, \tag{C.3}$$

$$\nabla \times \mathbf{B}(x, y, z, t) - \varepsilon_0 \mu_0 \frac{\partial}{\partial t}\mathbf{E}(x, y, z, t) = \mathbf{0}. \tag{C.4}$$

Here $\mathbf{E}$ is the electric field, $\mathbf{B}$ is the magnetic induction, $\varepsilon_0$ corresponds to the electrical permittivity of free space, $\mu_0$ corresponds to the magnetic permeability of free space. Note that both the $\varepsilon_0$ and the $\mu_0$ are constants in vacuum.The $\nabla$ denotes the three-dimensional gradient operator, and $\nabla\times$ denotes the three-dimensional curl operator. The $(x, y, z)$ represents the Cartesian coordinates of three-dimensional space, $t$ is time, and $\mathbf{0}$ is a zero-length vector. One should notice that Eq. (C.1) is the free-space form of Gauss' Law, which asserts the zero electric flux through all closed surfaces that do not contain any enclosed electric charge. Eq. (C.2) is the magnetic equivalent of Gauss' Law, with the exception that the right side is zero in the presence of matter, amounting to an assertion of the nonexistence of magnetic monopoles. Eq. (C.3) is Faraday's Law of induction, while Eq. (C.4) is the free-space form of Ampére's Law[163].

The above four equations are not independent from each other. Mathematically, one is able to derive the following equations for both electric field $\mathbf{E}$ and the magnetic induction $\mathbf{B}$

$$\left(\varepsilon_0\mu_0\frac{\partial^2}{\partial t^2} - \nabla^2\right)\mathbf{E}(x,y,z,t) = \mathbf{0}, \tag{C.5}$$

$$\left(\varepsilon_0\mu_0\frac{\partial^2}{\partial t^2} - \nabla^2\right)\mathbf{B}(x,y,z,t) = \mathbf{0}. \tag{C.6}$$

Since the electric field $\mathbf{E}$ and the magnetic induction $\mathbf{B}$ behave very similarly, we only focus on the electric field $\mathbf{E}$ in the following discussions. Assuming $\mathbf{E}$ can be represented as a product of a spatial wave function $\mathbf{E}(x,y,z)$ with a harmonic time factor $e^{-i\omega t}$, by substituting it back into Eq. (C.5), we get the well-known Helmholtz wave equation[162]

$$\left(\varepsilon_0\mu_0\omega^2 + \nabla^2\right)\mathbf{E}(x,y,z) = \mathbf{0}. \tag{C.7}$$

Consider one possible solution that a plane wave traveling in the $x$ direction[162],

$$\mathbf{E}(x,y,z) = \mathrm{E}_0 e^{ikx}, \tag{C.8}$$

we will find a relationship between the wave number $k$ and the frequency $\omega$

$$k = \sqrt{\varepsilon_0\mu_0}\,\omega. \tag{C.9}$$

Thus, we will get the phase velocity[162] of the electromagnetic fields that propagate in vacuum

$$c = \frac{1}{\sqrt{\varepsilon_0\mu_0}}. \tag{C.10}$$

**X-rays interacting with matter**

In the above, we've discussed the electromagnetic field propagation in vacuum using the Maxwell equations. For real applications, it is more important to study how the electromagnetic field propagates through mediums. In this condition, the Maxwell equations in Eq. (C.1) to Eq. (C.4) need to be modified as following

$$\nabla \cdot \mathbf{D}(x,y,z,t) = \rho(x,y,z,t), \tag{C.11}$$

$$\nabla \cdot \mathbf{B}(x,y,z,t) = 0, \tag{C.12}$$

$$\nabla \times \mathbf{E}(x,y,z,t) + \frac{\partial}{\partial t}\mathbf{B}(x,y,z,t) = \mathbf{0}, \tag{C.13}$$

$$\nabla \times \mathbf{H}(x,y,z,t) - \frac{\partial}{\partial t}\mathbf{D}(x,y,z,t) = \mathbf{J}(x,y,z,t). \tag{C.14}$$

Here, $\mathbf{D}$ is the electric displacement, $\mathbf{H}$ is the magnetic field, $\rho$ is the charge density, and $\mathbf{J}$ is the current density. In general, the electric displacement $\mathbf{D}$ and the magnetic induction $\mathbf{B}$ are functions of the electric field $\mathbf{E}$ and the magnetic field $\mathbf{H}$, respectively. To ease the discussion, we restrict to consider the mediums that are linearly isotropic, namely

$$\mathbf{D}(x,y,z,t) = \varepsilon(x,y,z,t)\mathbf{E}(x,y,z,t), \tag{C.15}$$

$$\mathbf{B}(x,y,z,t) = \mu(x,y,z,t)\mathbf{H}(x,y,z,t), \tag{C.16}$$

where $\varepsilon$ denotes the electrical permittivity, and $\mu$ denotes the magnetic permeability of the material. Notice that both the $\varepsilon$ and $\mu$ are functions of space and time in this assumption. For materials that have more complex electric and magnetic properties, there are no such simple relationships exist. To deal with these anisotropic materials, one needs to use more complicated mathematical representations: the tensor method[162].

For a certain isotropic material, it is reasonable to further assume that the electrical permittivity $\varepsilon$ and magnetic permeability $\mu$ are time independent, thus having $\varepsilon(x,y,z)$ and $\mu(x,y,z)$. If the medium considered is also a non-magnetic material, then $\mu(x,y,z) = \mu_0$. With all the approximations above, we now obtain

$$\left(\varepsilon(x,y,z)\mu_0\frac{\partial^2}{\partial t^2} - \nabla^2\right)\mathbf{E}(x,y,z,t) = \nabla\left(\frac{1}{\varepsilon(x,y,z)}\nabla\varepsilon(x,y,z)\cdot\mathbf{E}(x,y,z,t)\right), \tag{C.17}$$

$$\left(\varepsilon(x,y,z)\mu_0\frac{\partial^2}{\partial t^2} - \nabla^2\right)\mathbf{H}(x,y,z,t) = \frac{1}{\varepsilon(x,y,z)}\nabla\varepsilon(x,y,z)\times\left(\nabla\times\mathbf{H}(x,y,z,t)\right). \tag{C.18}$$

It is not easy to precisely quantify the electrical permittivity $\varepsilon$ on the right hand sides, therefore, another assumption is needed to proceed the derivation further. We assume that the internal structures of medium varies sufficiently slow over length scales comparable to the wavelength of the incident electromagnetic waves, hence, one may neglect the quantity on the right hand sides, leading to:

$$\left(\varepsilon(x,y,z)\mu_0\frac{\partial^2}{\partial t^2}-\nabla^2\right)\mathbf{E}(x,y,z,t)=0, \tag{C.19}$$

$$\left(\varepsilon(x,y,z)\mu_0\frac{\partial^2}{\partial t^2}-\nabla^2\right)\mathbf{H}(x,y,z,t)=0. \tag{C.20}$$

Obviously, Eq. (C.19) and (C.20) look very similar as Eq. (C.5) and (C.6). The only difference is the electrical permittivity. In vacuum, it is a constant $\varepsilon_0$, while in materials, it is a spatial related variable $\varepsilon(x, y, z)$.

Similarly as has been done for the vacuum circumstance, we only take Eq. (C.19) as the example to explain the following derivations. Idential results will be obtained if working on Eq. (C.20) with the same procedures. First of all, assuming

$$\mathbf{E}(x,y,z,t)=\mathbf{E}(x,y,z)e^{-i\omega t}. \tag{C.21}$$

Substituting it into Eq. (C.19) and get

$$\left(\varepsilon(x,y,z)\mu_0\omega^2+\nabla^2\right)\mathbf{E}(x,y,z)=\mathbf{0}. \tag{C.22}$$

With the help of $\omega = ck$, one is able to obtain the refractive index of the material

$$n(x,y,z)=\sqrt{\frac{\varepsilon(x,y,z)}{\varepsilon_0}}. \tag{C.23}$$

It is still very difficult to solve Eq. (C.22) directly so far. To proceed, we add another hypothesis on $\mathbf{E}(x, y, z)$ such that

$$\mathbf{E}(x,y,z)=\mathrm{E}_0(x,y,z)e^{ikz} \tag{C.24}$$

can be treated as an electromagnetic wave passing through the medium along the $z$ axis with a space dependent wave amplitude $\mathrm{E}_0(x, y, z)$. By doing so, Eq. (C.22) will be simplified as

$$\left((n^2(x,y,z)-1)k^2+2ik\frac{\partial}{\partial z}+\frac{\partial^2}{\partial x^2}+\frac{\partial^2}{\partial y^2}+\frac{\partial^2}{\partial z^2}\right)\mathrm{E}_0(x,y,z)=0. \tag{C.25}$$

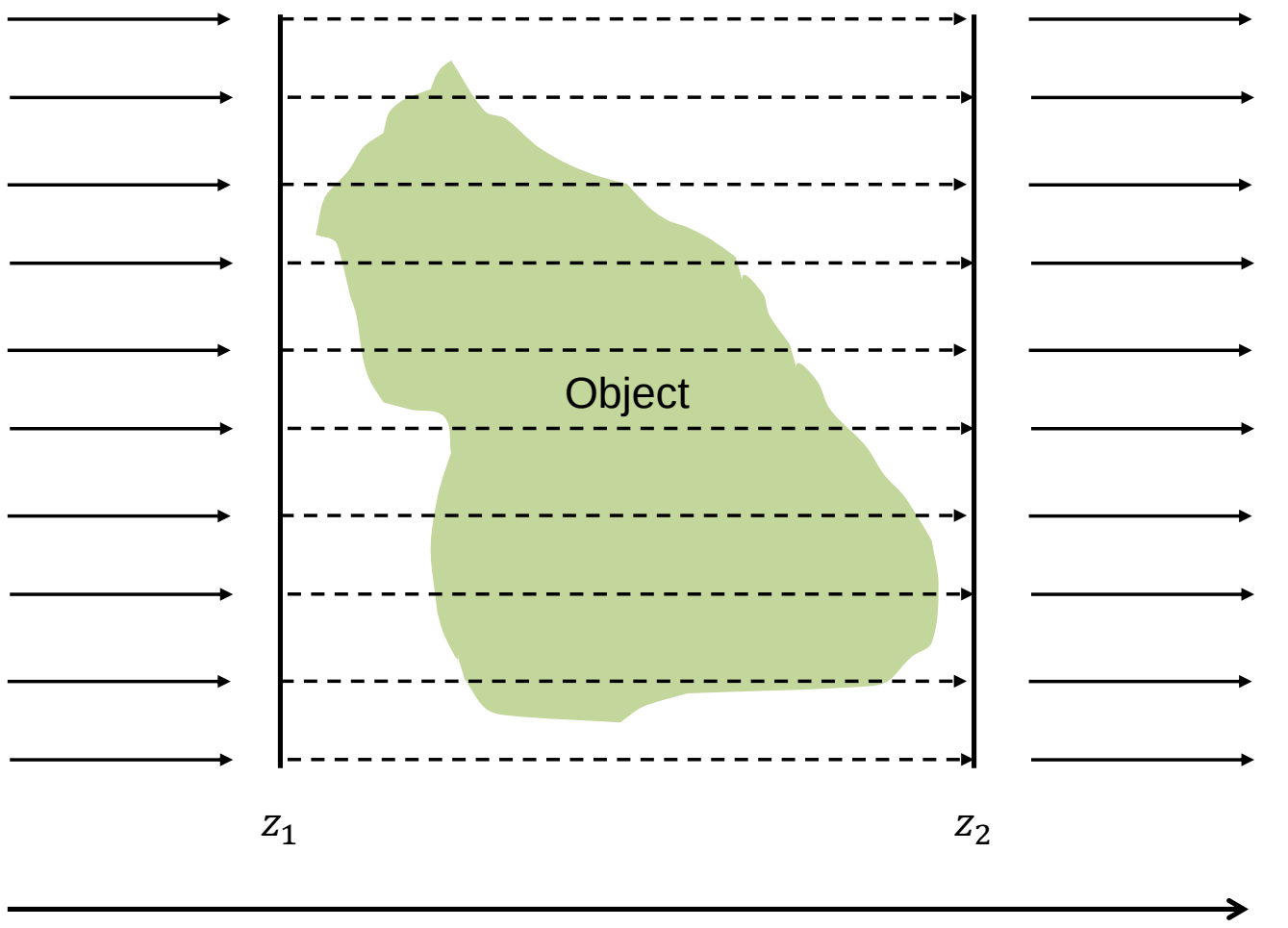

Figure C.1: Monochromatic plane wave propagating along the $z$-direction incidents upon the $z_1$ plane. The object locates between plane $z_1$ and $z_2$. The dotted line shows the path that an incident "ray" would take, in the absence of the scatterer.

It is reasonable to assume that $\mathrm{E}_0(x, y, z)$ varies slowly in space on scale length of the field wavelength. Therefore, the first derivatives will be more dominant compared to the second derivatives. Neglecting the second derivatives in Eq. (C.25), one may obtain

$$\left((n^2(x, y, z) - 1)k^2 + 2ik\frac{\partial}{\partial z}\right) \mathrm{E}_0(x, y, z) = 0, \tag{C.26}$$

whose solution is

$$\mathrm{E}_0(x, y, z) \mid_{z_2} = \mathrm{E}_0(x, y, z) \mid_{z_1} e^{\frac{k}{2i}\int_{z_1}^{z_2}\left[1-n^2(x,y,z)\right]dz}, \tag{C.27}$$

where locations $z_1$ and $z_2$ represent the left and right boundaries of the medium, see the diagram in Figure. C.1.

For X-ray energy band, the refraction index $n$ in Eq. (C.23) is usually written as a complex number[59, 162–164]

$$n(x, y, z) = 1 - \delta(x, y, z) + i\beta(x, y, z), , \tag{C.28}$$

in which the decremental factor $\delta$ is way small than 1. If substituting Eq. (C.28) into Eq. (C.27),

one will get

$$\mathrm{E}_0(x,y,z)\mid_{z_2} = \mathrm{E}_0(x,y,z)\mid_{z_1} e^{-k\int_{z_1}^{z_2}\beta(x,y,z)dz} e^{-ik\int_{z_1}^{z_2}\delta(x,y,z)dz}. \tag{C.29}$$

The Eq. (C.27) is important, because it indicates that one can still visualize "ray paths" as depicted by a dashed line in Figure. C.1, even though we start the entire mathematical derivations within the context of wave optics. This is more obvious if we were working within the framework of geometric optics. The essence of the above projection approximation is to assume that the scatterers are sufficiently weak so as to negligibly perturb the ray paths which would have existed in the volume occupied by the scatterer had the scatterer been absent. As can be seen in Eq. (C.29), the amplitude and phase of the disturbance at the exit $z_2$ plane can be expressed in terms of the amplitude and phase shifts accumulated as the disturbance traverses a given ray path connecting the entrance and exit surfaces, with the ray paths corresponding to those that would have existed if the scattering volume were replaced by vacuum[163].

By taking the squared modulus of Eq. (C.29) and identifying $|\mathrm{E}_0(x,y,z)|^2$ with the intensity $\mathrm{I}(x,y,z)$ of the wave-field, we obtain:

$$\mathrm{I}_0(x,y,z)\mid_{z_2} = \mathrm{I}_0(x,y,z)\mid_{z_1} e^{-2k\int_{z_1}^{z_2}\beta(x,y,z)dz}. \tag{C.30}$$

For the case of a scatterer composed of a single material, this formula becomes the well known Beer-Lambert's law of absorption[165], which relates the attenuation of light to the properties of the object and is formulated as:

$$I_{out} = I_{in}e^{-\int\mu(x,y,z)dz}, \tag{C.31}$$

Therefore, one can immediately get

$$\mu = \frac{4\pi\beta}{\lambda}. \tag{C.32}$$

Clearly, the material attenuation coefficient $\mu$ is linearly proportional to the imaginary part, $\beta$, of the refractive index $n$, while it is inversely proportional to the beam wavelength $\lambda$.

# D Fresnel diffraction

Optics studies the light propagation and detection. It was found that in certain regions the simple geometrical model of light propagation is sufficient and accurate enough to explain some physical phenomenons. However, due to its wave-particle duality nature, it becomes inadequate if only treating light propagation with geometrical approximations. One such exception is the diffraction effect. The deviations are manifested by the appearance of dark and bright bands, known as the diffraction fringes[59].

Precisely solving the diffraction problems are difficult in optics. The first rigorous solution was given by A. Sommerfeld in 1896. In his important paper, he discussed the diffraction of a plane wave by a perfectly conducting semi-infinite plane screen. Because of the mathematical difficulties, approximate methods have to be used for practical applications when solving the diffraction problems. By far, the most powerful and adequate approximation is the Huygens-Fresnel principle. The basic idea of the Huygens-Fresnel principle is that the light disturbance at a point P arises from the superposition of secondary waves that proceed from a surface situated between this point and the light source. Solid mathematical basis was derived by G. Kirchhoff. The following discussions will base on Kirchhoff's diffraction theory[59].

Consider a very simple scenario, where a monochromatic wave originated from a point source $P_0$ is propagating through an opening in a plane opaque screen, see Figure D.1. We are interested in determining the light disturbance at point $P$ which is located behind the screen. In Figure D.1(a), the opening is denoted as $A$, portion $B$ represents the nonilluminated side of the screen, and portion $C$ is a large sphere of radium of $R$, centered at $P$. The $A$, $B$, and $C$ together form a closed surface. Theoretical calculations show that the disturbance at point $P$ is equal to

$$\mathrm{U}(P) = -\frac{i\mathrm{U}_0}{2\lambda} \int \int_A \frac{e^{ik(r+s)}}{rs} \left[\cos(n, r) - \cos(n, s)\right] dS, \tag{D.1}$$

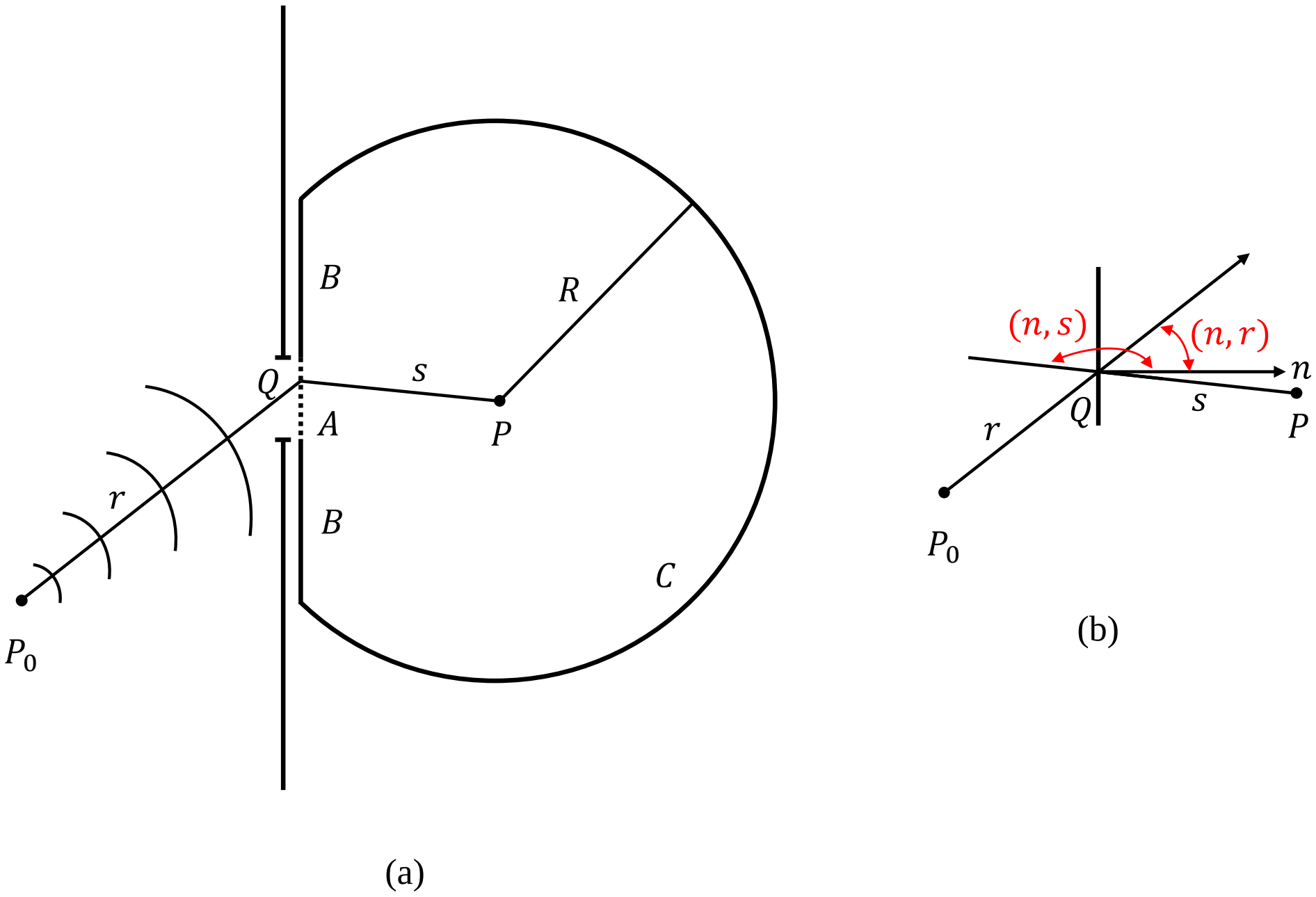

Figure D.1: (a) Illustration of the Fresnel-Kirchhoff diffraction. (b) Definition of angles.

where $\mathrm{U}_0$ is the amplitude of the disturbance at point $P_0$. Eq. (D.1) is known as the Fresnel-Kirchhoff diffraction formula[59].

If assuming the distance of the points $P_0$ and $P$ from the screen are large compared to the linear dimensions of the opening $A$, the factor $[\cos(n,r) - \cos(n,s)]$ can be approximated by $2\cos\theta$, where $\theta$ is the angle between the line $P_0P$ and the normal to the screen. Thus having,

$$\mathrm{U}(P) = -\frac{i\mathrm{U}_0}{\lambda}\int\int_A \frac{e^{ik(r+s)}}{rs}\cos\theta dS. \tag{D.2}$$

In the following discussions, a Cartesian reference system is assumed, of whose origin locates in the center of opening $A$ as illustrated in Figure D.2(a). Denote $(x_0, y_0, z_0)$ and $(x, y, z)$ as the coordinates of $P_0$ and of $P$ respectively, and $(\xi, \eta)$ the coordinates of an arbitrary point $Q$ on opening $A$. Therefore,

$$s = \sqrt{(x-\xi)^2 + (y-\eta)^2 + z^2} \approx z + \frac{(x-\xi)^2}{2z} + \frac{(y-\eta)^2}{2z}. \tag{D.3}$$

where we keep the second order of the Taylor expansion. Substituting Eq. (D.3) into Eq. (D.2), and

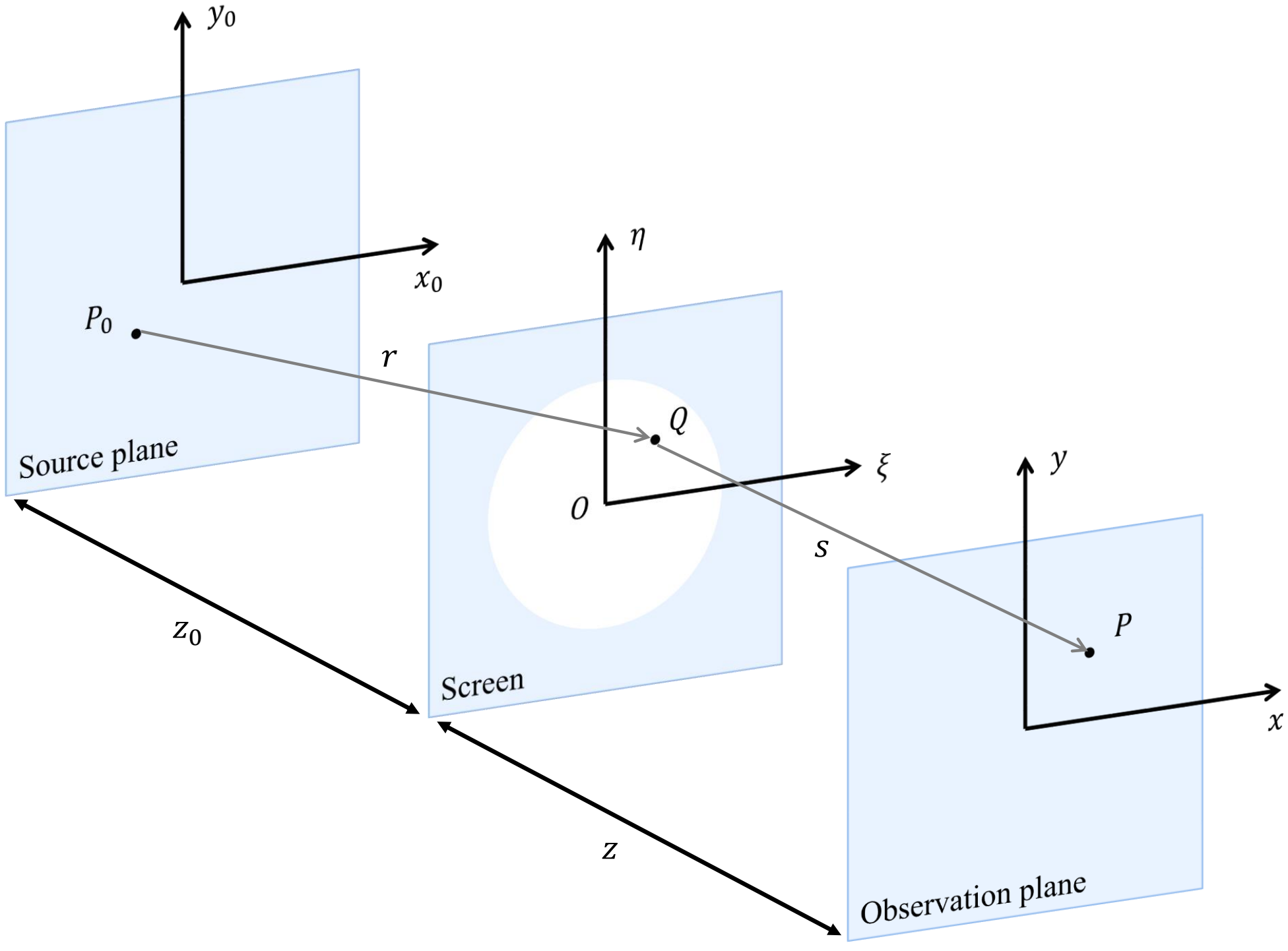

Figure D.2: Coordinates for the Fresnel-Kirchhoff diffraction integral.

approximate $\cos\theta$ by $z/s$, one gets

$$\mathrm{U}(x,y) = \frac{\mathrm{U}_0 e^{ikz}}{i\lambda z} \int\int_A \frac{e^{ikr}}{r} e^{ik\left[\frac{(x-\xi)^2}{2z} + \frac{(y-\eta)^2}{2z}\right]} d\xi d\eta. \tag{D.4}$$

Replace the point-source incident wavefront $\frac{e^{ikr}}{r}$ produced by a point source ($P$ in this discussion) with a general wavefront $u_{inc}$ produced by an extended source (modeled as a collection of point sources), and characterize the aperture by a transmission function $\tau$ to model the amplitude and/or phase changes due to diffraction gratings (or lens), the following formula can be obtained (only considering one point source located at $(x_0, y_0)$ of the extended source):

$$\mathrm{U}(x,y) = \frac{\mathrm{U}_0 e^{ikz}}{i\lambda z} \int\int_A \tau(\xi,\eta) u_{inc}(\xi,\eta) e^{ik\left[\frac{(x-\xi)^2}{2z} + \frac{(y-\eta)^2}{2z}\right]} d\xi d\eta. \tag{D.5}$$

Equation (D.5) shows the primary formula that can be used to estimate the light disturbance on the observation plane if given the source and the screen transmission function. Sometimes, the formula can also be to study the free-space wave propagations. The detected diffraction pattern generated in this way is known as the Fresnel diffraction. It is a near-field diffraction effect. In contrast, the diffraction pattern in the far field region is given by the Fraunhofer diffraction equation, which cuts off the previous Taylor expansion of $s$ in Eq. (D.3) only by the first order.

# E Talbot and Lau effect

### X-ray Talbot effect

The Talbot effect is a near-field diffraction effect first observed in 1836 by Henry Fox Talbot [166]. When a plane wave is incident upon a periodic diffraction grating, the image of the grating (also known as self-imaging effect) is repeated at regular distances away from the grating plane. The regular distance is called the Talbot length [167]. The Talbot effect and Talbot interferometry were studied extensively in the visible light region and used for wavefront sensing. Recently, the Talbot effect and the fractional Talbot effect in the hard X-ray energy region was also demonstrated with gratings [128, 137]. In this part, the Talbot effect will be discussed using the Fresnel diffraction theory.

To simplify the discussions, a plane wave is assumed to be incident on a 1D diffraction grating, and the observation plane is located $z$ distance downstream of the grating plane. By jointly using the grating transmission function $\mathrm{T}(x)$ defined in Eq. (8.2.2) and the Fresnel diffraction equation Eq. (D.5), one can get the wave disturbance at the observation plane as following:

$$\mathrm{U}(x,z) = \frac{e^{ikz}}{\sqrt{i\lambda z}} \sum_{n=-\infty}^{n=\infty} a_n \int_{-\infty}^{+\infty} e^{\frac{i2\pi n\xi}{p}} e^{i\frac{(x-\xi)^2}{\lambda z}} d\xi, \tag{E.1}$$

$$= e^{ikz} \sum_{n=-\infty}^{n=\infty} a_n e^{-\frac{i\pi\lambda n^2 z}{p^2}} e^{\frac{i2\pi nx}{p}}, \tag{E.2}$$

$$= e^{ikz} \sum_{n=-\infty}^{n=\infty} g_n e^{\frac{i2\pi nx}{p}}. \tag{E.3}$$

Here the modified Fourier coefficients $g_n$ are defined as follows,

$$g_n = a_n e^{-\frac{i\pi\lambda n^2 z}{p^2}}. \tag{E.4}$$

For some special observation distance $z$,

$$z = m \times \frac{2p^2}{\lambda}, \quad m = \text{positive integers}, \tag{E.5}$$

it is easy to demonstrate that the $g_n$ become identical with $a_n$, namely,

$$g_n = a_n e^{-i2\pi n^2 m} = a_n. \tag{E.6}$$

Thus Eq. (E.3) becomes

$$\mathrm{U}(x, z) = e^{ikz}\mathrm{T}(x), \tag{E.7}$$

and the observed beam intensity distribution at these specific $z$ positions can be found equal to

$$\mathrm{I}(x, z) = |\mathrm{U}(x, z)|^2 = |\mathrm{T}(x)|^2. \tag{E.8}$$

Interestingly, Eq. (E.8) shows that the image of the grating is repeated at regular distances away from the grating plane. The regular distance is called the Talbot distance, denoted as $z_T$,

$$z_T = \frac{2p^2}{\lambda}. \tag{E.9}$$

The repeated images are called self images, and this effect is known as the Talbot self-imaging effect. Notice that these self images are appeared at integer times of the Talbot distance, see Eq. (E.5).

**X-ray fractional Talbot effect**

This discussed Talbot self-imaging effect is only valid for an amplitude diffraction grating. For phase diffraction gratings, it is easy to prove that the observed intensity distribution is a flat one (because $|\mathrm{T}(x)| = 1$). Actually, for phase diffraction gratings, the self images with varied periods occur at fractional Talbot distances [168]. For instance, if a pure $\frac{\pi}{2}$ phase grating is utilized, the fractional Talbot distances at which self images occur are

$$z = \frac{z_T}{4} \times (2q - 1), \quad q = \text{positive integers}. \tag{E.10}$$

At these special observation locations, the period of the detected self images is the same as the $\frac{\pi}{2}$

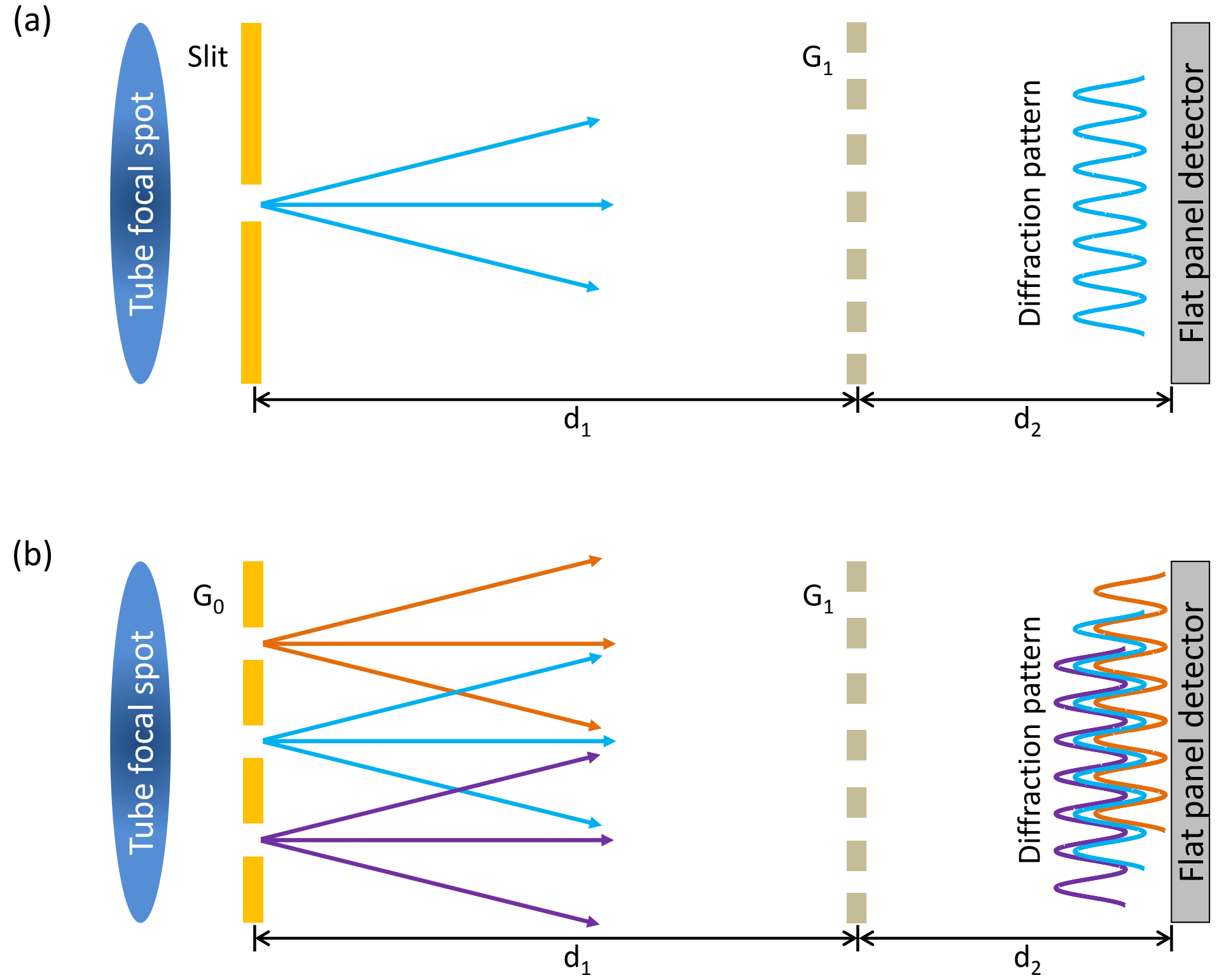

Figure E.1: (a) A slit with narrow opening is positioned close to the X-ray source to increase the beam spatial coherence. (b) A group of slits, namely a source grating $G_0$ is used to let more photons pass through. The diffraction pattern produced by each individual grating opening overlap constructively at the plane of detection.

phase grating. However, the fractional Talbot distances of a pure $\pi$ phase grating are equal to

$$z = \frac{z_T}{16} \times (2q-1), \quad q = \text{positive integers}. \tag{E.11}$$

At these particular positions, the period of the observed self images become half of the diffraction grating. For details of the mathematical derivations, please refer to Appendix F.

**The Lau effect**

To increase the spatial coherence of the X-ray beam generated from a medical grade X-ray tube (tube focal spot size ~1.0 mm), a narrow slit with very small opening size is always added close to the output window of the tube, as illustrated in Figure 8.4 and Figure E.1(a). The narrow slit opening makes the source act like a point source. Thus, the reduced source size thus can greatly increase the spatial coherence of the X-ray beam, and the needed interference pattern can be generated at some distance downstream of the diffraction grating. However, it is very energy inefficient by only using one single slit opening, because most of the X-ray beam flux are blocked and not used for imaging purposes. To improve the beam usage efficiency while still maintain the high spatial coherence of the beam, in practice one can align an array of slit openings, i.e., a grating, with appropriate pitch size. As shown in Figure E.1(b), the source grating $G_0$ enables more X-ray photons coming through. For each individual grating opening, the generated spatially coherent X-ray beam will form a certain diffraction pattern. The modulated interference patterns generated from different grating openings have the same period and are located at the same distance behind the $G_1$ grating. More importantly, the properly designed $G_0$ grating pitch size will help the interference pattern generated from every single slit opening constructively add upon each other. This principle is known as the Lau effect [169–171], and the interferometer system is called the Talbot-Lau interferometry.

# F Self imaging

First of all, define the transmission function $\mathrm{T}(x)$ of an 1D phase diffraction grating as

$$\mathrm{T}(x) = \begin{cases} 1 & \text{if } \operatorname{mod}(x,p) \in [0 \;\; p/2) \\ e^{i\varphi} & \text{if } \operatorname{mod}(x,p) \in [p/2 \;\; p) \end{cases} \tag{F.1}$$

and the corresponding Fourier coefficients $a_n$ are found equal to

$$a_n = \begin{cases} \frac{1+e^{i\varphi}}{2} & \text{if } n = 0 \\ 0 & \text{if } \operatorname{mod}(n,2) = 0, \text{ but } n \neq 0 \\ \frac{ie^{i\varphi}-i}{n\pi} & \text{if } \operatorname{mod}(n,2) = 1 \end{cases} \tag{F.2}$$

Using Eq. (E.3) and Eq. (E.4), the beam intensity detected at downstream distance $z$ is found to be

$$\mathrm{I}(x,z) = |\mathrm{U}(x,z)|^2 = \sum_{m=-\infty}^{m=\infty} \sum_{n=-\infty}^{n=\infty} g_m g_n^* e^{\frac{i2\pi x(m-n)}{p}}. \tag{F.3}$$

If assume $m - n = t$, then Eq. (F.3) becomes

$$\mathrm{I}(x,z) = \sum_{t=-\infty}^{t=\infty} \sum_{n=-\infty}^{n=\infty} g_{t+n} g_n^* e^{\frac{i2\pi xt}{p}}. \tag{F.4}$$

Depending on the value of $t$, the above summation can be divided into three main parts, namely,

$$\mathrm{I}(x,z) = \mathrm{A}_0 + \sum_{s=-\infty}^{s=\infty} \mathrm{A}_{odd} e^{\frac{i2\pi x(2s+1)}{p}} + \sum_{s=-\infty, s\neq 0}^{s=\infty} \mathrm{A}_{even} e^{\frac{i2\pi x(2s)}{p}}. \tag{F.5}$$

Notice that the summation for $t$ has been replaced by $s$. In it, the first part $A_0$ corresponds to $t = 0$ and is defined as

$$A_0 = \sum_{n=-\infty}^{n=\infty} g_n g_n^*. \tag{F.6}$$

Similarly, the second part $A_{odd}$ are related with odd values of $t = 2s + 1$,

$$A_{odd} = \sum_{n=-\infty}^{n=\infty} g_{2s+n+1} g_n^*. \tag{F.7}$$

Finally, the third part $A_{even}$ represents the even values of $t = 2s$

$$A_{even} = \sum_{n=-\infty}^{n=\infty} g_{2s+n} g_n^*. \tag{F.8}$$

By substituting Eq. (E.4) and Eq. (E.2) into Eq. (F.6), one can get

$$A_0 = \sum_{n=-\infty}^{n=\infty} g_n g_n^* = \sum_{n=-\infty}^{n=\infty} |a_n|^2 = 1. \tag{F.9}$$

Similarly, the Eq. (F.7) and Eq. (F.8) become

$$A_{odd} = \sum_{n=-\infty}^{n=\infty} g_{2s+n+1} g_n^*, \tag{F.10}$$

$$= \sum_{n=-\infty}^{n=\infty} a_{2s+n+1} a_n^* e^{-\frac{i\pi\lambda z}{p^2}(2s+1)(2n+2s+1)}, \tag{F.11}$$

$$= a_{2s+1} a_0^* e^{-\frac{i\pi\lambda z}{p^2}(2s+1)^2} + \sum_{n=-\infty, n\neq 0}^{n=\infty} a_{2s+n+1} a_n^* e^{-\frac{i\pi\lambda z}{p^2}(2s+1)(2n+2s+1)}, \tag{F.12}$$

$$= a_{2s+1} a_0^* e^{-\frac{i\pi\lambda z}{p^2}(2s+1)^2}, \tag{F.13}$$

$$= \frac{i(1 - e^{-i2\varphi})}{2\pi(2s+1)} e^{-\frac{i\pi\lambda z}{p^2}(2s+1)^2}. \tag{F.14}$$

$$\mathrm{A}_{even} = \sum_{n=-\infty}^{n=\infty} g_{2s+n} g_n^*, \tag{F.15}$$

$$= \sum_{l=-\infty}^{l=\infty} g_{2s+2l+1} g_{2l+1}^*, \tag{F.16}$$

$$= \frac{2}{\pi^2} \frac{\sin^2(\frac{\varphi}{2})}{s} e^{-\frac{i\pi\lambda z}{p^2}(2s)^2} \sum_{l=-\infty}^{l=\infty} \left( \frac{1}{2l+1} - \frac{1}{2l+2s+1} \right) e^{-\frac{i\pi\lambda z}{p^2} 4s(2l+1)}, \tag{F.17}$$

$$= \frac{-4i}{\pi^2} \frac{\sin^2(\frac{\varphi}{2})}{s} \sin\left( \frac{\pi\lambda z}{p^2}(2s)^2 \right) \mathrm{Q}, \tag{F.18}$$

where Q is defined as

$$\mathrm{Q} = \sum_{l=-\infty}^{l=\infty} \frac{1}{2l+1} e^{-\frac{i\pi\lambda z}{p^2} 4s(2l+1)}, \tag{F.19}$$

$$= -2i \sum_{l=1}^{l=\infty} \frac{\sin\left[ (2l-1)\frac{\pi\lambda z}{p^2} 4s \right]}{2l-1}, \tag{F.20}$$

$$= -\frac{i\pi}{2} \begin{cases} 1, & \mathrm{mod}(\frac{\pi\lambda z}{p^2}4s, 2\pi) \in [0, \pi) \\ -1, & \mathrm{mod}(\frac{\pi\lambda z}{p^2}4s, 2\pi) \in (-\pi, 0) \end{cases} . \tag{F.21}$$

**$\frac{\pi}{2}$ phase grating** In this case, $\varphi = \frac{\pi}{2}$. Moreover, the downstream observation distance $z$ is assumed to be

$$z = \frac{z_T}{4}(2q-1), \quad q = \text{positive integers}. \tag{F.22}$$

If substituting Eq. (F.22) into Eq. (F.14), one can get,

$$\mathrm{A}_{odd} = \frac{1}{\pi(2s+1)} e^{-\frac{\pi}{2}(2q-1)(2s+1)^2}, \tag{F.23}$$

$$= \frac{i}{\pi} \frac{(-1)^q}{2s+1}. \tag{F.24}$$

Similarly,

$$\sin\left( \frac{\pi\lambda z}{p^2}(2s)^2 \right) = \sin\left( 2\pi(2q-1)s^2 \right) = 0, \tag{F.25}$$

therefore, $A_{even} = 0$. In all, the Eq. (F.5) now becomes

$$I(x,z) = 1 + (-1)^q \sum_{s=-\infty}^{s=\infty} \frac{i}{\pi(2s+1)} e^{\frac{i2\pi x(2s+1)}{p}}, \tag{F.26}$$

$$= \begin{cases} 1+(-1)^q, & \mathrm{mod}(x,p) \in [0, \frac{p}{2}) \\ 1+(-1)^{q+1}, & \mathrm{mod}(x,p) \in [\frac{p}{2}, p) \end{cases}. \tag{F.27}$$

In conclusion, the Talbot self images of a $\frac{\pi}{2}$ phase grating can be found at some special $z$ values. At those observation planes, the period of the self images is the same as of the grating period $p$. However, depending on the value of $q$, the self images may be laterally shifted by half the width of the grating period.

$\pi$ **phase grating** In this case, $\varphi = \pi$, and the downstream observation distance $z$ is assumed to be

$$z = \frac{z_T}{16}(2q-1), \quad q = \text{positive integers}. \tag{F.28}$$

Since $\varphi = \pi$, it is easily to show that $A_{odd} = 0$. Whereas,

$$\sin\left(\frac{\pi\lambda z}{p^2}(2s)^2\right) = \sin\left(\pi\frac{2q-1}{2}s^2\right), \tag{F.29}$$

$$= \begin{cases} 0, & \mathrm{mod}(s,2) = 0 \\ (-1)^q, & \mathrm{mod}(s,2) = 1 \end{cases}. \tag{F.30}$$

Hence, only odd values of $s = 2u - 1$ will be considered in the following discussions. The value of Q also depends on the observation distance $z$ and $s$. Now having,

$$\frac{\pi\lambda z}{p^2}4s = \frac{\pi}{2}(2q-1)(2u-1). \tag{F.31}$$

It is easy to demonstrate that the value of Q can be written in terms of $q$ and $u$, i.e.,

$$Q = \frac{i\pi}{2}(-1)^{u+q+1}. \tag{F.32}$$

As a result, the Eq. (F.5) in this case becomes

$$\mathrm{I}(x,z) = 1 + \sum_{u=-\infty}^{u=\infty} \frac{2}{\pi s}(-1)^{u+1} e^{\frac{i2\pi x(2s)}{p}}, \tag{F.33}$$

$$= 1 - \frac{4}{\pi} \sum_{k=1}^{k=\infty} (-1)^{k-1} \frac{\cos\left[(2k-1)\frac{4\pi x}{p}\right]}{2k-1}, \tag{F.34}$$

$$= 1 - \frac{4}{\pi} \begin{cases} \frac{\pi}{4}, & \mathrm{mod}(x, \frac{p}{2}) \in (-\frac{p}{8}, \frac{p}{8}) \\ -\frac{\pi}{4}, & \mathrm{mod}(x, \frac{p}{2}) \in [\frac{p}{8}, \frac{3p}{8}) \end{cases}, \tag{F.35}$$

$$= \begin{cases} 0, & \mathrm{mod}(x, \frac{p}{2}) \in (-\frac{p}{8}, \frac{p}{8}) \\ 2, & \mathrm{mod}(x, \frac{p}{2}) \in [\frac{p}{8}, \frac{3p}{8}) \end{cases}. \tag{F.36}$$

In summary, the Talbot self images of a $\pi$ phase grating can be found at some special $z$ values. At those observation planes, the period of the self images becomes half of the grating period, i.e., $\frac{p}{2}$.

# G Talbot-Lau interferometry

To initiate the following analyses, an ideal point source is assumed. It emits spherical wavefront with single wavelength and propagates outwardly in the three-dimensional space. Moreover, the already obtained projection approximation, see Eq. (C.29), is used to describe the beam interactions inside the medium. Finally, the amplitude disturbance at the detection plane is analyzed via the previously discussed Fresnel diffraction equation Eq. (D.5). Be aware that only one $\pi$-phase grating is involved in the following analytical framework.

Figure G.1 diagrams two possible imaging system arrangements. As an example, only the imaging geometry in Figure G.1(a), in which the object is located in front of the $G_1$ grating is discussed with details in the following part. Using the same mathematical framework, the imaging geometry shown in Figure G.1(b) can also be investigated.

The wave front arriving upon the object plane A can be expressed as

$$\mathrm{U}_1(x_1, y_1) = \frac{\mathrm{U}_0 e^{ikd_1}}{d_1} e^{ik\left[\frac{(x_1-x_s)^2}{2d_1} + \frac{(y_1-y_s)^2}{2d_1}\right]}, \tag{G.1}$$

where the coordinates $(x_s, y_s)$ represent the point source location. This is the parabolic approximation of the spherical wavefront (details can be found in the Appendix section). For the considered 3D case, the beam intensity drops according to the inverse square law of the distance, therefore, the wave amplitude $\mathrm{U}_1$ inversely proportional to the distance $d_1$ between the source to incident plane A.

When shining upon the object, the X-ray wave starts to interact with the electrons and atomic nucleus that composite the object very complicatedly. However, with appropriate approximations, the exit wave can be estimated using the obtained projection approximation, where the primary direction of the X-ray beam does not change. Due to the presence of the object, the exit beam will change its amplitude and phase accordingly, see Eq. (C.29). As a result, we obtain the amplitude of

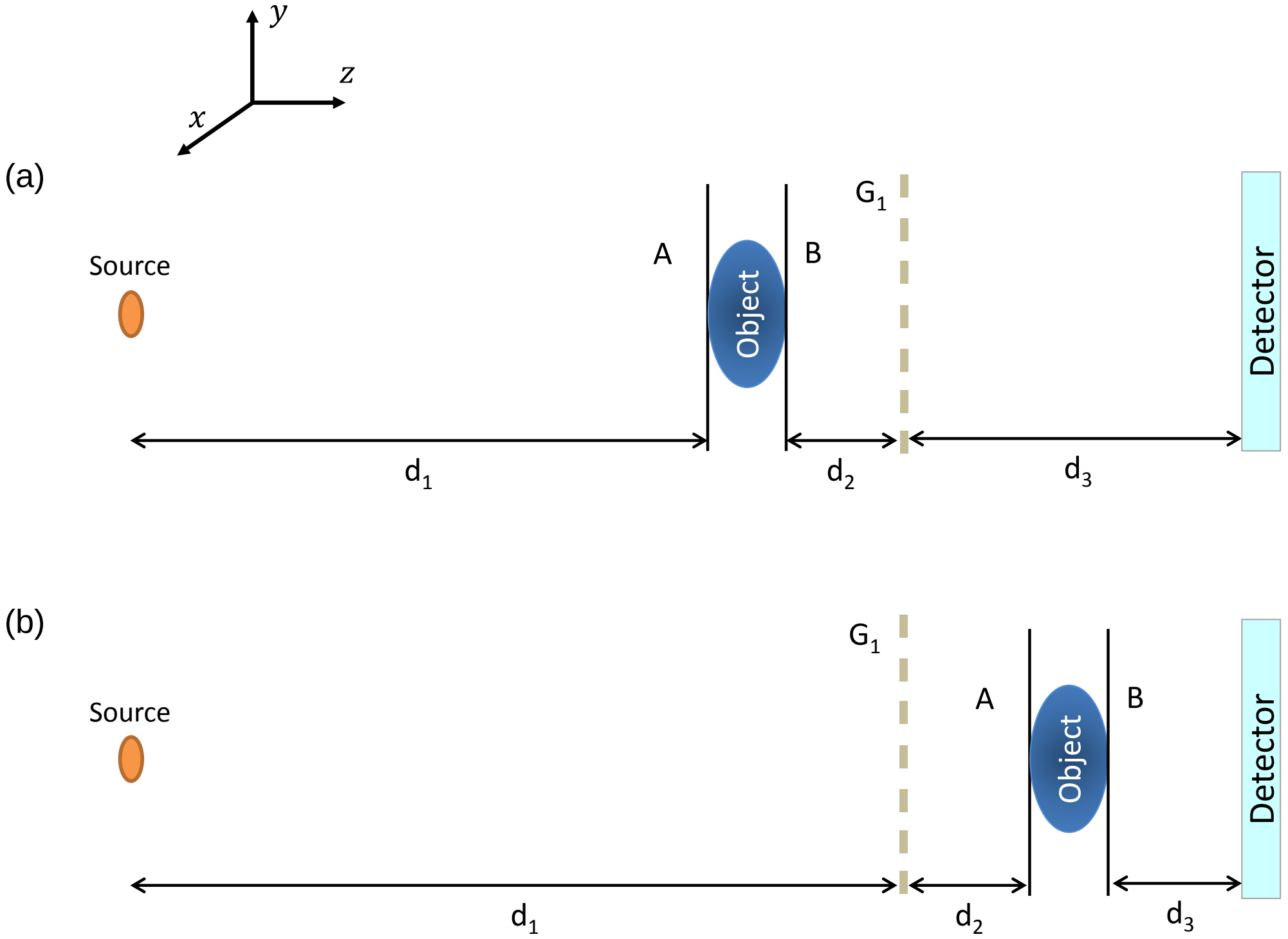

Figure G.1: Two system arrangements with one $\pi$-phase grating $G_1$: (a) the imaging object is put between the source and $G_1$; (b) the imaging object is put between $G_1$ and the detector.

the exiting wave field at the object plane B

$$\mathrm{U}_1^{'}(x_1, y_1) = \mathrm{U}_1(x_1, y_1) e^{-ik\left[\int \delta(x_1,y_1,z_1)dz_1 - i\int \beta(x_1,y_1,z_1)dz_1\right]}, \tag{G.2}$$

where the line integrals are performed along the $z$-axis direction. This is true only if the thickness of the object is relative thin compared to the total length of the imaging system.

Once leaving the object, the X-ray wave continues traveling towards the phase grating $G_1$ in free space. Right upon the surface of $G_1$ grating, the wave field is given as

$$\mathrm{U}_2(x_2, y_2) = \frac{e^{ikd_2}}{i\lambda d_2} \int\int \mathrm{U}_1^{'}(x_1, y_1) e^{ik\left[\frac{(x_2-x_1)^2}{2d_2} + \frac{(y_2-y_1)^2}{2d_2}\right]} dx_1 dy_1. \tag{G.3}$$

Immediately, the above wave front meets the $G_1$ grating and modified by the transmission function

(see Eq. (8.2.2))

$$\mathrm{U}_2^{'}(x_2, y_2) = \mathrm{U}_2(x_2, y_2)\mathrm{T}(x_2). \tag{G.4}$$

Equation (G.4) represents the wave field right behind the $\mathrm{G}_1$ grating. Using the Fresnel diffraction formula one more time, the wave field distribution that locates $d_3$ distance behind the $\mathrm{G}_1$ grating can be expressed as following

$$\mathrm{U}_3(x_3, y_3) = \frac{e^{ikd_3}}{i\lambda d_3} \int\int \mathrm{U}_2^{'}(x_2, y_2) e^{ik\left[\frac{(x_3-x_2)^2}{2d_3} + \frac{(y_3-y_2)^2}{2d_3}\right]} dx_2 dy_2. \tag{G.5}$$

With some derivations, one can get the final expression for $\mathrm{U}_3(x_3, y_3)$

$$\begin{aligned}\mathrm{U}_3(x_3, y_3) =& \frac{\mathrm{U}_0 e^{ik(d_1+d_2+d_3)}}{d_1+d_2+d_3} e^{ik\left[\frac{(x_3-x_s)^2}{2(d_1+d_2+d_3)} + \frac{(y_3-y_s)^2}{2(d_1+d_2+d_3)}\right]} \\ & e^{-k\int \beta\left[\frac{(d_2+d_3)x_s+d_1x_3}{d_1+d_2+d_3}, \frac{(d_2+d_3)y_s+d_1y_3}{d_1+d_2+d_3}, z_1\right]dz_1} e^{-ik\int \delta\left[\frac{(d_2+d_3)x_s+d_1x_3}{d_1+d_2+d_3}, \frac{(d_2+d_3)y_s+d_1y_3}{d_1+d_2+d_3}, z_1\right]dz_1} \\ & \sum_{n=-\infty}^{n=+\infty} a_n e^{-\frac{i2\pi n}{p}\frac{d_1d_3}{d_1+d_2+d_3}\frac{\partial}{\partial x_3}\int \delta\left[\frac{(d_2+d_3)x_s+d_1x_3}{d_1+d_2+d_3}, \frac{(d_2+d_3)y_s+d_1y_3}{d_1+d_2+d_3}, z_1\right]dz_1} \\ & e^{-\frac{i\pi n^2\lambda}{p^2}\frac{(d_1+d_2)d_3}{d_1+d_2+d_3}} e^{\frac{i2\pi n}{p}\frac{d_3x_s}{d_1+d_2+d_3}} e^{\frac{i2\pi n}{p}\frac{(d_1+d_2)x_3}{d_1+d_2+d_3}}.\end{aligned} \tag{G.6}$$

In reality, it is the signal intensity, $\mathrm{I}_3(x_3, y_3)$, that would be recorded during the signal detection procedures. In general, we have

$$\begin{aligned}\mathrm{I}_3(x_3, y_3) &= |\mathrm{U}_3(x_3, y_3)|^2 \\ &= \frac{|\mathrm{U}_0|^2}{(d_1+d_2+d_3)^2} e^{-2k\int \beta\left[\frac{(d_2+d_3)x_s+d_1x_3}{d_1+d_2+d_3}, \frac{(d_2+d_3)y_s+d_1y_3}{d_1+d_2+d_3}, z_1\right]dz_1} \\ & \sum_{n=-\infty}^{n=+\infty}\sum_{n'=-\infty}^{n'=+\infty} a_n a_{n'}^* e^{-\frac{i2\pi(n-n')}{p}\frac{d_1d_3}{d_1+d_2+d_3}\frac{\partial}{\partial x_3}\int \delta\left[\frac{(d_2+d_3)x_s+d_1x_3}{d_1+d_2+d_3}, \frac{(d_2+d_3)y_s+d_1y_3}{d_1+d_2+d_3}, z_1\right]dz_1} \\ & e^{-\frac{i\pi(n^2-n'^2)\lambda}{p^2}\frac{(d_1+d_2)d_3}{d_1+d_2+d_3}} e^{\frac{i2\pi(n-n')}{p}\frac{d_3x_s}{d_1+d_2+d_3}} e^{\frac{i2\pi(n-n')}{p}\frac{(d_1+d_2)x_3}{d_1+d_2+d_3}}.\end{aligned} \tag{G.7}$$

Details of the derivations please refer to the Appendix H. In the following discussions, further simplifications will be performed on Eq. (G.7) in order to get a general idea of the property of the beam intensity $\mathrm{I}_3(x_3, y_3)$. Define

$$\varphi(x_3, y_3, x_s, y_s) = -\frac{\partial}{\partial x_3}\int \delta\left[\frac{(d_2+d_3)x_s+d_1x_3}{d_1+d_2+d_3}, \frac{(d_2+d_3)y_s+d_1y_3}{d_1+d_2+d_3}, z_1\right]dz_1, \tag{G.8}$$

and

$$\mathrm{J}_3(x_3,y_3)\sum_{n=-\infty}^{n=+\infty}\sum_{n'=-\infty}^{n'=+\infty}a_n a^*_{n'}e^{\frac{i2\pi(n-n')}{p}\frac{d_1+d_2}{d_1+d_2+d_3}\left[x_3+\frac{d_1d_3}{d_1+d_2}\varphi(x_3,y_3,x_s,y_s)\right]}$$
$$e^{-\frac{i\pi(n^2-n'^2)\lambda}{p^2}\frac{(d_1+d_2)d_3}{d_1+d_2+d_3}}e^{\frac{i2\pi(n-n')}{p}\frac{d_3x_s}{d_1+d_2+d_3}}. \tag{G.9}$$

If assume $n = n' + s$, the Eq. (G.9) can be rewritten as

$$\mathrm{J}_3(x_3,y_3)=\sum_{s=-\infty}^{s=+\infty}\sum_{n'=-\infty}^{n'=+\infty}a_{n'+s}a^*_{n'}\mathrm{G}(x_3,y_3,x_s,y_s,s)\mathrm{Q}(\lambda,n',s)\mathrm{S}(x_s,s), \tag{G.10}$$

where

$$\mathrm{G}(x_3,y_3,x_s,y_s,s)=e^{\frac{i2\pi(n-n')}{p}\frac{d_1+d_2}{d_1+d_2+d_3}\left[x_3+\frac{d_1d_3}{d_1+d_2}\varphi(x_3,y_3,x_s,y_s)\right]}, \tag{G.11}$$

$$\mathrm{Q}(\lambda,n',s)=e^{-\frac{i\pi(2n's+s^2)\lambda}{p^2}\frac{(d_1+d_2)d_3}{d_1+d_2+d_3}}, \tag{G.12}$$

$$\mathrm{S}(x_s,s)=e^{\frac{i2\pi s}{p}\frac{d_3x_s}{d_1+d_2+d_3}}. \tag{G.13}$$

According to Eq. (8.2.8) and the plotting results in Fig. 8.8, the first several most dominant terms in Eq. (G.10) should correspond to small $s$ values. If only consider the first three most dominant terms, Eq. (G.10) can be approximated as below,

$$\mathrm{J}_3(x_3,y_3)\approx\sum_{n'=-\infty}^{n'=+\infty}a_{n'}a^*_{n'}+\sum_{n'=-\infty}^{n'=+\infty}a_{n'-2}a^*_{n'}\mathrm{G}(x_3,y_3,-2)\mathrm{Q}(\lambda,n',-2)\mathrm{S}(x_s,-2)$$
$$+\sum_{n'=-\infty}^{n'=+\infty}a_{n'+2}a^*_{n'}\mathrm{G}(x_3,y_3,2)\mathrm{Q}(\lambda,n',2)\mathrm{S}(x_s,2), \tag{G.14}$$
$$\approx 1+a_{-1}a^*_1\mathrm{G}(x_3,y_3,-2)\mathrm{Q}(\lambda,1,-2)\mathrm{S}(x_s,-2)$$
$$+a_1a^*_{-1}\mathrm{G}(x_3,y_3,2)\mathrm{Q}(\lambda,-1,2)\mathrm{S}(x_s,2). \tag{G.15}$$

In Eq. (G.14), only the terms related to $s = 0$, $s = -2$, and $s = 2$ are maintained. In Eq. (G.15), only the terms related to $n' = -1$, $n' = 1$ are maintained. In addition, since

$$\mathrm{G}(x_3,y_3,x_s,y_s,-2)=\mathrm{G}^*(x_3,y_3,x_s,y_s,2), \tag{G.16}$$

$$\mathrm{Q}(\lambda,1,-2)=\mathrm{Q}(\lambda,-1,2), \tag{G.17}$$

$$\mathrm{S}(x_s,-2)=\mathrm{S}^*(x_s,2), \tag{G.18}$$

one can further obtain

$$\mathrm{J}_3(x_3, y_3) \approx 1 - \frac{8}{\pi^2}\mathrm{Q}(\lambda, 1, -2)\cos\left[\frac{2\pi}{p'}\left(\frac{d_3 x_s}{d_1 + d_2} + x_3 + \frac{d_1 d_3}{d_1 + d_2}\varphi(x_3, y_3, x_s, y_s)\right)\right], \tag{G.19}$$

where

$$p' = \frac{p}{2} \times \frac{d_1 + d_2 + d_3}{d_1 + d_2}. \tag{G.20}$$

If assume the source size is small enough and the signal intensity observation plane is exactly the self-imaging plane, i.e., $x_s = 0$, $y_s = 0$, and $\mathrm{Q}(\lambda, n', s) = 1$, the Eq. (G.19) can be finally simplified into

$$\mathrm{J}_3(x_3, y_3) \approx 1 - \frac{8}{\pi^2}\cos\left[\frac{2\pi}{p'}\left(x_3 + \frac{d_1 d_3}{d_1 + d_2}\varphi(x_3, y_3)\right)\right]. \tag{G.21}$$

Therefore,

$$\begin{aligned}\mathrm{I}_3(x_3, y_3) \approx\ & \frac{|\mathrm{U}_0|^2}{(d_1 + d_2 + d_3)^2} e^{-2k\int \beta\left[\frac{(d_2+d_3)x_s + d_1 x_3}{d_1+d_2+d_3}, \frac{(d_2+d_3)y_s + d_1 y_3}{d_1+d_2+d_3}, z_1\right] dz_1} \\ & \left[1 - \frac{8}{\pi^2}\cos\left[\frac{2\pi}{p'}\left(x_3 + \frac{d_1 d_3}{d_1 + d_2}\varphi(x_3, y_3)\right)\right]\right]. \end{aligned} \tag{G.22}$$

The approximated results in Eq. (G.22) indicates that the detected beam intensity distribution map can be modeled by a sinusoidal equation. In principle, both the object attenuation information and the differential phase information can be retrieved from the detected intensity signals.

Remember that the results in Eq. (G.22) only considers the contributions from $a_{-1}$ and $a_1$. It also assumes that the source size is small enough and thus can be treated as a point source. Moreover, it is also assumed that the incident X-ray photons have one single energy (same $\lambda$). Admittedly, these approximations will have to be modified for real benchtop experimental systems. In practice, the source might have a finite size, the used X-ray beam might be polychromatic, and the fabricated gratings might not be perfect, and so on. All of these imperfect experimental conditions will greatly decrease the fringe visibility of the detected modulations.

As shown in Eq. (G.7) or Eq. (G.22), there is a decay factor,

$$e^{-2k\int \beta\left[\frac{(d_2+d_3)x_s + d_1 x_3}{d_1+d_2+d_3}, \frac{(d_2+d_3)y_s + d_1 y_3}{d_1+d_2+d_3}, z_1\right] dz_1}, \tag{G.23}$$

due to the interactions between the X-ray electromagnetic wave and matter. This term reduces the incident beam intensity and is recognized as the attenuation properties of the X-ray imaging object. Mathematically, Eq. (G.23) and Eq. (C.30) have the same form. Therefore, the relationship between the material attenuation coefficient $\mu$ and the imaginary part, $\beta$, of the refractive index $n$ should be the same as defined in Eq. (C.32).

Another important finding is about the geometrical magnification effect. Assuming that the source size is negligible, i.e., $x_s \approx 0$, and $y_s \approx 0$, then the geometrical magnification factor is exactly equal to $\frac{d_1+d_2+d_3}{d_1}$. Be aware that the imaging object is located at $d_1$ plane, and the image recording plane is positioned $d_1 + d_2 + d_3$ apart from the light source. For X-ray imaging systems with relatively small focal spot (way smaller than the imaging geometry), such magnification factor can still be used as a good approximation.

Essentially, the fundamental imaging principle of the X-ray Talbot-Lau interferometer has been interpreted via the Fresnel diffraction theory in the preceding sections. As obtained in Eq. (G.22), the final detected X-ray intensity distribution is partly modulated by $\varphi$, which is defined as the first derivative of the line integral of $\delta$ with respect to the $x_3$ coordinate, as can be seen in Eq. (G.8). Actually, $\varphi$ corresponds to the refraction angle of X-ray beam when it propagates through the object.

In the Talbot-Lau interferometer imaging system, this tiny beam refraction angle $\varphi$ has been greatly magnified to make it detectable,

$$\phi(x, y) = \frac{2\pi}{p} \frac{d_1 d_3}{d_1 + d_2} \varphi. \tag{G.24}$$

In the above equation, $\frac{2\pi}{p} \frac{d_1 d_3}{d_1+d_2}$ is known as the signal magnification factor (sometimes called as the interferometer sensitivity). For instance, if considering a $\pi$-phase grating with period of couple microns, and an observation distance of several tens of centimeters, such magnification factor could be as large as $10^5$. In fact, it is such dramatic signal magnification mechanism that makes the detection of the tiny X-ray refraction angle $\phi$ become capable in the X-ray Talbot-Lau interferometer system.

In theory, the decrement of the real part of the refractive index $n$ is related with the electron density distribution $\rho_e$ of the imaging object via

$$\delta(x, y, z, E) = \frac{h^2 c^2 r_e}{2\pi E^2} \rho_e(x, y, z), \tag{G.25}$$

in which $r_e$ denotes the classical electron radius, $h$ is the Planck's constant, and $c$ is the light speed in vacuum. Hence, the detection of phase information, also the differential phase information, reflects the internal electron density distribution.

# H How to derive $\mathrm{U}_3$

First of all, define

$$\Psi(x_1, y_1) = k \int \delta(x_1, y_1, z_1) dz_1 - ik \int \beta(x_1, y_1, z_1) dz_1, \tag{H.1}$$

and

$$F(x_1, y_1) = k \left[ \frac{(x_1 - x_s)^2}{2d_1} + \frac{(y_1 - y_s)^2}{2d_1} + \frac{(x_2 - x_1)^2}{2d_2} + \frac{(y_2 - y_1)^2}{2d_2} \right] - \Psi(x_1, y_1), \tag{H.2}$$

then, Eq. (G.3) can be rewritten as

$$\mathrm{U}_2(x_2, y_2) = \frac{\mathrm{U}_0 e^{ik(d_1+d_2)}}{i\lambda d_1 d_2} \int\int e^{-iF(x_1,y_1)} dx_1 dy_1. \tag{H.3}$$

If taking the first derivative of $F(x_1, y_1)$ with respect to $x_1$ and $y_1$ separately, one is able to get

$$\frac{\partial F(x_1, y_1)}{\partial x_1} = k \left[ \frac{x_1 - x_s}{d_1} + \frac{x_1 - x_2}{d_2} \right] - \frac{\partial \Psi(x_1, y_1)}{\partial x_1}, \tag{H.4}$$

$$\frac{\partial F(x_1, y_1)}{\partial y_1} = k \left[ \frac{y_1 - y_s}{d_1} + \frac{y_1 - y_2}{d_2} \right] - \frac{\partial \Psi(x_1, y_1)}{\partial y_1}. \tag{H.5}$$

By setting both equations equal to 0, the optimality conditions, denoted as $(x_1^0, y_1^0)$, that ensure $F(x_1, y_1)$ achieve its extremum are

$$x_1^0 = \frac{1}{k(d_1 + d_2)} \frac{\partial \Psi(x_1, y_1)}{\partial x_1} + \frac{d_2 x_s + d_1 x_2}{d_1 + d_2} \approx \frac{d_2 x_s + d_1 x_2}{d_1 + d_2}, \tag{H.6}$$

$$y_1^0 = \frac{1}{k(d_1 + d_2)} \frac{\partial \Psi(x_1, y_1)}{\partial y_1} + \frac{d_2 y_s + d_1 y_2}{d_1 + d_2} \approx \frac{d_2 y_s + d_1 y_2}{d_1 + d_2}. \tag{H.7}$$

In the above two equations, the x-ray wavelength $\lambda$ related terms are assumed to be small quantities and thus are ignored. Meanwhile, the Taylor expansion of $F(x_1, y_1)$ leads to (only consider the

quadratic orders)

$$F(x_1,y_1) \approx F(x_1^0,y_1^0) + (x_1-x_1^0)\frac{\partial F(x_1,y_1)}{\partial x_1}|_{x_1=x_1^0} + (y_1-y_1^0)\frac{\partial F(x_1,y_1)}{\partial y_1}|_{y_1=y_1^0}$$
$$+ \frac{(x_1-x_1^0)^2}{2}\frac{\partial^2 F(x_1,y_1)}{\partial (x_1)^2}|_{x_1=x_1^0} + \frac{(y_1-y_1^0)^2}{2}\frac{\partial^2 F(x_1,y_1)}{\partial (y_1)^2}|_{y_1=y_1^0}. \tag{H.8}$$

Because $(x_1^0, y_1^0)$ are assumed to be the optimality conditions of $F(x_1, y_1)$, therefore,

$$\frac{\partial F(x_1,y_1)}{\partial x_1}|_{x_1=x_1^0} = 0,$$
$$\frac{\partial F(x_1,y_1)}{\partial y_1}|_{y_1=y_1^0} = 0.$$

In addition,

$$\frac{\partial^2 F(x_1,y_1)}{\partial (x_1)^2} = \frac{k}{d_1} + \frac{k}{d_2} - \frac{\partial^2 \Psi(x_1,y_1)}{\partial (x_1)^2} \approx \frac{k}{d_1} + \frac{k}{d_2}, \tag{H.9}$$
$$\frac{\partial^2 F(x_1,y_1)}{\partial (y_1)^2} = \frac{k}{d_1} + \frac{k}{d_2} - \frac{\partial^2 \Psi(x_1,y_1)}{\partial (y_1)^2} \approx \frac{k}{d_1} + \frac{k}{d_2}, \tag{H.10}$$

therefore,

$$F(x_1,y_1) \approx F(x_1^0,y_1^0) + \left(\frac{k}{d_1} + \frac{k}{d_2}\right)\left[\frac{(x_1-x_s)^2}{2} + \frac{(y_1-y_s)^2}{2}\right]. \tag{H.11}$$

Substitute Eq. (H.11) into Eq. (H.3), the following result can be immediately obtained

$$\mathrm{U}_2(x_2,y_2) = \frac{\mathrm{U}_0 e^{ik(d_1+d_2)}}{i\lambda d_1 d_2} e^{-iF(x_1^0,y_1^0)} \int e^{i\left(\frac{k}{d_1}+\frac{k}{d_2}\right)\frac{(x_1-x_s)^2}{2}} dx_1 \int e^{i\left(\frac{k}{d_1}+\frac{k}{d_2}\right)\frac{(y_1-y_s)^2}{2}} dy_1. \tag{H.12}$$

It is known that

$$\int_{-\infty}^{+\infty} e^{ix^2} dx = \sqrt{i\pi}, \tag{H.13}$$

so, Eq. (H.12) can be simplified into

$$\mathrm{U}_2(x_2,y_2) = \frac{\mathrm{U}_0 e^{ik(d_1+d_2)}}{d_1+d_2} e^{-i\Psi\left[\frac{d_2 x_s + d_1 x_2}{d_1+d_2}, \frac{d_2 y_s + d_1 y_2}{d_1+d_2}\right]} e^{ik\left[\frac{(x_2-x_s)^2}{2(d_1+d_2)} + \frac{(y_2-y_s)^2}{2(d_1+d_2)}\right]}. \tag{H.14}$$

Clearly, the wave amplitude $\mathrm{U}_2(x_2, y_2)$ right upon the $\mathrm{G}_1$ phase grating surface has the similar form as of on the object incident plane $\mathrm{U}_1(x_1, y_1)$, besides that the wave front has been modified by the object accordingly.

Substituting Eq. (8.2.2), Eq. (H.14) into Eq. (G.5),

$$\mathrm{U}_3(x_3, y_3) = \frac{\mathrm{U}_0 e^{ik(d_1+d_2+d_3)}}{i\lambda(d_1+d_2)d_3} \sum_{-\infty}^{+\infty} a_n \int\int e^{\frac{i2\pi n x_2}{p}} e^{-i\Psi\left(\frac{d_2 x_s + d_1 x_2}{d_1+d_2}, \frac{d_2 y_s + d_1 y_2}{d_1+d_2}\right)} e^{ik\left[\frac{(x_2-x_s)^2}{2(d_1+d_2)} + \frac{(y_2-y_s)^2}{2(d_1+d_2)} + \frac{(x_3-x_2)^2}{2d_3} + \frac{(y_3-y_2)^2}{2d_3}\right]} dx_2 dy_2. \tag{H.15}$$

Define

$$G(x_2, y_2) = k\left[\frac{(x_2-x_s)^2}{2(d_1+d_2)} + \frac{(y_2-y_s)^2}{2(d_1+d_2)} + \frac{(x_3-x_2)^2}{2d_3} + \frac{(y_3-y_2)^2}{2d_3}\right] + \frac{2\pi n x_2}{p} - \Psi\left(\frac{d_2 x_s + d_1 x_2}{d_1+d_2}, \frac{d_2 y_s + d_1 y_2}{d_1+d_2}\right), \tag{H.16}$$

and then repeat the similar mathematical tricks implemented in the above discussions, one can obtain the optimality conditions for $G(x_2, y_2)$

$$x_2^0 = \frac{d_1 d_3}{k(d_1+d_2+d_3)}\frac{\partial\Psi}{\partial x_2} + \frac{d_3 x_s + (d_1+d_2)x_3}{d_1+d_2+d_3} - \frac{(d_1+d_2)d_3}{k(d_1+d_2+d_3)}\frac{2\pi n}{p} \approx \frac{d_3 x_s + (d_1+d_2)x_3}{d_1+d_2+d_3} - \frac{(d_1+d_2)d_3}{k(d_1+d_2+d_3)}\frac{2\pi n}{p}, \tag{H.17}$$

$$y_2^0 = \frac{d_1 d_3}{k(d_1+d_2+d_3)}\frac{\partial\Psi}{\partial y_2} + \frac{d_3 y_s + (d_1+d_2)y_3}{d_1+d_2+d_3} \approx \frac{d_3 y_s + (d_1+d_2)y_3}{d_1+d_2+d_3}. \tag{H.18}$$

Immediately,

$$\mathrm{U}_3(x_3, y_3) = \frac{\mathrm{U}_0 e^{ik(d_1+d_2+d_3)}}{d_1+d_2+d_3} \sum_{-\infty}^{+\infty} a_n e^{iG(x_2^0, y_2^0)}, \tag{H.19}$$

where

$$G(x_2^0, y_2^0) = \frac{k\left[(x_3-x_s)^2 + (y_3-y_s)^2\right]}{2(d_1+d_2+d_3)} + \frac{2\pi n}{p}\frac{(d_1+d_2)x_3 + d_3 x_s}{d_1+d_2+d_3} - \frac{2\pi n^2\lambda}{p^2}\frac{(d_1+d_2)d_3}{d_1+d_2+d_3} - \Psi\left(\frac{(d_2+d_3)x_s + d_1 x_3}{d_1+d_2+d_3} - \frac{d_1 d_3}{d_1+d_2+d_3}\frac{\lambda n}{p}, \frac{(d_2+d_3)y_s + d_1 y_3}{d_1+d_2+d_3}\right). \tag{H.20}$$

The $\Psi$ part, as defined in Eq. (H.1), can be further simplified via

$$\begin{aligned}
&\int \delta\left[\frac{(d_2+d_3)x_s+d_1x_3}{d_1+d_2+d_3}-\frac{d_1d_3}{d_1+d_2+d_3}\frac{\lambda n}{p},\frac{(d_2+d_3)y_s+d_1y_3}{d_1+d_2+d_3},z_1\right]dz_1\\
\approx&\int \delta\left[\frac{(d_2+d_3)x_s+d_1x_3}{d_1+d_2+d_3},\frac{(d_2+d_3)y_s+d_1y_3}{d_1+d_2+d_3},z_1\right]dz_1\\
&-\frac{d_1d_3}{d_1+d_2+d_3}\frac{\lambda n}{p}\frac{\partial}{\partial x_3}\int \delta\left[\frac{(d_2+d_3)x_s+d_1x_3}{d_1+d_2+d_3},\frac{(d_2+d_3)y_s+d_1y_3}{d_1+d_2+d_3},z_1\right]dz_1,
\end{aligned} \tag{H.21}$$

and

$$\begin{aligned}
&\int \beta\left[\frac{(d_2+d_3)x_s+d_1x_3}{d_1+d_2+d_3}-\frac{d_1d_3}{d_1+d_2+d_3}\frac{\lambda n}{p},\frac{(d_2+d_3)y_s+d_1y_3}{d_1+d_2+d_3},z_1\right]dz_1\\
\approx&\int \beta\left[\frac{(d_2+d_3)x_s+d_1x_3}{d_1+d_2+d_3},\frac{(d_2+d_3)y_s+d_1y_3}{d_1+d_2+d_3},z_1\right]dz_1.
\end{aligned} \tag{H.22}$$

By combining Eqs. (H.19)-(H.22), one can easily obtain the results shown below:

$$\begin{aligned}
\mathrm{U}_3(x_3,y_3)=&\frac{\mathrm{U}_0e^{ik(d_1+d_2+d_3)}}{d_1+d_2+d_3}e^{ik\left[\frac{(x_3-x_s)^2}{2(d_1+d_2+d_3)}+\frac{(y_3-y_s)^2}{2(d_1+d_2+d_3)}\right]}\\
&e^{-k\int \beta\left[\frac{(d_2+d_3)x_s+d_1x_3}{d_1+d_2+d_3},\frac{(d_2+d_3)y_s+d_1y_3}{d_1+d_2+d_3},z_1\right]dz_1}e^{-ik\int \delta\left[\frac{(d_2+d_3)x_s+d_1x_3}{d_1+d_2+d_3},\frac{(d_2+d_3)y_s+d_1y_3}{d_1+d_2+d_3},z_1\right]dz_1}\\
&\sum_{n=-\infty}^{n=+\infty}a_ne^{-\frac{i2\pi n}{p}\frac{d_1d_3}{d_1+d_2+d_3}\frac{\partial}{\partial x_3}\int \delta\left[\frac{(d_2+d_3)x_s+d_1x_3}{d_1+d_2+d_3},\frac{(d_2+d_3)y_s+d_1y_3}{d_1+d_2+d_3},z_1\right]dz_1}\\
&e^{-\frac{i\pi n^2\lambda}{p^2}\frac{(d_1+d_2)d_3}{d_1+d_2+d_3}}e^{\frac{i2\pi n}{p}\frac{d_3x_s}{d_1+d_2+d_3}}e^{\frac{i2\pi n}{p}\frac{(d_1+d_2)x_3}{d_1+d_2+d_3}}.
\end{aligned} \tag{H.23}$$